\documentclass[11pt, a4paper, oneside]{Thesis} % Paper size, default font size
\graphicspath{{Pictures/}} % Specifies the directory where pictures are stored

\usepackage[square,sort,compress,numbers]{natbib}
\usepackage{cite}
\usepackage{amsmath,latexsym} 
\usepackage{float}
\usepackage{epstopdf}
\usepackage{caption}
\usepackage{subcaption} 
\usepackage{graphicx} 
\usepackage{lscape}
\usepackage{hyperref}
\usepackage{enumitem}
\usepackage{physics}
\hypersetup{colorlinks,citecolor=red}
\hypersetup{colorlinks=true,urlcolor=black}
\title{\ttitle} % Defines the thesis title - don't touch this
\DeclareUnicodeCharacter{2212}{-}
\DeclareUnicodeCharacter{2212}{+}

\begin{document}

\setstretch{1.3} % Line spacing of 1.3

% Define page headers using FancyHdr package and set up for one-sided printing
\fancyhead{} % Clears all page headers and footers
\rhead{\thepage} % Sets the right side header to show the page number
\lhead{} % Clears the left side page header

% Input all the variables used in the document. Please fill out the
% variables.tex file with all your details.
%-------------------------------------------------------------------------------
%	DOCUMENT VARIABLES
%
%	Fill in the lines below to set the various variables for the document
%-------------------------------------------------------------------------------

%-------------------------------------------------------------------------------
% Your thesis title - this is used in the title and abstract
% Command: \ttitle
\thesistitle{Dynamical and Cosmological Aspects of Teleparallel and Extended Teleparallel Gravity}
%-------------------------------------------------------------------------------
% The document type: Thesis / report, etc.
% Command: \doctype
\documenttype{\textbf{THESIS}}
%-------------------------------------------------------------------------------
% Your supervisor's name - this is used in the title page
% Command: \supname
\supervisor{Prof. BIVUDUTTA MISHRA}
%-------------------------------------------------------------------------------
% The supervisor's position - Used on Certificate
% Command: \suppos
\supervisorposition{Professor}
%-------------------------------------------------------------------------------
% Supervisor's institute
% Command: \supinst
\supervisorinstitute{BITS-Pilani, Hyderabad Campus}
%-------------------------------------------------------------------------------
% Your Co-Supervisor's name
% Command: \cosupname
%\cosupervisor{Prof. Dipak Kumar Satpathi}
%-------------------------------------------------------------------------------
% Co-Supervisor's Position - Used on Certificate
% Command: \cosuppos
%\cosupervisorposition{Associate Professor}
%-------------------------------------------------------------------------------
% Co-Supervisor's Institute
% Command: \cosupinst
%\cosupervisorinstitute{BITS-Pilani, Hyderabad Campus}
%-------------------------------------------------------------------------------
% Your Examiner's name. Not currently used anywhere.
% Command: \examname
\examiner{}
%-------------------------------------------------------------------------------
% Name of your degree
% Command: \degreename
\degree{Ph.D. Research Scholar}
%-------------------------------------------------------------------------------
% The BITS Course Code for which this report is written
% COmmand: \ccode
\coursecode{\textbf{DOCTOR OF PHILOSOPHY}}
%\coursecode{BITS C799T}
%-------------------------------------------------------------------------------
% The name of the Course
% Command: \cname
\coursename{\textbf{Thesis}}
%-------------------------------------------------------------------------------
% Your name. Extend manually in case of multiple authors
% Command: \authornames
\authors{\textbf{KADAM SIDDHESHWAR ATMARAM}}
%-------------------------------------------------------------------------------
% Your ID Number - used on the Title page and abstract
% Command: \idnum
\IDNumber{\textbf{2020PHXF0002H}}
%-------------------------------------------------------------------------------
% Your address
% Command: \addressnames
\addresses{}
%-------------------------------------------------------------------------------
% Your subject area
% Command: \subjectname
\subject{}
%-------------------------------------------------------------------------------
% Keywords for this report.
% Command: \keywordnames
%-------------------------------------------------------------------------------
% University details
% Command: \univname
\university{\texorpdfstring{\href{http://www.bits-pilani.ac.in/} % URL
                {Birla Institute of Technology and Science, Pilani}} % University name
                {Birla Institute of Technology and Science, Pilani}}
%-------------------------------------------------------------------------------
% University details, in Capitals
% Command: \UNIVNAME
\UNIVERSITY{\texorpdfstring{\href{http://www.bits-pilani.ac.in/} % URL
                {\textbf{BIRLA INSTITUTE OF TECHNOLOGY AND SCIENCE, PILANI}}} % name in capitals
                {BIRLA INSTITUTE OF TECHNOLOGY AND SCIENCE, PILANI}}
%-------------------------------------------------------------------------------
% Campus Name
% Command: \campusname
%\campus{Hyderabad Campus}

%-------------------------------------------------------------------------------
% Campus Name, in capitals
% Command: \CAMPUSNAME
%\CAMPUS{HYDERABAD CAMPUS}

%-------------------------------------------------------------------------------
% Department Details
% Command: \deptname
\department{\texorpdfstring{\href{http://www.bits-pilani.ac.in/pilani/Mathematics/Mathematics} % Your department's URL
                {Mathematics}} % Your department's name
                {Mathematics}}
%-------------------------------------------------------------------------------
% Department details, in Capitals
% Command: \DEPTNAME
\DEPARTMENT{\texorpdfstring{\href{http://www.bits-pilani.ac.in/pilani/Mathematics/Mathematics} % Your department's URL
                {Mathematics}} % Your department's name in capitals
                {Mathematics}}
%-------------------------------------------------------------------------------
% Research Group Details
% Command: \groupname
\group{\texorpdfstring{\href{Research Group Web Site URL Here (include http://)}
                {Research Group Name}} % Your research group's name
                {Research Group Name}}
%-------------------------------------------------------------------------------
% Research Group Details, in Capitals
% Command: \GROUPNAME
\GROUP{\texorpdfstring{\href{Research Group Web Site URL Here (include http://)}
                {RESEARCH GROUP NAME (IN BLOCK CAPITALS)}}
                {RESEARCH GROUP NAME (IN BLOCK CAPITALS)}}
%-------------------------------------------------------------------------------
% Faculty details
% Command: \facname
\faculty{\texorpdfstring{\href{Faculty Web Site URL Here (include http://)}
                {Faculty Name}}
                {Faculty Name}}
%-------------------------------------------------------------------------------
% Faculty details, in Capitals
% Command: \FACNAME
\FACULTY{\texorpdfstring{\href{Faculty Web Site URL Here (include http://)}
                {FACULTY NAME (IN BLOCK CAPITALS)}}
                {FACULTY NAME (IN BLOCK CAPITALS)}}
%-------------------------------------------------------------------------------

%-------------------------------------------------------------------------------
%   NON-CONTENT PAGES
%-------------------------------------------------------------------------------
\maketitle

%-------------------------------------------------------------------------------
%	DEDICATION
%-------------------------------------------------------------------------------
\clearpage
 % Add a gap in the Contents, for aesthetics
\frontmatter
\Certificate
%\pagenumbering{roman}

\Declaration
%\frontmatter % Use roman numbering style (i, ii...) for the pre-content pages
%\pagenumbering{roman}

%\Quotation{Insert Random Quote here. Publish like a boss.}{Your Name}
\setstretch{1.3} % Return the line spacing back to 1.3

\pagestyle{empty} % Page style needs to be empty for this page
% Dedication text
\pagenumbering{gobble}

\Dedicatory{\bf \begin{LARGE}
In the Loving Memories of  
\end{LARGE} 
\\
\vspace{3cm}
 My Grandfather\\
 \vspace{1cm}
 }

\addtocontents{toc}{\vspace{2em}}

\begin{acknowledgements}

I would like to express my deep gratitude to all those who have provided me with inspiration and support as I work towards my doctorate.

I am sincerely grateful to my supervisor, \textbf{Prof. Bivudutta Mishra}, Professor of the Department of Mathematics at BITS-Pilani, Hyderabad Campus. His determination and unwavering enthusiasm have been a great source of motivation for me. Working with him has been a tremendous privilege and I have benefited greatly from his exceptional mathematical expertise, teaching skills and humanistic approach. I am truly indebted to him for his dedicated guidance and support, without which this dissertation would not have been possible.

I also thank my DAC members,\textbf{ Prof. Pradyumn Kumar Sahoo} and \textbf{Prof. Nirman Ganguly}, for their valuable insights, thoughtful comments and consistent feedback throughout my research.

I would like to convey my sincere gratitude to the Head of the Department, the DRC Convener and all the faculty members for their encouragement and support in my research endeavors. Also, I would like to thank the entire staff of the Department of Mathematics at BITS-Pilani Hyderabad Campus for their assistance, support and encouragement during my research.

I also thank the Associate Dean of AGRSD BITS-Pilani, Hyderabad Campus.

I am thankful to BITS-Pilani, Hyderabad Campus, for providing me with the necessary resources to conduct my research.

I express my gratitude to the University Grants Commission (UGC) for the financial support provided through the Senior Research Fellowship (UGC Ref. No. 191620205335). 

I want to take this opportunity to express my gratitude to \textbf{Prof. Sunil Kumar Tripathy}, Assistant Professor, Department of Physics, IGIT Saranga, Dhenkanal, Odisha and \textbf{Prof. Jackson Levi-Said}, \textbf{Dr. Gabriel Farrugia} Institute of Space Sciences and Astronomy, University of Malta, for the opportunity to collaborate with them. I have gained valuable knowledge from our collaboration, for which I am truly thankful.

I would like to express my gratitude to my research colleagues Dr. Pratik, Dr. Sankarsan, Dr. Amar, Santosh, Lokesh, Shubham, Rahul, Kalpana, Priyobatra and Shivam and extend special thanks to all my other friends and family members.

Most importantly, I want to thank my parents (\textbf{Mr. Atmaram Kadam and Mrs. Panchafula Kadam}), brothers, family members and friends for their love, care and support in my personal life. I am deeply grateful to my wife, \textbf{Mrs. Megha Kadam} and daughter, \textbf{Shreeja Kadam}, for the light they brought to my life.

\end{acknowledgements}

\clearpage

\begin{abstract}       
This thesis explores the dynamics of modified teleparallel gravity models that include a scalar field and boundary terms, focusing on their cosmological implications, especially to describe late-time cosmic acceleration. Teleparallel gravity an alternative to General Relativity, describes gravitation through torsion rather than curvature. The inclusion of a scalar field may allow the description of a broader range of cosmological scenarios. The most generalised teleparallel gravity formalisms with a scalar field, known as the teleparallel analog of the Horndeski theory and its dynamical system approach have been presented. The dynamical system analysis in the recently developed teleparallel scalar field gravity model $f(T,\phi)$ gravity contains two different potential functions are presented. This study also explores the phase space, the conditions under which different cosmological epochs, such as radiation-dominated, matter-dominated and late-time acceleration, emerge and the transition from early to late-time are also examined.

Inclusion of boundary terms, such as the teleparallel boundary term $B$ and the teleparallel Gauss-Bonnet term $T_G$, enriches the gravitational interaction. The teleparallel gravity theories with boundary terms like $f(T, B)$, $f(T, T_G)$ and one of the most general teleparallel formalisms without a scalar field, i.e., $f(T, B, T_G, B_G)$ gravity, are explored. In these works, cosmologically viable models have been demonstrated to frame the autonomous dynamical system. A comprehensive dynamical system analysis of these models is performed to examine the stability of critical points and the corresponding cosmological scenarios, including accelerated expansion phases and potential late-time attractors.

Furthermore, the role of the non-canonical scalar field coupled to the boundary term $B$ in driving the evolution of the Universe and influencing observational parameters is investigated. Here, findings for $H(z)$ and $\mu(z)$ using cosmological Hubble data and Supernovae Ia and are compared. The analysis demonstrates that modified teleparallel gravity models provide a viable framework for explaining both early phases and late-time cosmic acceleration, offering new insights into the dynamics of the Universe and potential extensions to standard cosmological models. The analysis reveals rich dynamical structures, with the boundary terms playing a crucial role in shaping the phase space and altering the evolution of the Universe. The results of this thesis show that the inclusion of non-canonical fields and complex boundary terms interactions can describe novel or new cosmological scenarios, offering a versatile framework for addressing current and upcoming challenges in the study of cosmic evolution.
\end{abstract}

\clearpage
%\chapter{Preface}

%-------------------------------------------------------------------------------
%	LIST OF CONTENTS/FIGURES/TABLES PAGES
%-------------------------------------------------------------------------------

%\lhead{\emph{Contents}} % Set the left side page header to "Contents"
%\tableofcontents % Write out the Table of Contents
\addtocontents{toc}{\vspace{1em}}
% Set the left side page header to "List of Figures"
 % Write out the List of Figures
\tableofcontents % Write out the Table of Contents
\addtocontents{toc}{\vspace{1em}}
% Set the left side page header to "List of Tables"
\lhead{\emph{List of Tables}}
\listoftables % Write out the List of Tables
\addtocontents{toc}{\vspace{1em}}
% Set the left side page header to "List of Figures"
\lhead{\emph{List of Figures}}
\listoffigures % Write out the List of Figures
\addtocontents{toc}{\vspace{1em}}
 
%-------------------------------------------------------------------------------
%	ABBREVIATIONS
%-------------------------------------------------------------------------------

%\clearpage % Start a new page

 % Set the line spacing to 1.5, this makes the following tables easier to read
\setstretch{1.5}
\lhead{\emph{List of Symbols and Abbreviations}}
% Set the left side page header to "Abbreviations"
\listofsymbols{ll}{
\begin{tabular}{cp{0.5\textwidth}}
$\mathcal{L}_{m}$ & : Matter Lagrangian\\
$H(z)=\frac{\dot a}{a}$& : Hubble parameter\\
$a(t)$ & : Scale factor\\
$q$ & : Deceleration parameter\\
$g_{\mu\nu}$ &: Metric tensor\\
 $g$ &: Determinant of $g_{\mu\nu}$  \\ 
$G_{\mu \nu} \equiv R_{\mu \nu} - \frac{1}{2} g_{\mu \nu} R$ & : Einstein tensor\\
$\Lambda$ & : Cosmological constant\\
 $\Gamma^{\sigma}_{\mu \nu}$  &: Teleparallel  Weitzenb\"{o}ck connection \\ 
 %$\{^{k} _{ij}\}$ &: Levi-Civita connection\\  
%$\nabla_{i}$ &: Co-variant derivative \\ %w.r.t. Levi- Civita connection\\
%$(ij)$&: Symmetrization over the indices $i$ and $j$\\
%$[ij]$ &: Anti-symmetrization over the indices $i$ and $j$\\
$R^{k}_{\sigma \mu \nu}$ &: Riemann tensor \\
$R_{\mu \nu}$ &: Ricci tensor \\
$R$ &: Ricci scalar \\
$S_{m}, S_{r}$ & : Matter, radiation action\\
$T^{\alpha}_{\ \ \mu \nu}$ & : Torsion tensor\\
$T$ & : Torsion scalar \\
%$ \Delta^{\lambda}_{ij}$ & : Hypermomentum\\
$S_{\rho}^{~~~\mu \nu}$ & : Superpotential\\
$e_{a}^{~~\nu}$ &: Tetrad\\
$e$ & : Determinant of tetrad\\
%$\hbar=c=k_B=1$ &: Natural units\\
$\kappa^2\equiv 8\pi G_N= M_{\rm Pl}^{-2}$ &: Gravitational constant\\
$(-\,+\,+\,+)$ &: Metric signature\\
$\rho_{m}$, $\rho_{DE}$ & : Energy density of dark matter and dark energy \\
$p_{m}$, $p_{DE}$&: Pressure of dark matter and dark energy \\
\end{tabular}
}

\clearpage

\noindent
\begin{center}
\begin{tabular}{cp{0.5\textwidth}}
$\Omega_m$, $\Omega_{\rm DE}$ & : Standard density parameters for dark matter and dark energy\\
  \textbf{.} &: Derivative w. r. to time\\
  $\lor$ &: Or \\
  $\land$ &: And\\
  \textbf{GR}&: \textbf{G}eneral \textbf{R}elativity\\
\textbf{TEGR}&: \textbf{T}eleparallel \textbf{E}quivalent of \textbf{G}eneral \textbf{R}elativity\\
\textbf{DE}&: \textbf{D}ark \textbf{E}nergy\\
\textbf{DM}&: \textbf{D}ark \textbf{M}atter\\
\textbf{EoS}&: \textbf{E}quation \textbf{o}f \textbf{S}tate\\
\textbf{$\Lambda$CDM}&: \textbf{L}ambda-\textbf{C}old-\textbf{D}ark-\textbf{M}atter\\
\textbf{FLRW}&: \textbf{F}riedmann \textbf{L}ema\^{i}tre \textbf{R}obertson \textbf{W}alker \\
\textbf{CMB}&: \textbf{C}osmic \textbf{M}icrowave \textbf{B}ackground 
\end{tabular}
\end{center}

%-------------------------------------------------------------------------------
%	PHYSICAL CONSTANTS/OTHER DEFINITIONS
%-------------------------------------------------------------------------------

%\clearpage % Start a new page
%
%% Set the left side page header to "Physical Constants"
%\lhead{\emph{Physical Constants}}
%
% % Include a list of Physical Constants (a four column table)
%\listofconstants{lrcl}
%{
%Speed of Light & $c$ & $=$ & $2.997\ 924\ 58\times10^{8}\ \mbox{ms}^{-\mbox{s}}$ (exact)\\
%% Constant Name & Symbol & = & Constant Value (with units) \\}

%-------------------------------------------------------------------------------
%	SYMBOLS
%-------------------------------------------------------------------------------

\clearpage % Start a new page

\lhead{\emph{Glossary}} % Set the left side page header to "Symbols"

%\listofnomenclature % List the nomenclature. (We use the glossaries package)
%{The standard nomenclatures used in this thesis are listed below. \\
 
%$\mu$ \& $\nu$\\
%$\Gamma_{\mu \nu}^{\lambda}$}
%-------------------------------------------------------------------------------
%	DEDICATION
%-------------------------------------------------------------------------------

%-------------------------------------------------------------------------------
%	THESIS CONTENT - CHAPTERS
%-------------------------------------------------------------------------------

\mainmatter % Begin numeric (1,2,3...) page numbering
\setstretch{1.2} % Line spacing of 1.3
\pagestyle{fancy} % Return the page headers back to the "fancy" style

% Include the chapters of the thesis as separate files from the Chapters folder
% Uncomment the lines as you write the chapters

% Chapter 1

\chapter{Introduction} % Main chapter title
\label{Chapter1}
%\ref{Chapter1}  % For referencing the chapter elsewhere, use \ref{Chapter1} 

\lhead{Chapter 1. \emph{Introduction}} % This is for the header on each page - perhaps a shortened title

%\blindtext
%----------------------------------------------------------------------------------------
%\clearpage

%----------------------------------------------------------------------------------------
\newpage 
\section{Introduction}\label{ch1FDE}
This thesis titled \textbf{Dynamical and Cosmological Aspects of
Teleparallel and Extended 
Teleparallel Gravity} aims to study the late-time cosmic acceleration issue with the well-known and universally adopted technique, known as Dynamical System Analysis. This method allows us to systematically analyse the stability of critical points corresponding to different cosmological epochs, such as radiation-dominated, matter-domination, and DE domination. The phase space analysis highlights the conditions under which cosmological trajectories lead to physically viable solutions, rich structures, and dynamics. In this study, the teleparallel gravity models are considered to demonstrate different phases of the evolution of the Universe. The reader may surely find interesting results that may help to describe the late-time cosmic acceleration issue. The main focus is to describe the evolution of the Universe by including different teleparallel boundary terms like $B$, $T_G$, and the scalar field $\phi$ in the form of different potential functions.   

The observational findings in 1998 \cite{Riess:1998cb,SupernovaCosmologyProject:1998vns} indicate that the expansion of Universe is accelerating. This discovery drastically transformed our understanding of the Universe. The accelerating expansion was attributed to a mysterious form of energy known as DE \cite{Hinshaw:2013, DIVALENTINO:2016}. Although GR is widely accepted as an accurate description of gravity for large masses at the scale of the solar system \cite{misner1973gravitation,Capozziello:2011et,Faraoni:2008mf}, the theory requires the introduction of a cosmological constant $\Lambda$ to account for the accelerating expansion of the Universe. This cosmological constant behaves like a fluid with negative pressure, which can explain the violation of certain energy conditions and the emergence of a repulsive gravitational force. However, due to the lack of theoretical understanding of the cosmological constant $\Lambda$ \cite{sahni2000case}, researchers are exploring alternative explanations for the Universe accelerating behaviour at late times. Hence, to study the late time cosmic evolution, one of the main avenues being pursued is the development of modified theories of gravity. This involves altering or modifying GR to account for cosmological observations. The desire to explain DM and inflation serves as further motivation for modifying GR \cite{Arbey_2021}.

We know that the GR uses the Levi-Civita connection, which is symmetric and torsionless. An alternative approach is to replace this connection with the Weitzenböck connection, which is skew-symmetric and curvatureless and has a non-zero torsion tensor \cite{Weitzenbock1923,bahamonde:2021teleparallel}. The teleparallel gravity and its field equations are similar to that of GR, so this formalism is referred to as the TEGR. Both GR and TEGR offer an equal explanation of gravity. While TEGR makes the same observational predictions as GR, it has different physical and mathematical interpretations. It is important to note that modifications to TEGR are not the same as those to GR. This thesis sets out to investigate modifications to TEGR using different scalars, such as the teleparallel boundary term and the teleparallel Gauss-Bonnet invariant. This study will also emphasise the scalar field coupling function, which offers a wider range of viable cosmic models to explore cosmic evolution. This study will primarily focus on the Universe late-time accelerating behaviour through the lens of modified teleparallel gravity.

Though modifications of GR are frequently used to address theoretical challenges, the suggestion of modifications to teleparallel gravity is relatively recent. This sparks an inquiry into whether modified teleparallel gravity theories could effectively resolve or alleviate cosmological issues like DE and DM. Are there any benefits of modified teleparallel gravity in cosmology compared to standard modifications? In what way does higher-order teleparallel gravity contribute to the study of the late-time cosmic acceleration in the evolution of the Universe? Do these theories offer a means to examine the history of the evolution of the Universe? This thesis will present some of the modified teleparallel theories to address these queries.

This thesis is organised as a brief discussion of standard
Friedman-Lema\^{i}tre-Robertson-Walker (FLRW) Universe, the prerequisite fundamental formulas for formulating TEGR. A basic overview of late-time cosmic acceleration is summarized in the remaining section of chapter \ref{Chapter1}, it will also elaborate on the cosmological implications of the dynamical system analysis, which is a helpful tool to select and analyse the cosmologically viable models followed by chapter \ref{Chapter2}, in which we present the comprehensive study of one of the most recent and general formalisms of the teleparallel scalar field gravity that is the teleparallel Horndeski theory of gravity. \\

Chapter \ref{Chapter3} will give a description of the formulation and implication of the autonomous dynamical system in the non-minimally coupled scalar field $f(T,\phi)$ theory. The role of teleparallel boundary term $B$ and the teleparallel Gauss-Bonnet invariant $T_G$ is analysed in detail in chapter \ref{Chapter4}. Moreover, chapter \ref{Chapter5} sums up both the scalars $B, T_G$ and elaborates the cosmological implications of the inclusion of scalars $T, B, T_G, B_G$ in a more general way in the study of $f(T, B, T_G, B_G)$ gravity formalism. In chapter \ref{Chapter6}, the role of nonminimal boundary couplings has been investigated. To match the recent observational data sets the error bar plot has been presented using Hubble and the Supernova data sets.

\section{The history of the evolution of the Universe}\label{cosmichistroy}
The concept of the Big Bang theory is widely embraced in cosmology as it explains the beginning and development of the Universe. Following the Big Bang, the Universe underwent rapid expansion during inflation. Subsequently, it transitioned through a phase where radiation dominated, then matter, and it currently seems to be in a phase dominated by DE.

\subsection{Inflation}

In 1981, Alan Guth \cite{AGuth1981} proposed the concept of the inflationary phase, which was later supported by John D. Barrow, Michael S. Turner, and Andrei D. Linde \cite{barrow1981inflatio,LINDE1982389}. During the inflationary phase, the Universe undergoes a rapid and exponential expansion, increasing its volume by a factor of $e^{60}$ to $e^{70}$ compared to its initial size before inflation. This expansion takes place over a very short period, estimated to last from $10^{-36}$ seconds to approximately $10^{-33}$ to $10^{-32}$ seconds following the Big Bang. The inflation phenomena describe the early Universe epoch during which the scale factor exponentially expanded in just a fraction of a second ($\ddot{a}>0$). When this condition is applied to the acceleration equation, it violates the strong energy condition $(\rho+3p)<0$. In cosmology, it is always found that the energy density $\rho>0$, and therefore, to fulfill the above condition, the pressure must be negative.

\subsection{Radiation-dominated era} The Universe, in its early stages, went through a phase called the Radiation-dominated epoch. This phase began at the Planck epoch, which was around $10^{-43}$ seconds after the Big Bang. The radiation-dominated epoch continued until approximately 47,000 years after the Big Bang when the Universe shifted to being matter-dominated after cooling enough for matter to become dominant. During the early stages of the Universe, the dominant energy density consisted mainly of photons and other relativistic particles. The scale factor $a(t)$, which indicates the Universe size over time, followed the behaviour of $a(t) \propto t^{1/2}$. This behaviour suggested rapid expansion that gradually decelerated. The Universe experienced a cooling process due to its expansion, as it had extremely high temperatures at the outset of the radiation-dominated era, around $10^{10}$ K. This era established the initial conditions for the formation of large-scale structures by introducing minor irregularities in the CMB, which acted as the starting points for future galaxy and structure formation \cite{sciama1990impact}. The transition from the radiation-dominated epoch to matter domination left marks in the CMB, providing valuable insights into the early Universe. In essence, the radiation-dominated epoch played a crucial role in cosmology, laying the groundwork for the Universe present-day observed state \cite{bellini2022gravitational}. 

\subsection{Matter dominated era}
The era dominated by matter began approximately 47,000 years after the Big Bang and lasted about 9.8 billion years until DE took over. During this period, the Universe energy density was mainly driven by matter, including baryonic matter and DM. The scale factor \( a(t) \) continued to follow \( a(t) \propto t^{2/3} \), indicating a slowing down of the expansion rate compared to the radiation-dominated era \cite{ryden2017cosmology, silk2000big}. The Universe continued to cool as it expanded, leading to a significant drop in temperatures. Additionally, this era facilitated the gravitational collapse of residual density fluctuations from the radiation-dominated era, ultimately leading to the formation of cosmic structures such as galaxies and galaxy clusters. Notably, the large-scale structure of the Universe transitioned from a uniform distribution of matter to a complex, web-like structure that can be formulated \cite{peebles1980large}. The redshift \( z \) was related to the scale factor \( a(t) \) through the equation \( 1 + z = \frac{1}{a(t)} \). Distant objects displayed an increasing redshift as the Universe expanded, indicating their recession. The matter-dominated era significantly influenced the distribution of matter and the CMB radiation, playing a crucial role in shaping the large-scale structure of the Universe. Essentially, this era represented a substantial shift in the evolution of the Universe from a radiation-dominated, hot and dense state to a cooler, more organised cosmos governed by the gravitational influence of matter, laying the groundwork for the complex structure and dynamics observed in the Universe today.

\subsection{Late-time cosmic acceleration era}

The issue of late-time cosmic acceleration pertains to the observations in cosmology that confirm the Universe expansion rate has been increasing in the past few billion years. In 1998, two separate research teams \cite{Riess:1998cb,SupernovaCosmologyProject:1998vns} concluded by their studies of Supernova of Type Ia that the Universe expansion was accelerating, which is contrary to expectations based on the standard cosmological model, which predicts that the expansion rate should be decelerating ($\ddot{a}<0$) due to gravitational attraction between all matter and energy in the Universe. 
Late-time cosmic acceleration is believed to be driven by an unidentified form of energy that counteracts the gravitational pull of matter. This enigmatic energy is known as DE. Subsequent cosmological observations, such as those related to the CMB, baryon acoustic oscillations (BAO), and large-scale structure (LSS), have provided additional evidence for the Universe accelerated expansion in recent times. Using observations of the CMB radiation (CMBR) by the PLANCK satellite experiment, the current energy density of DE is estimated to be 0.69 \cite{Planck:2018vyg}. This phenomenon can be addressed by examining observational evidence, such as Supernova Type Ia, CMB, BAO, large Scale Structure, and many more. In addition, the theoretical frameworks of modified theories of gravity are proposed in the upcoming topics to explain it.
%\begin{figure}[!htb]
 %   \centering
  %  \includegraphics[width=10cm]{Figures/Chapter-1/LCDM.jpg}
   % \caption{$\Lambda$CDM Model of Cosmology [Credit: NASA Goddard Space Flight Center]}
    %\label{fig:galaxy}
%\end{figure}
\section{Friedmann--Lema\^{i}tre--Robertson--Walker Universe}\label{FLRWUniverse}
To understand the evolution in the Universe, one must solve the Einstein field equations, which require a metric that accurately depicts the Universe large-scale geometry. In standard cosmology, the cosmological principle is a fundamental assumption, asserting the Universe’s homogeneity (translation invariance) and isotropy (rotational invariance). There is conflicting research challenging the cosmological principle \cite{Colin:2019opb}. In such a Universe, there are three reliable types of topological geometries: i) flat, ii) spherical (closed), and iii) hyperbolic (open).

The mathematical implication regarding the spacetime metric is that all points in spacetime adhere to the equation determined by the specific geometry. For instance, upon inserting a 2D sphere in 3D, the equation describes all points within the sphere by $r^2=a^2$. Moreover, one can generalise it and consider the 3D embedding into 4D; in this case, the equation described is $t^2 \pm r^2 = \pm a^2$. 
In cosmology, the plus sign $(+)$ denotes a closed, spherical Universe, while the minus sign $(-)$ represents an open, hyperbolic Universe. The metric is then defined accordingly as, carrying the same sign convention, one can construct the Friedmann--Lema\^{i}tre--Robertson--Walker (FLRW) metric as,
 \begin{align}
     d\textit{s}^2 = -dt^2 +a^2 (t) \left(\frac{1}{ 1 - k r^2} dr^2+r^2 d\theta^ 2+r^2 sin^ 2\theta d\phi^2 \right)\,,\label{FLRWW}
\end{align}
where we have used the flat form of the above metric with the $ (-, +, +, +)$ signature, carried throughout our analysis, and can be presented as, 
\begin{equation}
    ds^2 = -dt^2+a(t)^2(dx^2+dy^2+dz^2)\,.\label{FLATFLRW}
\end{equation}

Moreover, this form generally appears in the special theory of relativity. 
In studying the Universe large-scale properties, the scale factor, represented by $a(t)$, is a significant parameter that indicates the Universe size relative to a reference point. It is a quantity that changes with time, and it is conventionally set such that the scale factor at present ($t=t_0$) is equal to 1. It is essential to understand that the FLRW metric uses comoving coordinates. This means that if the Universe expansion were to cease today, the physical coordinates would align with the comoving coordinates. To determine the physical distance at any other time, one must multiply the comoving coordinates by the scale factor. For example, $a(t) r$ provides the radial distance to a point at time t. Another crucial parameter is $k$, referred to as the curvature parameter. It can take values of $-1$ for a hyperbolic Universe, 0 for a flat Universe, and $+1$ for a spherical Universe. When $k=0$, the FLRW metric simplifies to the flat metric in spherical coordinates.

\subsection{Fundamental equations of cosmology}\label{FunCosmo}
Einstein's equations describe the relation between the geometry of space-time and the distribution of matter within it. The field equations for GR can be written as,
\begin{align}
G_{\mu\nu} = 8\,\pi\,G\,T_{\mu\nu} \,,\label{GRFE}
\end{align}
where 
$G_{\mu \nu}=R_{\mu\nu} - \frac{1}{2}R g_{\mu\nu}$ is the Einstein tensor, 
\(g_{\mu \nu}\) is the metric tensor, 
\(T_{\mu \nu}\) is the stress-energy tensor,  
\(8 \pi G\) is the Einstein gravitational constant.

Now, applying two simplifying assumptions: (i) the Universe contains matter that can be modeled as a perfect fluid, and (ii) the Universe is equipped with the FLRW metric, we can reduce Einstein's ten equations to only two, referred to as the Friedmann equations.

The first assumption of perfect fluid leads to vanishing off-diagonal elements of energy-momentum tensor and the pressure will remain the same in all three spatial directions, can be presented as,
\[
T^{\mu}_{\nu} =
\begin{pmatrix}
-\rho & 0 & 0 & 0 \\
0 & p & 0 & 0 \\
0 & 0 & p & 0 \\
0 & 0 & 0 & p
\end{pmatrix}\,,
\]
consequently, this equation can be written in the tensor form as 
\begin{equation}
T^{\mu \nu} = (\rho + p)u^{\mu}u^{\nu} + p g^{\mu \nu}\,.\label{stressenergytensor}
\end{equation}
Here $u^{\mu}$ is the four-velocity vector. Now, with the second consideration of the FLRW metric, we find for the components $\mu,\nu=0$ (time coordinates), and the remaining spatial coordinates will generate the Friedmann equations as follows,
\begin{align}
    3H^2=8\pi G \rho\,,\nonumber\\
    3H^2+2\dot{H}=-8\pi G p\,,\label{FriedmanEQ}
\end{align}
 where $H = \frac{\dot{a}}{a}$ is the Hubble parameter. The fluid equation is another differential equation that plays a crucial role in describing how the energy density of matter evolves, which is also called the continuity equation,
\begin{align}
\dot{\rho}+3H\left(\rho+p\right)=0\,.\label{inConservationEq}
\end{align}
One thing to notice is that the equation for continuity can also be obtained from the first law of thermodynamics, where $dQ = dE + pdV$ \cite{ryden2017cosmology}. This equation can be represented in terms of the scale factor $a(t)$, initially, by rephrasing the Hubble parameter as $\frac{dlna}{dt}$, on further simplification, one can obtain,
\begin{align}
    \rho \propto a^{-3(1+\omega)}\,.
\end{align}
Through this solution of the continuity equation, we determined the relationship between the energy density and the scale factor. The symbol $\omega$ represents EoS, the constant of proportionality in the thermodynamic EoS, $p=\omega\rho$. The value of the EoS parameter $\omega$ varies based on the type of matter being considered. For standard matter, $\omega$ has been established as 0. However, for gas of photons, the EoS is $p=\frac{\rho}{3}$, indicating that $\omega$ equals $\frac{1}{3}$ for radiation.

\subsection{Dynamics of the scale factor in a flat Universe}
The astronomical evidence indicates a flat Universe \cite{LINDE1982389}. Therefore, it is beneficial to conduct a more thorough examination of the change in the scale factor for a flat geometry. With $k = 0 $ for a flat Universe, the initial Friedmann equation Eq. \eqref{FriedmanEQ} becomes simplified to
\begin{align}
    H^2 = \left(\frac{\dot{a}}{a}\right)^2 = \frac{8\pi G}{3} \rho\,,
\end{align}
which implies 
\begin{align}
    \left(\frac{\dot{a}}{a}\right)^2 \propto a^{-3(1+\omega)} \implies \dot{a} \propto a^{-\frac{1}{2}(1+3w)} \implies a \propto t^{\frac{2}{3(1+\omega)}}\,.
\end{align}
Now, as we are aware of $H=\frac{\dot{a}}{a}$ i.e. $Hdt= \frac{1}{a} dt$, on integrating both side will generate 
$a \propto t^{\frac{2}{3(1+\omega)}}$.
We can infer that a Universe governed by a cosmological constant would result in rapid expansion. This phenomenon is commonly referred to as de Sitter expansion. Another scenario to consider is an empty Universe with $\rho = 0$, in this situation, the Friedmann equation simplifies to $\dot{a} = 0$, which means the scale factor a remains constant. To summarize the results, the scale factor for a flat Universe changes in the following manner:
\begin{align}
a(t) \propto
\begin{cases}
t^{\frac{2}{3(1+w)}} & w \neq -1 \\[8pt]
e^{Ht} & w = -1 \\[8pt]
\tau_{0} & \rho = 0
\end{cases}\,.
\end{align}
Where $\tau_{0}$ is a constant. We have gathered the outcomes for various dominance scenarios in Table \ref{scalefactor}. From this, it is evident that only a Universe dominated by a cosmological constant results in accelerated expansion. This observation has led to the inference that the ``DE" thought to drive this acceleration is attributed to the cosmological constant $\Lambda$ \cite{Lin:2018kjm}
\begin{table}[H]
     % title of Table
    \centering % used for centering table
    \scalebox{0.8}{
    \begin{tabular}{|c|c|c|c|} % centered columns (5 columns)
    \hline %inserts double horizontal lines
    \parbox[c][1.3cm]{1.3cm}{$a(t)$}& Name of an era & $\rho(a)$ & $\omega$ \\  % inserts table %heading
    \hline\hline % inserts single horizontal line
    \parbox[c][1.3cm]{1.3cm}{$t^{\frac{1}{2}}$ } & Radiation dominated & $a^{-4}$ & $\frac{1}{3}$\\
    \hline
    \parbox[c][1.3cm]{1.3cm}{$t^{\frac{2}{3}}$ } & Matter dominated & $a^{-3}$ & 0\\
    \hline
   \parbox[c][1.3cm]{1.3cm}{$e^{Ht}$ }   &Cosmological constant $\Lambda$ & $a^{0}$ & $-1$  \\
    \hline
    \end{tabular}}
    \caption{Evolution of a scale factor representing different important eras of the evolution.}
    % is used to refer to this table in the text
    \label{scalefactor}
\end{table}

\section{Brief overview of \texorpdfstring{$\Lambda$}{}CDM model}\label{lambdaoverview}
The prevailing model that best matches observational data and explains the evolution of the Universe is referred to as the standard model of cosmology. It is also called the Benchmark model \cite{ryden2017cosmology}. The standard model can be viewed as a fusion of two theories: $\Lambda$CDM (lambda-Cold Dark Matter) and the Hot Big Bang, although these are at times used interchangeably and even as synonyms for the standard model itself. For this thesis, we will treat these two theories as distinct and as a part of the standard model. The major difference between the $\Lambda$CDM and the Hot-Big Bang Model is the $\Lambda$CDM model characterizes the distribution of fluid in the Universe, while the Hot Big Bang model is the theory that proposes the Universe developed from a hot, dense state from which matter was produced. In this section, we will further discuss the $\Lambda$CDM model.

When discussing how matter is spread out in the Universe, it is typical to use density parameters, which can be established based on the Friedmann equations when they are reformulated. The Friedmann first equation for a flat Universe is as presented in Eq. \eqref{FriedmanEQ}. At the current time, it can be rewritten as,
\begin{align}
    \rho =\frac{3H_{0}^2}{8\pi G}=\rho_{crit} \,,\label{rhocrit}
\end{align}
where $\rho_{crit}$ is the critical current density and $H_0$ is the Hubble parameter at present epoch. The density at which the scale factor reaches an extremum was historically defined as being a maximum in a closed Universe. It is possible to separate the energy density and pressure into various contributions. For instance, $\rho_{\Lambda}$ and $\rho_{m}$ represent the energy density contribution from DE and DM, in that order. Along with the critical density definition in Eq. \eqref{rhocrit}, we can now establish the expressions for the standard density parameters as,
\begin{align}\label{densityparameters}
    \Omega_{k}=\frac{\rho_{k}}{3H^2}\,, \quad \Omega_{m}=\frac{\rho_{m}}{3H^2}\,,\quad \Omega_{\Lambda}=\frac{\rho_{\Lambda}}{3H^2}\,. 
\end{align}
With this definition, one can construct the Friedmann equations in a different form as,
\begin{align}
\left(\frac{H}{H_0}\right)^2 = \Omega_i a^{-3(1+w_i)} + \Omega_\kappa a^{-2} \,.\label{ageoftheun}
\end{align}
 Evaluating the Friedmann equations at the present time $t = t_0$, where we have $H = H_0$ and the scale factor at the present time $a = a_0= 1$, which leads to the cosmic sum and can be presented as,
 \begin{align}\label{CONSTRAINEQ}
\Omega_{k}+\Omega_{m}+\Omega_{\Lambda}=1\,.
 \end{align}
We must ensure that all the standard density parameters adhere to the aforementioned relationship. In Table \ref{parametersbestfit}, we have compiled all density parameters obtained from the CMB by the Planck satellite. It is important to note that we have not accounted for contributions from radiation neutrinos or any other fluid for now, as the majority consists of DM and DE. DM is an unidentified component known to cause flat rotation curves in galaxies and accounts for most of the matter contribution \cite{ryden2017cosmology}. This collection of findings is referred to as $\Lambda$CDM, where the $\Lambda$ suggests the assumption that the cosmological constant is responsible for DE.
\begin{table}[h!]
\centering
\begin{tabular}{lcc}
\toprule
Parameter & Value & Uncertainty \\
\midrule
$\Omega_m$   & 0.3111 & $\pm 0.0056$ \\
$\Omega_\Lambda$   & 0.6889 & $\pm 0.0056$ \\
$\Omega_\kappa$   & 0 & $\pm 0.011$ \\
Sum & 1 & \\
$H_0$ (km/s/Mpc) & 67.66 & $\pm 0.42$ \\
\bottomrule
\end{tabular}
\caption{Present values of the cosmological parameters}\label{parametersbestfit}
\end{table}

Data from the most recent Planck satellite release \cite{Planck:2018vyg} shows the following parameters: $H_0$, which represents the Hubble constant; $\Omega_{m}$, symbolizing matter (including both baryonic and DM); $\Omega_{\Lambda}$, standing for DE; and $\Omega_{k}$, which denotes the curvature parameter. If we assume that the Universe has been expanding since the beginning of time, we can calculate the age of the Universe by projecting backward until the scale factor reaches zero. Once we have assessed the density parameters, this task becomes surprisingly straightforward.

\section{Understanding the need to modify GR or TEGR}\label{NeedofTEGR}
 In this section, we will explore the motivations and physical basis for considering modified gravity as a potential approach to addressing certain problems that are not properly addressed using GR or TEGR. For a comprehensive review of modified gravity, refer to the review articles \cite{Cai:2015emx,bahamonde:2021teleparallel,Bamba:2012cp,Clifton:2011jh}. Since the TEGR and GR have the same set of field equations, they might have the same physical motivation, as TEGR faces similar limitations as GR. When the Einstein-Hilbert (or TEGR) action is modified or extended, it is often called modified gravity. This modification can be made by changing the matter or energy section, which is generally presented on the right-hand side of the GR's equations, or by obtaining the change in the section where the geometrical part lies, which is on the left-hand side of the GR field equations.  For instance, the quintessence and phantom models introduce a canonical or non-canonical scalar field to explain the Universe late-time accelerating behaviour.

 To begin, GR requires a cosmological constant, which serves as a fluid with a negative pressure $p_{\Lambda} = -\rho_{\Lambda}$, in order to account for the current accelerating behaviour of the Universe. In the absence of evoking a cosmological constant, GR can generate this scenario by incorporating additional fields like a canonical or non-canonical scalar field. The observed value of the cosmological constant differs from the expected value derived from considering quantum and classical aspects associated with vacuum energy. This discrepancy is known as the cosmological constant problem \cite{Bernardo:2021izq,Henneaux:1989zc,Lombriser:2019jia}, and some researchers argue that there is no solution to this problem without either modifying GR, adding extra scalar fields, or altering the standard model of physics. When contemplating modifications to GR, certain theories have the potential to portray an accelerating late-time behaviour of the Universe expansion without resorting to cosmological constant. Moreover, the acceleration is attributed to the new terms arising from the modifications. This stands as one of the primary motivations behind modified gravity.
 
 During the early stages of the Universe, there was a period of incredibly quick expansion. This expansion cannot be explained by a cosmological constant because it is followed by a period of radiation, and a cosmological constant would not cause the acceleration to stop. The accelerated expansion can be accounted for GR using a scalar field known as inflation. However, GR does not provide an explanation for the origin or nature of inflation, and its predictions have issues with fine-tuning parameters. Examples of this include the initial conditions for inflation and the slow-roll approximations. Modified gravity can describe both the early inflationary period and the current DE era in a unified manner \cite{Arbey_2021,Nojiri:2017ncd}.

 Another important issue that cannot be resolved with GR could also be tackled with modified theories of gravity. The current energy density of matter (including baryonic and DM) is comparable to the energy density of DE. What is the reason for these values being almost identical today? Is there a physical explanation for the similarity of these quantities? This is referred to as the coincidence problem. Some physicists argue that this is merely a coincidence and not a problem, but there is no initial theoretical justification for this. Certain modifications of GR may offer a solution to this problem \cite{Nojiri:2017ncd}.
 
In conclusion, studying modified gravity theories provides an opportunity to better understand GR or TEGR. Modified gravity can be viewed as an extension of this basic structure of gravities; this will allow for the discovery of properties that may not be readily apparent when solely studying GR or TEGR. There are various ways to modify the gravity sector. One perspective is that while GR performs well at the scale of the solar system, it may require modifications at larger cosmological scales. Thus, any modification of GR must accurately predict the same observations at the solar system scale. This can be achieved by incorporating an Einstein-Hilbert action and introducing additional terms derived from the modifications. Alternatively, one can develop a modified theory that departs from a GR background and incorporates screening mechanisms to ensure accurate predictions at the solar system scale. Nevertheless, a conclusive modified theory of gravity that resolves all current issues has not yet been established.

\section{TEGR and its modifications}\label{TEGRMODIFICATION}

After Einstein formulated the GR field equations, alternative versions of GR were developed and discussed. One of these descriptions, TEGR, is particularly interesting. The final equations of motion in TEGR are the same as those in GR, with the only difference being a total derivative term in their actions. These two theories are experimentally indistinguishable. In TEGR, there is no geodesic equation, unlike in GR. Instead, force equations are used to describe the movement of particles under the influence of gravity. The tetrad is the dynamical variable in TEGR instead of the metric in GR. These concepts were initially introduced by Einstein in 1928 \cite{einstein2005riemann}. The fundamental mathematical finding of this method can be traced back to Weitzenbock's work in 1923. This section is dedicated to presenting the fundamental principles of TEGR, formulated as a gauge theory of translations. Ref. \cite{Aldrovandi:2013wha} provides a more detailed explanation of TEGR.

Any theory of gravity is formulated on a manifold, where at each point, a tangent space is defined as a Minkowski spacetime with a metric $\eta_{ab} = \text{diag} (-1, +1, +1, +1)$. In this work, Greek indices $\alpha, \beta, \gamma,...$ represent spacetime indices, while Latin indices $a, b, c,...$ represent tangent space indices. The local bases for vector fields are defined as $\partial_{\mu} = \frac{\partial}{\partial{x^\mu}}$ for a spacetime coordinate $x^{\mu}$, and $\partial_{a} = \frac{\partial}{\partial{x^a}}$ for a tangent space coordinate $x^{a}$. The primary quantity in teleparallel gravity is the tetrad field or vierbein, which forms the basis for vectors in the tangent space. In this work, the tetrad is denoted as $e^{a}_{~\mu}$ and its inverse as $e^{~\mu}_{a}$. For flat FLRW space-time it will take the form as,
\begin{align}\label{FLRWTETRAD}
e^{a}_{~\mu}=(1,a(t),a(t),a(t))\,.
\end{align}
Thus, the basis can be expressed as $e^{a}=e^{a}_{~\mu} dx^{\mu}$ and $e_{a}=e^{~\mu}_{a} \partial_{\mu}$. It is important to note that tetrads can always be defined if the manifold is assumed to be differentiable. The tetrads satisfy the orthonormal conditions, as described below,
\begin{align}
 e^{~\nu}_{m}e^m_{~\mu}= \delta^{\nu}_{\mu} \,, \quad
    e^{~\mu}_{m}e^n_{~\mu}= \delta^{n}_{m}\,.
\end{align}
The relationship between the Minkowski tangent spacetime metric $\eta_{ab}$ and the Minkowski spacetime metric $\eta_{\mu\nu}$ is provided by linear frames,
\begin{align}
    \eta_{\mu\nu} =e^{a}_{~\mu} e^{b}_{~\nu} \eta_{ab}\,,
\end{align}
which are known to be a trivial frame, In those frames, a pseudo-Riemannian metric and its inverse can be written with the tetrads as follows,
\begin{align}
    g_{\mu\nu} = e^a_{~\mu} e^b_{~\nu} \eta_{ab}\,, \quad
    g^{\mu\nu} = e^{~\mu}_{a}e^{~\nu}_{b}\eta^{ab} \,.
\end{align}
One can easily notice that the determinant of the tetrad is related to the determinant of the metric by $e=\sqrt{-g}=det(e^{a}_{~\mu})$.

The quantities known as spin connections, denoted by $\omega^{a}_{\, \ \, b \mu}$, it depict the inertial effects of the tetrad frame. An additional characteristic of the spin connection is its antisymmetry in the first two indices $(\omega_{ab\mu} = -\omega_{ba\mu})$. This connection can also be mathematically represented, as shown in the references \cite{Aldrovandi:2013wha, Krssak:2015oua}, and is defined as,
    \begin{align}
        \omega^{a}_{\, \, \, b\mu}=-\Lambda_{b}^{c}\partial_{\mu} \Lambda_{c}^{a} \,,
    \end{align}
where $\Lambda_{c}^{a}$ is the Lorentz metric. Now the connection coefficients $\Gamma^{\theta}_{\, \, \,\nu \sigma }$ of these connections were originally introduced to preserve the covariance of the covariant derivative. However, they also serve to describe the geometric properties of spacetime. It is said that the connection describes the motion of freely falling particles, while the metric determines the causal structure \cite{Cai:2015emx,Capozziello:2011et}. To differentiate between possible different geometries, the Riemann tensor, torsion tensor, and metric compatibility are all defined, and they all depend on the form of the connection. The Riemann tensor is initially defined to be \cite{Krssak:2015oua},
\begin{equation}
    R^\theta_{\ \sigma\mu\nu} = \partial_\mu \Gamma^\theta_{\nu\sigma} - \partial_\nu \Gamma^\theta_{\mu\sigma} + \Gamma^\theta_{\mu\alpha} \Gamma^\alpha_{\nu\sigma} - \Gamma^\theta_{\nu\alpha} \Gamma^\alpha_{\mu\sigma}\,,\label{Riemanntensor}
\end{equation}
it gives a concept of curvature. This can be viewed as a way to measure the change in orientation when a vector is moved around a closed curve. Thus, we consider a connection flat (no curvature) when $R^\theta_{\ \sigma\mu\nu}=0$. The torsion tensor is specifically defined as,
\begin{equation}
    T^\theta_{~\mu\nu} = \Gamma^\theta_{\nu\mu} - \Gamma^\theta_{\mu\nu}\,,
\end{equation}
and is providing a measure of the non-closure of parallelograms, as vectors are parallel transported. From this, it can be seen that a connection is said
 to be torsionless if $T^\theta_{~~\mu\nu}=0$. In the framework of GR, it is assumed that the connection is without torsion and satisfies metric compatibility, signifying that spacetime is entirely characterized by curvature. As a result, this gives rise to a distinctive connection known as the Levi-Civita connection, which is purely defined in relation to the metric tensor \cite{carroll2004spacetime} as,
 \begin{equation}
    \Gamma^\mu_{\nu\sigma} = \frac{1}{2} g^{\mu\alpha} \left( \partial_\nu g_{\alpha\sigma} + \partial_\sigma g_{\alpha\nu} - \partial_\alpha g_{\nu\sigma} \right)\,.
\end{equation}
Teleparallel gravity, conversely, requires the connection to be flat and for metric compatibility to be maintained. In this category, the torsion is the only non-zero and the surviving quantity. The Weitzenböck connection is an example of a connection with these characteristics and can be presented as \cite{Weitzenbock1923},
\begin{equation}
\widehat{\Gamma}^\theta_{\nu\mu} = e^{~\theta}_a \partial_\mu e^a_{~\nu} + e^{~\theta}_{a} \omega^{a}_{~b\mu} e^{b}_{~\nu}\,.\label{Weitzenbockeq}
\end{equation}
The contortion tensor calculates the difference between the torsional and curvature components of the connection \cite{Aldrovandi:2013wha},
\begin{equation}
    K^{\mu}_{~\alpha \beta} = \widehat{\Gamma}^\mu_{\alpha \beta}-\Gamma^\mu_{\alpha \beta}\,.\label{kotorsiontensor}
\end{equation}
This expression can also be expressed in terms of the torsion tensor as,
\begin{equation}
     K^{\mu}_{~\alpha \beta} =\frac{1}{2}\left
     (T^{~\mu}_{\alpha ~~ \beta} + T^{~~\mu}_{\beta ~~ \alpha}- T^{\mu}_{~\alpha \beta}\right)\,.
\end{equation}
The connection between curvature (gravity) and mass (energy) is explained by a series of field equations. These field equations can be derived using the principle of least action and by introducing the notion of an action of some Lagrangian density, initially developed by Hilbert \cite{Hilbert:1915tx}. The Lagrangian contains a scalar, and since gravity is depicted by a manifestation of curvature, this scalar must be created from the Riemann tensor. In the case of GR, the Ricci scalar $R = R^{\mu}_{~\mu}$ gives the description, where $R_{\mu \nu} = R^{\alpha}_{~\mu \alpha \nu}$ is the Ricci tensor. Now, to include the contributions from matter, the matter Lagrangian $\mathcal{L}_{\text{m}}$, which includes all the information about the matter fields, is added to the total Lagrangian. Consequently, the combined gravitational and matter action takes the form,
\begin{equation}
    S_{EH} = \frac{1}{2} \int d^4x \, \sqrt{-g} \, R + \int d^4x \, \sqrt{-g} \, \mathcal{L}_{\text{m}} \,.
\end{equation}
 Taking variations of the above action with respect to the metric field yields the Einstein field equations presented in Eq. \eqref{GRFE}, and the stress of the energy-momentum tensor is presented in Eq. \eqref{stressenergytensor}.
 
After checking the wide range of physical acceptance of the argument of GR, a similar thought process is repeated from a teleparallel point of view. Here, the tetrad takes the place of the metric as the fundamental gravitational variable. Rather than the Ricci scalar $R$, a new torsional scalar quantity called the torsion scalar $T$ is constructed as,
\begin{equation}\label{TORSIONSCALAR}
    T = S_\theta^{\ \mu\nu} T^\theta_{\ \mu\nu}=\frac{1}{4}T^{\theta\mu\nu} T_{\theta\mu\nu} +\frac{1}{2} T^{\theta\mu\nu} T_{\nu\mu\theta} -  T_{\theta\mu}^{~~~\theta} T^{\nu\mu}_{~~~\nu}\,,
\end{equation}
where \( S_\theta^{\ \mu\nu} \) is the superpotential tensor, defined as,
\begin{equation}
    S_\theta^{\ \mu\nu} = \frac{1}{2} \left( K^{\mu\nu}_{\ \ \theta} + \delta^\mu_\theta T^{\sigma \nu}_{\ \ \ \sigma} - \delta^\nu_\theta T^{\sigma \mu}_{\ \ \ \sigma} \right)\,,
\end{equation}
and \( T^\theta_{\ \mu\nu} \) is the torsion tensor, and \( K^{\mu\nu}_{\ \ \theta} \) is the contorsion tensor. The expression of gravity is replaced by torsion instead of curvature. This means that the TEGR gravitational action is provided,
\begin{equation}
    S_{TEGR} = \frac{1}{2} \int d^4x \, e \, T + S_m\,.
\end{equation}
Where $S_m=\int d^4x \sqrt{-g} \mathcal{L}_{m}$ represents the matter action, on varying this action with respect to the tetrad field, the general field equations for TEGR can be obtained as \cite{Aldrovandi:2013wha},
\begin{equation}
    e^{-1} \partial_{\nu} (e S_{a}^{~~\mu\nu})+\frac{1}{4} e^{~\mu}_{a} T -T^{b}_{~~\nu a} S_{b}^{~~\nu\mu}+\omega^{b}_{~~a\nu} S_{b}^{~~\nu\mu}=\frac{1}{2} \Theta^{~~\mu}_{a}\,.
\end{equation}
It is important to note that the connection between TEGR and GR comes from the correspondence between the curvature-based quantities (Riemann tensor) and the torsional ones (torsion tensor). This correspondence is established by combining the definition of the Riemann tensor in Eq. \eqref{Riemanntensor} with the definition given by the contorsion tensor in Eq. \eqref{kotorsiontensor}. As detailed in Ref. \cite{Aldrovandi:2013wha}, this simplifies the teleparallel field equations to that of GR. The relationship between curvature and torsion results in an inherent connection between the Ricci and torsion scalars as,
\begin{equation}
    R = -T + B=-T - \frac{2}{e} \partial_{\mu} \left( e T^{\nu \mu}_{~~~\nu} \right)\,.
\end{equation}
The term $B$ is known as the boundary term. This term is not present in the GR and TEGR Lagrangians, which is expected because it results in a total derivative. At the action level, the term $B$ is set to vanish at the boundary, which results in preserving the equivalence even at the Lagrangian level \cite{Aldrovandi:2013wha}. The contribution of this term is useful in distinguishing between the non-equivalence of the curvature and teleparallel extensions $f(R)$ and $f(T)$ gravity, respectively.

\subsection{\texorpdfstring{$f(T)$}{} gravity}
The initial straightforward extension of the TEGR Lagrangian includes a general function of the torsion scalar $f(T)$, which can be seen as the torsional equivalent of the $f(R)$ gravity, which is the extension of GR, curvature gravity. The gravitational action for $f(T)$ gravity can be formulated as,
\begin{equation}
    S_{f(T)} = \frac{1}{2} \int d^4x \, e \, f(T) + S_m\,.
\end{equation}
The general $f(T)$ gravity field equations can be obtained as \cite{Ferraro:2006jd,Krssak:2015oua},
\begin{equation}
   f_{TT} S_{a}^{~~\mu\nu} \partial_{\nu} T+\frac{1}{4} e_{a}^{~~\mu} f +f_{T} e^{-1} \partial_{\nu} (e S_{a}^{~~\nu \mu})-f_{T} T^{b}_{~~\nu a} S_{b}^{~~\nu\mu}+f_{T}\omega^{b}_{~~a\nu} S_{b}^{~~\nu\mu}=\frac{1}{2} \Theta_{a}^{~~\mu}\,.\label{f(T)FE}
\end{equation}
The field equations are different from those found in $f(R)$ gravity since $f(R)=f(-T+B)=f(T)$. Indeed, $f(R)$ gravity is a subset of $f(T, B)$ gravity, as we'll demonstrate shortly. Additionally, the equations are second order in the tetrad field, contrasting the fourth-order equations derived in $f(R)$ gravity. The theory's local Lorentz invariance is an important point to consider and has been the subject of investigation in many studies (see Ref. \cite{Cai:2015emx}). $f(T)$ gravity is noted to break this invariance due to the non-local Lorentz invariance of the torsion scalar. In the context of TEGR, this is not problematic, as the field equations still uphold local Lorentz invariance. However, for $f(T)$ gravity, the equations would no longer have local Lorentz invariance \cite{Li:2010cg,Krssak:2015oua}.

\subsection{\texorpdfstring{$f(T, B)$}{} gravity}\label{ftbformulasec}
As discussed in the previous section, it was noted that $f(T)$ gravity and $f(R)$ gravity are not equivalent because of the difference in their boundary terms. Thus, it seems appropriate to introduce a gravitational Lagrangian that depends on both torsion and boundary term, known as $f(T, B)$ gravity, which was proposed in 
 Ref. \cite{Bahamonde:2015zma}, where $B$ represents a total divergence term and is defined as
\begin{equation}\label{eq:boundary_term_def}
    B = \frac{2}{e}\partial_{\theta}\left(e T^{\mu \ \ \theta}_{\ \ \mu}\right)\,.
\end{equation}
In essence, the theory is built upon gravitational action as follows,
\begin{equation}
    S_{f(T, B)} = \frac{1}{2} \int d^4x \, e \, f(T, B) + S_m\label{f(TBaction)}\,.
\end{equation}
On varying this action formula with respect to the tetrad field, the field equations are obtained to be, \cite{Bahamonde:2015zma},
\begin{eqnarray}
e_a{}^{\mu} \square f_B -  e_a {}^{\nu} \nabla ^{\mu} \nabla_{\nu} f_B +
	\frac{1}{2} B f_B e_a{}^{\mu} - \left(\partial _{\nu}f_B + \partial
	_{\nu}f_{T} \right)S_a{}^{\mu\nu} -\frac{1}{e} f_{T}\partial_{\nu} (e S_a{}^{\mu\nu})
	\nonumber\\+ f_{T} T^{b}{}_{\nu a}S_{b}{}^{\nu\mu}- f_T \omega ^b{}_{a\nu}
	S_b{}^{\nu\mu} -\frac{1}{2}  f e_{a}{}^{\mu} = \kappa ^2  \Theta _a{}^{\mu} \,.\label{f(T,B)FE}
\end{eqnarray}
The field equations for $f(R)$ can be derived when $f(T, B) = f(-T + B) = f(R)$ by equating $f_B$ with $f_R$ and $f_T$ with $f_R$. Consequently, gravity is regained in all analyses conducted within the $f(R)$ gravity framework. Various studies have examined $f(T, B)$ gravity theory with more general Lagrangians. Some of the relevant studies can be seen in Refs. \cite{Franco:2021,Franco:2020lxx}. The autonomous dynamical system is framed, and this formalism is analysed in detail in the later section.

\subsection{\texorpdfstring{$f(T, T_G)$}{} gravity}
The effects of gravity, which result in the field equations and describe curvature, have been traditionally attributed to the Ricci scalar. However, it is important to note that there are two other scalar quantities capable of quantifying curvature that originate from the Riemann tensor: the Kretschmann scalar $R_{\alpha\beta\gamma\eta} R^{\alpha\beta\gamma\eta}$, and another from the Ricci scalar $R_{\alpha\beta} R^{\alpha\beta}$. These two quantities introduce terms that are fourth order in the metric in the field equations, unlike the second-order behaviour found in GR. As a result, the system requires more degrees of freedom than GR, as it necessitates more information regarding the behaviour of the higher-order derivatives of the metric \cite{DeFelice:2010aj}. In 1971, Lovelock \cite{Lovelock:1971yv} developed a specific combination of quadratic curvature terms to address this undesired fourth-order effect, ultimately leaving the field equations as second-order. This combination is known as the Gauss-Bonnet invariant, and it is defined to be,
\begin{equation}
    G =R^2-4R_{\mu\nu} R^{{\mu\nu}} +R_{\alpha\beta\gamma\eta} R^{\alpha\beta\gamma\eta} \,.
\end{equation}
Despite this appealing characteristic, in 4-dimensional space, this term does not affect the field equations as it results in a total derivative unless higher order dimensions are taken into account \cite{DeFelice:2010aj}. Because of this, two potential alternatives have been explored to preserve the contributions of the Gauss-Bonnet term in the field equations: (a) coupling it with a scalar field, or (b) considering a general function $f(G)$ \cite{DeFelice:2010aj,Felicef(G)}. In either scenario, the equations are no longer second order, which means that the theory still encompasses more degrees of freedom than GR.

Focusing on the latter scenario, extensive research has been conducted on this model in both localized and cosmological environments. In the former case, the Newtonian approximation remains effectively unchanged, ensuring that the theory remains consistent with constraints derived from the solar ystem \cite{Cognola_2006}. From a cosmological perspective, these models can describe an expanding Universe, and can also produce shifts from deceleration to acceleration \cite{Cognola_2006,Felicef(G)}.

Motivated by these characteristics, researchers have explored the teleparallel counterpart of this scalar in Ref. \cite{Kofinas:2014owa}. The expression of $T_G$ can be presented as,
\begin{align}\label{eq:T_G_def}
    T_G &= \Big(K_{a \ \ e}^{\ \ i} K_{b}^{\ \ ej}K_{c \ \ f}^{\ \ k} K_{d}^{\ \ fl} - 2K_{a}^{\ \ ij} K_{b \ \ e}^{\ \ k} K_{c \ \ f}^{\ \ e} K_{d}^{\ \ fl}  + 2K_{a}^{\ \ ij}K_{b \ \ e}^{\ \ k}K_{f}^{\ \ el}K_{d \ \ c}^{\ \ f} \nonumber\\&+2K_{a}^{\ \ ij}K_{b \ \ e}^{\ \ k} K_{c,d}^{\ \ \ \ el} \Big)\delta_{ijlk}^{abcd} \,,
\end{align}
where $\delta_{ijkl}^{abcd} = \epsilon^{abcd}\epsilon_{ijkl}$ is the generalized Kronecker delta function \cite{Bahamonde:2016kba}. Similarly, just as the relation between the Ricci scalar and the torsion scalar differs by a boundary contribution, the curvature Gauss-Bonnet scalar $G$ and its teleparallel equivalent, $T_G$, also differ by a boundary term $B_G$, denoted as $G = T_G + B_G$, where,
\begin{equation}
    \frac{1}{e}\delta^{abcd}_{ijkl}\partial_{a} \left[K_{b}^{\ \ ij}\left(K_{c\ \ ,d}^{\ \ kl} + K_{d \ \ c}^{\ \ m}K_{m}^{\ \ kl}\right)\right]=B_G  \,.
\end{equation}
In this context, partial differentiation with respect to the index is indicated by commas, where $K_{c,d}^{~~~ el} =\partial_{d}(K_{c}^{~~ el})$. Therefore, it is possible to examine any functional dependence on the torsion and teleparallel equivalence of the Gauss-Bonnet scalars, such as $f(T, T_G)$, influence the gravitational source. In other words, this implies that the gravitational (and matter) action forms such a functional dependence as \cite{Kofinas:2014daa, Kofinas:2014owa}
\begin{equation}
    \mathcal{S}_{f(T,T_G)}^{} =  \frac{1}{2\kappa^2}\int \mathrm{d}^4 x\; e\,f(T,T_G) + \int \mathrm{d}^4 x\; e\mathcal{L}_{\text{m}}\,.\label{f(TTG)action}
\end{equation}
The \( f(T, T_G) \) field equations are given by,
\begin{eqnarray}
    2 \left(H^{[ac] b}+
    H^{[ba] c} -H^{[cb] a} \right)_{,c}+ 2\left(H^{[ac] b}+H^{[ba] c}-H^{[cb] a}\right)C^{d}_{~~dc}\nonumber\\
    +\left(2H^{[ac] b}+H^{dca}\right) C^{b}_{~~cd}+4H^{[db]c} C_{dc}^{~~a}+\left(T^{a}_{~~cd}+2\omega^{a}_{~~[cd]}\right) H^{cdb}\nonumber\\
    -h^{ab}+\left(f-T f_{T}-T_{G}f_{T_G}\right)\eta^{ab}=\Theta^{ab}\,,
\end{eqnarray}
where
\begin{eqnarray}
H^{abc}&=&f_{T}\left(\eta^{ac}K^{bd}_{~~d}-K^{bca}\right)\nonumber\\&+&f_{T_G}\Big[ \epsilon^{cmrt}K^{nf}_{~~t} \Big(2\epsilon^{a}_{~~dkf}K^{bk}_{~~m} K^{d}_{~~nr}+\epsilon_{ndkf} K^{ak}_{~~m} K^{bd}_{~~r}
+\epsilon^{ab}_{~~kf} K^{k}_{~~dm} K^{d}_{~~nr}\Big)  \nonumber\\
&+&\epsilon^{cmrt}\epsilon^{fd}_{~~md}K^{fd}_{~~n}\Big(K^{k}_{~~fr,t}-\frac{1}{2}K^{k}_{~~en} C^{n}_{~~tr}+\omega^{k}_{~~nt} K^{n}_{~~fr} +\omega^{n}_{~~fr}K^{k}_{~~nt}\Big)\nonumber\\
&+&\epsilon^{cmrt} \epsilon^{ak}_{~~df} K^{df}_{~m}\Big(K^{b}_{~~kr,t}-\frac{1}{2} K^{b}_{~~kn}C^{n}_{~~tr}+\omega^{b}_{~~nt} K^{n}_{~~kr}+\omega^{n}_{~~kr} K^{b}_{~~nt}\Big)\Big]\nonumber\\
&+&\epsilon^{cmrt}\epsilon^{a}_{~kdf}\Big[\Big(f_{T_G} K^{bk}_{~~m} K^{df}_{~~r}\Big)_{,t}+\frac{1}{2}f_{T_G} C^{n}_{~~mt} K^{bk}_{~~n}K^{df}_{~~r}-\frac{1}{2}f_{T_G} C^{n}_{~~mt} K^{bk}_{~~r} K^{df}_{~~n}\nonumber\\
&+&f_{T_G}\Big(\omega^{b}_{nm}K^{nk}_{~~r}+\omega^{k}_{nm}K^{bn}_{~~r}\Big) \Big)K^{df}_{~~t}+f_{T_G}\Big(\omega^{d}_{nm}K^{nf}_{~~t}+\omega^{f}_{nm}K^{dn}_{~~t}\Big)K^{bk}_{~~r}\Big]\,,\nonumber\\
h^{ab}&=&f_{T}\epsilon^{a}_{~~kcd}\epsilon^{bmnd}K^{k}_{~~fm} K^{fc}_{~~n}\,,\nonumber\\
C^{c}_{~~ab}&=&e^{~~\mu}_{a}e^{~~\nu}_{b}\Big(\partial_{\nu}e^{c}_{~~\mu}-\partial_{\mu}e^{c}_{~~\nu}\Big)\,.
\end{eqnarray}
Where $C^c_{~~
ab}$ is the structure coefficients function. Like its curvature counterpart, this theory of gravity has successfully explained both early (inflation) and late-time acceleration without the need for DE \cite{Kadam2023,Kofinas:2014aka,Kofinas:2014daa}.
\subsection{\texorpdfstring{$f(T, B, T_G, B_G)$}{} gravity}\label{1.6.4}
At this stage, it is crucial to make an important observation. To fully consider all the degrees of freedom in a particular gravitational theory, one must take into account every conceivable geometric invariant at a specific level.  In Ref. \cite{BOGDANOS2010236}, this discussion is considered for the fourth
order gravity, in this work the $f(R)$ gravity is further modified to $f(R, P, Q)$ gravity where $P = R_{\mu\nu}R^{\mu\nu}$ and
$Q=R_{\alpha\beta\mu\nu}R^{\alpha\beta\mu\nu}$. An equivalent theory can be expressed as $f(R, G)$ where $G = R^2−4P +Q$ represents the Gauss-Bonnet topological invariant, imposing a constraint on the other curvature invariant. A similar method is also feasible in TEGR. In Ref. \cite{Kofinas:2014aka}, a broader theory was presented, incorporating a teleparallel Gauss-Bonnet term. In the curvature approach, in addition to the definition above, it is a well-established fact that in 4 dimensions, the Gauss-Bonnet scalar $G$ becomes a boundary term. These scalars with detailed formalism have been presented in the previous section. In this section, a general teleparallel Gauss-Bonnet formalism has been presented,

The action formula can be presented as,
\begin{equation}
    \mathcal{S}_{f(T, B, T_G, B_G)}^{} =  \frac{1}{2\kappa^2}\int \mathrm{d}^4 x\; e\,f(T, B, T_G, B_G) + \int \mathrm{d}^4 x\; e\mathcal{L}_{\text{m}}\,.\label{ftbtgaction}
\end{equation}
Variation of this action formula with respect to the  tetrad gives,
\begin{equation}
    \delta \mathcal{S} =\int \Big[\frac{1}{2\kappa} \Big(f \delta e + e f_B \delta B + e f_T \delta T
    + e f_{T_G} \delta T_G + e f_{B_G} \delta B_G \Big)+ \delta \left
    ( e \mathcal{L}_m \right) \Big]d^4x\,,
\end{equation}
where,
\begin{eqnarray}
    e f_T \delta T &=& -4\Big[e (\partial_\mu f_T) S_a^{\ \mu\beta} + \partial_\mu (e S_a^{\ \mu\beta}) f_T 
    - e f_{T} T^\sigma_{~\ \mu a} S_\sigma^{\ \beta\mu}\Big] \delta e^a_\beta \,,\nonumber\\
    e f_B \delta B &=& \Big[2e e^\nu_{\ a} \nabla^\beta \nabla_\mu f_B - 2e e^\beta_{\ a} \Box f_B
    - B e f_B e^\beta_{\ a} - 4e (\partial_\mu f_B) S_a^{\ \mu\beta} \Big]\delta e^a_{~~\beta}\,, \nonumber\\
    e f_{T_G} \delta T_G &=& \Big[\partial_\mu\Big( e^\mu_{\ s} e^\beta_{\ b} \left( Y^{b~~s}_{~a} - Y^{s~~b}_{~a} + Y_{a}^{~ [bs]} \right)
    + T^{i}_{~~ab} e^\beta_{\ s} \left( Y^{b~~s}_{i} - Y^{s~~b}_{i} + Y_i^{\ [bs]} \right)\nonumber\\
    &-& 2e f_{T_G} \delta^{mbcd}_{\ ijkl}  e^\beta_{\ d} K_m^{\ ij} K_{b~~e}^{k} \partial_a (K^{~~el}_{\ c})\Big] \delta e^a_\beta, \nonumber\\
    e f_{B_G} \delta B_G &=& -\Big[\partial_\mu \left( P^{b~~s}_{\ ~a} - P^{s~~b}_{\ a~} + P_{i}^{~~[bs]} \right) e^\mu_{\ s} e^\beta_{\ b}\Big)
    \nonumber\\ 
    &+& T^i_{\ ab} E^\beta_{\ s} \left( P^{b~~s}_{~i} - P^{s~~b}_{~i} + P_i^{\ [bs]} \right)\nonumber\\
    &-& \delta^{mbcd}_{\ ijkl} e e^\beta_{\ d} \partial_m (f_{B_G}) K_b^{\ ij} (\partial_a K_{c}^{\ kl}) + e \partial_\mu (f_{B_G})
    \times \left( e^\beta_{\ a} B^\mu_{\ G} - e^\mu_{\ a} B^\beta_{\ G} \right)\nonumber\\
    &+& e f_{B_G} B_G e^\beta_{\ a}\Big] \delta e^a_\beta\,, \nonumber\\
    f \delta e &=& e f e^\beta_{\ a} \delta e^a_\beta \,.
\end{eqnarray}
The motivation here is to study the general case in which the torsional Gauss-Bonnet invariant and the boundary term can broadly be regarded in the $f(T, B, T_G, B_G)$ gravity formalism. This formalism allows us to study the effect of both the boundary term and the teleparallel Gauss-Bonnet invariant \cite{Bahamonde:2016kba}. These models are cosmologically viable and tested firmly using the Noether symmetry approach \cite{bahamonde2019noether}. Hence, we aim to analyse these models in studying the evolutionary epoch of the Universe and the contribution of Gauss-Bonnet torsion scalar $T_G$ and the boundary term $B$ in the prism of dynamical system analysis presented in a further section.

\subsection{\texorpdfstring{$f(T, \phi)$}{} gravity}

We studied the modified gravity models using different boundary terms in the previous sections. Starting from the second-order modification, we accumulate one of our study's most general teleparallel Gauss-Bonnet formalisms. Although these are the different extensions of TEGR with the inclusion of boundary terms, the scalar field modifications can also be considered. Similar to the modifications made in GR with the scalar field \cite{Capozziello:2011et}, the same has been introduced in the TEGR \cite{Cai:2015emx,Bahamonde:2017ize}. It is important to note that although TEGR and GR are dynamically equivalent, a scalar-torsion theory with non-minimal coupling is not equivalent to its curvature-based counterpart. A scalar-torsion theory with a non-minimal coupling term, $\xi \phi^2 T$, where $\phi$ is the dynamical scalar field, was initially applied to DE in Ref. \cite{Geng:2011aj,Geng_2012}. These modifications can be more generalised by introducing the general form of the non-minimally coupled scalar field form $F(\phi)G(T)$. These generalisations are assumed to be an extension of $f(T)$ gravity in terms of the gravitational sector \cite{Ferraro:2008ey,Linder:2010py}. Motivated by the study of scaling solutions presented in Ref, \cite{Uzan1999,Amendola1999}, in TEGR, the $f(T,\phi)$ gravity \cite{Gonzalez-Espinoza:2021mwr,Gonzalezreconstruction2021,Gonzalez-Espinoza:2020jss} can be considered. 

The action for $f(T,\phi)$ gravity is \cite{Gonzalez-Espinoza:2021mwr},
\begin{equation}\label{ActionEqf(Tphi)}
S =\int d^{4}xe[f(T,\phi)+P(\phi)X]+S_{r}+S_{m} \,,
\end{equation}
where $X= -\partial_{\mu} \phi \partial^{\mu} \phi/2$ represents the kinetic term of the field, $S_{m}, S_{r}$ represents the action contribution from the matter and the radiation part.

On varying this action with respect to the tetrads, we get the following set of field equations,
\begin{equation}
    f_{,T} G_{\mu\nu} + S_{\mu\nu}^{\ \ \ \theta} \partial_\theta f_{,T} + \frac{1}{4} g_{\mu\nu} \left( f - T f_{,T} \right) + \frac{P}{4} \left( g_{\mu\nu} X + \partial_\mu \phi \, \partial_\nu \phi \right) = -\frac{1}{4} T_{\mu\nu},
\end{equation}
where \( G^{\mu}_{~\nu} = e_a^{\ \mu} G^a_{\ \nu} \) is the Einstein tensor \cite{Aldrovandi:2013wha}, with,
\begin{equation}
    G^{\mu}_{\ a} \equiv e^{-1} \partial_\nu \left( e e^{~~\sigma}_{a} S_{\sigma}^{~~\mu\nu} \right) - e^{~~\sigma}_{\ a} T^{\alpha}_{\ \theta\sigma} S^{~~\theta\mu}_{\alpha}
    + e^{~~\alpha}_{\ b} S^{~~\theta\mu}_{\ \alpha} \omega^{b}_{ ~~a\theta} + \frac{1}{4} e^{~~\mu}_{a} T \,.
\end{equation}
The autonomous dynamical system is constructed with different potential coupling functions, which are discussed in detail in the following section. 
\subsection{Teleparallel Horndeski gravity}\label{teleparllelhorndesky}

In the work presented in chapter \ref{Chapter2}, we explore the dynamical systems of a particular class of scalar-tensor models that uniquely appear in teleparallel gravity. To form the broader class of scalar-tensor extensions in TG, we first consider the irreducible pieces of the torsion tensor \cite{Hayashi:1979qx,Bahamonde:2017wwk}
\begin{align}
    a_{\mu} & =\frac{1}{6}\epsilon_{\mu\nu\alpha\theta}T^{\nu\alpha\theta}\,,\\
    v_{\mu} & ={T}^{\alpha}_{~~\alpha\mu}\,,\\
    t_{\alpha\mu\nu} & =\frac{1}{2}\left(T_{\alpha\mu\nu}+T_{\mu\alpha\nu}\right)+\frac{1}{6}\left(g_{\nu\alpha}v_{\mu}+g_{\nu\mu}v_{\alpha}\right)-\frac{1}{3}g_{\alpha\mu}v_{\nu}\,,
\end{align}
which are respectively the axial, vector, and purely tensorial parts, and where $\epsilon_{\mu\nu\alpha\theta}$ is the totally antisymmetric Levi-Civita tensor in four dimensions. The axial, vector and purely tensorial parts can be used to construct the scalar invariant \cite{Bahamonde:2015zma}
\begin{align}
    T_{\text{ax}} & = a_{\mu}a^{\mu} = -\frac{1}{18}\left(T_{\alpha\mu\nu}T^{\alpha\mu\nu}-2T_{\alpha\mu\nu}T^{\mu\alpha\nu}\right)\,,\\
    T_{\text{vec}} & =v_{\mu}v^{\mu}=T^{\alpha}_{~~\alpha\mu}T_{\theta}^{~~\theta\mu}\,,\\
    T{_{\text{ten}}} & =t_{\alpha\mu\nu}t^{\alpha\mu\nu}=\frac{1}{2}\left(T_{\alpha\mu\nu}T^{\alpha\mu\nu}+T_{\alpha\mu\nu}T^{\mu\alpha\nu}\right)-\frac{1}{2}T^{\alpha}_{~~\alpha\mu}T_{\theta}^{~~\theta\mu}\,.
\end{align}
These three scalars form the most general parity preserving scalars that are quadratic in contractions of the torsion tensor, and even reproduce the torsion scalar $T=\frac{3}{2}T_{\text{ax}}+\frac{2}{3}T_{\text{ten}}-\frac{2}{3}T{_{\text{vec}}}$. Recently, this has led to the proposal of a teleparallel analogue of Horndeski gravity \cite{Bahamonde:2019shr,Bahamonde:2019ipm,Bahamonde:2020cfv,Dialektopoulos:2021ryi,Bahamonde:2021dqn,Bernardo:2021izq,Bernardo:2021bsg}, also called Bahamonde-Dialektopoulos-Levi Said (BDLS) theory. As in curvature-baed gravity, this is grounded on the Lovelock theorem \cite{Lovelock:1971yv,Gonzalez:2015sha,Gonzalez:2019tky}, and leads to the linear torsion scalar contraction scalar invariants \cite{Bahamonde:2019shr}
\begin{equation}
    I_2 = v^{\mu} \phi_{;\mu}\,,\label{eq:lin_contrac_scalar}
\end{equation}
where $\phi$ is the scalar field, and while for the quadratic scenario, we find
\begin{align}
    J_{1} & =a^{\mu}a^{\nu}\phi_{;\mu}\phi_{;\nu}\,,\label{eq:quad_contrac_scal1}\\
    J_{3} & =v_{\sigma}t^{\sigma\mu\nu}\phi_{;\mu}\phi_{;\nu}\,,\\
    J_{5} & =t^{\sigma\mu\nu}t_{\sigma~~~\nu}^{~~\alpha}\phi_{;\mu}\phi_{;\alpha}\,,\\
    J_{6} & =t^{\sigma\mu\nu}t_{\sigma}^{~~\alpha\beta}\phi_{;\mu}\phi_{;\nu}\phi_{;\alpha}\phi_{;\beta}\,,\\
    J_{8} & =t^{\sigma\mu\nu}t_{\sigma\mu}^{~~\alpha}\phi_{;\nu}\phi_{;\alpha}\,,\\
    J_{10} & ={\epsilon}^{\mu}_{~~\nu\sigma\theta}a^{\nu}t^{\alpha\theta\sigma}\phi_{;\mu}\phi_{;\alpha}\,,\label{eq:quad_contrac_scal10}
\end{align}
where semicolons represent covariant derivatives with respect to the Levi-Civita connection. The Levi-Civita connection enters into the scalar field sector through the minimal coupling prescription of TG (See Ref.~\cite{Bahamonde:2019shr} for further details). Naturally, the regular Horndeski terms from curvature-based gravity also appear in this framework \cite{Horndeski:1974wa}
\begin{align}
    \mathcal{L}_{2} & =G_{2}(\phi,X)\,,\label{eq:LagrHorn1}\\
    \mathcal{L}_{3} & =G_{3}(\phi,X)\mathring{\Box}\phi\,,\\
    \mathcal{L}_{4} & =G_{4}(\phi,X)\left(-T+B\right)+G_{4,X}(\phi,X)\left(\left(\mathring{\Box}\phi\right)^{2}-\phi_{;\mu\nu}\phi^{;\mu\nu}\right)\,,\\
    \mathcal{L}_{5} & =G_{5}(\phi,X)\mathring{G}_{\mu\nu}\phi^{;\mu\nu}-\frac{1}{6}G_{5,X}(\phi,X)\left(\left(\mathring{\Box}\phi\right)^{3}+2{\phi}_{;\mu}^{~~\nu}{\phi}_{;\nu}^{~~\alpha}{\phi}_{;\alpha}^{~~\mu}-3\phi_{;\mu\nu}\phi^{;\mu\nu}\,\mathring{\Box}\phi\right)\,,\label{eq:LagrHorn5}
\end{align}
where the kinetic term is defined as $X=-\frac{1}{2}\partial^{\mu}\phi\partial_{\mu}\phi$. BDLS theory simply adds the further Lagrangian component \cite{Bahamonde:2019shr}
\begin{equation}
    \mathcal{L}_{\text{{\rm Tele}}}= G_{\text{{\rm Tele}}}\left(\phi,X,T,T_{\text{ax}},T_{\text{vec}},I_2,J_1,J_3,J_5,J_6,J_8,J_{10}\right)\,.
\end{equation}
This results in the BDLS action given by
\begin{equation}\label{action}
    \mathcal{S}_{\text{BDLS}} = \frac{1}{2\kappa^2}\int d^4 x\, e\mathcal{L}_{\text{{\rm Tele}}} + \frac{1}{2\kappa^2}\sum_{i=2}^{5} \int d^4 x\, e\mathcal{L}_i+ \int d^4x \, e\mathcal{L}_{\rm m}\,.
\end{equation}
We will make detailed analysis of two general models with their two particular cases in detail in chapter \ref{Chapter2}. 
    
\section{Dynamical system analysis fundamental formulation}\label{dmsoverview}

A dynamical system can be anything, ranging from something as simple as a single pendulum to something as complex as the human brain and the entire Universe. In general, a dynamical system can be considered as any abstract system comprising (a) a space (state space or phase space) and (b) a mathematical rule that explains the development of any point in that space. The importance of the second point cannot be overstated. For example, discovering a mathematical rule that explains the development of information at any neuron in the human brain is likely impossible \cite{B_hmer_2016}. Therefore, we require a mathematical rule as an input, and finding one might be quite challenging. There are two primary types of dynamical systems. The first type consists of continuous dynamical systems, whose development is determined by a set of ordinary differential equations (ODEs), and the other type consists of time-discrete dynamical systems, which are determined by a map or difference equations. In the field of cosmology, we study the Einstein field equations, which, for a homogeneous and isotropic space, lead to a set of ODEs. Consequently, we are solely concerned with continuous dynamical systems and will not address time-discrete dynamical systems from now on. 

Let's start by establishing the collection of variables $(x_1, ..., x_n) \in X \subseteq \mathbb{R}^n$, which denote positions in an $n$-dimensional phase space. Let's formally define our independent variable, $t \in \mathbb{R}$, which may not necessarily signify time. As a general structure, a series of ODE that describe a dynamic system can be expressed as
\begin{equation}
\begin{aligned}
    \dot{x}_1 &= f_1(x_1, \dots  \dots, x_n) \\
    &\vdots \\
    \dot{x}_n &= f_n(x_1, \dots  \dots, x_n)\,.
\end{aligned}\label{_GDS_}
\end{equation}
Here, $f_i$ are the mapping function $f_i: X \rightarrow X$. The dynamical system is considered autonomous if the equations do not explicitly depend on the independent variable $t$. The behaviour of the system presented in Eq. \eqref{_GDS_} can be represented by the equivalent equation $\dot{x} = f(x)$, where $\dot{x}$ represents the derivative with respect to time. In this context, the vector $x = (x_1, \dots, x_n)$, and the function $f(x) = (f_1(X), \dots, f_n(X))$ can be interpreted as a vector field on $\mathbb{R}^n$.

We discuss the scenario where $f(x)$ is smooth and has real values. Any specific solution to Eq. \eqref{_GDS_} with an initial condition $x_0$ will correspond to a point that moves along a curve in the phase space. This curve or solution is referred to as the trajectory or orbit. Therefore, the phase space contains trajectories beginning from different initial conditions. By using this visual representation, we can gather information about the system simply by observing the paths of trajectories in the phase space.

We define the point $x_0$ as the critical point as points $x_0 \in U$ where $f(x_0) = 0$ and consider the stability of these critical points as discussed in \cite{wiggins2003introduction}. It is a way to determine if a trajectory starting near $x_0$ moves towards or away from $x_0$. Specifically, we outline two stability types, a critical point $(x, y) = (x_0, y_0)$ is considered stable (also known as Lyapunov stable) if all solutions $x(t)$ that start close to it remain nearby, and it is considered asymptotically stable if it is stable and the solutions approach the critical point for all nearby initial conditions. If the point is unstable, then the solutions will move away from it. The stability or instability of the fixed points can also be determined through linearization. Once we have identified a critical point or a set of critical points for a system during the practical application, our objective is to comprehend the stability of each point and categorize its behaviour. 

\subsection{Linear stability theory}

For a dynamical system with $\dot{x} = f(x)$ and a critical point at $x = x_0$, the first step in linearizing the system is to perform a Taylor expansion of the system,  
\begin{eqnarray}
f(x) = f(x_0) + \frac{f'(x_0)}{1!}(x - x_0) + \frac{f''(x_0)}{2!}(x - x_0)^2  + \cdots + \frac{f^{(n)}(x_0)}{n!}(x - x_0)^n + \cdots\,.
\end{eqnarray}
In this case, we ignore the second-order or the higher-order derivative terms and, using the definition of critical point, we have 
\begin{equation}
   \dot{x} = f^{'}(x_0)(x - x_0)\,.
\end{equation}
In this setup, we can describe the stability of the critical points as (1) it is stable if $f^{'}(x_0) < 0$,
(2) is unstable if $f^{'}(x_0) > 0$,
(3) and the stability is not known that is, the linear stability theory fails if $f^{'}(x_0) = 0$. If the linearization leads to the outcomes mentioned in case (3) above, then it is necessary to carry out a non-linear stability analysis. The system in the previous statement was one-dimensional. For systems with more dimensions, examining the eigenvalues of the Jacobian matrix of the system at critical points would provide insights into their stabilities. 

For the higher dimension dynamical systems, the Jacobian matrix which is also known as the stability matrix, is to be derived and can be stated as
\begin{eqnarray}
    J =
\begin{pmatrix}
\frac{\partial f_1}{\partial x_1} & \cdots \cdots & \frac{\partial f_1}{\partial x_n} \\
 \vdots & \ddots &  \vdots \\
\frac{\partial f_n}{\partial x_1} & \cdots \cdots & \frac{\partial f_n}{\partial x_n}
\end{pmatrix}\,.\label{jacobean}
\end{eqnarray}
The eigenvalues at the critical point of this Jacobian matrix play a crucial role in deciding the stability of the critical point. The Jacobian is a matrix of size $n \times n$ containing $n$ eigenvalues, which may be complex (taking into account repeated eigenvalues). Linear stability theory has limitations that prompt the introduction of hyperbolic points.
\begin{itemize}
    \item  (Hyperbolic point): Consider $x = x_0 \in X \subseteq \mathbb{R}^n $ as a fixed point (critical point) of the system $\dot{x} = f(x).$ Then $x_0$ is called hyperbolic if none of the eigenvalues of the Jacobian matrix $J(x_0)$ have a real part equal to zero. Otherwise, the point is referred to as non-hyperbolic.
\end{itemize}
When classifying fixed points for cosmological dynamical systems, it is important to distinguish three broad cases. Firstly, if all eigenvalues of the Jacobian matrix have positive real parts, trajectories are repelled from the fixed point, and in this case, it is referred to as an unstable point, repeller, or repelling node. Secondly, if all eigenvalues have negative real parts, the point would attract all nearby trajectories, and it is considered stable, sometimes referred to as an attractor or attracting node. Lastly, if at least two eigenvalues have real parts with opposite signs, the corresponding fixed point is called a saddle point, which attracts trajectories in some directions but repels them along others. In 2 and 3 dimensions, it is possible to classify all possible critical points, but in more than 3 dimensions, this task becomes very difficult. For the majority of models in cosmology, the above broad classifications are sufficient for all practical purposes. However, since we have been working in the higher-order teleparallel gravity formalism here, we utilize other techniques to assess the stability of the critical point, such as the central manifold method, which is further discussed in detail. One thing one should notice here is that two theoretical methods are generally used to obtain the stability of the non-hyperbolic critical points, both are discussed in detail as,

\subsection{Lyapunov’s method}
A wide range of literature in applied mathematics explores dynamical systems beyond linear stability theory. While this study primarily focuses on the application of linear stability theory, it is important to note a highly effective technique that can demonstrate both local and global stability, even at non-hyperbolic points. This method, developed by Lyapunov does not depend on linear stability and can be directly applied to the dynamical system, making it very powerful. However, it has a drawback, finding a Lyapunov function is necessary, and there is no systematic method for doing so. We avoid the use of this technique but use the central manifold method, which is presented in detail in the following Sec \ref{centralmanifoldtheory}.

\subsection{Central manifold theory (CMT)}\label{centralmanifoldtheory}
Linear stability theory cannot assess the stability of critical points when the Jacobian has eigenvalues with zero real parts. By utilizing the method of center manifold theory, it is possible to reduce the dimensionality of the dynamical system and subsequently examine the stability of this reduced system. The stability of the critical points of the full system is determined by the stability properties of the reduced system. The following discussion covers the fundamental theorems of center manifold theory. We follow \cite{wiggins2003introduction,carr1981applications,Bahamonde:2017ize} to discuss the basics of central manifold theory. Let us consider the dynamical system of the form,
\begin{equation}
    \dot{z}=F(z)\,,\label{ddsm}
\end{equation}
where $F$ is a regular function of $z \in \mathbb{R}^n$. Suppose the system possesses a fixed point $z_0$. Using the linear stability method, we can linearise the system near this point by employing Eq. \eqref{jacobean}. By defining $z_{\ast} = z - z_0$, we can express Eq. \eqref{ddsm} at the linear level in the following manner,
\begin{equation}
    \dot{z}_{\ast}= J z_{\ast}\,.
\end{equation}
As $J$ is an $n \times n$ matrix, it will also possess $n$ eigenvalues (taking into account repeated eigenvalues). This fact leads to the establishment of the subsequent three spaces. The space $\mathbb{R}^n$ can be a direct sum into three subspaces denoted as $E^s$, $E^u$, and $E^c$, where the superscripts denote (s) stable, (u) unstable, and (c) center, respectively. The `stable' space $E^s$ comprises the eigenvectors of $J$ associated with eigenvalues having a negative real part. The `unstable' space $E^u$ encompasses the eigenvectors of $J$ associated with eigenvalues having a positive real part, and $E^c$ contains the eigenvectors of $J$ associated with eigenvalues having a zero real part. The behaviour of the phase space trajectories in $E^s$ and $E^u$ can be determined using linear stability theory, while the center manifold theory enables us to decide the dynamics of trajectories in $E^c$. If the unstable space is non-empty, i.e., if $J$ has at least one eigenvalue with a positive real part, then the corresponding critical point cannot be stable, regardless of whether it is hyperbolic or non-hyperbolic. If on the other hand, the unstable space of a non-hyperbolic critical point is empty, i.e., if $J$ has no eigenvalues with a positive real part, then stability can be determined by employing center manifold techniques. In this latter scenario, there always exists a coordinate transformation that permits us to express the dynamical system Eq. \eqref{ddsm} in the following form:
\begin{eqnarray}
    \dot{x}&=& Ax + f(x, y), \label{dsa1}\\
    \dot{y}&=& By + g(x, y), \label{dsa2}
\end{eqnarray}
here $(x, y) \in \mathbb{R}^c\times \mathbb{R}^s$
where $c$ is the dimension of $E^{c}$, and s is the dimension of $E^{s}$. The two functions $f$ and $g$ will satisfy,
\begin{eqnarray}
f(0, 0) &=& 0, \quad \grad f(0, 0) = 0\,,\nonumber\\ 
g(0, 0) &=& 0, \quad \grad g(0, 0) = 0\,,
\end{eqnarray}
here in Eq. \eqref{dsa1}, $A$ is a matrix having dimensions $c\times c$ with the eigenvalues with zero real parts, while in Eq. \eqref{dsa2} $B$ is a matrix having dimensions 
$s \times s$ whose eigenvalues have negative real parts. 
\begin{itemize}
    \item Central manifold:
    A space with geometry becomes a center manifold for Eq. \eqref{dsa1} and Eq. \eqref{dsa2} if it can be locally presented as,
    \begin{equation}
W^{c}(0) = \{(x, y) \in \mathbb{R}^c \times \mathbb{R}^s \mid y = h(x), \, |x| < \delta, \, h(0) = 0, \, \grad h(0) = 0\} \,.\label{cmtdef}
\end{equation}
\end{itemize}
For $\delta$ sufficiently small and for a function $h(x)$ on $\mathbb{R}^s$ that is sufficiently regular, the conditions $h(0) = 0$ and $\grad h(0) = 0$ from the definition imply that the space $W^c (0)$ is tangent to the eigenspace $E^c$ at the critical point $(x, y) = (0, 0)$. The following section presents three theorems fundamental to center manifold theory, which will be useful in applications to cosmology. The first theorem demonstrates the existence of the center manifold, while the second deals with stability. The final theorem explains how to construct the actual center manifold locally and that this local construction is sufficient to investigate stability. Interested readers are directed to \cite{carr1981applications} for proofs of the following theorems.
\begin{itemize}
    \item Existence theorem:  There exists a center manifold for Eq. \eqref{dsa1} and Eq. \eqref{dsa2}. The system dynamics for these equations on the center manifold can be represented as
   \begin{equation}
       \dot{u} = Au + f(u, h(u)) \,,\label{dsa22}
   \end{equation}
    where $u \in \mathbb{R}^c$ is sufficiently small.
    \item Stability theorem: If the zero solution of Eq. \eqref{dsa22} is stable (asymptotically stable or unstable), then the zero solution of Eq. \eqref{dsa1} and Eq. \eqref{dsa2} is also stable (asymptotically stable or unstable). Moreover, if $(x(t), y(t))$ is a solution of Eq. \eqref{dsa1} and Eq. \eqref{dsa2} with $(x(0), y(0))$ sufficiently small, there is a solution u(t) of Eq. \eqref{dsa22} such that,
     \begin{eqnarray}
       x(t) &=& u(t) + \mathrm{O}( e^{- \gamma t} )\,,\nonumber\\
       y(t) &=& h(u(t)) + \mathrm{O}( e^{- \gamma t} )\,,
    \end{eqnarray}
    as $t\rightarrow \infty$, where $\gamma > 0$ is a constant.
\end{itemize}
The validity of these theorems depends on the understanding of the function $h(x)$, which must be determined. We will now establish a differential equation for the function $h(x).$ According to definition Eq. \eqref{cmtdef}, it can be stated that $y = h(x)$. By differentiating this with respect to time and subsequently applying the chain rule, we obtain the following:
\begin{equation}
    \dot{y} = \grad h(x) \dot{x} \,.
\end{equation}
Given that $W^c (0)$ relies on the dynamics produced by the system Eq. \eqref{dsa1} and Eq. \eqref{dsa2}, we can replace $\dot{x}$ with the right-hand side of Eq. \eqref{dsa1} and $\dot{y}$ with the right-hand side of Eq. \eqref{dsa2}. As a result, we obtain the following:
\begin{equation}
Bh(x) + g(x, h(x)) = \grad h(x) \cdot [Ax + f(x, h(x))]\,.
\end{equation}
We also utilized the information that $y = h(x)$. The latter equation can be rearranged to form the quasilinear partial differential equation, as depicted below.
\begin{equation}
\mathcal{N}(h(x)) = \grad h(x) [Ax + f(x, h(x))] - Bh(x) - g(x, h(x)) = 0\,.\label{cmt1}
\end{equation}
Here, $h(x)$ must satisfy this condition to be considered as the center manifold. Typically, it is challenging, and sometimes impossible, to find a solution to this equation, even for relatively uncomplicated systems. The third and final theorem clarifies that it is unnecessary to have knowledge of the entire function.
\begin{itemize}
    \item Approximation theorem: Consider a mapping $\phi : \mathbb{R}^c \rightarrow \mathbb{R}^s$ where $\phi(0) = \grad \phi(0) = 0$ and $\mathcal{N} (\phi(x)) = O(|x|^q )$ as $x \rightarrow 0$ for some $q > 1$. Then
\begin{equation}
|h(x) - \phi(x)| = O(|x|^q) \quad \text{as} \quad x \to 0\,.
\end{equation}
\end{itemize}
The key idea here is that an estimated understanding of the center manifold provides the same stability information as the precise solution of Eq. \eqref{cmt1}. Additionally, an approximation for the center manifold can frequently be obtained simply by assuming a series expansion of $h$. The coefficients in this series are then determined by satisfying Eq. \eqref{cmt1} for each order. We have used this technique to check the stability of the non-hyperbolic critical point.

\section{Conclusion}\label{conclusionch1}
In this section, we present the required prerequisites for our study. This chapter briefly describes the history of cosmic acceleration in Sec \ref{cosmichistroy}, starting from the inflationary era and then the radiation, matter-dominated era, and, at last, the current epochs of cosmic acceleration. The fundamental equations of cosmology obtained using the flat FLRW space-time and are presented in Sec \ref{FLRWUniverse}
like the Friedmann equation, the conservation equation, and the dynamics of the scale factor are discussed in this section. The brief overview of the $\Lambda$CDM model, along with the description of its key point, is presented in Sec \ref{lambdaoverview}. The importance and need of both of the fundamental theories GR and TEGR are discussed in Sec \ref{NeedofTEGR}. In the next section Sec \ref{TEGRMODIFICATION}, it has been explained the basic formalism of the modified teleparallel gravity models that are studied in the next few chapters of this thesis. Moreover, at last the detailed overview of the dynamical system analysis technique, which is discussed in different modified teleparallel gravity models in this thesis, is presented in Sec \ref{dmsoverview}. In this overview, the major highlighted points are the formation method of the autonomous dynamical system, some important definitions like the critical point and hyperbolic critical points, and some methods like Lypunov stability, central manifold method of checking the stability of this critical point are mentioned in detail. 

In the following chapters, we will analyse the role of the dynamical system analysis to describe different phases of the evolution of the Universe. The different teleparallel modifications, which include the teleparallel scalars like teleparallel Gauss-Bonnet scalar $T_G$, teleparallel boundary term $B$, and canonical scalar field $\phi$, have been analysed in detail. Moreover, one of the most general teleparallel formalisms, teleparallel analog to the Horndeski gravity, has been considered and studied in its role in demonstrating the dynamics of the evolution of the Universe.

% Chapter 2

\chapter{Teleparallel scalar-tensor gravity through cosmological dynamical systems} % Main chapter title

\label{Chapter2} % For referencing the chapter elsewhere, use \ref{Chapter1} 

\lhead{Chapter 2. \emph{Teleparallel scalar-tensor gravity through cosmological dynamical systems}} % This is for the header on each page - perhaps a shortened title

\vspace{10 cm}
*The work in this chapter is covered by the following publication:\\

\textbf{S.A. Kadam}, B. Mishra, and Jackson Levi-Said, ``Teleparallel scalar-tensor gravity through cosmological dynamical systems", \textit{European Physical Journal C}, \textbf{82}, 680 (2022).

%----------------------------------------------------------------------------------------
%\section{Summary of Results}
\clearpage
\section{Introduction}\label{Introchapt_1}  
This chapter is dedicated to construct and to study the autonomous dynamical system in one of the most general formalisms, representing a teleparallel analog of Horndeski gravity \cite{Bahamonde:2019shr,Bahamonde:2019ipm,Bahamonde:2020cfv,Dialektopoulos:2021ryi,Bahamonde:2021dqn,Bernardo:2021izq,Bernardo:2021bsg}. This work considers two models and describes their cases for $\alpha=1, 2$. The Friedmann equations are obtained for the general form of the models and the particular cases. The critical points are discussed in relation to the values of deceleration and EoS parameters. Further, the eigenvalues are obtained to examine the stability of the critical points. 

\section{Teleparallel scalar-tensor flat FLRW cosmology} \label{sec:backgroundexpressions}
In addition to the fundamental formalism presented in Sec \ref{teleparllelhorndesky}, we consider a flat isotropic and homogeneous background cosmology through the FLRW metric as presented in Eq. \eqref{FLATFLRW} \cite{misner1973gravitation} and the tetrad is as presented in Eq. \eqref{FLRWTETRAD} which is consistent with the Weitzenb\"{o}ck gauge described in Sec \ref{TEGRMODIFICATION}. We take the standard definition of the Hubble parameter $H= \frac{\dot{a}}{a}$, where dots refer to derivatives with respect to cosmic time. We also consider the EoS for matter $\omega_{m}=\frac{p_{\rm m}}{\rho_{\rm m}} = 0$ and radiation $\omega_{r}=\frac{p_{r}}{\rho_{\rm r}} = 1/3$, which will both contribute to our representation of cosmology. In this work, we consider the class of models in which
\begin{align}
    G_2 &= X - V(\phi)\,,\\
    G_3 &= 0 = G_5\,,\\
    G_4 &= 1/2\kappa^2\,,
\end{align}
where we take a generalization of a canonical scalar field together with a TEGR term. This then lets us probe different forms of the $G_{\rm Tele}$ term in action Eq. (\ref{action}). As discussed, we aim to probe the nature of power-law couplings with the kinetic term. To that end, we consider two models that embody nonvanishing terms for an FLRW background cosmology, which are
\begin{align}
    G_{{\rm Tele}_1} &= X^{\alpha} T\,,\label{model1}\\
    G_{{\rm Tele}_2} &= X^{\alpha} I_2\,,
\end{align}
where the other terms effectively do not contribute to the Friedmann equations \cite{Bahamonde:2019shr}, and where
\begin{align}
    T &= 6H^2\,,\label{torsionscalar__}\\
    I_2 &= 3H\dot{\phi}\,.\label{eq:I_2_scalar}
\end{align}
Thus, we can write the effective Friedmann equations as
\begin{align}
    \dfrac{3}{\kappa^{2}}H^{2} &= \rho_{\rm m} + \rho_{\rm r} + X + V + 6H\dot{\phi} G_{{\rm Tele},I_2} + 12H^2 G_{{\rm Tele},T} + 2X G_{{\rm Tele},X} - G_{{\rm Tele}}\,,\label{eq:30}\\
    -\dfrac{2}{\kappa^{2}}\dot{H} &= \rho_{\rm m} + \frac{4}{3}\rho_{\rm r} + 2X + 3H\dot{\phi}G_{{\rm Tele},I_2} + 2XG_{{\rm Tele},X} - \frac{d}{dt}\left(4HG_{{\rm Tele},T} + \dot{\phi} G_{{\rm Tele},I_2}\right) \,,\label{eq:31}
\end{align}
and where the scalar field equation is given by
\begin{equation}
    \frac{1}{a^3}\frac{d}{dt}\left[a^3 \dot{\phi} \left( 1+ G_{{\rm Tele},X}\right)\right] = -V'(\phi) - 9H^2 G_{{\rm Tele},I_2} + G_{{\rm Tele},\phi} - 3\frac{d}{dt}\left(HG_{{\rm Tele},I_2}\right)\,.\label{eq:kg_eq}
\end{equation}
This effective fluid satisfies the energy-conservation equation in Eq. \eqref{inConservationEq}.

\section{ Kinetic term coupled with torsion scalar}\label{generalmodelI}

The action for the model presented in Eq. \eqref{model1} is given by a coupling between a power-law-like term and the torsion scalar represented by
\begin{equation}\label{eq:action_model_1}
    S = \int d^4x e [X-V(\phi)-\frac{T}{2\kappa^{2}}+X^{\alpha}T]+S_{m}+S{r}\,,
\end{equation}
where $V(\phi)$ is the scalar potential, $S_m$ represents the action for matter and $S_r$ describes the action for the radiation component. We shall perform the dynamical system analysis for the general case $\alpha$, followed by two examples with $\alpha=1$, $\alpha=2$. Taking background FLRW cosmology and the action above, we can obtain the Friedmann equations in Eqs. \eqref{eq:44} to \eqref{eq:45}, whereas the Klein-Gordon equation can be obtained in Eq. (\ref{eq:46}), altogether giving
\begin{align}
    \frac{T}{2\kappa^{2}}-V(\phi)-X-X^{\alpha}T-2\alpha X^{\alpha}T =& \rho_{\rm m}+\rho_{\rm r}\,,\label{eq:44}\\
    -V(\phi)+\frac{T}{2\kappa^{2}}+ X-X^{\alpha}T+\frac{2\dot{H}}{\kappa^{2}}-4X^{\alpha}\dot{H}-8\alpha X^{\alpha}H\frac{\ddot{\phi}}{\dot{\phi}} =& -p_{r}\,,\label{eq:45}\\
    V^{'}(\phi)+3H\dot{\phi}+\frac{6X^{\alpha}T\alpha H}{\dot{\phi}}+\frac{4\alpha X^{\alpha}T \dot{H}}{\dot{\phi} H}+\ddot{\phi}[1-\alpha X^{\alpha-1}T+\alpha^{2}X^{\alpha-2}T] =& 0\,.\label{eq:46}
\end{align}
Now using the background expressions defined in Sec \ref{sec:backgroundexpressions}, we can obtain the expression for energy density and pressure of effective DE as
\begin{align}
    \rho_{\rm DE} &= X^{\alpha}T+2\alpha X^{\alpha}T+V(\phi)+X\,,\label{eq:47}\\
    p_{\rm DE} &= -V(\phi)+X-X^{\alpha}T-4X^{\alpha}\dot{H}-\frac{8\alpha X^{\alpha} H \ddot{\phi}} {\dot{\phi}}\,.\label{eq:48}
\end{align}
To study the phases of cosmic evolution, the autonomous dynamical system for the above set of cosmological expressions can be defined using the following dynamical variables.
\begin{align}\label{eq:49}
    x=\dfrac{\kappa\dot{\phi}}{\sqrt{6}H}\,,\quad y=\frac{\kappa\sqrt{V}}{\sqrt{3}H}\,,\quad u=2 X^\alpha \kappa^2\,,\quad \rho= \frac{\kappa\sqrt{\rho_{\rm r}}}{\sqrt{3}H}\,,\quad \lambda=\frac{-V^{'}(\phi)}{\kappa V(\phi)}\,,\quad \Gamma=\dfrac{V(\phi)V^{''}(\phi)}{V^{'}(\phi)^{2}}\,,
\end{align}
where the constraint equation for the dynamical variables can be obtained as,
\begin{align}\label{eq:50}
    x^{2}+y^{2}+(1+2\alpha)u+\Omega_{m}+\rho^{2}=1\,.
\end{align}
Now the autonomous dynamical system can be defined by differentiating the dynamical variables with respect to $N=log(a)$ as,
\begin{align}
    \dfrac{dx}{dN} &= \frac{x \left(-x^2 \left(\rho ^2+3 (\alpha  (2 \alpha +5)+1) u-3 y^2-3\right)+\sqrt{6} \lambda  x y^2 (2 \alpha  u+u-1)-3 x^4\right)}{2 (u-1) x^2-2 \alpha  u (2 \alpha  (u+1)+u-1)}\nonumber\\
    & - \frac{\alpha  u x \left(2 \alpha  \left(\rho ^2+3\right)+\rho ^2+(6 \alpha +3) u-3 (2 \alpha +1) y^2-3\right)}{2 (u-1) x^2-2 \alpha  u (2 \alpha  (u+1)+u-1)}\,,\label{eq:dx_dN_model_1}\\
    \dfrac{dy}{dN} &= \frac{-y \left(x^2 \left(\rho ^2+\left(6 \alpha ^2+9 \alpha -3\right) u-3 y^2+3\right)-2 \sqrt{6} \alpha  \lambda  u x y^2+3 x^4\right)}{2 (u-1) x^2-2 \alpha  u (2 \alpha  (u+1)+u-1)} \nonumber\\
    & + \frac{\alpha  u (-y) \left((6 \alpha +3) u-(2 \alpha -1) \left(-\rho ^2+3 y^2-3\right)\right)}{2 (u-1) x^2-2 \alpha  u (2 \alpha  (u+1)+u-1)}-y\sqrt{\frac{3}{2}}  \lambda  x \,,\\
    \dfrac{du}{dN} &= \frac{\alpha  u \left(2 \alpha  u \left(\rho ^2+3 x^2-3 y^2\right)+(u-1) x \left(6 x-\sqrt{6} \lambda y^2\right)\right)}{\alpha u (2 \alpha (u+1)+u-1)-(u-1) x^2}\,,\\
    \frac{d\rho}{dN} &= \frac{\rho \left(-x^2 \left(\rho ^2+6 \alpha ^2 u+9 \alpha  u+u-3 y^2-1\right)+2 \sqrt{6} \alpha \lambda  u x y^2-3 x^4\right)}{2 (u-1) x^2-2 \alpha  u (2 \alpha  (u+1)+u-1)} \nonumber\\
    & + \frac{\alpha \rho u \left(2 \alpha  u+u+(2 \alpha -1) \left(-\rho ^2+3 y^2+1\right)\right)}{2 (u-1) x^2-2 \alpha  u (2 \alpha  (u+1)+u-1)}\,,\\
    \frac{d\lambda}{dN} &= \sqrt{6}(\Gamma-1)\lambda^{2}x\,.\label{eq:dl_dN_model_1}
\end{align}
Unless the parameter $\Gamma$ is known, the dynamical systems presented in this work are not autonomous systems. Now onwards, we will focus on the exponential potential $V(\phi) =V_{0}e^{-\tau \kappa \phi}$  with $\tau$ is a dimensionless constant. This particular form of potential function leads to $\Gamma = 1$ and can show the accelerating Universe. We obtain the critical points (or fixed points) for autonomous dynamical system presented in Eqs. \eqref{eq:dx_dN_model_1} to \eqref{eq:dl_dN_model_1} by imposing conditions $\frac{dx}{dN}=0$, $\frac{dy}{dN}=0$, $\frac{du}{dN}=0$, $\frac{d\rho}{dN}=0$. The critical points are titled with capital letters and presented in corresponding tables. The tables also present the value of the deceleration parameter and the value for the total EoS $(\omega_{tot})$ to study the cosmological implications. From the Table \ref{CHITABLEI} observation, it can be concluded that parameter $\alpha$ contributes in the co-ordinates of critical points $D$ and $E$ and represents the de Sitter solution for the dynamical system presented in Eqs. \eqref{eq:dx_dN_model_1} to \eqref{eq:dl_dN_model_1}. While analyzing the critical points for the case general $\alpha$, there is a chance to get more critical points than for a particular value of $\alpha$. The critical points $L$, $M$, $F$, $G$ and $A$ represent same cosmological implication. These critical points show deceleration parameter value $q=\frac{1}{2}$ and $\omega_{tot}=0$, hence explaining the cold DM-dominated era. Similarly, the critical points $B$, $C$, and $N$ represent the same phase of evolution with the value of $\omega_{tot} =1$; hence, it cannot describe the current accelerated phase of evolution and behave as stiff matter. The critical points $J$ and $K$ are defined at $\lambda=2$ and show value of $\omega_{tot}=\frac{1}{3}$ hence  describe the radiation dominated era. Since the critical points $H$ and $I$ represent deceleration parameter value, $q=-1+\frac{\lambda^2}{2}$, these critical points can describe the current acceleration of the Universe for any real value of $\lambda$ and are compatible with the current observational data.
\begin{table}[H]
 % title of Table
\centering % used for centering table
\renewcommand{\arraystretch}{1.2} % Adjust row height
\scalebox{0.9}{
\begin{tabular}{|c|c|c|c|c|c|c|c|} % centered columns (7 columns)
\hline\hline %inserts double horizontal lines
\parbox[c]{2.5cm}{\centering Critical points} & 
\parbox[c]{1.2cm}{\centering $x_{c}$} & 
\parbox[c]{1.2cm}{\centering $y_{c}$} & 
\parbox[c]{1.2cm}{\centering $u_{c}$} & 
\parbox[c]{1.2cm}{\centering $\rho_{c}$} &
\parbox[c][1.2cm]{2.5cm}{\centering Existence condition} & 
\parbox[c]{1.2cm}{\centering $q$} & 
\parbox[c]{1.2cm}{\centering $\omega_{tot}$} \\ [0.5ex] % inserts table heading
\hline\hline % inserts single horizontal line
$A$ & 
0 & $0$ & $\tau$ & $0$ & \parbox[c][2.3cm]{2.3cm}{\centering $2 \alpha ^2 \tau^2$\\ $+\alpha  \tau^2+2 \alpha ^2 \tau-\alpha  \tau\neq 0 $}&$\frac{1}{2}$ &$0$\\
\hline
$B$ & $1$ & $0$ & $0$ & $0$ & -& $2$ & $1$\\
\hline
$C$ & $-1$ & $0$ & $0$ & $0$ & -&$2$ & $1$\\
\hline
$D$  & $\zeta$ & 
$\frac{\sqrt{(\alpha +1) \zeta ^2+\alpha }}{\sqrt{\alpha }}$ & 
$-\frac{\zeta ^2}{\alpha }$ & $0$ & \parbox[c][2.2cm]{2.2cm}{\centering $(\alpha +1)\zeta^2 \neq (\alpha -1)\alpha$, $\lambda=0$ }& $-1$ & $-1$\\
\hline
$E$& $\zeta$ & 
-$\frac{\sqrt{(\alpha +1) \zeta ^2+\alpha }}{\sqrt{\alpha }}$ & 
$-\frac{\zeta ^2}{\alpha }$ & $0$ &\parbox[c][2.2cm]{2.2cm}{\centering $(\alpha +1)\zeta^2 \neq (\alpha -1)\alpha$ , $\lambda=0$ }&$-1$ & $-1$\\
\hline
$F$ & $\frac{\sqrt{\frac{3}{2}}}{\lambda }$ & $\sqrt{\frac{3}{2}} \sqrt{\frac{1}{\lambda ^2}}$ & $0$ & $0$ &-&$\frac{1}{2}$ & $0$\\
\hline
$G$ & $\frac{\sqrt{\frac{3}{2}}}{\lambda }$ & $-\sqrt{\frac{3}{2}} \sqrt{\frac{1}{\lambda ^2}}$ & $0$ & $0$ &-&$\frac{1}{2}$ & $0$\\
\hline
$H$ & $\frac{\lambda }{\sqrt{6}}$ & $\sqrt{1-\frac{\lambda ^2}{6}}$ & $0$ & $0$&-&$\frac{1}{2} (\lambda ^2-2)$ & $-1+\frac{\lambda^2}{3}$\\
\hline
$I$ & $\frac{\lambda }{\sqrt{6}}$ & $-\sqrt{1-\frac{\lambda ^2}{6}}$ & $0$ & $0$&-&$\frac{1}{2} (\lambda ^2-2)$ & $-1+\frac{\lambda^2}{3}$\\
\hline
$J$ & $\sqrt{\frac{2}{3}}$ & $\sqrt{\frac{1}{3}}$ & $0$ & $0$ & $\lambda=2$ & $1$ & $\frac{1}{3}$\\
\hline
$K$ & $\sqrt{\frac{2}{3}}$ & $-\sqrt{\frac{1}{3}}$ & $0$ & $0$ & $\lambda=2$ & $1$ & $\frac{1}{3}$\\
\hline
$L$ & $\frac{\sqrt{\frac{3}{2}}}{\lambda }$ & $\frac{\sqrt{\frac{3}{2}}}{\lambda }$ & $\chi$ & $0$ & $\chi -1\neq 0$ &$\frac{1}{2}$ & $0$\\
\hline
$M$ & $\frac{\sqrt{\frac{3}{2}}}{\lambda }$ & $-\frac{\sqrt{\frac{3}{2}}}{\lambda }$ & $\chi$ & $0$ & $\chi -1\neq 0$ &$\frac{1}{2}$ & $0$\\
\hline
$N$ & $\delta$ & $0$ & $1-\delta^2$ & $0$ & $ \delta\neq 0,\alpha =0$ &$2$ & $1$\\
[1ex] % [1ex] adds vertical space
\hline %inserts single line
\end{tabular}}
\caption{Critical points for Model \ref{generalmodelI}, for general $\alpha$. }
\label{CHITABLEI}
\end{table}
The stability of critical points can be studied by obtaining eigenvalues of linear perturbation matrix at critical points. The stability techniques are well discussed and presented in detail in Sec \ref{dmsoverview}. The eigenvalues and stability conditions for the dynamical system in Eqs. \eqref{eq:dx_dN_model_1} to \eqref{eq:dl_dN_model_1} are presented in Table \ref{CHITABLE-II}. The existence of positive and negative eigenvalues for the permutation matrix at the critical points $A$, $B$, $C$, $J$, $K$, and $N$ describe saddle point behaviour at these critical points; hence these critical points are unstable. However, the value of the deceleration parameter at these critical points clarifies that these critical points can not explain the accelerated expansion phase of the Universe. Critical points $F$ and $G$ ensure stability in the range of parameter $\alpha >0$ and $\left(-2 \sqrt{\frac{6}{7}}\leq \lambda <-\sqrt{3}\lor \sqrt{3}<\lambda \leq 2 \sqrt{\frac{6}{7}}\right)$ these critical points explain the standard matter-dominated era. The critical points $D$ and $E$ show stable behaviour and explain the de Sitter solution.

Note: During this study, for non-hyperbolic critical points, we have a dimension of the set of eigenvalues one which is equal to the number of vanishing eigenvalues, the set of eigenvalues is normally hyperbolic, and the critical point corresponding to the set is stable, but they can not be a global attractor \cite{coley2003dynamical}.
\begin{table}[H]
\small\addtolength{\tabcolsep}{-8pt}
 % title of Table
\centering % used for centering table
\renewcommand{\arraystretch}{1.2} % Adjust row height
\scalebox{0.9}{
\begin{tabular}{|c|c|c|} % centered columns (3 columns)
\hline %inserts double horizontal lines
\hline

\parbox[c][1.2cm]{2.8cm}{\centering Critical points } & 
\parbox[c]{6.5cm}{\centering Eigenvalues} & 
\parbox[c]{4.0cm}{\centering Stability} \\ [0.5ex] % inserts table heading

\hline % inserts single horizontal line
\hline

$A$ & $ \Bigl\{\frac{3}{2},\frac{3}{2},-\frac{1}{2},0 \Bigl\}$ & Unstable\\
\hline

$B$ & $ \Bigl\{3,1,-6 \alpha ,\frac{1}{2} \left(6-\sqrt{6} \lambda \right) \Bigl\}$ & Unstable\\
\hline

$C$ & $ \Bigl\{3,1,-6 \alpha ,\frac{1}{2} \left(6+\sqrt{6} \lambda \right) \Bigl\}$ & Unstable\\
\hline

$D$ & $\{0,-3,-3,-2\}$ & Stable \\
\hline

$E$ &$ \Bigl\{0,-3,-3,-2 \Bigl\}$ & Stable \\
\hline

$F$ & \parbox[c][2cm]{6.5cm}{\centering $ \Bigl\{-\frac{1}{2},-3 \alpha ,\frac{3 \left(-\lambda ^2-\sqrt{24 \lambda ^2-7 \lambda ^4}\right)}{4 \lambda ^2}, $\\$ \frac{3 \left(\sqrt{24 \lambda ^2-7 \lambda ^4}-\lambda ^2\right)}{4 \lambda ^2} \Bigl\}$} & \parbox[c]{4.0cm}{\centering Stable for  $\alpha >0$\\ $\land  \Bigl\{-2 \sqrt{\frac{6}{7}}\leq \lambda <-\sqrt{3}\lor \sqrt{3}<\lambda \leq 2 \sqrt{\frac{6}{7}} \Bigl\}$}\\
\hline

$G$ & \parbox[c][2cm]{6.5cm}{\centering $ \Bigl\{-\frac{1}{2},-3 \alpha ,\frac{3 \left(-\lambda ^2-\sqrt{24 \lambda ^2-7 \lambda ^4}\right)}{4 \lambda ^2},$\\$\frac{3 \left(\sqrt{24 \lambda ^2-7 \lambda ^4}-\lambda ^2\right)}{4 \lambda ^2} \Bigl\}$} & \parbox[c]{4.0cm}{\centering Stable for  $\alpha >0$\\ $\land  \Bigl\{-2 \sqrt{\frac{6}{7}}\leq \lambda <-\sqrt{3} $\\$\lor \sqrt{3}<\lambda \leq 2 \sqrt{\frac{6}{7}} \Bigl\}$}\\
\hline

$H$ & \parbox[c][1.6cm]{6.5cm}{\centering $ \Bigl\{-\alpha  \lambda ^2,\frac{1}{2} \left(\lambda ^2-6\right),\frac{1}{2} \left(\lambda ^2-4\right),\lambda ^2-3 \Bigl\}$} &\parbox[c]{4.0cm}{\centering  Stable for $\alpha >0 $\\$ \land  \Bigl\{-\sqrt{3}<\lambda <0\lor 0<\lambda <\sqrt{3} \Bigl\}$}\\
\hline

$I$ & \parbox[c][1.6cm]{6.5cm}{\centering $ \Bigl\{-\alpha  \lambda ^2,\frac{1}{2} \left(\lambda ^2-6\right),\frac{1}{2} \left(\lambda ^2-4\right),\lambda ^2-3 \Bigl\}$} & \parbox[c]{4.0cm}{\centering  Stable for $\alpha >0 $\\$\land  \Bigl\{-\sqrt{3}<\lambda <0\lor 0<\lambda <\sqrt{3} \Bigl\}$}\\
\hline

$J$ & $ \Bigl\{-1,1,0,-4 \alpha  \Bigl\}$ & Unstable \\
\hline

$K$ & $ \Bigl\{-1,1,0,-4 \alpha  \Bigl\}$ & Unstable \\
\hline

$L$ & \parbox[c][2.1cm]{6.5cm}{\centering $ \Bigl\{0,-\frac{1}{2},\frac{3}{4} \left(\frac{\sqrt{-\lambda ^2 (\chi -1) \left(7 \lambda ^2 (\chi -1)+24\right)}}{\lambda ^2 (\chi -1)}-1\right), $\\$-\frac{3 \sqrt{-\lambda ^2 (\chi -1) \left(7 \lambda ^2 (\chi -1)+24\right)}}{4 \lambda ^2 (\chi -1)}-\frac{3}{4}  \Bigl\}$} & \parbox[c][2.1cm]{6.5cm}{\centering Stable for $\lambda \in \mathbb{R} $\\$  \Bigl\{\land \lambda \neq 0\land \frac{7 \lambda ^2-24}{7 \lambda ^2}\leq \chi <\frac{\lambda ^2-3}{\lambda ^2} \Bigl\}$} \\
\hline

$M$ & \parbox[c][2.1cm]{6.5cm}{\centering $ \Bigl\{0,-\frac{1}{2},\frac{3}{4} \left(\frac{\sqrt{-\lambda ^2 (\chi -1) \left(7 \lambda ^2 (\chi -1)+24\right)}}{\lambda ^2 (\chi -1)}-1\right), $\\$-\frac{3 \sqrt{-\lambda ^2 (\chi -1) \left(7 \lambda ^2 (\chi -1)+24\right)}}{4 \lambda ^2 (\chi -1)}-\frac{3}{4}  \Bigl\}$} & \parbox[c][2.1cm]{6.5cm}{\centering  Stable for $\lambda \in \mathbb{R}$\\$ \Bigl\{\land \lambda \neq 0  \land \frac{7 \lambda ^2-24}{7 \lambda ^2}\leq \chi <\frac{\lambda ^2-3}{\lambda ^2} \Bigl\}$}\\
\hline

$N$ & $ \Bigl\{0,1,3,\frac{1}{2} \left(6-\sqrt{6} \delta  \lambda \right) \Bigl\}$ &  Unstable\\
[1ex] % [1ex] adds vertical space
\hline %inserts single line
\end{tabular}}
\caption{Eigenvalues and stability of critical points for model \ref{generalmodelI}.}
 % is used to refer this table in the text
\label{CHITABLE-II}
\end{table}
For any real value of $\lambda$, the critical points $H$ and $I$ represent the DE-dominated era. This point shows stable behaviour for the parameters that obey the value $\alpha >0$ and $\Big(-\sqrt{3}<\lambda <0\lor 0<\lambda <\sqrt{3}\Big)$. During the study of stability conditions, for critical points $F$, $G$, $H$ and $I$, we get condition on $\alpha$, $\alpha>0$. This implies that the stability of critical points can be assessed for a positive value of $\alpha$ for model \ref{generalmodelI}. All the eigenvalues at critical points $L$ and $M$ are less than zero for $\lambda \in \mathbb{R}\land \lambda \neq 0\land \frac{7 \lambda ^2-24}{7 \lambda ^2}\leq \chi <\frac{\lambda ^2-3}{\lambda ^2}$, hence show stable behaviour at these parametric values.

\begin{figure}[H]
    \centering
    \includegraphics[width=60mm]{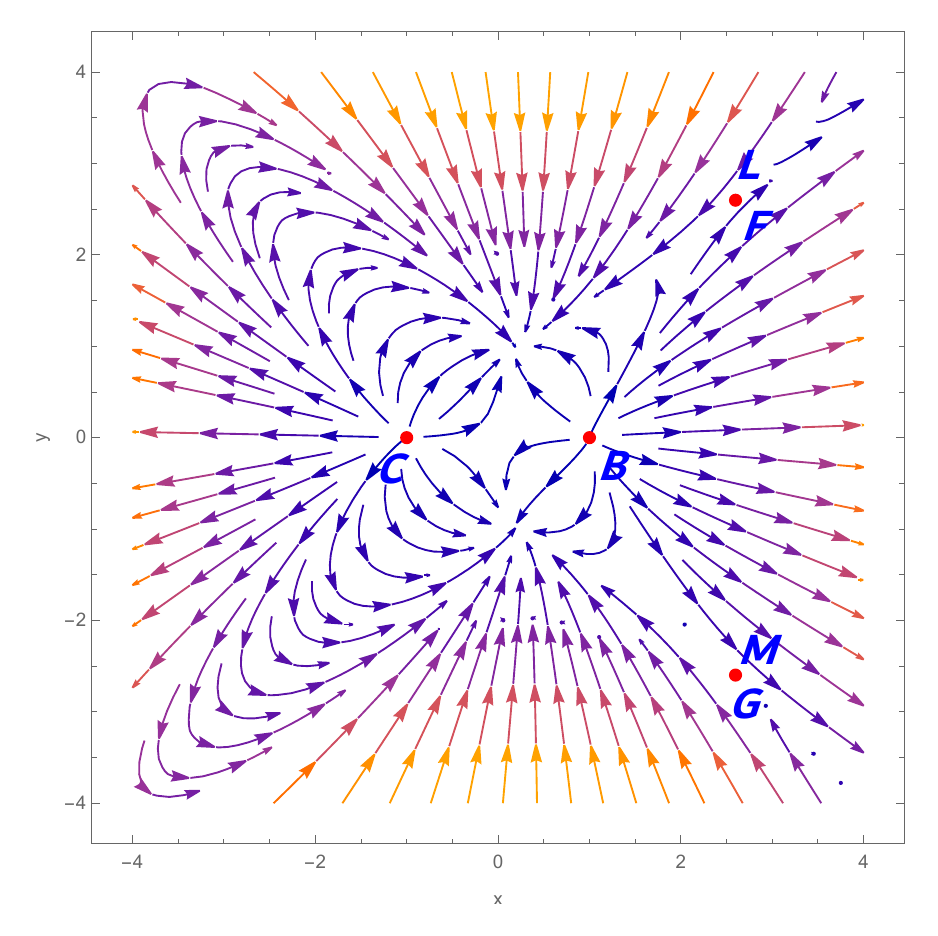}
    \includegraphics[width=60mm]{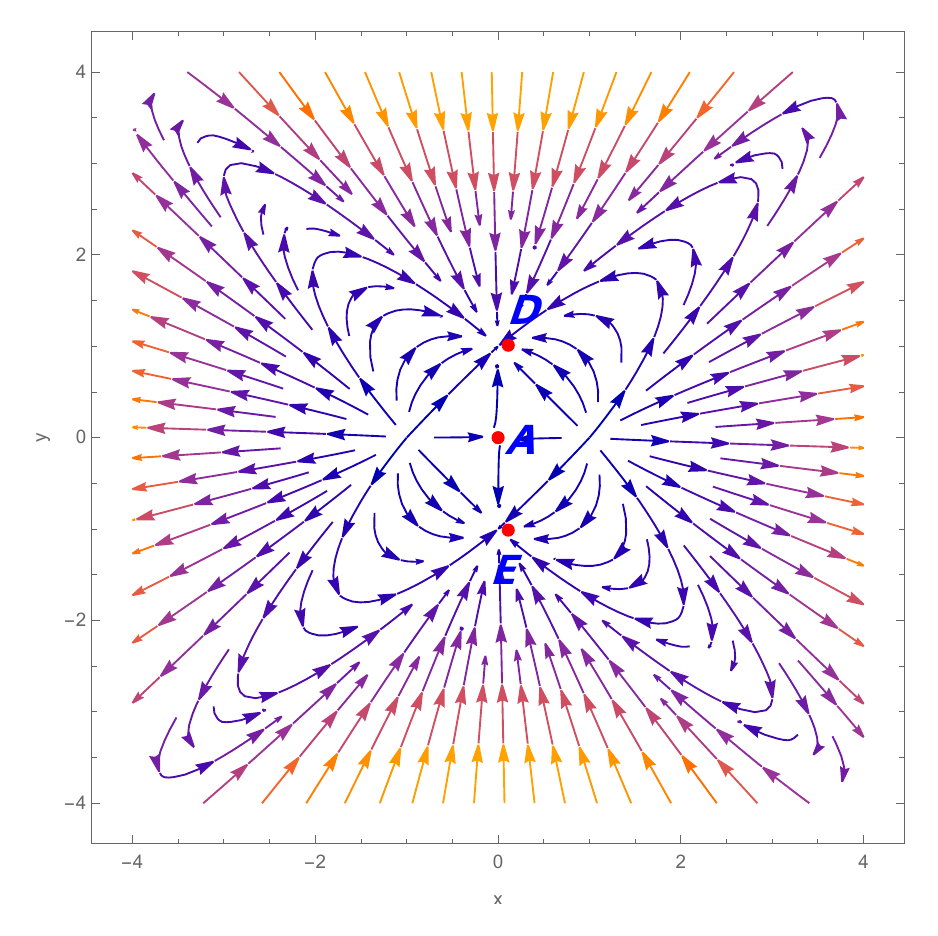}
    \includegraphics[width=60mm]{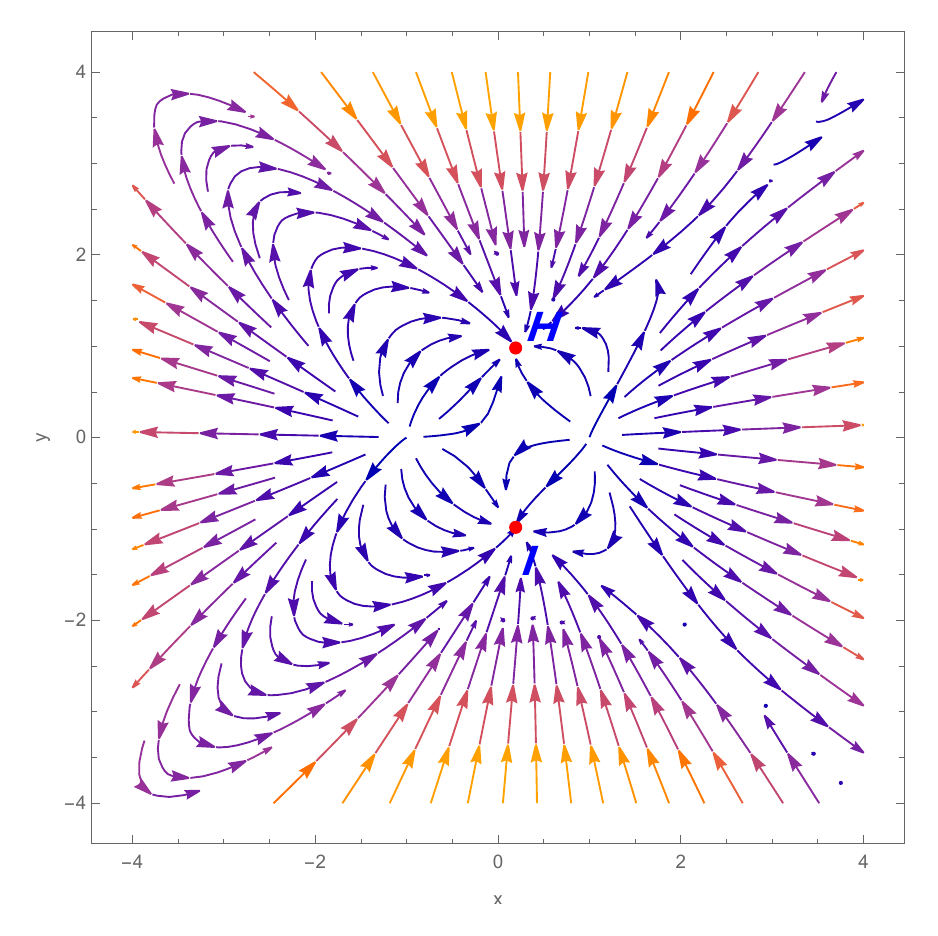}
    \includegraphics[width=60mm]{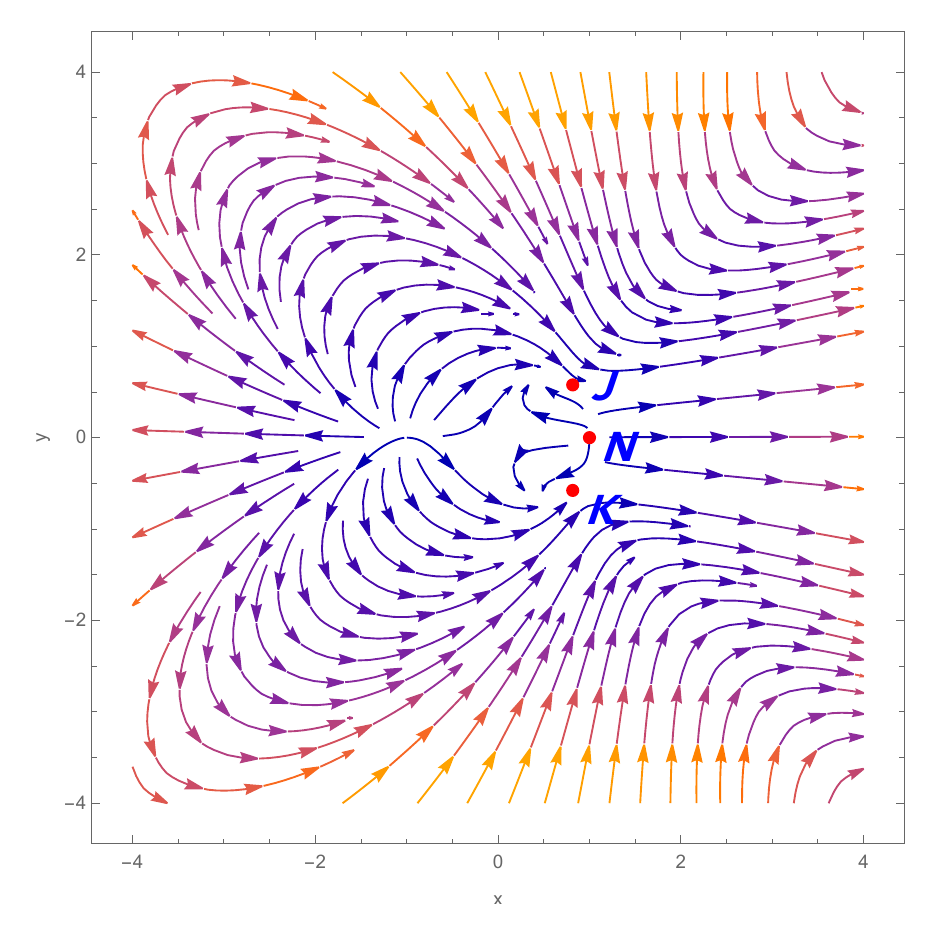}
    \caption{2D phase portrait for model \ref{generalmodelI} for general $\alpha$. } \label{Ch1Fig1}
\end{figure}

In Fig. \ref{Ch1Fig1}, the model parameters and the dynamical variables take the values the upper left plot is for $u=0$, $\rho=0$, $\lambda=\sqrt{\frac{2}{9}}$. The upper right plot having parameter values $u=0$, $\rho=0$, $\zeta =\frac{1}{9}$, $\tau=1$, $\alpha=1.1$, lower left phase portrait is for $u=0$, $\rho=0$, $\lambda=\sqrt{\frac{2}{9}}$, lower right phase portrait is for $u=0$, $\rho=0$, $\delta=1$. We have analysed the phase space for all the critical points by fixing some parameters to an appropriate value. The phase space plots for the dynamical system described in Eqs. \eqref{eq:dx_dN_model_1} to \eqref{eq:dl_dN_model_1} are presented in Fig. \ref{Ch1Fig1}. From the phase space diagram for model \ref{generalmodelI}, we can conclude that the phase space trajectories for critical points $L$, $F$, $M$, $G$, $A$, $J$, $N$, and $K$ are moving away from the critical point hence confirm saddle point behaviour. Critical points $F$ and $G$ show stability for $-2 \sqrt{\frac{6}{7}}\leq \lambda <-\sqrt{3}$ or $\sqrt{3}<\lambda \leq 2 \sqrt{\frac{6}{7}}$ but we choose $\lambda=\sqrt{\frac{2}{9}}$ hence these are showing saddle points nature in phase diagram. 
For the critical points $L$ and $M$, the stability conditions depend on $\chi$, which represents $u$ co-ordinate, and the phase plots are analysed in the xy-axis plane; hence critical points $L$ and $M$ may show saddle point behaviour. From the phase diagram, it can be observed that critical points, $H$, $I$, $D$, and $E$ trajectories are attracted towards the critical point, hence describing the attracting behaviour of these critical points. Also, these critical points can explain the DE-dominated Universe. Although eigenvalues at $B$ and $C$ contain both positive and negative signatures, from the phase portrait critical points $B$ and $C$ represent an unstable node leading to the positive eigenvalues only (due to consideration of $u=0$, $\rho=$ may the negative eigenvalues are not contributing in the phase space plot). We shall present below two examples of this model for $\alpha=1$ and $\alpha=2$. 
\subsection{Case A: \texorpdfstring{$\alpha=1$}{}}\label{sec:model_1_case_1}

In this case, we have considered $\alpha=1$ in the action equation Eq. (\ref{eq:action_model_1}. The evolution equations can be obtained by limiting Eqs. \eqref{eq:44} to \eqref{eq:46} using $\alpha=1$. To study cosmic evolution through a dynamical system analysis approach, the set of dynamical variables associated with the above set of cosmological equations can be defined as follows \cite{Gonzalez-Espinoza:2020jss}
\begin{equation}\label{eq:56}
    x=\dfrac{\kappa\dot{\phi}}{\sqrt{6}H}\,,\quad y=\frac{\kappa\sqrt{V}}{\sqrt{3}H}\,, u=3\dot{\phi}^2 \kappa^2\,,\quad \rho= \frac{\kappa\sqrt{\rho_{\rm r}}}{\sqrt{3}H}\,,\quad \lambda=\frac{-V^{'}(\phi)}{\kappa V(\phi)}\,,\quad \Gamma=\dfrac{V(\phi)V^{''}(\phi)}{V^{'}(\phi)^{2}}\,.
\end{equation}
In this case the dynamical variables are selected such that they can linked with each other in following constraint equation form, note that in this expression u is considered as it is (without any scalar multiplier).
\begin{align}\label{eq:57}
    x^{2}+y^{2}+u+\Omega_{m}+\rho^{2}=1\,.
\end{align}
Using these dynamical variables the cosmological expressions in this case can be written in terms of autonomous dynamical system as follow, 
\begin{align}
    \dfrac{dx}{dN} &= \frac{3 x y^2 \left(\sqrt{6} \lambda  (u-1) x+3 u+3 x^2\right)-3 x \left(u^2+u \left(\rho ^2+8 x^2+1\right)+x^2 \left(\rho ^2+3 x^2-3\right)\right)}{2 (u-3) x^2-2 u (u+1)}\,,\label{eq:dx_dN_model_1_alpha_1}\\
    \dfrac{dy}{dN} &= -y \left(\frac{-3 u^2-u \left(\rho ^2+12 x^2+3\right)+u y^2 \left(2 \sqrt{6} \lambda x+3\right)-3 x^2 \left(\rho ^2+3 x^2-3 y^2+3\right)}{2 u (u+1)-2 (u-3) x^2}+\sqrt{\frac{3}{2}} \lambda  x\right)\,,\\
    \dfrac{du}{dN} &= \frac{2 u \left(\rho ^2 u+(6 u-9) x^2\right)-u y^2 \left(\sqrt{6} \lambda  (u-3) x+6 u\right)}{u (u+1)-(u-3) x^2}\,,\\
    \frac{d\rho}{dN} &=\frac{\rho \left(-u^2+u \left(\rho ^2+16 x^2-y^2 \left(2 \sqrt{6} \lambda  x+3\right)-1\right)+3 x^2 \left(\rho ^2+3 x^2-3 y^2-1\right)\right)}{2 u (u+1)-2 (u-3) x^2}\,,\\
    \dfrac{d\lambda}{dN} &=-\sqrt{6}(\Gamma-1)\lambda^{2}x\label{eq:dl_dN_model_1_alpha_1}\,.
\end{align}

From Table \ref{CHITABLE-III} observations, we can conclude that, at critical points $A$, $F$, $G$ the value of deceleration parameter is $\frac{1}{2}$ with $\omega_{tot}=0$. These critical points do not represent the accelerating Universe but the cold, DM-dominated era. The critical points $B$ and $C$ behave as stiff matter showing $\omega_{tot}=1$. The critical points $J$ and $K$ represent the radiation-dominated solutions. The critical points $D$, $E$, $H$, and $I$ show the value of the deceleration parameter negative; these critical points can represent the accelerating behaviour of the Universe. The critical points $D$ and $E$ are the de Sitter solutions with the value of $\omega_{tot}=-1$ and can be obtained only at a particular value of $\lambda=0$.  At the critical points $H$ and $I$ deceleration parameter shows negative value for $-\sqrt{2}$ $<$ $\lambda$ $<$ $\sqrt{2}$, these points explains DE, matter-dominated Universe. To analyse the stability behaviour of all these critical points, the eigenvalues and stability conditions are presented in Table \ref{CHITABLE-IV}. 

\begin{figure}[H]
    \centering
    \includegraphics[width=60mm]{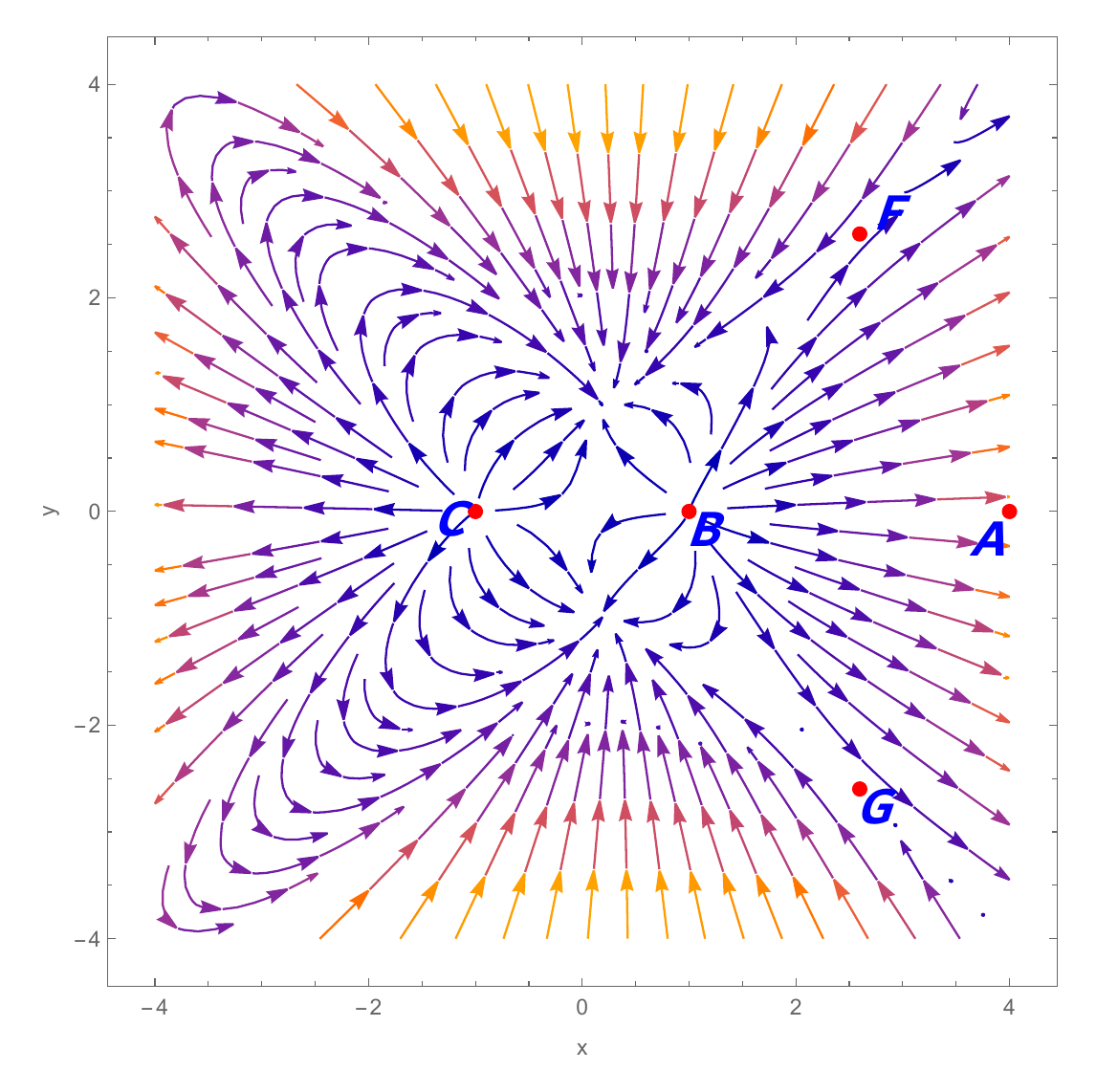}
    \includegraphics[width=60mm]{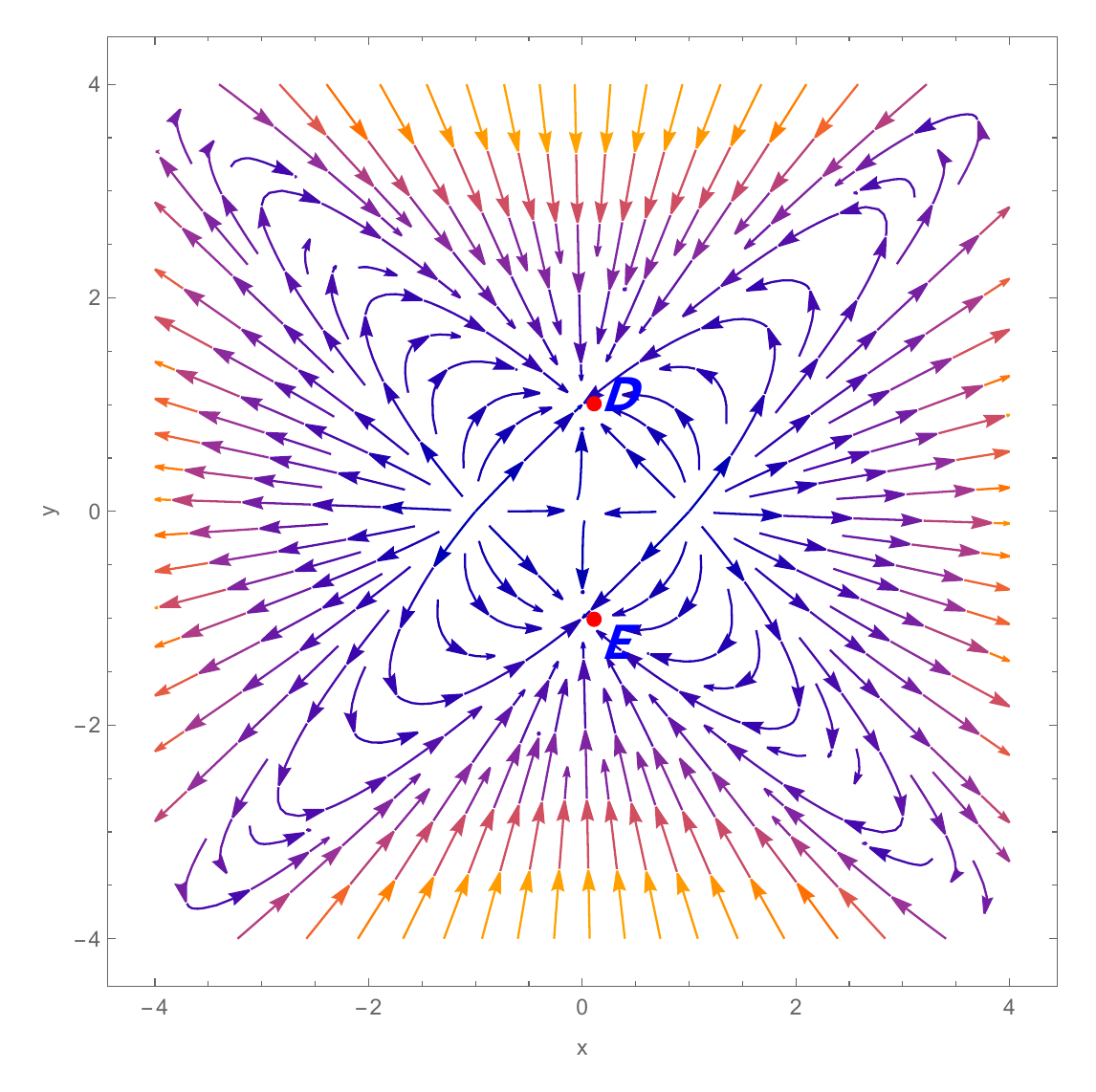}
    \includegraphics[width=60mm]{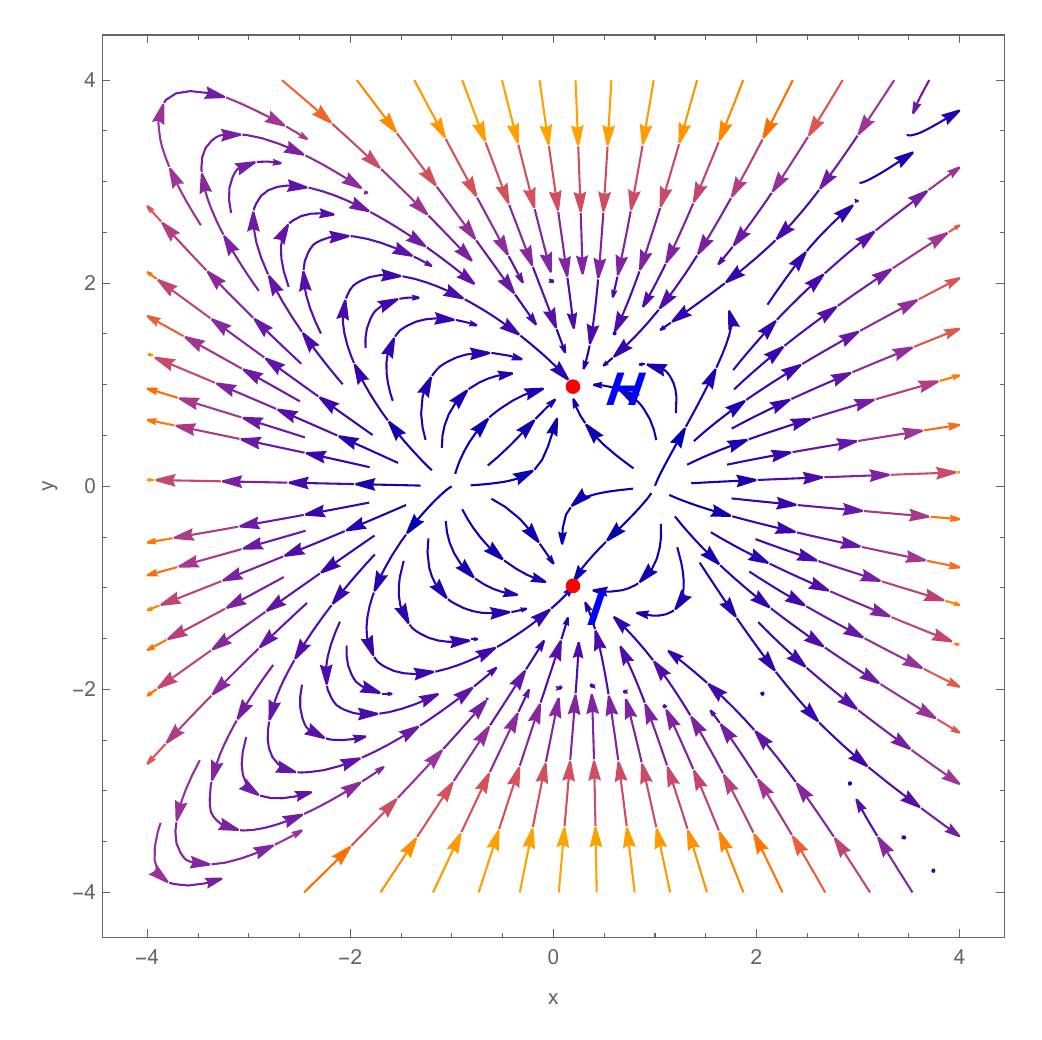}
    \includegraphics[width=60mm]{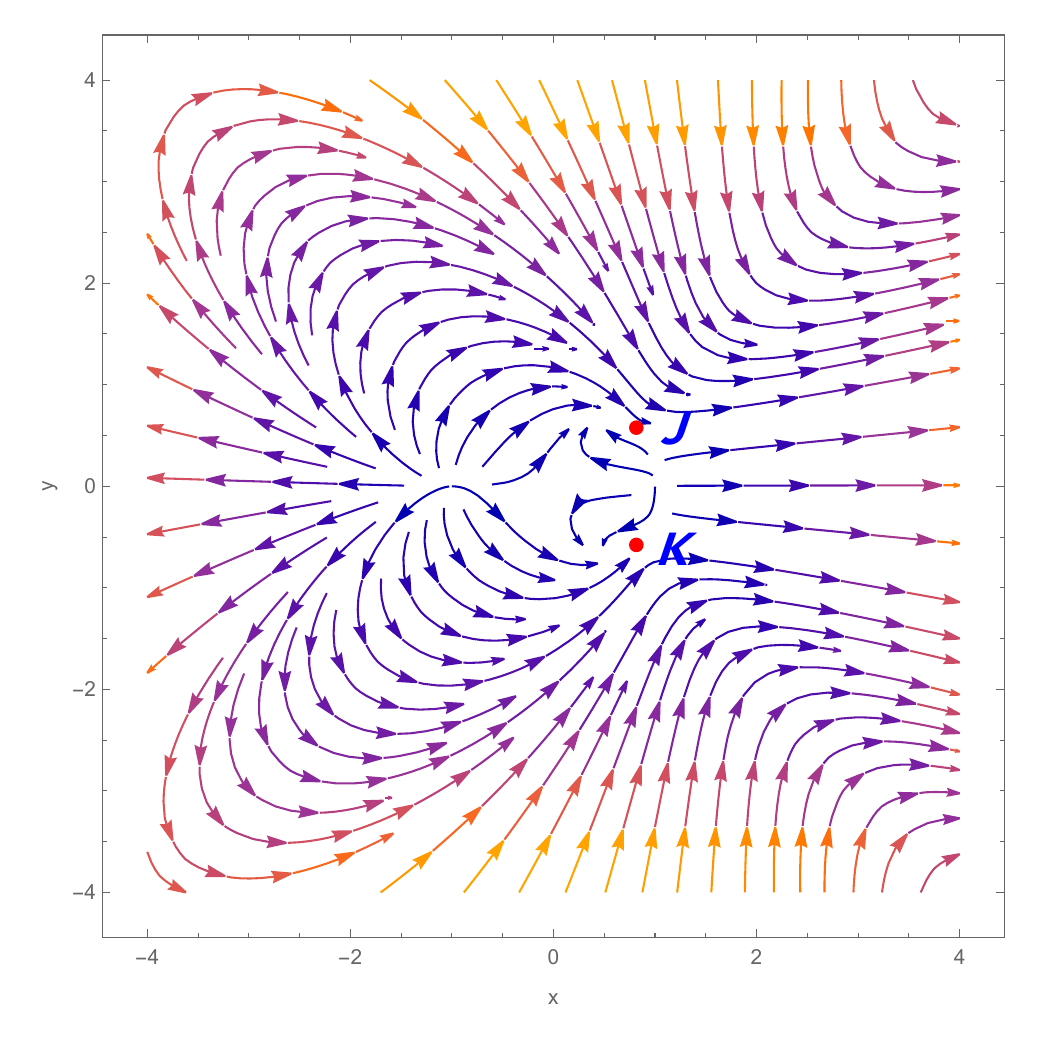}
    \caption{2D phase portrait for model \ref{generalmodelI}, $\alpha=1$.} \label{Ch1Fig2}
\end{figure}

From Table \ref{CHITABLE-IV}, we can conclude that, the critical points $D$, $E$ show stable behaviour and these critical points can attract the Universe at late time. At the critical points $H$ and $I$ eigenvalues show stability at $-\sqrt{3}<\lambda <0$ or $0<\lambda <\sqrt{3}$ these points explain DE domination at late time. The critical points $A$ to $C$ are saddle points, hence unstable for all values of $\lambda$. However, these points cannot describe the current accelerated expansion of the Universe. The radiation-dominated representation belongs to the critical points $J$ and $K$; these points are also saddle points for any value of $\lambda$ and, hence, unstable. Although critical points $F$ and $G$ represent cold DM-dominated Universe, these critical points obey stability at $-2 \sqrt{\frac{6}{7}}\leq \lambda <-\sqrt{3}$ or $\sqrt{3}<\lambda \leq 2 \sqrt{\frac{6}{7}}$.
\begin{table}[H]
\small\addtolength{\tabcolsep}{-5pt}
\centering % used for centering table
\renewcommand{\arraystretch}{1.2} % Adjust row height
\scalebox{0.95}{
\begin{tabular}{|c|c|c|c|c|c|c|c|} % centered columns (7 columns)
\hline\hline %inserts double horizontal lines

\parbox[c][1.2cm]{2.5cm}{\centering Critical points} & 
\parbox[c]{1.5cm}{\centering $x_{c}$} & 
\parbox[c]{1.5cm}{\centering $y_{c}$} & 
\parbox[c]{1.5cm}{\centering $u_{c}$} & 
\parbox[c]{1.5cm}{\centering $\rho_{c}$} & 
\parbox[c]{2.0cm}{\centering Existence conditions} &
\parbox[c]{2.0cm}{\centering $q$} & 
\parbox[c]{2.0cm}{\centering $\omega_{tot}$} \\ [0.5ex] % inserts table heading

\hline\hline % inserts single horizontal line

$A$ & $\tau$ & $0$ & $0$ & $0$ & $\tau ^2+\tau \neq 0$ & $\frac{1}{2}$ & $0$\\
\hline

$B$ & $1$ & $0$ & $0$ & $0$ &-&$2$ & $1$\\
\hline

$C$ & $-1$ & $0$ & $0$ & $0$ &-&$2$ & $1$\\
\hline

$D$  & $\zeta$ & $\sqrt{2 \zeta ^2+1}$ & $-3 \zeta ^2$ & $0$ & $\lambda=0, \zeta \neq 0$  &$-1$ & $-1$\\
\hline

$E$  & $\zeta$ & $-\sqrt{2 \zeta ^2+1}$ & $-3 \zeta ^2$ & $0$ &  $\lambda=0, \zeta \neq 0$ & $-1$ & $-1$\\
\hline

$F$ & $\frac{\sqrt{\frac{3}{2}}}{\lambda }$ & $\sqrt{\frac{3}{2}} \sqrt{\frac{1}{\lambda ^2}}$ & $0$ & $0$ &-& $\frac{1}{2}$ & $0$\\
\hline

$G$ & $\frac{\sqrt{\frac{3}{2}}}{\lambda }$ & $-\sqrt{\frac{3}{2}} \sqrt{\frac{1}{\lambda ^2}}$ & $0$ & $0$ &-&$\frac{1}{2}$ & $0$\\
\hline

$H$ & $\frac{\lambda }{\sqrt{6}}$ & $\sqrt{1-\frac{\lambda ^2}{6}}$ & $0$ & $0$&-& $\frac{1}{2} \left(\lambda ^2-2\right)$ & $-1+\frac{\lambda^2}{3}$\\
\hline

$I$ & $\frac{\lambda }{\sqrt{6}}$ & $-\sqrt{1-\frac{\lambda ^2}{6}}$ & $0$ & $0$&-&$\frac{1}{2} \left(\lambda ^2-2\right)$ & $-1+\frac{\lambda^2}{3}$\\
\hline

$J$ & $\sqrt{\frac{2}{3}}$ & $\frac{1}{\sqrt{3}}$ & $0$ & $0$ &  $\lambda=2$  & $1$ & $\frac{1}{3}$\\
\hline

$K$  & $\sqrt{\frac{2}{3}}$ & $-\frac{1}{\sqrt{3}}$ & $0$ & $0$ & $\lambda=2$   & $1$ & $\frac{1}{3}$\\
[1ex] % [1ex] adds vertical space
\hline %inserts single line
\end{tabular}}
\caption{Critical points corresponding to model \ref{generalmodelI}, $\alpha=1$.}
\label{CHITABLE-III}
\end{table}
From the Fig. \ref{Ch1Fig2} observation, it can be conclude that critical points $H$, $I$ are attractors and can be analysed by setting $\lambda=\sqrt{\frac{2}{9}}$ which belongs to the stability range of $\lambda$ (that is $-\sqrt{3}<\lambda <0\lor 0<\lambda <\sqrt{3}$ ). Since $F$ and $G$ show stability  for $-2 \sqrt{\frac{6}{7}}\leq \lambda <-\sqrt{3}$ or $\sqrt{3}<\lambda \leq 2 \sqrt{\frac{6}{7}}$ and here, we have analyse phase plots at $\lambda=\sqrt{\frac{2}{9}}$ which does not belong to stability range of $\lambda$, the critical points $F$ and $G$ show saddle point behaviour. The critical points $A$, $J$, and $K$ have eigenvalues with both positive and negative signs; hence, these are saddle points. The parametric value $\lambda=2$ helps us analyse the stability at $J$ and $K$. Also, from phase space analysis, it can be observed that phase space trajectories are moving away at these critical points, showing the saddle point behaviour. The de Sitter solution is represented by critical points $D$ and $E$; these critical points exist only for parametric value $\lambda=0$, and from the phase space analysis, it can be concluded that these points represent an attractive solution. The critical points $A$ to $K$ can obtained in particular case $\alpha=1$ similar to general $\alpha$, but the critical points  $L$, $M$, and $N$ discussed in general $\alpha$ case are not contributing in the case $\alpha=1$.
\begin{table}[H]
\small\addtolength{\tabcolsep}{-6pt}
 % title of Table
\centering % used for centering table
\renewcommand{\arraystretch}{1.2} % Adjust row height
\scalebox{0.95}{
\begin{tabular}{|c|c|c|} % centered columns (3 columns)
\hline %inserts double horizontal lines

\parbox[c][1.2cm]{2.5cm}{\centering Critical points} & 
\parbox[c]{5.5cm}{\centering  Eigenvalues} & 
\parbox[c]{2.5cm}{\centering Stability} \\ [0.5ex] % inserts table heading

\hline % inserts single horizontal line

$A$ & $ \Bigl\{\frac{3}{2},\frac{3}{2},-\frac{1}{2},0 \Bigl\}$ & Unstable\\
\hline

$B$ & $ \Bigl\{-6,3,1,\frac{1}{2} \left(6-\sqrt{6} \lambda \right) \Bigl\}$ & Unstable\\
\hline

$C$ & $ \Bigl\{-6,3,1,\frac{1}{2} \left(6+\sqrt{6} \lambda \right) \Bigl\}$ & Unstable\\
\hline

$D$ & $ \Bigl\{0,-3,-3,-2 \Bigl\}$ & Stable\\
\hline

$E$ &$ \Bigl\{0,-3,-3,-2 \Bigl\}$ & Stable\\
\hline

$F$ & \parbox[c][2cm]{4.5cm}{\centering $ \Bigl\{-3,-\frac{1}{2},\frac{3 \left(-\lambda ^4-\sqrt{24 \lambda ^6-7 \lambda ^8}\right)}{4 \lambda ^4},$\\$\frac{3 \left(\sqrt{24 \lambda ^6-7 \lambda ^8}-\lambda ^4\right)}{4 \lambda ^4} \Bigl\}$} & \parbox[c][2.1cm]{4.5cm}{\centering Stable for\\ $ \Bigl\{-2 \sqrt{\frac{6}{7}}\leq \lambda <-\sqrt{3}\lor \sqrt{3}<\lambda \leq 2 \sqrt{\frac{6}{7}} \Bigl\}$} \\
\hline

$G$ & \parbox[c][2cm]{4.5cm}{\centering $\Bigl\{-3,-\frac{1}{2},\frac{3 \left(-\lambda ^4-\sqrt{24 \lambda ^6-7 \lambda ^8}\right)}{4 \lambda ^4},$\\$\frac{3 \left(\sqrt{24 \lambda ^6-7 \lambda ^8}-\lambda ^4\right)}{4 \lambda ^4} \Bigl\}$} & \parbox[c][2.1cm]{4.5cm}{\centering Stable for \\ $ \Bigl\{-2 \sqrt{\frac{6}{7}}\leq \lambda <-\sqrt{3}\lor \sqrt{3}<\lambda \leq 2 \sqrt{\frac{6}{7}} \Bigl\}$} \\
\hline

$H$ & $ \Bigl\{-\lambda ^2,\frac{1}{2} \left(\lambda ^2-6\right),\frac{1}{2} \left(\lambda ^2-4\right),\lambda ^2-3 \Bigl\}$ & \parbox[c][1.5cm]{4.5cm}{\centering Stable for $ \Bigl\{-\sqrt{3}<\lambda <0\lor 0<\lambda <\sqrt{3} \Bigl\}$} \\
\hline

$I$ & $\Bigl\{-\lambda ^2,\frac{1}{2} \left(\lambda ^2-6\right),\frac{1}{2} \left(\lambda ^2-4\right),\lambda ^2-3 \Bigl\}$ & \parbox[c][1.5cm]{4.5cm}{\centering Stable for $\Bigl\{-\sqrt{3}<\lambda <0\lor 0<\lambda <\sqrt{3}\Bigl\}$}\\
\hline

$J$ & $\Bigl\{-4,-1,1,0 \Bigl\}$ & Unstable\\
\hline

$K$ & $ \Bigl\{-4,-1,1,0 \Bigl\}$ & Unstable\\
[1ex] % [1ex] adds vertical space

\hline %inserts single line
\end{tabular}}
\caption{Eigenvalues and stability of critical points for model \ref{generalmodelI}, $\alpha=1$.}
\label{CHITABLE-IV}
\end{table}
 In the phase portrait presented in Fig. \ref{Ch1Fig2} , the upper left plot is for the parametric value $u=0$, $\rho=0$, $\tau=1$ and $\lambda=\sqrt{\frac{2}{9}}$. The upper right plot is for the parameteric values $u=0$, $\rho=0$ and $\zeta =\frac{1}{9}$. The lower left phase portrait is for $u=0$, $\rho=0$ and $\lambda=\sqrt{\frac{2}{9}}$ and lower right phase portrait is for $u=0$ and $\rho=0$.

\subsection{Case B: \texorpdfstring{$\alpha=2$}{}}\label{sec:model_1_case_2}

In this case, we have analysed cosmological implications by using a dynamical system approach for particular value $\alpha=2$ in action Eq. (\ref{eq:action_model_1}). In this case, the set of dynamical variables to obtain an autonomous dynamical system can be defined as follows,
\begin{equation}\label{eq:63}
    x=\dfrac{\kappa\dot{\phi}}{\sqrt{6}H}\,,\quad y=\frac{\kappa\sqrt{V}}{\sqrt{3}H}\,,\quad u=\frac{5}{2}\kappa^2\dot{\phi}^{4}\,,\quad \rho= \frac{\kappa\sqrt{\rho_{\rm r}}}{\sqrt{3}H}\,,\quad \lambda=\frac{-V^{'}(\phi)}{\kappa V(\phi)}\,,\quad \Gamma=\dfrac{V(\phi)V^{''}(\phi)}{V^{'}(\phi)^{2}}\,.
\end{equation}

One can observe that the dynamical variables defined in the study of these scalar-tensor models are not the same in \cite{Gonzalez-Espinoza:2020jss}. Still, these variables are usually used to obtain viable critical points in cosmology. The dynamical variables defined in Eq. (\ref{eq:63}) also satisfy  the constraint equation Eq. (\ref{eq:57}), and the dynamical system in this case can be defined as follows,
\begin{align}
    \frac{dx}{dN} &= \frac{x \left(-5 \rho ^2 \left(2 u+x^2\right)+5 y^2 \left(\sqrt{6} \lambda  (u-1) x+6 u+3 x^2\right)+3 (5-19 u) x^2-6 u (u+3)-15 x^4\right)}{2 (u-5) x^2-4 u (u+3)}\,,\label{eq:dx_dN_model_1_alpha_2}\\
    \frac{dy}{dN} &= -y \left(\frac{-6 u^2-3 u \left(2 \rho ^2+13 x^2+6\right)+2 u y^2 \left(2 \sqrt{6} \lambda  x+9\right)-5 x^2 \left(\rho ^2+3 x^2-3 y^2+3\right)}{4 u (u+3)-2 (u-5) x^2}\right)\nonumber\\
    &-y\sqrt{\frac{3}{2}} \lambda  x\,,\\
    \frac{du}{dN} &= \frac{4 u \left(2 \rho ^2 u+3 (3 u-5) x^2\right)-2 u y^2 \left(\sqrt{6} \lambda  (u-5) x+12 u\right)}{2 u (u+3)-(u-5) x^2}\,,\\
    \frac{d\rho}{dN} &= \frac{\rho  \left(-2 u^2+u \left(6 \rho ^2+43 x^2-2 y^2 \left(2 \sqrt{6} \lambda  x+9\right)-6\right)+5 x^2 \left(\rho ^2+3 x^2-3 y^2-1\right)\right)}{4 u (u+3)-2 (u-5) x^2}\,,\\
    \frac{d\lambda}{dN} &= -\sqrt{6}(\Gamma-1)\lambda^{2}x \,.\label{eq:dl_dN_model_1_alpha_2}
\end{align}

Studying dynamical systems at critical points obtained from cosmological evolution equations is very important to studying cosmological implications. Critical points for autonomous dynamical system presented in Eqs. \eqref{eq:dx_dN_model_1_alpha_2} to \eqref{eq:dl_dN_model_1_alpha_2} are presented in Table \ref{CHITABLE-V}. From the table, it can be concluded that, although critical points have different co-ordinates than those in the Table \ref{CHITABLE-III} (for $\alpha=1$) case, the cosmological implications are almost similar. When we observe critical points in Table \ref{CHITABLE-III} and Table \ref{CHITABLE-V}, we can easily see that the deceleration parameter ($q$) and $\omega_{tot}$ for critical points with the same name are the same. From Table \ref{CHITABLE-V}, it can be observed that critical points $D$, $E$, $H$, and $I$ can show deceleration parameter value in the negative range, and hence these critical points can deal with the DE, matter-dominated era. For critical points $H$ and $I$, we get accelerating behaviour for $-\sqrt{2}$ $<$ $\lambda$ $<$ $\sqrt{2}$ and critical points $D$ and $E$ are defined only for parametric value $\lambda=0$ and represent de Sitter solution for the system. The other critical points do not give a negative value for the deceleration parameter and hence define the non-accelerating phase of evolution. The critical points $A$, $F$ and $G$ represents cold DM-dominated era with $\omega_{tot}=0$. In this case (for $\alpha=2$) also, we are getting critical points $B$ and $C$ representing stiff matter. The critical points $J$ and $K$ are defined for $\lambda=2$ and deliver value for $\omega_{tot}=\frac{1}{3}$, hence represent radiation-dominated era.

The stability conditions for critical points corresponding to the dynamical system in Eqs. \eqref{eq:dx_dN_model_1_alpha_2} to \eqref{eq:dl_dN_model_1_alpha_2} are presented in Table \ref{CHITABLE-VI}. The signature of eigenvalues confirms the stability of the corresponding critical point.  From the table observations, we can conclude that critical points $A$, $B$, and $C$ are unstable for all values of $\lambda$ and are saddle points. At the critical points, $H$ and $I$ eigenvalues show stability at $-\sqrt{3}<\lambda <0$ or $0<\lambda <\sqrt{3}$ and these points explain DE domination at late time. Critical points $D$ and $E$ show stable behaviour, and in further analysis, it is noticed that they can attract the Universe at a late time. The critical points $J$ and $K$ represent radiation dominated, and from the signature of eigenvalues, these points are saddle points for any value of $\lambda$, hence unstable. Critical points $F$ and $G$ represent cold DM-dominated Universe and these critical point obey stability at $-2 \sqrt{\frac{6}{7}}\leq \lambda <-\sqrt{3}$ or $\sqrt{3}<\lambda \leq 2 \sqrt{\frac{6}{7}}$. 

\begin{table}[H]
\small\addtolength{\tabcolsep}{-6pt}
 % title of Table
\centering % used for centering table
\renewcommand{\arraystretch}{1.2} % Adjust row height
\scalebox{0.8}{
\begin{tabular}{|c|c|c|c|c|c|c|c|} % centered columns (7 columns)
\hline\hline %inserts double horizontal lines
\parbox[c][1.6cm]{2.5cm}{\centering Critical points} & 
\parbox[c]{1.5cm}{\centering $x_{c}$} & 
\parbox[c]{1.5cm}{\centering $y_{c}$} & 
\parbox[c]{1.5cm}{\centering $u_{c}$} & 
\parbox[c]{2.0cm}{\centering $\rho_{c}$} & 
\parbox[c]{2.5cm}{\centering Existence conditions} &
\parbox[c]{2.5cm}{\centering $q$} & 
\parbox[c]{2.0cm}{\centering $\omega_{tot}$} \\ [0.5ex] % inserts table heading

\hline\hline % inserts single horizontal line

$A$ & $\tau$ & $0$ & $0$ & $0$ & $\tau ^2+3\tau \neq 0$ &$\frac{1}{2}$ & $0$\\
\hline

$B$ & $1$ & $0$ & $0$ & $0$ &-&$2$ & $1$\\
\hline

$C$ & $-1$ & $0$ & $0$ & $0$ &-&$2$ & $1$\\
\hline

$D$  & \parbox[c][0.2cm]{1.2cm}{\centering $\zeta$} & $\sqrt{\frac{3 \zeta ^2}{2}+1}$ & $-\frac{1}{2} \left(5 \zeta ^2\right)$ & $0$ &\parbox[c][1.4cm]{2.0cm}{\centering $\lambda=0$, \\ $3 \zeta ^3-2 \zeta \neq 0$} & $-1$ & $-1$\\
\hline

$E$ & \parbox[c][0.2cm]{1.2cm}{\centering $\zeta$} & $-\sqrt{\frac{3 \zeta ^2}{2}+1}$ & $-\frac{1}{2} \left(5 \zeta ^2\right)$ & $0$ &\parbox[c][1.4cm]{2.0cm}{\centering $\lambda=0$, \\ $3 \zeta ^3-2 \zeta \neq 0$} & $-1$ & $-1$\\
\hline

$F$ & $\frac{\sqrt{\frac{3}{2}}}{\lambda }$ & $\sqrt{\frac{3}{2}} \sqrt{\frac{1}{\lambda ^2}}$ & $0$ & $0$ &-& $\frac{1}{2}$ & $0$\\
\hline

$G$ & $\frac{\sqrt{\frac{3}{2}}}{\lambda }$ & $-\sqrt{\frac{3}{2}} \sqrt{\frac{1}{\lambda ^2}}$ & $0$ & $0$ &-&$\frac{1}{2}$ & $0$\\
\hline

$H$ & $\frac{\lambda }{\sqrt{6}}$ & $\sqrt{1-\frac{\lambda ^2}{6}}$ & $0$ & $0$ &-&$\frac{1}{2} \left(\lambda ^2-2\right)$ & $-1+\frac{\lambda^2}{3}$\\
\hline

$I$ & $\frac{\lambda }{\sqrt{6}}$ & $-\sqrt{1-\frac{\lambda ^2}{6}}$ & $0$ & $0$ &-&$\frac{1}{2} \left(\lambda ^2-2\right)$ & $-1+\frac{\lambda^2}{3}$\\
\hline

$J$& $\sqrt{\frac{2}{3}}$ & $\sqrt{\frac{1}{3}}$ & $0$ & $0$ & $\lambda=2$ & $1$ & $\frac{1}{3}$\\
\hline

$K$ & $\sqrt{\frac{2}{3}}$ & $-\sqrt{\frac{1}{3}}$ & $0$ & $0$ & $\lambda=2$ &$1$ & $\frac{1}{3}$\\
[1ex] % [1ex] adds vertical space

\hline %inserts single line
\end{tabular}}
\caption{Critical points for  model \ref{generalmodelI}, $\alpha=2$.}
\label{CHITABLE-V}
\end{table}

\begin{table}[H]
\small\addtolength{\tabcolsep}{-6pt}
 % title of Table
\centering % used for centering table
\renewcommand{\arraystretch}{1.2} % Adjust row height
\scalebox{0.8}{
\begin{tabular}{|c|c|c|} % centered columns (3 columns)
\hline % inserts double horizontal lines

\parbox[c]{2.5cm}{\centering Critical points} & 
\parbox[c]{7.5cm}{\centering Eigenvalues} & 
\parbox[c]{2.5cm}{\centering Stability} \\ [0.5ex] % inserts table heading

\hline % inserts single horizontal line

$A$ & $\Bigl\{\frac{3}{2},\frac{3}{2},-\frac{1}{2},0\Bigl\}$ & Unstable \\
\hline

$B$ & $\Bigl\{-12,3,1,\frac{1}{2} \left(6-\sqrt{6} \lambda \right)\Bigl\}$ & Unstable \\
\hline

$C$ & $\Bigl\{-12,3,1,\frac{1}{2} \left(\sqrt{6} \lambda +6\right)\Bigl\}$ & Unstable \\
\hline

$D$ & $\Bigl\{0,-3,-3,-2\Bigl\}$ & Stable \\
\hline

$E$ & $\Bigl\{0,-3,-3,-2\Bigl\}$ & Stable \\
\hline

$F$ & \parbox[c]{7.5cm}{\centering $\Bigl\{-6,-\frac{1}{2},\frac{3 \left(-\lambda ^2-\sqrt{24 \lambda ^2-7 \lambda ^4}\right)}{4 \lambda ^2},\frac{3 \left(\sqrt{24 \lambda ^2-7 \lambda ^4}-\lambda ^2\right)}{4 \lambda ^2}\Bigl\}$} & \parbox[c][2.1cm]{4.5cm}{\centering Stable for \\ $\Bigl\{ -2 \sqrt{\frac{6}{7}}\leq \lambda <-\sqrt{3}\lor \sqrt{3}<\lambda \leq 2 \sqrt{\frac{6}{7}}\Bigl\}$} \\
\hline

$G$ & \parbox[c]{7.5cm}{\centering $\Bigl\{-6,-\frac{1}{2},\frac{3 \left(-\lambda ^2-\sqrt{24 \lambda ^2-7 \lambda ^4}\right)}{4 \lambda ^2},\frac{3 \left(\sqrt{24 \lambda ^2-7 \lambda ^4}-\lambda ^2\right)}{4 \lambda ^2}\Bigl\}$} &  \parbox[c][2.1cm]{4.5cm}{\centering Stable for \\ $\Bigl\{-2 \sqrt{\frac{6}{7}}\leq \lambda <-\sqrt{3}\lor \sqrt{3}<\lambda \leq 2 \sqrt{\frac{6}{7}}\Bigl\}$} \\
\hline

$H$ & \parbox[c]{7.5cm}{\centering $\Bigl\{-2 \lambda ^2,\frac{1}{2} \left(\lambda ^2-6\right),\frac{1}{2} \left(\lambda ^2-4\right),\lambda ^2-3\Bigl\}$} & \parbox[c][1.8cm]{4.5cm}{\centering Stable for\\ $\Bigl\{-\sqrt{3}<\lambda <0\lor 0<\lambda <\sqrt{3}\Bigl\}$} \\
\hline

$I$ & \parbox[c]{7.5cm}{\centering $\Bigl\{-2 \lambda ^2,\frac{1}{2} \left(\lambda ^2-6\right),\frac{1}{2} \left(\lambda ^2-4\right),\lambda ^2-3\Bigl\}$} & \parbox[c][1.8cm]{4.5cm}{\centering Stable for\\ $\Bigl\{-\sqrt{3}<\lambda <0\lor 0<\lambda <\sqrt{3}\Bigl\}$} \\
\hline

$J$ &$\Bigl\{-8,-1,1,0\Bigl\}$ & Unstable \\
\hline

$K$ & $\Bigl\{-8,-1,1,0\Bigl\}$ & Unstable \\
[1ex] % [1ex] adds vertical space

\hline %inserts single line
\end{tabular}}
\caption{Eigenvalues and stability of critical points for model \ref{generalmodelI}, $\alpha=2$.}
\label{CHITABLE-VI}
\end{table}

From Table \ref{CHITABLE-IV} and Table \ref{CHITABLE-VI}, it can be observed that the stability conditions for the case ($\alpha=1$) and ($\alpha=2$) show a similar nature to explain the evolution of the Universe.

\begin{figure}[H]
    \centering
    \includegraphics[width=60mm]{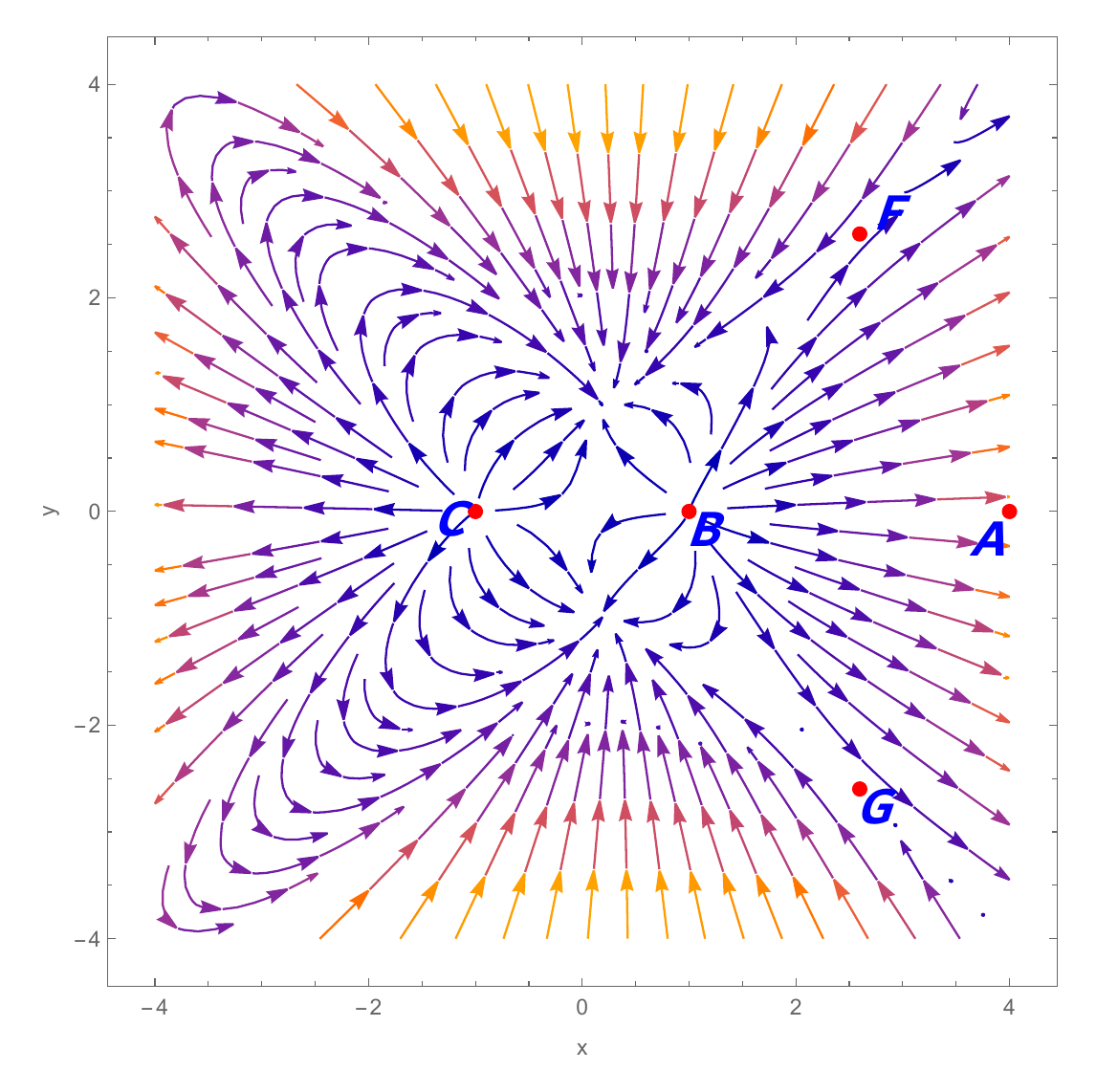}
    \includegraphics[width=60mm]{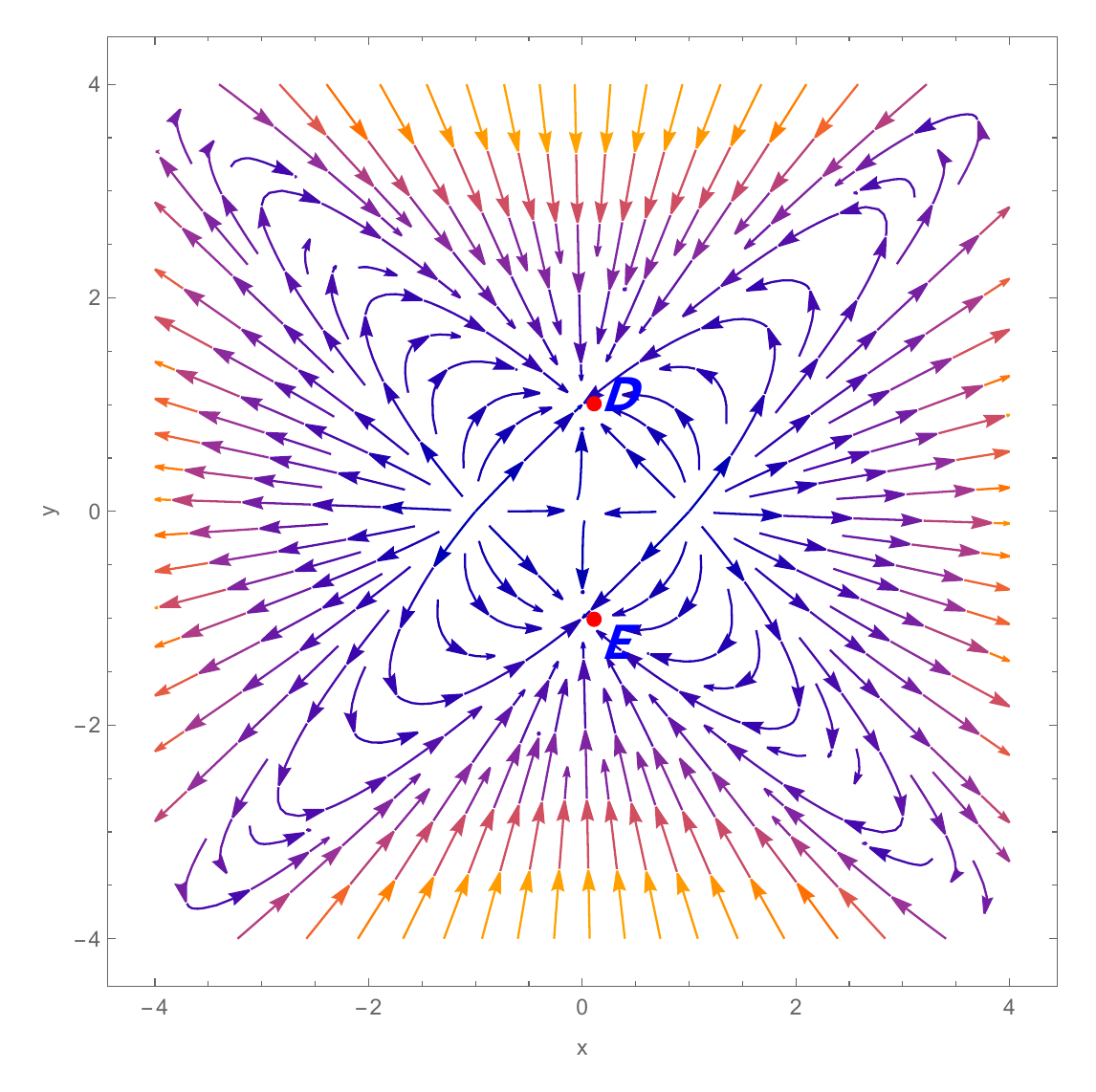}
    \includegraphics[width=60mm]{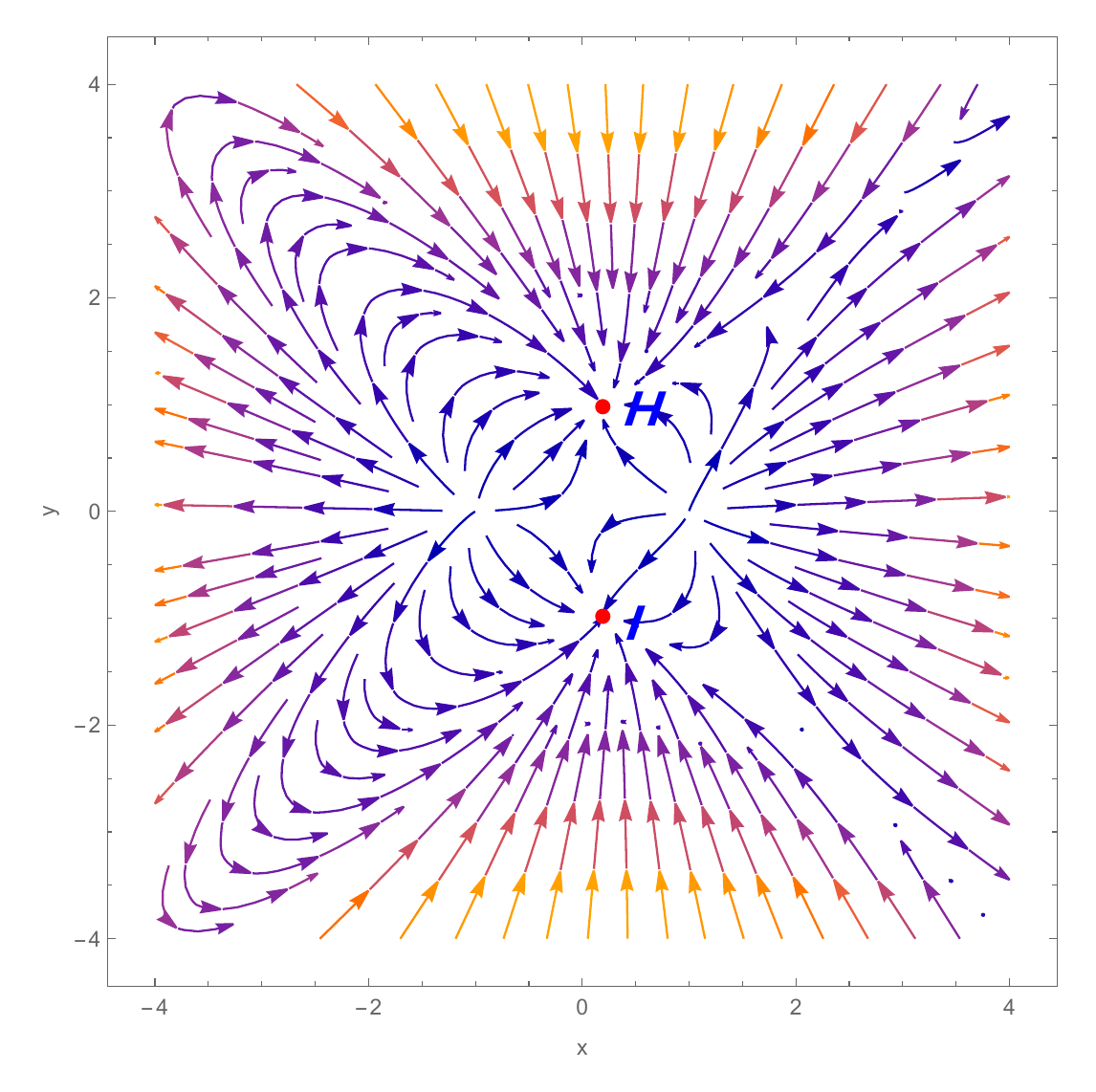}
    \includegraphics[width=60mm]{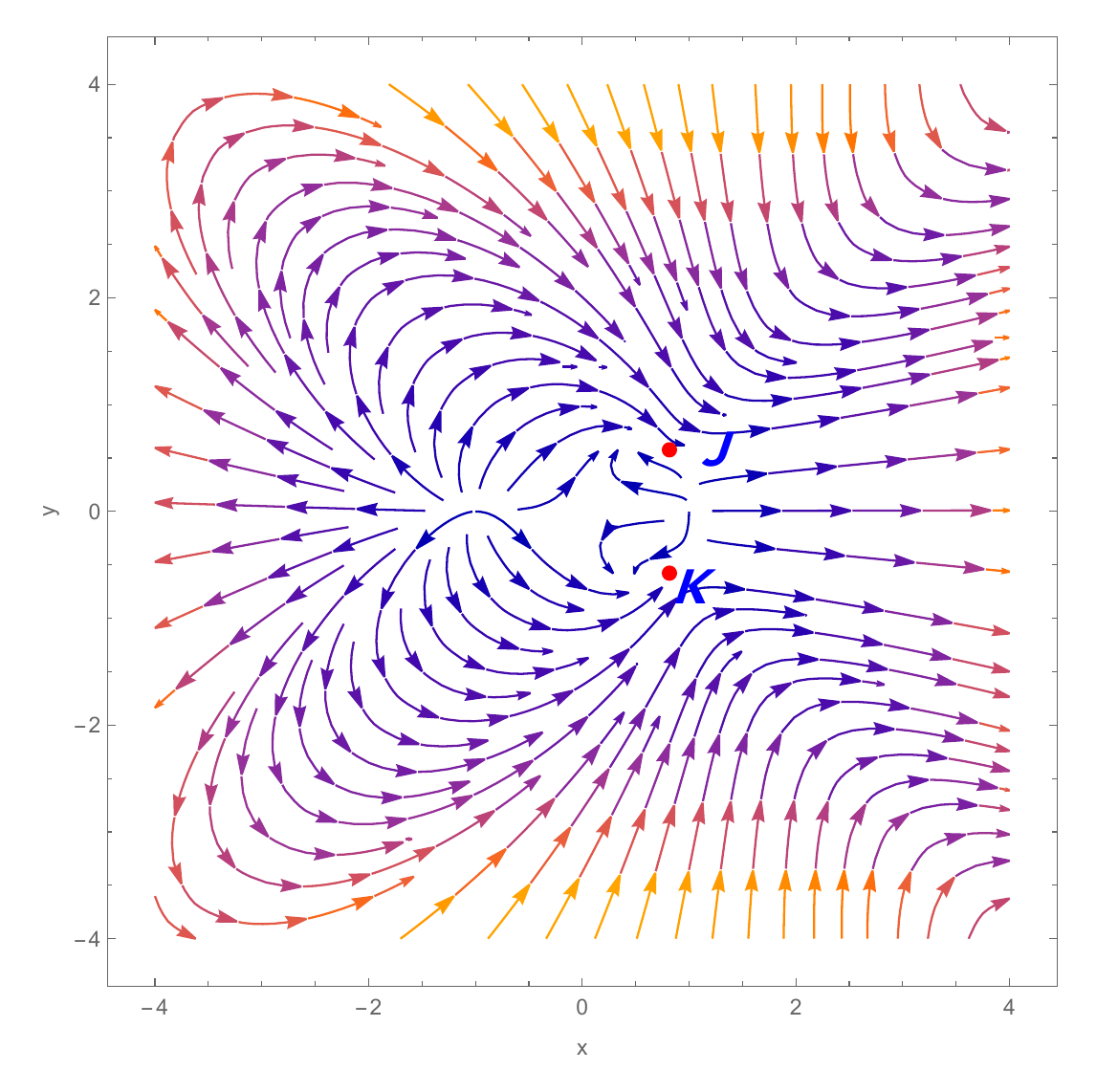}
    \caption{2D phase portrait for model \ref{generalmodelI}, $\alpha=2$. } \label{Ch1Fig31}
\end{figure}

The phase space diagram is presented in Fig. \ref{Ch1Fig31} for the parametric values $\lambda=\sqrt{\frac{2}{9}}$ which belongs to the stability range for $\lambda$ for critical points $H$ and $I$ and other parameters $\tau=1$, $\zeta=\frac{1}{9}$ are chosen such that the phase space diagram explain the stability conditions for the corresponding critical points. Critical points $D$ and $E$ represent the de Sitter solution. The phase space analysis confirms the attracting nature of these critical points. From the Fig. \ref{Ch1Fig2} observations, it can be concluded that due to different co-ordinates, the critical point $A$ is presented in Fig. {\ref{Ch1Fig31}} moves in positive X-axis than in Fig. \ref{Ch1Fig2}. Still, it does not impact its stable nature. Critical points $F$ and $G$ show stability  for $-2 \sqrt{\frac{6}{7}}\leq \lambda <-\sqrt{3}$ or $\sqrt{3}<\lambda \leq 2 \sqrt{\frac{6}{7}}$ but we choose $\lambda=\sqrt{\frac{2}{9}}$ hence these are a saddle points. The phase space diagram lets us conclude that critical points $J$ and $K$ are saddle points.

The upper left plot in Fig. \ref{Ch1Fig31} for $u=0$, $\rho=0$, $\tau=1$, $\lambda=\sqrt{\frac{2}{9}}$, the upper right plot having parametric values $u=0$, $\rho=0$, $\zeta =\frac{1}{9}$. The lower left phase portrait is for $u=0$, $\rho=0$, $\lambda=\sqrt{\frac{2}{9}}$ and the lower right phase portrait is for $u=0$, $\rho=0$.

\section{Kinetic term coupled with \texorpdfstring{$I_2$}{}}\label{sec:model-II}

The action equation consists of a coupling between the $X^\alpha$ term with $I_2$, the action for model \ref{sec:model-II} is described below with
\begin{equation}\label{eq:action_model_2}
    S = \int d^4x e [X-V(\phi)-\frac{T}{2 \kappa^{2}}+X^{\alpha}I_{2}]+S_{m}+S{r}\,.
\end{equation}
In the action equation, we have  $I_2=3H\dot{\phi}$, and other notations are the same as in the first model. The expression for the sum of energy density for matter and radiation and the negative of radiation pressure can be obtained on varying action equations for model \ref{sec:model-II} with respect to the tetrad field presented in Eqs. \eqref{eq:70} to \eqref{eq:71} respectively. The motion equation, in this case, can be obtained by taking a variation of the action equation with respect to $\phi$, presented in Eq. (\ref{eq:72}).
\begin{eqnarray}
    \frac{T}{2\kappa^{2}}-V(\phi)-X-X^{\alpha}I_2-2\alpha X^{\alpha}I_2 &=& \rho_{\rm m}+\rho_{\rm r} \,,\label{eq:70}\\
    -V(\phi)+\frac{T}{2\kappa^{2}}+ X+\frac{2\dot{H}}{\kappa^{2}}-X^{\alpha}\ddot{\phi}-2\alpha X^{\alpha}\ddot{\phi} &=& -p_{r} \,,\label{eq:71}\\
    V^{'}(\phi)+3H\dot{\phi}+(3+6\alpha)X^{\alpha}\left(\frac{T}{2}+\dot{H}\right)+\ddot{\phi}[1+\frac{\alpha X^{\alpha}6H}{\dot{\phi}}+\frac{\alpha^{2} X^{\alpha}12H}{\dot{\phi}}] &=& 0 \,.\label{eq:72}
\end{eqnarray}
Using background expressions discussed in Sec \ref{sec:backgroundexpressions}, we can obtain expressions for energy density and pressure for effective DE are presented in Eq. (\ref{eq:73}) and Eq. (\ref{eq:74}) respectively.
\begin{align}
    \rho_{\rm DE} &= V(\phi)+X+X^{\alpha}I_2+2\alpha X^{\alpha}I_2 \,,\label{eq:73}\\
    p_{\rm DE} &= -V(\phi)+X-(1+2\alpha)X^{\alpha}\ddot{\phi} \,,\label{eq:74}
\end{align}
The dynamical variables defined in this case are as follows,
\begin{equation}\label{eq:model_2_dynamic_var}
    x=\dfrac{\kappa\dot{\phi}}{\sqrt{6}H},\quad y=\frac{\kappa\sqrt{V}}{\sqrt{3}H}, \quad
    u=\frac{\kappa^2 X^\alpha \dot{\phi} }{H},\quad
    \rho= \frac{\kappa\sqrt{\rho_{\rm r}}}{\sqrt{3}H}, \quad \lambda=\frac{-V^{'}(\phi)}{\kappa V(\phi)}, \quad \Gamma=\dfrac{V(\phi)V^{''}(\phi)}{V^{'}(\phi)^{2}} \,.
\end{equation}
The critical points for the dynamical system described in Eqs. \eqref{eq:model_2_dx_dN} to \eqref{eq:model_2_dl_dN} are presented in Table \ref{CHITABLE-VII}. From table, it can be concluded that critical points $J$ to $O$ represent similar cosmological implications in terms of deceleration parameter $q=1$ and $\omega_{tot}=\frac{1}{3}$ describe the radiation-dominated phase of the Universe. Amongst these critical points the critical points $N$ and $O$ are defined for $\alpha=\frac{1}{2}$. The critical points $F$, $G$, and $P$ also describe the same cosmological implication and represent a cold DM-dominated era with $\omega_{tot}=0$. The parameter $\alpha$ plays a role in the coordinate representation of critical points $D$ and $E$, these critical points describe de Sitter solution and are defined for $\lambda=0$. The critical points $H$ and $I$ represent the value for the deceleration parameter $q=-1+\frac{\lambda^2}{2}$; these critical points can explain the DE-dominated Universe. Critical points $A$, $B$, and $C$ deliver the same value for $q$ and $\omega_{tot}$ hence these critical points also represent similar phases of Universe evolution and behave as stiff matter.

The dynamical system can be obtained as presented below,
\begin{align}
    \dfrac{dx}{dN} &= \frac{x \left((2 \alpha +1) u \left(2 \alpha  \rho ^2+6 \alpha +\rho ^2+6 \alpha  x^2+9 x^2-y^2 \left(6 \alpha +\sqrt{6} \lambda  x+3\right)-3\right)\right)}{(2 \alpha +1) u (2 \alpha  (u+2)+u)+4 x^2} \nonumber \\
    & +\frac{x \left(3 (2 \alpha  u+u)^2+2 x \left(3 x^3+x \left(\rho ^2-3 y^2-3\right)+\sqrt{6} \lambda  y^2\right)\right)}{(2 \alpha +1) u (2 \alpha  (u+2)+u)+4 x^2}\,,\label{eq:model_2_dx_dN}\\
    \dfrac{dy}{dN} &= -\sqrt{\frac{3}{2}} \lambda  x y+\frac{y \left(3 (2 \alpha  u+u)^2+2 x^2 \left(\rho ^2+3 x^2-3 y^2+3\right)\right)}{(2 \alpha +1) u (2 \alpha  (u+2)+u)+4 x^2}\nonumber \\
    & +\frac{(2 \alpha +1) u y \left(6 (\alpha +1) x^2-\sqrt{6} \lambda  x y^2+2 \alpha  \left(\rho ^2-3 y^2+3\right)\right)}{(2 \alpha +1) u (2 \alpha  (u+2)+u)+4 x^2}\,, \\ 
    \dfrac{du}{dN} &= \frac{u \left(3 (2 \alpha  u+u)^2+(2 \alpha +1) u \left(4 \alpha  \rho ^2+\rho ^2+12 \alpha  x^2+9 x^2-y^2 \left(12 \alpha +\sqrt{6} \lambda  x+3\right)-3\right)\right)}{(2 \alpha +1) u (2 \alpha  (u+2)+u)+4 x^2}\nonumber \\ 
    & + \frac{2 u x \left(3 x^3+x \left(\rho ^2-3 \left(4 \alpha +y^2+1\right)\right)+\sqrt{6} (2 \alpha +1) \lambda  y^2\right)}{(2 \alpha +1) u (2 \alpha  (u+2)+u)+4 x^2}\,, \\
    \frac{d\rho}{dN} &= \frac{\rho  \left((2 \alpha  u+u)^2+2 x^2 \left(\rho ^2+3 x^2-3 y^2-1\right)\right)}{(2 \alpha +1) u (2 \alpha  (u+2)+u)+4 x^2}\nonumber\\ 
    & +\frac{\rho  u \left((2 \alpha +1) \left(6 (\alpha +1) x^2-\sqrt{6} \lambda  x y^2+2 \alpha  \left(\rho ^2-3 y^2-1\right)\right)\right)}{(2 \alpha +1) u (2 \alpha  (u+2)+u)+4 x^2}\,, \\
    \frac{d\lambda}{dN} &= \sqrt{6}(\Gamma-1)\lambda^{2}x\,.\label{eq:model_2_dl_dN}
\end{align}

Table \ref{CHITABLE-VIII} discusses the stability conditions of the critical points. The critical points $J$ to $O$ present a radiation-dominated Universe, with eigenvalues for the linear perturbation matrix at these critical points having at least one eigenvalue with a positive signature, hence showing unstable behaviour. The critical points $F$, $G$ show eigenvalues in negative range for  $\alpha >0\land \left(-2 \sqrt{\frac{6}{7}}\leq \lambda <-\sqrt{3}\lor \sqrt{3}<\lambda \leq 2 \sqrt{\frac{6}{7}}\right)$ and critical point $P$ is stable at $\left(\lambda <0\land \sigma <\frac{\sqrt{\frac{3}{2}}}{\lambda }\right)\lor \left(\lambda >0\land \sigma >\frac{\sqrt{\frac{3}{2}}}{\lambda }\right)$, hence show stability within this range, also these points express $\omega_{tot}=0$, hence addressing cold DM phase of the Universe evolution. The critical points $B$ and $C$ represent stiff matter era and possess at least one eigenvalue with a positive signature; hence, they are unstable, whereas critical point $A$ shows stable behaviour in the parametric range, same as critical point $P$. The critical points $D$, $E$, $H$ and $I$ represent DE-dominated era, where $D$ and $E$ are non-hyperbolic critical points but are stable and critical points $H$ and $I$ show their stability in the parametric range $\alpha >0\land \left(-\sqrt{3}<\lambda <0\lor 0<\lambda <\sqrt{3}\right)$. The stability conditions of critical points $F$, $G$, $H$, and $I$ show $\alpha>0$ condition, which implies stability of critical points representing cold DM and DE, matter-dominated era can be obtained for the positive value of $\alpha$.

\begin{figure}[H]
    \centering
    \includegraphics[width=60mm]{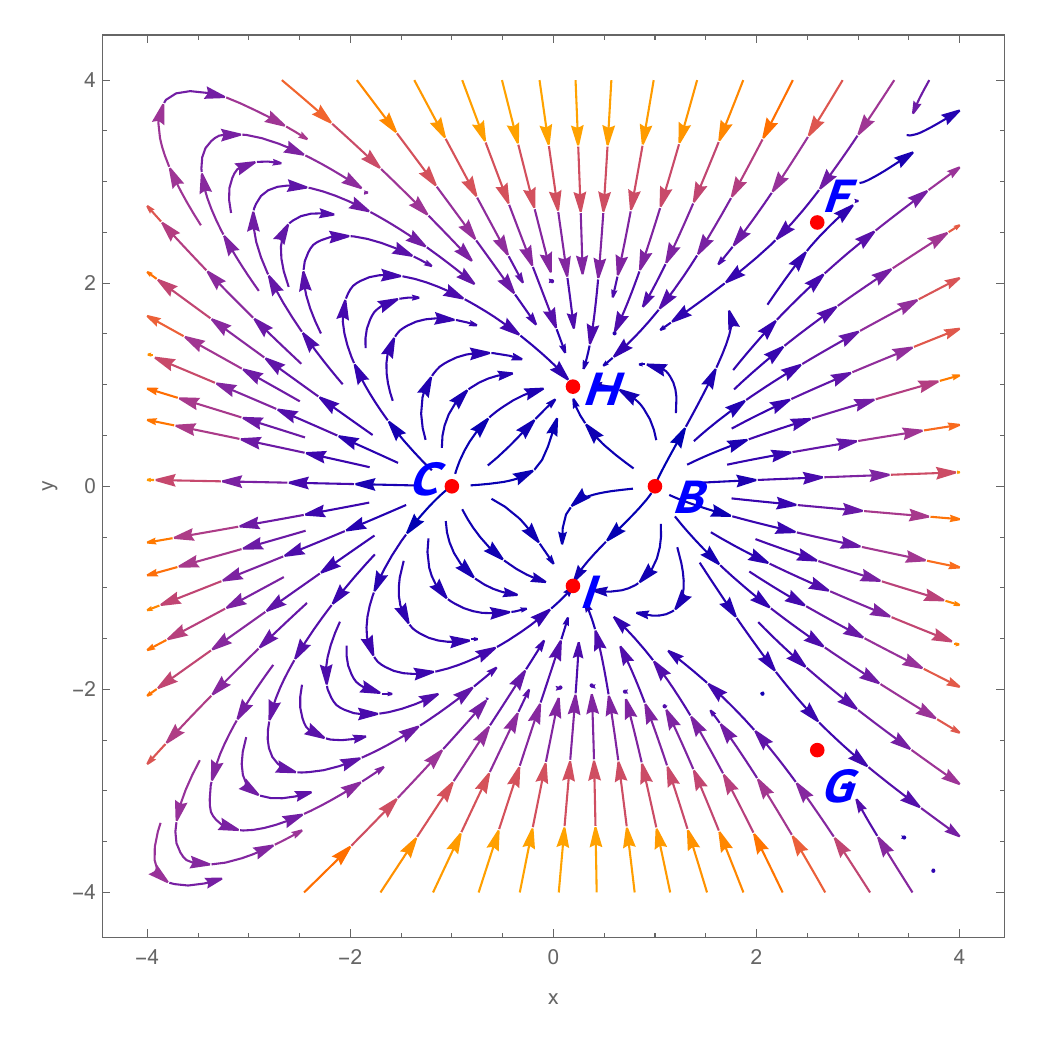}
    \includegraphics[width=60mm]{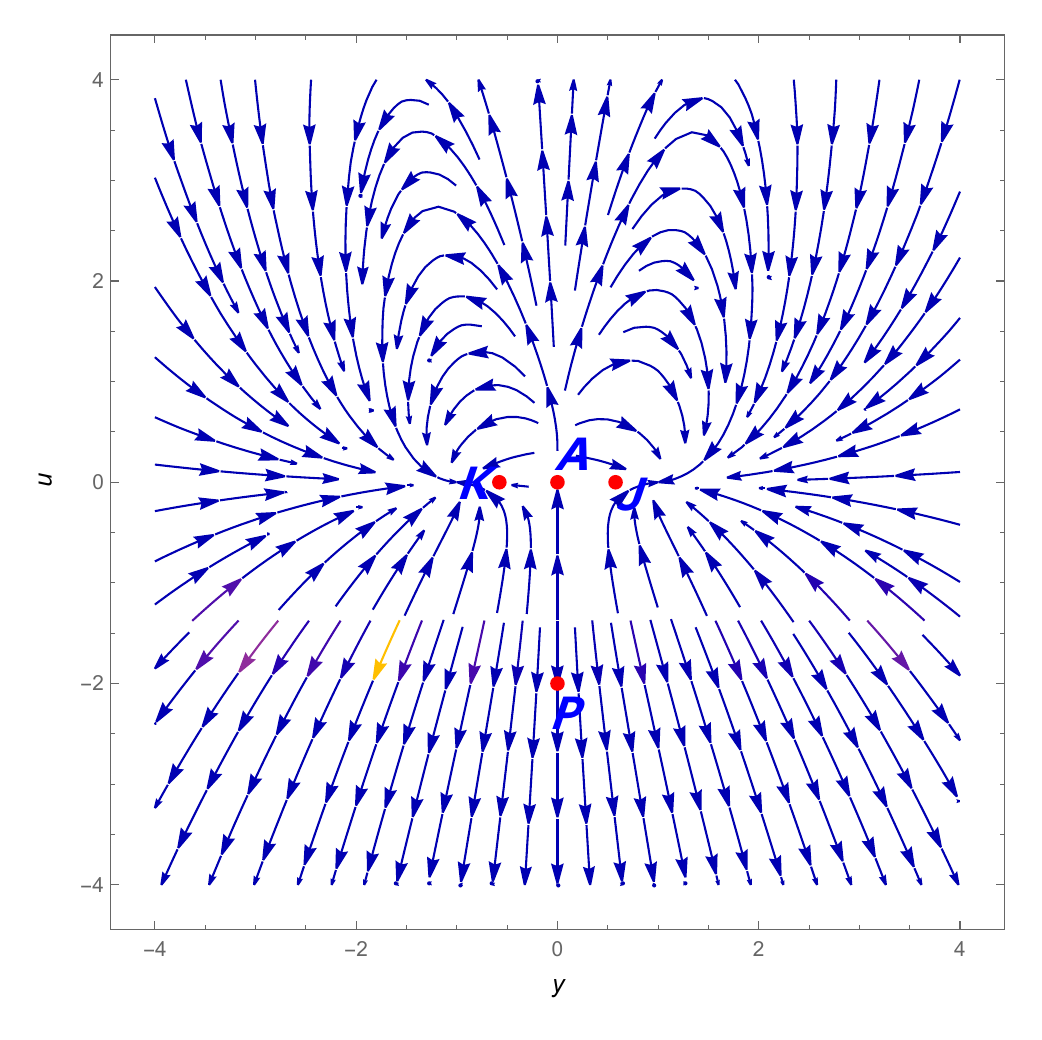}
    \caption{2D phase portrait for model  \ref{sec:model-II}, for general $\alpha$.} \label{CHIFig5}
\end{figure}

The critical points are plotted in the phase diagram Fig. \ref{CHIFig5}. These phase plots are plotted for the dynamical system presented in Eqs. \eqref{eq:model_2_dx_dN} to \eqref{eq:model_2_dl_dN}. The lower right plot shows that the phase space trajectories are moving away from critical points $L$, $M$, $N$, and $O$; hence these points represent instability with saddle point behaviour. The critical points $N$ and $O$ are defined for $\alpha=\frac{1}{2}$. The critical points $D$ and $E$ are de Sitter solutions, and the phase diagram clarifies that these points behave as attracting solutions. The upper left phase plot describes the phase space trajectories behaviour of critical points $H$, $I$, $B$, $C$, $F$, and $G$. We can observe that the critical points $B$ and $C$ which is unstable saddle points but show unstable node behaviour, which leads to the positive eigenvalues at critical points $B$ and $C$.
\begin{table}[H]
\small\addtolength{\tabcolsep}{-6pt}
% title of Table
\centering % used for centering table
\renewcommand{\arraystretch}{1.2} % Adjust row height
\scalebox{0.9}{
\begin{tabular}{|c|c|c|c|c|c|c|c|} % centered columns (7 columns)
\hline\hline %inserts double horizontal lines
\parbox[c]{2.4cm}{\centering Critical points} & 
\parbox[c][1.3cm]{2.5cm}{\centering $x_{c}$} & 
\parbox[c]{1.5cm}{\centering $y_{c}$} & 
\parbox[c]{1.1cm}{\centering $u_{c}$} & 
\parbox[c]{1.5cm}{\centering $\rho_{c}$} & 
\parbox[c]{2.4cm}{\centering Existence condition} & 
\parbox[c]{1.5cm}{\centering $q$} & 
\parbox[c]{1.5cm}{\centering $\omega_{tot}$} \\ [0.5ex] % inserts table heading

\hline\hline % inserts single horizontal line

$A$  & $\sigma$ & $0$ & $1-\sigma ^2$ & $0$ & \parbox[c][1.3cm]{2.5cm}{\centering $3 \mu  \sigma ^2-\mu +$\\$2 \sigma ^4-6 \sigma ^2 \neq 0$} & $2$ & $1$ \\
\hline

$B$ & $1$ & $0$ & $0$ & $0$ &-&$2$ & $1$ \\
\hline

$C$ & $-1$ & $0$ & $0$ & $0$ &-&$2$ & $1$ \\
\hline

$D$ & $\varphi$ & $\sqrt{\varphi ^2+1}$ & $-\frac{2 \varphi ^2}{2 \alpha +1}$ & $0$ & \parbox[c][2.2cm]{2.5cm}{\centering$\lambda=0$, $2 \alpha +1\neq 0$\\$-2 \alpha  \varphi +\varphi ^3+\varphi \neq 0$} & $-1$ & $-1$ \\ [1.5ex] 
\hline

$E$ & $\varphi$ & $-\sqrt{\varphi ^2+1}$ & $-\frac{2 \varphi ^2}{2 \alpha +1}$ & $0$ &\parbox[c][2.2cm]{2.5cm}{\centering $\lambda=0$, $2 \alpha +1\neq 0$\\$-2 \alpha  \varphi +\varphi ^3+\varphi \neq 0$} & $-1$ & $-1$ \\
\hline

$F$ & $\frac{\sqrt{\frac{3}{2}}}{\lambda }$ & $\sqrt{\frac{3}{2}} \sqrt{\frac{1}{\lambda ^2}}$ & $0$ & $0$ &-&$\frac{1}{2}$ & $0$ \\
\hline

$G$ & $\frac{\sqrt{\frac{3}{2}}}{\lambda }$ & $-\sqrt{\frac{3}{2}} \sqrt{\frac{1}{\lambda ^2}}$ & $0$ & $0$ &-&$\frac{1}{2}$ & $0$ \\
\hline

$H$ & $\frac{\lambda }{\sqrt{6}}$ & $\sqrt{1-\frac{\lambda ^2}{6}}$ & $0$ & $0$ &-&$\frac{1}{2} \left(\lambda ^2-2\right)$ & $-1+\frac{\lambda^2}{3}$ \\
\hline

$I$ & $\frac{\lambda }{\sqrt{6}}$ & $-\sqrt{1-\frac{\lambda ^2}{6}}$ & $0$ & $0$ &-&$\frac{1}{2} \left(\lambda ^2-2\right)$ & $-1+\frac{\lambda^2}{3}$ \\
\hline

$J$  & $\sqrt{\frac{2}{3}}$ & $\sqrt{\frac{1}{3}}$ & $0$ & $0$ &$\lambda=2$&$1$ & $\frac{1}{3}$ \\
\hline

$K$ & $\sqrt{\frac{2}{3}}$ & $-\sqrt{\frac{1}{3}}$ & $0$ & $0$ &$\lambda=2$&$1$ & $\frac{1}{3}$ \\
\hline

$L$ & $\eta$ & $0$ & $-2 \eta ^2$ & $\sqrt{\eta ^2+1}$ &$ \eta ^3+\eta \neq 0$&$1$ & $\frac{1}{3}$ \\
\hline

$M$ & $\eta$ & $0$ & $-2 \eta ^2$ & $-\sqrt{\eta ^2+1}$ &$\eta ^3+\eta \neq 0$& $1$ & $\frac{1}{3}$ \\
\hline

$N$ & $0$ & $0$ & \parbox[c]{2.5cm}{\centering $\epsilon$} & $\sqrt{1-2 \epsilon }$ & \parbox[c]{2.5cm}{\centering$\alpha=\frac{1}{2}$,\\$\epsilon ^2+\epsilon \neq 0$} & $1$ & $\frac{1}{3}$ \\
\hline

$O$ & $0$ & $0$ & \parbox[c]{2.5cm}{\centering $\epsilon$} & $-\sqrt{1-2 \epsilon }$ &\parbox[c]{2.5cm}{\centering$\alpha=\frac{1}{2}$\\ $\epsilon ^2+\epsilon \neq 0$} &$1$ & $\frac{1}{3}$ \\
\hline

$P$ & $\sigma$ & $0$ & $-2 \sigma ^2$ & $0$ &-&$\frac{1}{2}$ & $0$ \\
[1ex] % [1ex] adds vertical space

\hline % inserts single line
\end{tabular}}
\caption{Critical points for  model \ref{sec:model-II}, for general $\alpha$.} \label{CHITABLE-VII}
\end{table}
 The points $H$ and $I$ show attracting point behaviour; these critical points represent a DE-dominated era with stability as described in Table \ref{CHITABLE-VIII}. Critical points $F$ and $G$ represent cold DM-dominated era; we have plotted plots for $\lambda=\sqrt{\frac{2}{9}}$ which is not in the stability range of $F$ and $G$. The upper right diagram represents phase space trajectories at critical points $K$, $J$, and $P$; since the phase space trajectories are moving away from these critical points, these critical points are showing saddle point behaviour and hence unstable. Since the parametric value $\sigma=1$ does not follow the stability range of critical point $A$, it is getting saddle point behaviour.
\begin{table}[H]
%\small\addtolength{\tabcolsep}{-6pt}
 % title of Table
\centering % used for centering table
\scalebox{0.9}{
\begin{tabular}{|c|c|c|} % centered columns
\hline % inserts double horizontal lines

\parbox[c][1.3cm]{1.5cm}{\centering Critical points} & 
\parbox[c]{7cm}{\centering Eigenvalues} & 
\parbox[c]{7cm}{\centering Stability} \\ [0.5ex] % inserts table heading

\hline % inserts single horizontal line

$A$ & 
\parbox[c][1.3cm]{7.5cm}{\centering $\Bigl\{0,-\frac{3}{2},-\frac{1}{2},\frac{1}{2} \left(3-\sqrt{6} \lambda  \sigma \right)\Bigl\}$} & 
\parbox[c]{8cm}{\centering Stable for \\ $\Bigl\{\lambda <0 \land \sigma <\frac{\sqrt{\frac{3}{2}}}{\lambda }\Bigl\}\lor \Bigl\{\lambda >0\land \sigma >\frac{\sqrt{\frac{3}{2}}}{\lambda }\Bigl\}$} \\

\hline

$B$ & 
\parbox[c]{7.5cm}{\centering $\Bigl\{3,1,-6 \alpha ,\frac{1}{2} \left(6-\sqrt{6} \lambda \right)\Bigl\}$} & 
\parbox[c]{8cm}{\centering Unstable} \\

\hline

$C$ & 
\parbox[c]{7.5cm}{\centering $\Bigl\{3,1,-6 \alpha ,\frac{1}{2} \left(6+\sqrt{6} \lambda \right)\Bigl\}$} & 
\parbox[c]{8cm}{\centering Unstable} \\

\hline

$D$ & 
\parbox[c]{7.5cm}{\centering $\Bigl\{0,-3,-3,-2\Bigl\}$} & 
\parbox[c]{8cm}{\centering Stable} \\

\hline

$E$ & 
\parbox[c]{7.5cm}{\centering $\Bigl\{0,-3,-3,-2\Bigl\}$} & 
\parbox[c]{8cm}{\centering Stable} \\

\hline

$F$ & 
\parbox[c]{7.5cm}{\centering $\Bigl\{-\frac{1}{2},-3 \alpha ,\frac{3 \left(-\lambda ^2-\sqrt{24 \lambda ^2-7 \lambda ^4}\right)}{4 \lambda ^2},\frac{3 \left(\sqrt{24 \lambda ^2-7 \lambda ^4}-\lambda ^2\right)}{4 \lambda ^2}\Bigl\}$} & 
\parbox[c]{7.5cm}{\centering Stable for $\alpha >0$\\ $\land \Bigl\{-2 \sqrt{\frac{6}{7}}\leq \lambda <-\sqrt{3}\lor \sqrt{3}<\lambda \leq 2 \sqrt{\frac{6}{7}}\Bigl\}$} \\

\hline

$G$ & 
\parbox[c]{7.5cm}{\centering $\Bigl\{-\frac{1}{2},-3 \alpha ,\frac{3 \left(-\lambda ^2-\sqrt{24 \lambda ^2-7 \lambda ^4}\right)}{4 \lambda ^2},\frac{3 \left(\sqrt{24 \lambda ^2-7 \lambda ^4}-\lambda ^2\right)}{4 \lambda ^2}\Bigl\}$} & 
\parbox[c]{7.5cm}{\centering Stable for $\alpha >0$\\ $\land \Bigl\{-2 \sqrt{\frac{6}{7}}\leq \lambda <-\sqrt{3}\lor \sqrt{3}<\lambda \leq 2 \sqrt{\frac{6}{7}}\Bigl\}$} \\

\hline

$H$ & 
\parbox[c]{7.5cm}{\centering $\Bigl\{-\alpha  \lambda ^2,\frac{1}{2} \left(\lambda ^2-6\right),\frac{1}{2} \left(\lambda ^2-4\right),\lambda ^2-3\Bigl\}$} & 
\parbox[c]{7.5cm}{\centering Stable for $\alpha >0 $\\$\land \Bigl\{-\sqrt{3}<\lambda <0\lor 0<\lambda <\sqrt{3}\Bigl\}$} \\

\hline

$I$ & 
\parbox[c]{7.5cm}{\centering $\Bigl\{-\alpha  \lambda ^2,\frac{1}{2} \left(\lambda ^2-6\right),\frac{1}{2} \left(\lambda ^2-4\right),\lambda ^2-3\Bigl\}$} & 
\parbox[c]{8cm}{\centering Stable for $\alpha >0$\\$\land \Bigl\{-\sqrt{3}<\lambda <0\lor 0<\lambda <\sqrt{3}\Bigl\}$} \\

\hline

$J$ & 
\parbox[c]{7.5cm}{\centering $\Bigl\{-1,1,0,-4 \alpha \Bigl\}$} & 
\parbox[c]{8cm}{\centering Unstable} \\

\hline

$K$ & 
\parbox[c]{7.5cm}{\centering $\Bigl\{-1,1,0,-4 \alpha \Bigl\}$} & 
\parbox[c]{7.5cm}{\centering Unstable} \\

\hline

$L$ & 
\parbox[c]{7.5cm}{\centering $\Bigl\{0,1,\frac{1}{2} \left(4-\sqrt{6} \lambda  \eta \right),-\frac{\eta ^3+\eta }{\eta  \left(\eta ^2+1\right)}\Bigl\}$} & 
\parbox[c]{7.5cm}{\centering Unstable} \\

\hline

$M$ & 
\parbox[c]{7.5cm}{\centering $\Bigl\{0,1,\frac{1}{2} \left(4-\sqrt{6} \lambda  \eta \right),-\frac{\eta ^3+\tau }{\eta  \left(\eta ^2+1\right)}\Bigl\}$} & 
\parbox[c]{8cm}{\centering Unstable} \\

\hline

$N$ & 
\parbox[c]{7.5cm}{\centering $\Bigl\{\frac{1-2 \epsilon }{\epsilon +1},\frac{3 \epsilon }{\epsilon +1},1,2\Bigl\}$} & 
\parbox[c]{8cm}{\centering Unstable} \\

\hline

$O$ & 
\parbox[c]{7.5cm}{\centering $\Bigl\{\frac{1-2 \epsilon }{\epsilon +1},\frac{3 \epsilon }{\epsilon +1},1,2\Bigl\}$} & 
\parbox[c]{8cm}{\centering Unstable} \\

\hline

$P$ & 
\parbox[c]{7.5cm}{\centering $\Bigl\{0,-\frac{3}{2},-\frac{1}{2},\frac{1}{2} \left(3-\sqrt{6} \lambda  \sigma \right)\Bigl\}$} & 
\parbox[c]{7.5cm}{\centering Stable for \\$\Bigl\{\lambda <0\land \sigma <\frac{\sqrt{\frac{3}{2}}}{\lambda }\Bigl\}\lor \Bigl\{\lambda >0\land \sigma >\frac{\sqrt{\frac{3}{2}}}{\lambda }\Bigl\}$} \\
\hline
\end{tabular}}
\caption{Eigenvalues and stability for model \ref{sec:model-II}, for general $\alpha$.}
\label{CHITABLE-VIII}
\end{table}

\begin{figure}[H]
    \centering
    \includegraphics[width=60mm]{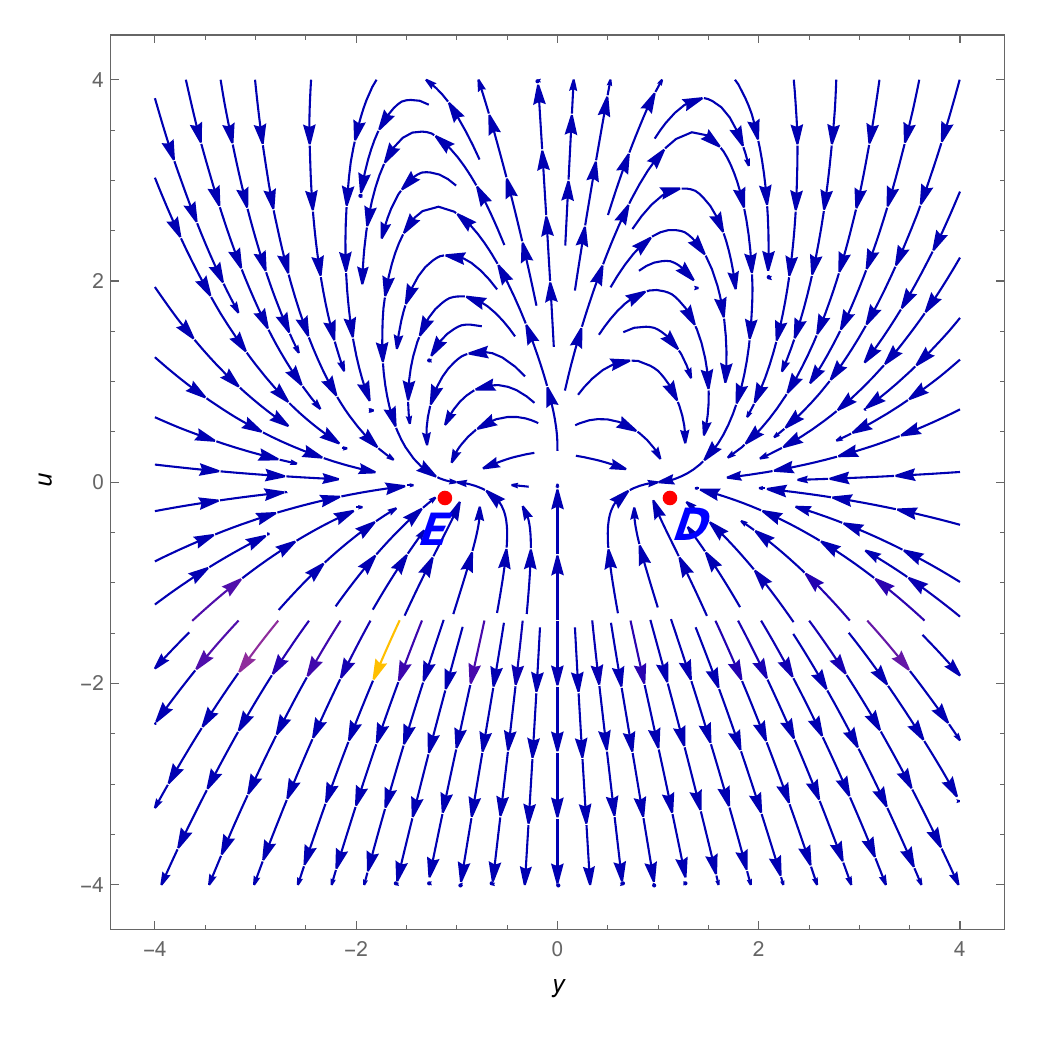}
    \includegraphics[width=60mm]{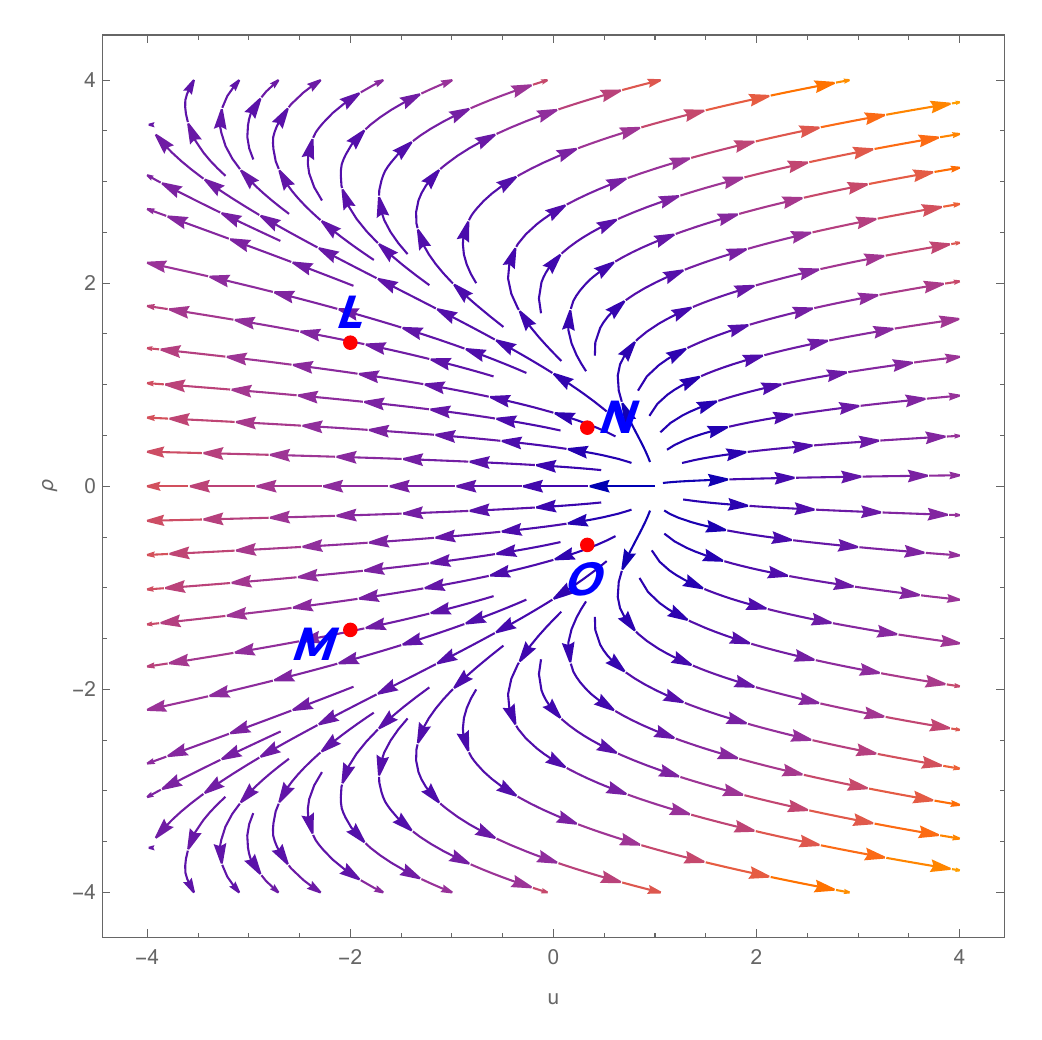}
    \caption{2D phase portrait for model  \ref{sec:model-II}, for general $\alpha$.} \label{CHIFig51}
\end{figure}

In above Fig. \ref{CHIFig5} and in Fig. \ref{CHIFig51}, the upper left plot is for $u=0$, $\rho=0$, $\lambda=\sqrt{\frac{2}{9}}$ and the upper right plot having parameter values $x=0$, $\rho=0$, $\sigma=1$, $\alpha=1.1$. The lower left phase portrait is for $x=0$ , $\rho=0$, $\alpha=1.1$, $\varphi=\frac{1}{2}$, the lower right phase portrait is for $u=0$, $\rho=0$, $\eta=1$, $\epsilon=\frac{1}{3}$.

\subsection{Case A: \texorpdfstring{$\alpha=1$}{}}

We have considered $\alpha=1$ in the action equation Eq. (\ref{eq:action_model_2}) to discuss the role of $\alpha$ in getting a number of critical points and their participation in describing different phases of Universe evolution. In this case, the evolution equations can be obtained by limiting the general $\alpha$ to $\alpha=1$ from Eqs. \eqref{eq:70} to \eqref{eq:72} the set of dynamical variables can be defined as follows
\begin{equation}\label{eq:65}
    x=\dfrac{\kappa\dot{\phi}}{\sqrt{6}H}\,,\quad y=\frac{\kappa\sqrt{V}}{\sqrt{3}H}\,,\quad u=\frac{3\kappa^2\dot{\phi}^3}{2H}\,,\quad \rho= \frac{\kappa\sqrt{\rho_{\rm r}}}{\sqrt{3}H}\,, \quad \lambda=\frac{-V^{'}(\phi)}{\kappa V(\phi)}\,,\quad \Gamma=\dfrac{V(\phi)V^{''}(\phi)}{V^{'}(\phi)^{2}}\,.
\end{equation}
These dynamical variables satisfy constraint equation Eq. (\ref{eq:57}), and the dynamical system in this case can be defined as follow,
\begin{align}
    \dfrac{dx}{dN} &= \frac{x \left(3 u^2+3 \rho ^2 u-y^2 \left(\sqrt{6} \lambda  (u-2) x+9 u+6 x^2\right)+15 u x^2+3 u+2 \rho ^2 x^2+6 x^4-6 x^2\right)}{u (u+4)+4 x^2}\,,\label{eq:model_2_csae_1_dx_dN}\\
    \dfrac{dy}{dN} &= -y \left(\frac{-3 u^2-2 u \left(\rho ^2+6 x^2+3\right)+u y^2 \left(\sqrt{6} \lambda  x+6\right)-2 x^2 \left(\rho ^2+3 x^2-3 y^2+3\right)}{u (u+4)+4 x^2}+\sqrt{\frac{3}{2}} \lambda  x\right)\,,\\
    \dfrac{du}{dN} &= \frac{u \left(3 u^2+5 \rho ^2 u-y^2 \left(\sqrt{6} \lambda  (u-6) x+15 u+6 x^2\right)+21 u x^2-3 u+2 \rho ^2 x^2+6 x^4-30 x^2\right)}{u (u+4)+4 x^2}\,,\\
    \frac{d\rho}{dN} &= \frac{\rho  \left(u^2+2 u \left(\rho ^2+6 x^2-1\right)-u y^2 \left(\sqrt{6} \lambda  x+6\right)+2 x^2 \left(\rho ^2+3 x^2-3 y^2-1\right)\right)}{u (u+4)+4 x^2}\,,
     \end{align}
    \begin{align}
    \dfrac{d\lambda}{dN} &= -\sqrt{6}(\Gamma-1)\lambda^{2}x\,.\label{eq:model_2_csae_1_dl_dN}
\end{align}

The critical points with the value of the deceleration parameter and $\omega_{tot}$ are presented in Table \ref{CHITABLE-IX}. From the table we can conclude that we get less number of critical points than the general case. The critical points $J$ and $K$ show deceleration parameter value $q=1$, hence describing the radiation-dominated era. The critical points $F$ and $G$ show similar cosmological implications in terms of the value of the deceleration parameter and the value of $\omega_{tot}$, representing a cold DM-dominated era. Amongst all the critical points, an accelerated phase of evolution can be described by the critical points $D$, $E$, $H$, and $I$. The critical points $D$ and $E$ describe de Sitter solution with $q=-1$ and critical points $H$ and $I$ deliver deceleration parameter value $q=-1+\frac{\lambda^2}{2}$ hence represent the DE, matter-dominated era. The critical points $B$ and $C$ give $q=2$; these points can describe stiff matter. The critical point $A$ shows a deceleration parameter value in the positive range; it can not describe the accelerated phase of the evolution.

Eigenvalues for all critical points are presented in Table \ref{TABLE-X} to analyse the stability conditions. The critical points $F$, $G$ show stability for the parametric values $-2 \sqrt{\frac{6}{7}}\leq \lambda <-\sqrt{3}\lor \sqrt{3}<\lambda \leq 2 \sqrt{\frac{6}{7}}$ and describe cold DM-dominated era. The critical points $H$ and $I$ show stable behaviour for the parametric values $-\sqrt{3}<\lambda <0\lor 0<\lambda <\sqrt{3}$ these critical points represent DE, matter-dominated era for any real values of $\lambda$. The critical points $J$ and $K$ are defined at $\lambda=2$, and the presence of eigenvalues with both positive and negative signatures leads to saddle points, hence unstable. The critical points $B$ and $C$ is also show saddle point behaviour and unstable. In this case, we get critical point $A$ with deceleration parameter value $q=\frac{4}{5}$ with the existence of positive eigenvalues at linear perturbation matrix, hence unstable in nature.

\begin{table}[H]
%\small\addtolength{\tabcolsep}{-4pt}
\centering
\scalebox{0.8}{\begin{tabular}{|c|c|c|c|c|c|c|c|} 
\hline\hline
\parbox[c][1.5cm]{2.5cm}{\centering Critical points} & 
\parbox[c]{2cm}{\centering $x_{c}$} & 
\parbox[c]{2cm}{\centering $y_{c}$} & 
\parbox[1.1]{1.1cm}{\centering $u_{c}$} & 
\parbox[c]{1.1cm}{\centering $\rho_{c}$} & 
\parbox[c]{2cm}{\centering Existence condition} & 
\parbox[c]{2.5cm}{\centering $q$} & 
\parbox[c]{2cm}{\centering $\omega_{tot}$} \\ [0.5ex] 

\hline\hline

$A$ & 
\parbox[c]{2cm}{\centering $0$} & 
\parbox[c]{2cm}{\centering $0$} & 
\parbox[c]{2cm}{\centering $1$} & 
\parbox[c]{1.1cm}{\centering $0$} &-&
\parbox[c]{2.5cm}{\centering $\frac{4}{5}$} & 
\parbox[c]{2cm}{\centering $\frac{1}{5}$} \\

\hline

$B$ & 
\parbox[c]{2cm}{\centering $1$} & 
\parbox[c]{2cm}{\centering $0$} & 
\parbox[c]{2cm}{\centering $0$} & 
\parbox[c]{1.1cm}{\centering $0$} &-&
\parbox[c]{2.5cm}{\centering $2$} & 
\parbox[c]{2cm}{\centering $1$} \\

\hline

$C$ & 
\parbox[c]{2cm}{\centering $-1$} & 
\parbox[c]{2cm}{\centering $0$} & 
\parbox[c]{2cm}{\centering $0$} & 
\parbox[c]{1.1cm}{\centering $0$} & -&
\parbox[c]{2.5cm}{\centering $2$} & 
\parbox[c]{2cm}{\centering $1$} \\

\hline

$D$ & 
\parbox[c]{2cm}{\centering $\sigma$} & 
\parbox[c]{2cm}{\centering $\sqrt{\sigma^2+1}$} & 
\parbox[c]{2cm}{\centering $-2 \sigma^2$} & 
\parbox[c]{1.1cm}{\centering $0$} & \parbox[c]{2cm}{\centering$\lambda=0$\\  $\sigma^3 - \sigma \neq 0$}& 
\parbox[c]{2.5cm}{\centering $-1$} & 
\parbox[c]{2cm}{\centering $-1$} \\

\hline

$E$,  & 
\parbox[c]{2cm}{\centering $\sigma $} & 
\parbox[c]{2cm}{\centering $-\sqrt{\sigma^2+1}$} & 
\parbox[c]{2cm}{\centering $-2 \sigma^2$} & 
\parbox[c]{1.1cm}{\centering $0$} &\parbox[c]{2cm}{\centering$\lambda=0$ \\ $\sigma^3 - \sigma \neq 0$}&
\parbox[c]{2.5cm}{\centering $-1$} & 
\parbox[c]{2cm}{\centering $-1$} \\

\hline

$F$ & 
\parbox[c]{2cm}{\centering $\frac{\sqrt{\frac{3}{2}}}{\lambda}$} & 
\parbox[c]{2cm}{\centering $\sqrt{\frac{3}{2}} \sqrt{\frac{1}{\lambda^2}}$} & 
\parbox[c]{2cm}{\centering $0$} & 
\parbox[c]{1.1cm}{\centering $0$} &-&
\parbox[c]{2.5cm}{\centering $\frac{1}{2}$} & 
\parbox[c]{2cm}{\centering $0$} \\

\hline

$G$ & 
\parbox[c]{2cm}{\centering $\frac{\sqrt{\frac{3}{2}}}{\lambda}$} & 
\parbox[c]{2cm}{\centering $-\sqrt{\frac{3}{2}} \sqrt{\frac{1}{\lambda^2}}$} & 
\parbox[c]{2cm}{\centering $0$} & 
\parbox[c]{1.1cm}{\centering $0$} & -&
\parbox[c]{2.5cm}{\centering $\frac{1}{2}$} & 
\parbox[c]{2cm}{\centering $0$} \\

\hline

$H$ & 
\parbox[c]{2cm}{\centering $\frac{\lambda}{\sqrt{6}}$} & 
\parbox[c]{2cm}{\centering $\sqrt{1-\frac{\lambda^2}{6}}$} & 
\parbox[c]{2cm}{\centering $0$} & 
\parbox[c]{1.1cm}{\centering $0$} & -&
\parbox[c]{2.5cm}{\centering $\frac{1}{2}\left(\lambda^2-2\right)$} & 
\parbox[c]{2cm}{\centering $-1+\frac{\lambda^2}{3}$} \\

\hline

$I$ & 
\parbox[c]{2cm}{\centering $\frac{\lambda}{\sqrt{6}}$} & 
\parbox[c]{2cm}{\centering $-\sqrt{1-\frac{\lambda^2}{6}}$} & 
\parbox[c]{2cm}{\centering $0$} & 
\parbox[c]{1.1cm}{\centering $0$} & -&
\parbox[c]{2.5cm}{\centering $\frac{1}{2}\left(\lambda^2-2\right)$} & 
\parbox[c]{2cm}{\centering $-1+\frac{\lambda^2}{3}$} \\

\hline

$J$ & 
\parbox[c]{2cm}{\centering $\sqrt{\frac{2}{3}}$} & 
\parbox[c]{2cm}{\centering $\frac{1}{\sqrt{3}}$} & 
\parbox[c]{2cm}{\centering $0$} & 
\parbox[c]{1.1cm}{\centering $0$} &$\lambda=2$ &
\parbox[c]{2.5cm}{\centering $1$} & 
\parbox[c]{2cm}{\centering $\frac{1}{3}$} \\

\hline

$K$ & 
\parbox[c]{2cm}{\centering $\sqrt{\frac{2}{3}}$} & 
\parbox[c]{2cm}{\centering $-\frac{1}{\sqrt{3}}$} & 
\parbox[c]{2cm}{\centering $0$} & 
\parbox[c]{1.1cm}{\centering $0$} &$\lambda=2$&
\parbox[c]{2.5cm}{\centering $1$} & 
\parbox[c]{2cm}{\centering $\frac{1}{3}$} \\
\hline
\end{tabular}}
\caption{Critical points for model \ref{sec:model-II}, $\alpha=1$.  }\label{CHITABLE-IX}
\end{table}

\begin{table}[H]
\small\addtolength{\tabcolsep}{-6pt}
\centering
\renewcommand{\arraystretch}{1.3} % Adjust row height
\scalebox{0.9}{
\begin{tabular}{|c|c|c|} 
\hline

\parbox[c][1.2cm]{2.5cm}{\centering Critical points} & 
\parbox[c]{9cm}{\centering  Eigenvalues} & 
\parbox[c]{4cm}{\centering Stability} \\ 

\hline

$A$ & 
\parbox[c]{9cm}{\centering $\Bigl\{\frac{9}{5}, \frac{6}{5}, \frac{3}{5}, -\frac{1}{5} \Bigl\}$} & 
\parbox[c]{3cm}{\centering Unstable} \\

\hline

$B$ & 
\parbox[c]{9cm}{\centering $\Bigl\{-12, 3, 1, \frac{1}{2} \Big(6 - \sqrt{6} \lambda \Big)\Bigl\}$} & 
\parbox[c]{3cm}{\centering Unstable} \\

\hline

$C$ & 
\parbox[c]{9cm}{\centering $\Bigl\{-12, 3, 1, \frac{1}{2} \Big(\sqrt{6} \lambda + 6 \Big)\Bigl\}$} & 
\parbox[c]{3cm}{\centering Unstable} \\

\hline

$D$ & 
\parbox[c]{9cm}{\centering $\Bigl\{0, -3, -3, -2 \Bigl\}$} & 
\parbox[c]{3cm}{\centering Stable} \\

\hline

$E$ & 
\parbox[c]{9cm}{\centering $\Bigl\{0, -3, -3, -2 \Bigl\}$} & 
\parbox[c]{3cm}{\centering Stable} \\

\hline

$F$ & 
\parbox[c]{9cm}{\centering $\Bigl\{-3, -\frac{1}{2}, \frac{3 \Big(-\lambda^2 - \sqrt{24 \lambda^2 - 7 \lambda^4}\Big)}{4 \lambda^2}, \frac{3 \Big(\sqrt{24 \lambda^2 - 7 \lambda^4} - \lambda^2 \Big)}{4 \lambda^2}\Bigl\}$} & 
\parbox[c]{3cm}{\centering Stable for\\ $-2 \sqrt{\frac{6}{7}} \leq \lambda < -\sqrt{3} \lor \sqrt{3} < \lambda \leq 2 \sqrt{\frac{6}{7}}$} \\

\hline

$G$ & 
\parbox[c]{9cm}{\centering $\Bigl\{-3, -\frac{1}{2}, \frac{3 \Big(-\lambda^2 - \sqrt{24 \lambda^2 - 7 \lambda^4}\Big)}{4 \lambda^2}, \frac{3 \Big(\sqrt{24 \lambda^2 - 7 \lambda^4} - \lambda^2 \Big)}{4 \lambda^2}\Bigl\}$} & 
\parbox[c]{3cm}{\centering Stable for \\$-2 \sqrt{\frac{6}{7}} \leq \lambda < -\sqrt{3} \lor \sqrt{3} < \lambda \leq 2 \sqrt{\frac{6}{7}}$} \\

\hline

$H$ & 
\parbox[c]{9cm}{\centering $\Bigl\{-\lambda^2, \frac{1}{2} \Big(\lambda^2 - 6 \Big), \frac{1}{2} \Big(\lambda^2 - 4 \Big), \lambda^2 - 3\Bigl\}$} & 
\parbox[c]{3cm}{\centering Stable for \\$-\sqrt{3} < \lambda < 0 \lor 0 < \lambda < \sqrt{3}$} \\

\hline

$I$ & 
\parbox[c]{9cm}{\centering $\Bigl\{-\lambda^2, \frac{1}{2} \Big(\lambda^2 - 6 \Big), \frac{1}{2} \Big(\lambda^2 - 4 \Big), \lambda^2 - 3\Bigl\}$} & 
\parbox[c]{3cm}{\centering Stable for \\$-\sqrt{3} < \lambda < 0 \lor 0 < \lambda < \sqrt{3}$} \\

\hline

$J$ & 
\parbox[c]{9cm}{\centering $\Bigl\{-4, -1, 1, 0 \Bigl\}$} & 
\parbox[c]{3cm}{\centering Unstable} \\

\hline

$K$ & 
\parbox[c]{9cm}{\centering $\Bigl\{-4, -1, 1, 0 \Bigl\}$} & 
\parbox[c]{3cm}{\centering Unstable} \\
\hline
\end{tabular}}
\caption{Eigenvalues and stability for  model \ref{sec:model-II}, $\alpha=1$.}
\label{TABLE-X}
\end{table}

\begin{figure}[H]
    \centering
    \includegraphics[width=60mm]{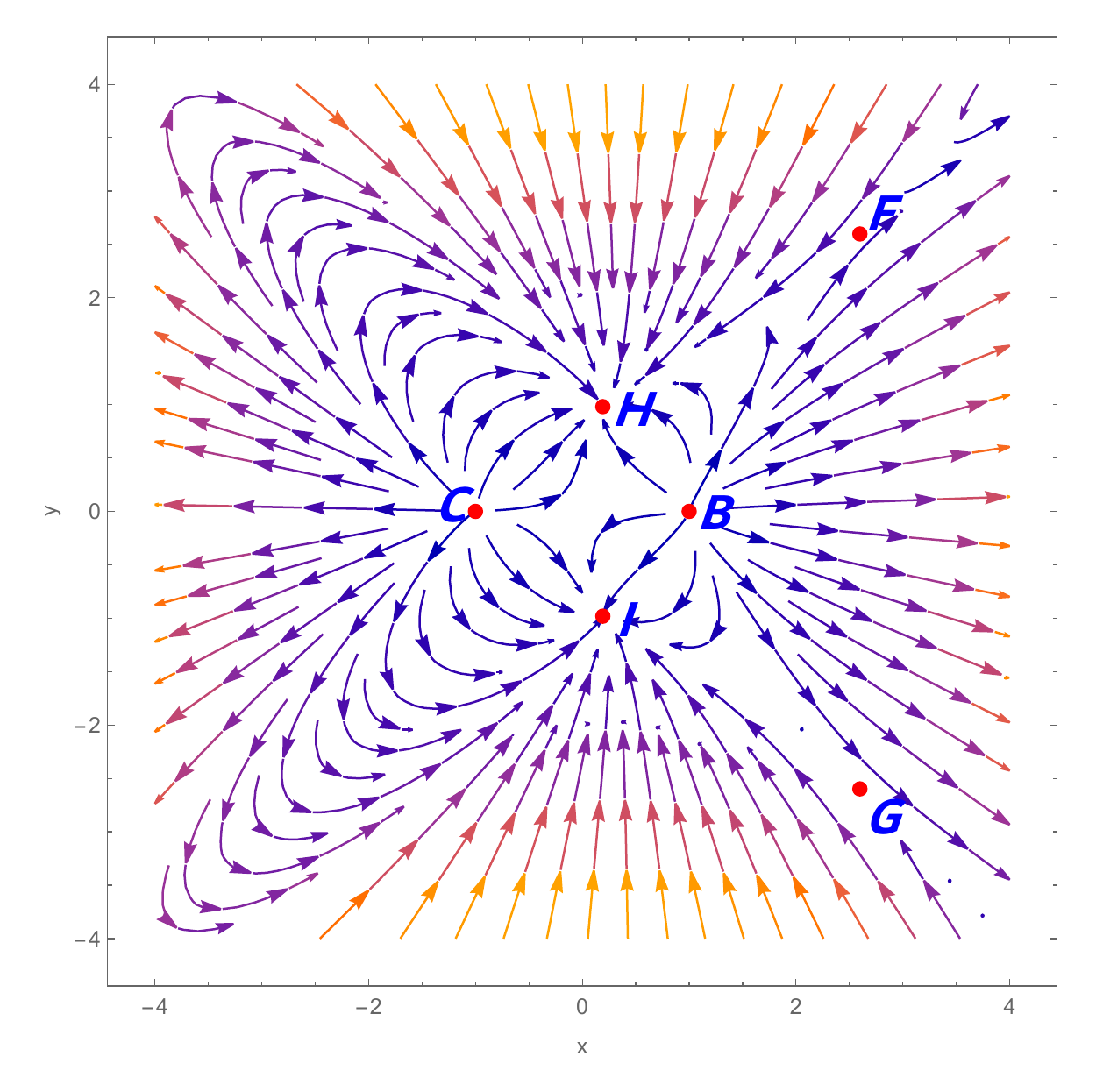}
    \includegraphics[width=60mm]{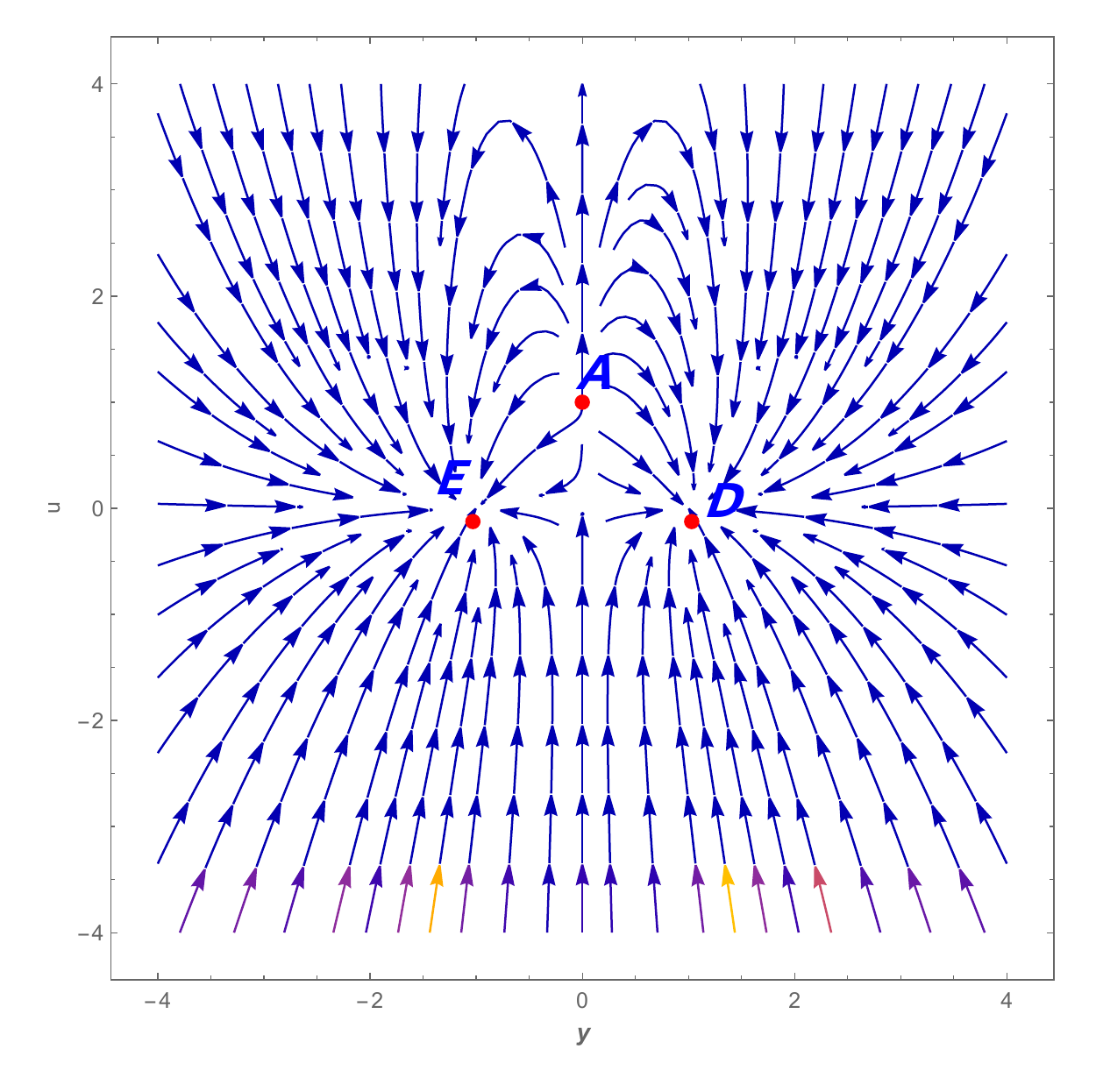}
    \caption{2D phase portrait for model \ref{sec:model-II}, $\alpha=1$.} \label{CHIFig6}
\end{figure}

\begin{figure}[H]
    \centering
    \includegraphics[width=60mm]{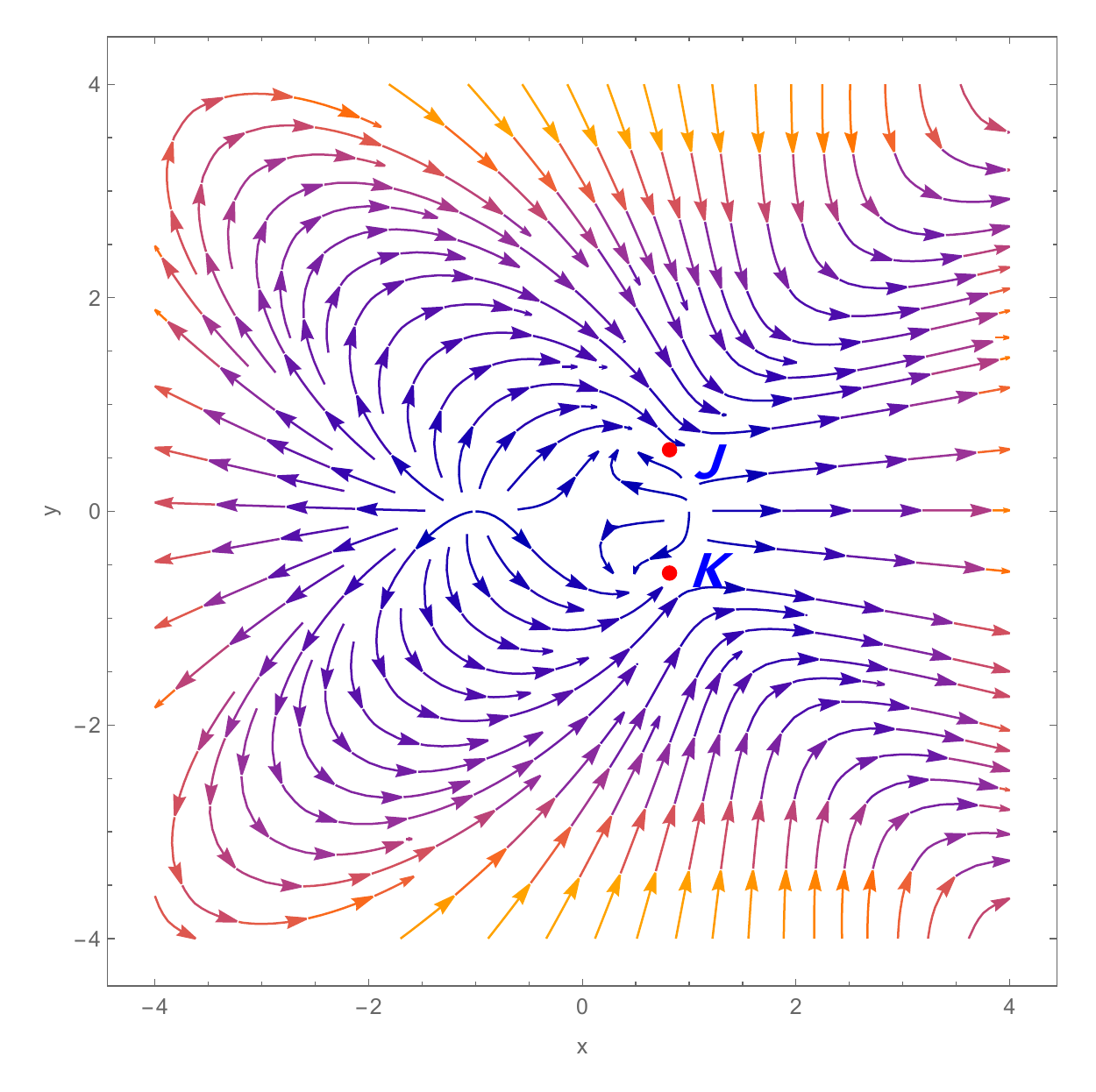}
    \caption{2D phase portrait for model \ref{sec:model-II} $\alpha=1$.}  \label{Fig7}
\end{figure}

Here, the left phase portrait is plotted for the parametric values $u=0$, $\rho=0$ and $\lambda=\sqrt{\frac{2}{9}}$ and the right side plot parametric values are  $x=0$, $\rho=0$, $\sigma=\frac{1}{4}$ for right side figure.
The phase space plots for the dynamical system presented in Eq. (\ref{eq:model_2_csae_1_dx_dN}--\ref{eq:model_2_csae_1_dl_dN}) are described in Figs. (\ref{CHIFig6}) and (\ref{Fig7}). The phase space analysis concludes that critical points $H$ and $I$ show attracting nature; the accelerating expansion of the Universe can be described at these critical points. The phase trajectories at critical points $B$ and $C$ are moving away from the critical point, hence showing unstable node behaviour leading to positive eigenvalues. The phase space plot for critical points $F$ and $G$ describe saddle point behaviour and represent the cold DM-dominated era. The phase plots at critical points $D$ and $E$ show attracting behaviour, and these critical points describe the de Sitter solution. The critical point $A$ is a saddle point and can be confirmed by observing phase trajectories at $A$. The plot in Fig. \ref{Fig7} also describes that phase space trajectories are moving away from critical points, hence showing instability.

\subsection{Case B: \texorpdfstring{$\alpha=2$}{}}

In this case, we have discussed dynamical system analysis for model \ref{sec:model-II}, $\alpha=2$. The evolution expressions can be obtained by limiting Eqs. \eqref{eq:70} to \eqref{eq:72} using $\alpha=2$. The dynamical variables to obtain an autonomous dynamical system can be defined as follows
\begin{equation}\label{eq:87}
    x=\dfrac{\kappa\dot{\phi}}{\sqrt{6}H}\,,\quad y=\frac{\kappa\sqrt{V}}{\sqrt{3}H}\,,\quad u=\dfrac{5\kappa^{2}\dot{\phi}^{5}}{4H}\,,\quad \rho= \frac{\kappa\sqrt{\rho_{\rm r}}}{\sqrt{3}H}\,,\quad \lambda=\frac{-V^{'}(\phi)}{\kappa V(\phi)}\,,\quad \Gamma=\dfrac{V(\phi)V^{''}(\phi)}{V^{'}(\phi)^{2}}\,.
\end{equation}
These dynamical variables satisfy constraint equation presented in Eq. (\ref{eq:57}), the dynamical system in this case is as follow
\begin{align}
    \frac{dx}{dN} &= \frac{x \left(3 u^2+5 \rho ^2 u-y^2 \left(\sqrt{6} \lambda  (u-2) x+15 u+6 x^2\right)+21 u x^2+9 u+2 \rho ^2 x^2+6 x^4-6 x^2\right)}{u (u+8)+4 x^2}\,,\label{eq:model_2_case_2_dx_dN}\\
    \frac{dy}{dN} &= -y \left(\frac{-3 u^2-2 u \left(2 \rho ^2+9 x^2+6\right)+u y^2 \left(\sqrt{6} \lambda  x+12\right)-2 x^2 \left(\rho ^2+3 x^2-3 y^2+3\right)}{u (u+8)+4 x^2}+\sqrt{\frac{3}{2}} \lambda  x\right)\,,\\
    \frac{du}{dN} &= \frac{u \left(3 u^2+9 \rho ^2 u-y^2 \left(\sqrt{6} \lambda  (u-10) x+27 u+6 x^2\right)+33 u x^2-3 u+2 \rho ^2 x^2+6 x^4-54 x^2\right)}{u (u+8)+4 x^2}\,,
    \end{align}
    \begin{align}
    \frac{d\rho}{dN} &= \frac{\rho  \left(u^2+2 u \left(2 \rho ^2+9 x^2-2\right)-u y^2 \left(\sqrt{6} \lambda  x+12\right)+2 x^2 \left(\rho ^2+3 x^2-3 y^2-1\right)\right)}{u (u+8)+4 x^2}\,,\\
    \frac{d\lambda}{dN} &= -\sqrt{6}(\Gamma-1)\lambda^{2}x\,.\label{eq:model_2_case_2_dl_dN}
\end{align}

The critical points with value of deceleration parameter and $\omega_{tot}$ for dynamical system in Eqs. \eqref{eq:model_2_case_2_dx_dN} to \eqref{eq:model_2_case_2_dl_dN} are presented in Table \ref{CHITABLE-XI}. From model \ref{sec:model-II} for critical point $A$ we get different positive deceleration parameter value for different values of $\alpha$. For $\alpha=2$ we are getting $q=\frac{2}{3}$ and $\omega_{tot}=\frac{1}{9}$. The critical points $D$ and $E$ represent the de Sitter solution to the system and are defined for $\lambda=0$. While critical points $H$ deliver deceleration parameter value $q=-1+\frac{\lambda^2}{2}$, which may describe the DE-dominated era of the Universe. Critical points $F$ and $G$ represent cold DM-dominated era with $\omega_{tot}=0$. The critical points $J$ and $K$ are defined for $\lambda=2$ and describe the radiation-dominated era of the Universe evolution. The critical points $B$ and $C$ behave as stiff matter with $\omega_{tot}=1$.

The stability conditions for this case $\alpha=2$ are presented in Table \ref{TABLE-XII}. Table observations conclude that critical points $H$ and $I$ show stability for parametric condition $-\sqrt{3}<\lambda <0\lor 0<\lambda <\sqrt{3}$ and can describe DE, matter-dominated era. The critical points $F$ and $G$ show stability at $-2 \sqrt{\frac{6}{7}}\leq \lambda <-\sqrt{3}\lor \sqrt{3}<\lambda \leq 2 \sqrt{\frac{6}{7}}$ and describe cold DM-dominated era. The critical points $A$, $B$, and $C$ with deceleration parameters in the positive range are unstable since eigenvalues have both positive and negative signatures. The critical points $D$ and $E$ represent the de Sitter solution and show stable behaviour. The critical points $J$ and $K$ represent the radiation-dominated era at $\lambda=2$ and are unstable in nature.
\begin{figure}[H]
    \centering
    \includegraphics[width=60mm]{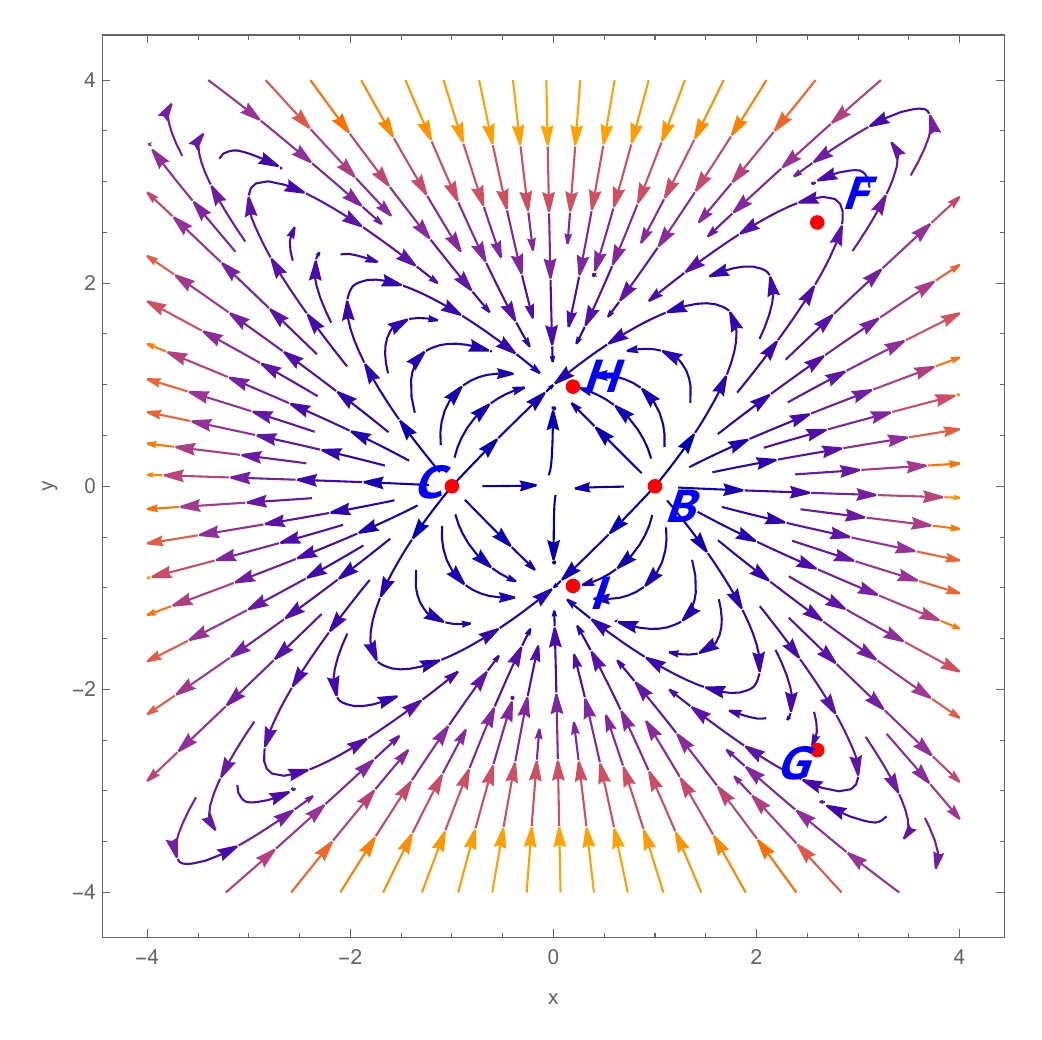}
    \includegraphics[width=60mm]{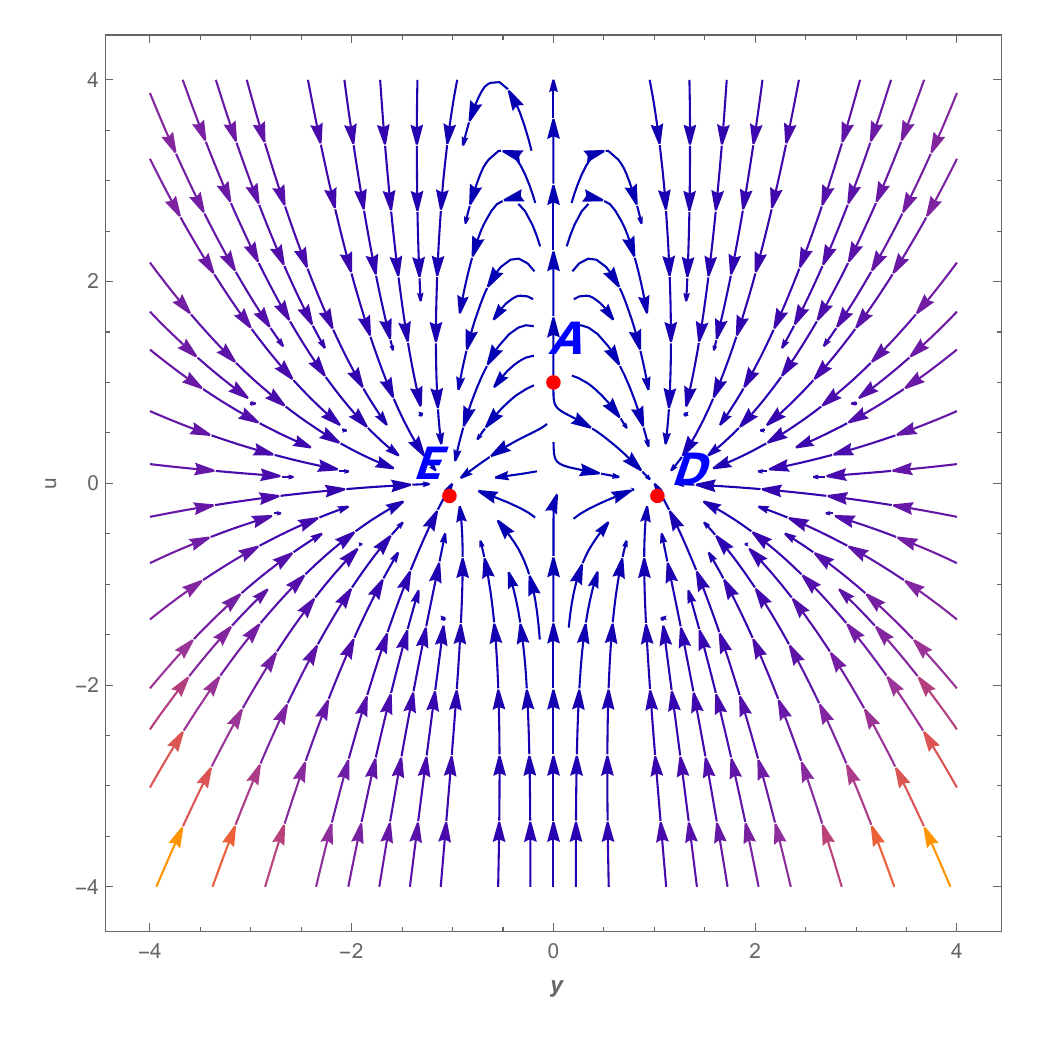}
    \caption{2D phase portrait for model \ref{sec:model-II}, $\alpha=2$.} \label{Fig8}
\end{figure}
Here in Fig. \ref{Fig8} left is plotted for the parametric values $u=0$, $\rho=0$ and $\lambda=\sqrt{\frac{2}{9}}$ and for right side plot parametric values are  $x=0$, $\rho=0$, $\sigma=\frac{1}{4}$. The phase portrait in Fig. \ref{Fig8-2}, for the parametric values $u=0$, $\rho=0$, $\lambda=2$.

The phase space diagram for dynamical system Eqs. \eqref{eq:model_2_case_2_dx_dN} to \eqref{eq:model_2_case_2_dl_dN} are plotted in Fig. \ref{Fig8} and \ref{Fig8-2}. The critical points $D$ and $E$ show attracting stable de Sitter solution and represent a DE-dominated era. While at critical point $A$, phase space trajectories are moving away, hence unstable addressing saddle point behaviour.
\begin{figure}[H]
    \centering
    \includegraphics[width=60mm]{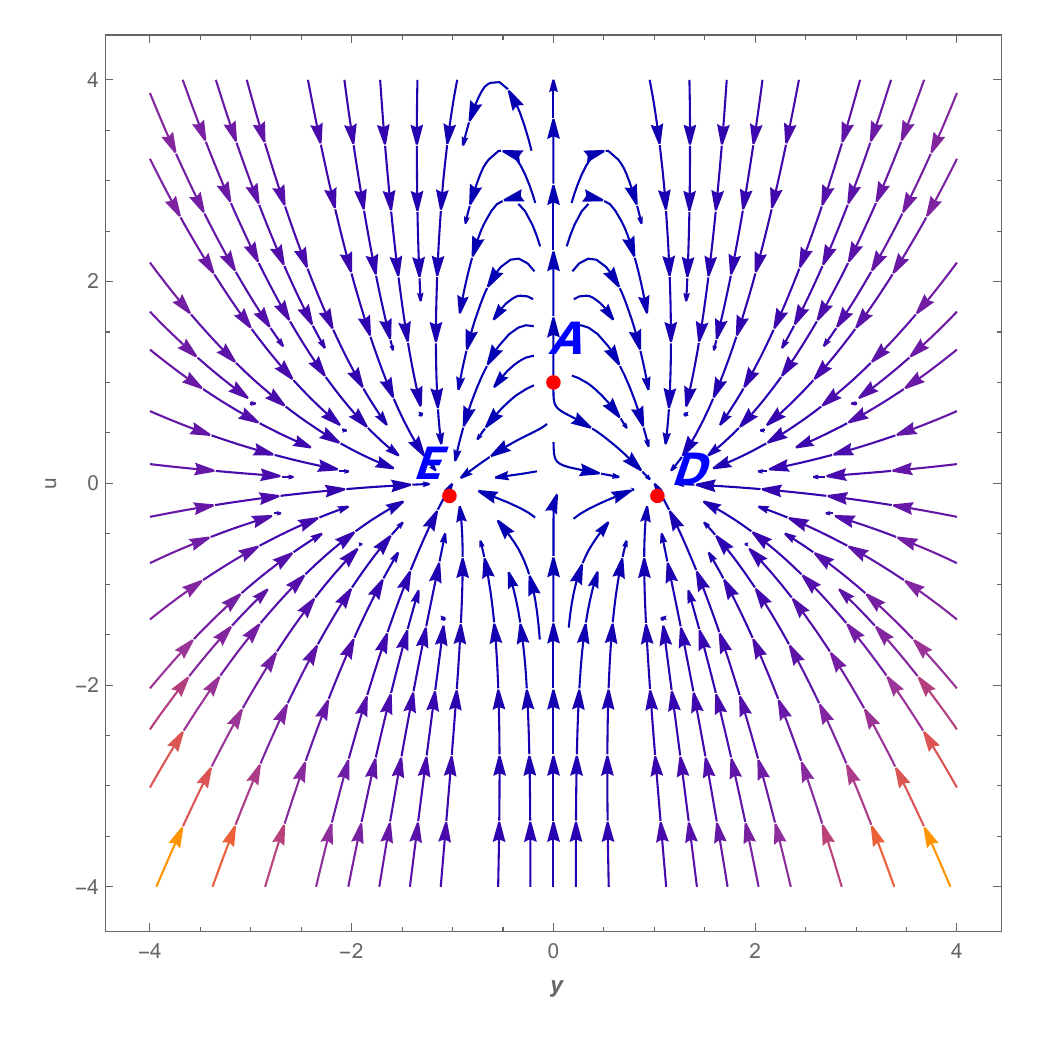}
    \caption{2D phase portrait for model \ref{sec:model-II}, $\alpha=2$.} \label{Fig8-2}
\end{figure}
\begin{table}[H]
%\small\addtolength{\tabcolsep}{-3pt} 
\centering
\scalebox{0.85}{\begin{tabular}{|c|c|c|c|c|c|c|c|} 
\hline\hline
\parbox[c][1.8cm]{2cm}{\centering Critical points} & 
\parbox[c]{2cm}{\centering $x_{c}$} & 
\parbox[c]{2cm}{\centering $y_{c}$} & 
\parbox[c]{2cm}{\centering $u_{c}$} & 
\parbox[c]{2cm}{\centering $\rho_{c}$} & 
\parbox[c]{2cm}{\centering Existence condition} & 
\parbox[c]{2cm}{\centering $q$} & 
\parbox[c]{2cm}{\centering $\omega_{tot}$} \\ [0.5ex]

\hline\hline

$A$ & 
\parbox[c]{2cm}{\centering $0$} & 
\parbox[c]{2cm}{\centering $0$} & 
\parbox[c]{2cm}{\centering $1$} & 
\parbox[c]{2cm}{\centering $0$} & -&
\parbox[c]{2cm}{\centering $\frac{2}{3}$} & 
\parbox[c]{2cm}{\centering $\frac{1}{9}$} \\

\hline

$B$ & 
\parbox[c]{2cm}{\centering $1$} & 
\parbox[c]{2cm}{\centering $0$} & 
\parbox[c]{2cm}{\centering $0$} & 
\parbox[c]{2cm}{\centering $0$} & -&
\parbox[c]{2cm}{\centering $2$} & 
\parbox[c]{2cm}{\centering $1$} \\

\hline

$C$ & 
\parbox[c]{2cm}{\centering $-1$} & 
\parbox[c]{2cm}{\centering $0$} & 
\parbox[c]{2cm}{\centering $0$} & 
\parbox[c]{2cm}{\centering $0$} & -&
\parbox[c]{2cm}{\centering $2$} & 
\parbox[c]{2cm}{\centering $1$} \\

\hline

$D$  & 
\parbox[c]{2cm}{\centering $\sigma$} & 
\parbox[c]{2cm}{\centering $\sqrt{\sigma^2+1}$} & 
\parbox[c]{2cm}{\centering $-2\sigma^2$} & 
\parbox[c]{2cm}{\centering $0$} &\parbox[c]{2cm}{\centering$\lambda=0$,\\$\sigma^3-3\sigma\neq0$}&
\parbox[c]{2cm}{\centering $-1$} & 
\parbox[c]{2cm}{\centering $-1$} \\

\hline

$E$ & 
\parbox[c]{2cm}{\centering $\sigma$} & 
\parbox[c]{2cm}{\centering $-\sqrt{\sigma^2+1}$} & 
\parbox[c]{2cm}{\centering $-2\sigma^2$} & 
\parbox[c]{2cm}{\centering $0$} &\parbox[c]{2cm}{\centering $\lambda=0, $\\$\sigma^3-3\sigma\neq0$}&
\parbox[c]{2cm}{\centering $-1$} & 
\parbox[c]{2cm}{\centering $-1$} \\

\hline

$F$ & 
\parbox[c]{2cm}{\centering $\frac{\sqrt{\frac{3}{2}}}{\lambda}$} & 
\parbox[c]{2cm}{\centering $\sqrt{\frac{3}{2}}\sqrt{\frac{1}{\lambda^2}}$} & 
\parbox[c]{2cm}{\centering $0$} & 
\parbox[c]{2cm}{\centering $0$} & -&
\parbox[c]{2cm}{\centering $\frac{1}{2}$} & 
\parbox[c]{2cm}{\centering $0$} \\

\hline

$G$ & 
\parbox[c]{2cm}{\centering $\frac{\sqrt{\frac{3}{2}}}{\lambda}$} & 
\parbox[c]{2cm}{\centering $-\sqrt{\frac{3}{2}}\sqrt{\frac{1}{\lambda^2}}$} & 
\parbox[c]{2cm}{\centering $0$} & 
\parbox[c]{2cm}{\centering $0$} & -&
\parbox[c]{2cm}{\centering $\frac{1}{2}$} & 
\parbox[c]{2cm}{\centering $0$} \\

\hline

$H$ & 
\parbox[c]{2cm}{\centering $\frac{\lambda}{\sqrt{6}}$} & 
\parbox[c]{2cm}{\centering $\sqrt{1-\frac{\lambda^2}{6}}$} & 
\parbox[c]{2cm}{\centering $0$} & 
\parbox[c]{2cm}{\centering $0$} & -&
\parbox[c]{2cm}{\centering $\frac{1}{2}\Big(\lambda^2-2\Big)$} & 
\parbox[c]{2cm}{\centering $-1 + \frac{\lambda^2}{3}$} \\

\hline

$I$ & 
\parbox[c]{2cm}{\centering $\frac{\lambda}{\sqrt{6}}$} & 
\parbox[c]{2cm}{\centering $-\sqrt{1-\frac{\lambda^2}{6}}$} & 
\parbox[c]{2cm}{\centering $0$} & 
\parbox[c]{2cm}{\centering $0$} & -&
\parbox[c]{2cm}{\centering $\frac{1}{2}\Big(\lambda^2-2\Big)$} & 
\parbox[c]{2cm}{\centering $-1 + \frac{\lambda^2}{3}$} \\

\hline

$J$& 
\parbox[c]{2cm}{\centering $\sqrt{\frac{2}{3}}$} & 
\parbox[c]{2cm}{\centering $-\frac{1}{\sqrt{3}}$} & 
\parbox[c]{2cm}{\centering $0$} & 
\parbox[c]{2cm}{\centering $0$} &$\lambda=2$ &
\parbox[c]{2cm}{\centering $1$} & 
\parbox[c]{2cm}{\centering $\frac{1}{3}$} \\

\hline

$K$  & 
\parbox[c]{2cm}{\centering $\sqrt{\frac{2}{3}}$} & 
\parbox[c]{2cm}{\centering $\frac{1}{\sqrt{3}}$} & 
\parbox[c]{2cm}{\centering $0$} & 
\parbox[c]{2cm}{\centering $0$} &$\lambda=2$ &
\parbox[c]{2cm}{\centering $1$} & 
\parbox[c]{2cm}{\centering $\frac{1}{3}$} \\
\hline
\end{tabular}}
\caption{Critical points for model \ref{sec:model-II}, $\alpha=2$.  }
\label{CHITABLE-XI}
\end{table}
Similarly, critical points $J$ and $K$ are showing unstable behaviour representing a radiation-dominated era. The upper right plot is presented for critical points $B$, $C$, $H$, $I$, $F$ and $G$. The critical points $B$ and $C$ behave as unstable nodes with respect to the existence of positive eigenvalues. The phase space trajectories are moving away from critical points $F$ and $G$; these critical points represent unstable behaviour, and critical points $H$ and $I$ display attracting behaviour of phase space trajectories and show consistency with current observational studies.

\begin{table}[H]
%\small\addtolength{\tabcolsep}{-6pt}
\centering
\renewcommand{\arraystretch}{1.3} % Adjust row height
\scalebox{0.8}{
\begin{tabular}{|c|c|c|}
\hline
\parbox[c]{3cm}{\centering Critical points} & 
\parbox[c]{7cm}{\centering Eigenvalues} & 
\parbox[c]{3cm}{\centering Stability} \\ [0.5ex]

\hline

$A$ & 
\parbox[c]{7cm}{\centering $\Bigl\{\frac{5}{3}, \frac{4}{3}, -\frac{1}{3}, \frac{1}{3}\Bigl\}$} & 
\parbox[c]{3cm}{\centering Unstable} \\

\hline

$B$ & 
\parbox[c]{7cm}{\centering $\Bigl\{-12, 3, 1, \frac{1}{2} \Big(6 - \sqrt{6} \lambda\Big)\Bigl\}$} & 
\parbox[c]{3cm}{\centering Unstable} \\

\hline

$C$ & 
\parbox[c]{7cm}{\centering $\Bigl\{-12, 3, 1, \frac{1}{2} \Big(\sqrt{6} \lambda + 6\Big)\Bigl\}$} & 
\parbox[c]{3cm}{\centering Unstable} \\

\hline

$D$ & 
\parbox[c]{7cm}{\centering $\Bigl\{0, -3, -3, -2\Bigl\}$} & 
\parbox[c]{3cm}{\centering Stable} \\

\hline

$E$ & 
\parbox[c]{7cm}{\centering $\Bigl\{0, -3, -3, -2\Bigl\}$} & 
\parbox[c]{3cm}{\centering Stable} \\

\hline

$F$ & 
\parbox[c]{7cm}{\centering $\Bigl\{-6, -\frac{1}{2}, \frac{3\Big(-\lambda^2 - \sqrt{24 \lambda^2 - 7 \lambda^4}\Big)}{4 \lambda^2},$\\$ \frac{3 \Big(\sqrt{24 \lambda^2 - 7 \lambda^4} - \lambda^2\Big)}{4 \lambda^2}\Bigl\}$} & 
\parbox[c]{3cm}{\centering Stable for $-2\sqrt{\frac{6}{7}} \leq \lambda < -\sqrt{3} \lor \sqrt{3} < \lambda \leq 2\sqrt{\frac{6}{7}}$} \\

\hline

$G$ & 
\parbox[c]{7cm}{\centering $\Bigl\{-6, -\frac{1}{2}, \frac{3\Big(-\lambda^2 - \sqrt{24 \lambda^2 - 7 \lambda^4}\Big)}{4 \lambda^2},$\\$ \frac{3 \Big(\sqrt{24 \lambda^2 - 7 \lambda^4} - \lambda^2\Big)}{4 \lambda^2}\Bigl\}$} & 
\parbox[c]{3cm}{\centering Stable for $-2\sqrt{\frac{6}{7}} \leq \lambda < -\sqrt{3} \lor \sqrt{3} < \lambda \leq 2\sqrt{\frac{6}{7}}$} \\

\hline

$H$ & 
\parbox[c]{7cm}{\centering $\Bigl\{-2 \lambda^2, \frac{1}{2} \Big(\lambda^2 - 6\Big), \frac{1}{2} \Big(\lambda^2 - 4\Big), \lambda^2 - 3\Bigl\}$} & 
\parbox[c]{3cm}{\centering Stable for $-\sqrt{3} < \lambda < 0 \lor 0 < \lambda < \sqrt{3}$} \\

\hline

$I$ & 
\parbox[c]{7cm}{\centering $\Bigl\{-2 \lambda^2, \frac{1}{2} \Big(\lambda^2 - 6\Big), \frac{1}{2} \Big(\lambda^2 - 4\Big), \lambda^2 - 3\Bigl\}$} & 
\parbox[c]{3cm}{\centering Stable for $-\sqrt{3} < \lambda < 0 \lor 0 < \lambda < \sqrt{3}$} \\

\hline

$J$ & 
\parbox[c]{7cm}{\centering $\Bigl\{-8, -1, 1, 0\Bigl\}$} & 
\parbox[c]{3cm}{\centering Unstable} \\

\hline

$K$ & 
\parbox[c]{7cm}{\centering $\Bigl\{-8, -1, 1, 0\Bigl\}$} & 
\parbox[c]{3cm}{\centering Unstable} \\
\hline
\end{tabular}}
\caption{Eigenvalues and stability for \ref{sec:model-II}, $\alpha=2$.}
\label{TABLE-XII}
\end{table}
 
\section{Conclusion}\label{conclusionch2}
In this chapter, we have explored the cosmological dynamics of DE through the prism of scalar-torsion gravity \cite{Paliathanasis:2021nqa,Leon:2022oyy} in the context of power-law couplings with the kinetic term. In particular, we have studied one nontrivial extension of the recently proposed teleparallel analogue of Horndeski gravity 

For our FLRW background cosmology, we explore these models in the presence of both radiation and cold DM through the density parameters in Eq. (\ref{densityparameters}). The Friedmann equations in Eqs. \eqref{eq:30}, \eqref{eq:31} and Klein-Gordon equation in Eq. (\ref{eq:kg_eq}) directly lead to a set of autonomous equations for each of the models under investigation. These are then used in each case to derive the critical points of the particular cosmologies from which we can expose the model behaviour using the dynamical analysis in the parameter phase space. This also opens the doorway to understand the stability of the models in question. In the first model for the action in Eq. (\ref{eq:action_model_1}), we utilize the dynamical variables defined in Eq. (\ref{eq:49}), which using the constraint in Eq. (\ref{eq:50}) together with the equations of motion, are then used to derive the system of autonomous equations given in Eqs. \eqref{eq:dx_dN_model_1} to \eqref{eq:dl_dN_model_1} which express the behaviour of the model in phase space. The critical points are then arrived at by imposing that each of these derivatives vanishes. These first order equations of motion of the dynamical variables are represented as derivatives with respect to $N = \ln a$ which shows the behaviour of the system in a more direct way. The result of this analysis is shown in Table \ref{CHITABLEI}. In this table, we show the values of the dynamical variables at which these critical points occur together with the value of the deceleration and EoS parameters, which already show an indication of the cosmological behaviour at those points in the evolution of the cosmological model. Each critical point is then further analysed for stability in Table \ref{CHITABLE-II}. In some circumstances, stability occurs for a smaller change of parameter values, as described in the last column of the table. For transparency, we also show the corresponding Eigenvalues at each critical point. In Fig. \ref{Ch1Fig1}, we show the phase portraits of this model for four specific examples of representative parameter values. In these plots, the nature of the critical points is further exposed through their impact on the contours of evolution. We further probe the behaviour of this model in the specific cases of the kinetic term being linear and quadratic, which represent the first cases that a Taylor expansion would open to. We do this in Sec \ref{sec:model_1_case_1} and Sec \ref{sec:model_1_case_2} respectively. We also show the phase portraits for these cases in Fig. \ref{Ch1Fig2} and Fig.~\ref{Ch1Fig31}, for the two respective cases. Finally, we show the nuanced critical points for both cases in Tables  \ref{CHITABLE-III}, \ref{CHITABLE-V} and their respective stability in Tables.~\ref{CHITABLE-IV}, \ref{CHITABLE-VI}.

In the second model, we explore the coupling between a power-law kinetic term and $I_2$ scalar written in Eq. (\ref{eq:I_2_scalar}). This scalar represents the only nonvanishing term that is linear in its contractions with the torsion tensor. In this case, we take the action Eq. (\ref{eq:action_model_2}), where TEGR and the canonical scalar field are complemented by this new coupling term together with the matter and radiation contributions. This leads directly to the Friedmann equations in Eq. (\ref{eq:70}, \ref{eq:71}), and the Klein-Gordon equation in Eq. (\ref{eq:72}). Now, by defining the dynamical variables in Eq. (\ref{eq:model_2_dynamic_var}) we can explore the dynamical system that the background cosmology represents through the system of linear autonomous differential equations given in Eqs. \eqref{eq:model_2_dx_dN} to \eqref{eq:model_2_dl_dN}. By performing a similar critical point analysis as in the first case, we find the critical points as listed in Table \ref{CHITABLE-VII}. These are then further analysed for their stability nature in Table \ref{CHITABLE-VIII}. As in the first case, we find a rich structure of critical points for the various model parameter values, which we show through the phase portrait in Fig. \ref{CHIFig5} and in Fig. \ref{CHIFig51}. As in the first model, we consider the cases where the power index takes linear and quadratic forms in the model action. For the linear case, the dynamical system is represented by Eqs. \eqref{eq:model_2_csae_1_dx_dN} to \eqref{eq:model_2_csae_1_dl_dN} which produce the critical points in Table \ref{CHITABLE-IX} with natures shown in Table \ref{TABLE-X}. This interesting scenario produces the phase portraits given in Figs. (\ref{CHIFig6}) and (\ref{Fig7}). On the other hand, the quadratic case is represented through the dynamical system given in Eqs. \eqref{eq:model_2_case_2_dx_dN} to \eqref{eq:model_2_case_2_dl_dN}. The corresponding critical point analysis produces Table \ref{CHITABLE-XI} with stability conditions described in Table \ref{TABLE-XII}. The final phase portraits for this case are then shown in Fig. \ref{Fig8} and \ref{Fig8-2}.

If we glance at a study made in Ref. \cite{Gonzalez-Espinoza:2020jss}, we can easily compare the cosmological implications and stability conditions of critical points from Ref. \cite{Gonzalez-Espinoza:2020jss} and Table \ref{CHITABLEI} for the model-I general $\alpha$ case. From this comparison, we can note that we get four more critical points than in Ref. \cite{Gonzalez-Espinoza:2020jss}, this may be possible because of the model construction and background formalism in the teleparallel Horndeski theory. This comparison allows us to know more about minor differences; in our study, we get four critical points that can describe the DE era that is critical points  $H, I, D, E$. 

% Chapter 3

\chapter{Dynamical system analysis for scalar field potential in teleparallel gravity} % Main chapter title

\label{Chapter3} % For referencing the chapter elsewhere, use \ref{Chapter1} 

\lhead{Chapter 3. \emph{Dynamical system analysis for scalar field potential in teleparallel gravity}} % This is for the header on each page - perhaps a shortened title

\vspace{10 cm}
*The work in this chapter is covered by the following publication:\\

 \textbf{S. A. Kadam}, Ananya Sahu, S. K. Tripathy, B. Mishra, ``Dynamical system analysis for scalar field potential in teleparallel gravity", \textit{European Physical Journal C}, \textbf{84}, 1088 (2024).

%----------------------------------------------------------------------------------------
%\section{Summary of Results}
\clearpage
\section{Introduction}\label{CH3Intro}
In the previous chapter, chapter \ref{Chapter2}, we have seen the dynamical system approach in detail for one of the most general teleparallel analogous to the Horndeski theory of gravity formalisms. There we have seen interesting points describing different phases of the evolution of the Universe. Continuing further, this work highlights the contribution of the teleparallel power law model of exponential and the power law scalar field potentials in the extension of teleparallel gravity.  Indeed, when the contribution from $F(\phi)=0$,
the models reduce to the standard TEGR with a scalar field. In our analysis, we explicitly considered the case $F\rightarrow
0$. While this limit simplifies the model and some auxiliary variables vanish, the overall consistency and dynamics of the remaining system were carefully examined. The reduction to TEGR with a scalar field does not introduce any new complications; our results and equations derived are consistent in this limit \cite{copelandLiddle}. However, we focused on how the presence of $F\ne0$ enriches the dynamics, which is the core novelty of the present work.

\section{\texorpdfstring{$f(T,\phi)$}{} gravity field equations}\label{Backgroundformalism}
Varying the action in Eq.~\eqref{ActionEqf(Tphi)} with respect to the tetrad field presented in Eq. \eqref{FLRWTETRAD} and the scalar field $\phi$, the field equations of $f(T,\phi)$ gravity are obtained as \cite{Hohmann:2018rwf},
\begin{eqnarray}
f(T,\phi)-P(\phi)X-2Tf,_{T}&=&\rho_{m}+\rho_{r}\,, \label{FE1}\\
f(T,\phi)+P(\phi)X-2Tf,_{T}-4\dot{H}f,_{T}-4H\dot{f},_{T} &=& -p_{r}\,,\label{FE2}\\
-P,_{\phi}X-3P(\phi)H\dot{\phi}-P(\phi)\ddot{\phi}+f,_{\phi}&=&0\label{KleinGordon1}\,. 
\end{eqnarray}

For brevity, we denote $f(T,\phi) \equiv f$ and $\frac{\partial f}{\partial T}=f_{, T}$.  $\rho_{m}$, $\rho_{r}$, and $p_{r}$ respectively represent the matter-energy density, radiation energy density, and radiation pressure.  We refer the non-minimal coupling function $f(T,\phi)$ in the form \cite{Hohmann:2018rwf},

\begin{equation}\label{Generalmodel}
f(T,\phi)=-\frac{T}{2\kappa^{2}}-F(\phi)G(T)-V(\phi)\,,
\end{equation}
where $V(\phi)$ is the scalar potential. $F(\phi)$ and $G(T)$ are respective suitable functions of the scalar field and torsion scalar $T$. 
With such a choice of the functional $f(T,\phi)$, the field Eqs. \eqref{FE1} and \eqref{KleinGordon1} reduce to
\begin{align}
\frac{3}{\kappa^2}H^2& =P(\phi)X+V(\phi)-2TG_{,T} F(\phi)+G(T) F(\phi)+\rho_m+\rho_r,\label{FEmr1}\\
-\frac{2}{\kappa^2}\dot{H}& =2P(\phi)X+4\dot{H} G_{,T}F(\phi) +4HG_{,TT} \dot{T} F(\phi) +4HG_{,T} \dot{F}+\rho_m+\frac{4}{3}\rho_r,\label{FEmr2}\\
&P(\phi)\ddot{\phi}+3P(\phi)H\dot{\phi} +P_{,\phi} X +G(T) F_{,\phi} +V_{,\phi}=0\label{KleinGordon2}.
\end{align}
Here, we have used the equations of state of the matter-dominated era ($p_m=0$) and that of the radiation-dominated era ($p_r=\frac{1}{3}\rho_r$).

Using the Friedmann equations in Eq. \eqref{FriedmanEQ},
we retrieve the energy density  and pressure  for the DE sector respectively as,
\begin{align}
\rho_{DE} &= P(\phi)X+V(\phi)+(G(T)-2TG_{,T}) F(\phi),\label{FEDE1ch3}\\
p_{DE} &= P(\phi) X-V(\phi)-G(T) F(\phi)+2TG_{,T} F(\phi) +8TG_{,TT} F(\phi)\dot{H}\nonumber\\
&+4 G_{,T} F(\phi)\dot{H} +4HG_{,T} F_{,\phi} \dot{\phi}.\label{FEDE2ch3}
\end{align}
Here, we consider $P(\phi)$ = 1. 

The fluid equations are expressed through the continuity equations as in Eq. \eqref{inConservationEq}, and
the standard density parameters for the matter, radiation, and that of the DE sector satisfy Eq. \eqref{densityparameters}.
The late-time cosmic acceleration issue is a recent phenomenon in modern cosmology. Several cosmological models have been proposed in the literature to find possible reasons for this strange behavior of the Universe. Amidst different approaches and models proposed, the stability of the models raises certain questions. In this context,  stability analysis of the models through dynamical system analysis has emerged as an effective tool \cite{Bahamonde:2017ize,Wu:2010xk}. Keeping these things in view, we study the dynamical system analysis with the torsion-based gravitational theory coupled with a scalar field. In this process, we require a certain form of $G(T)$ to be incorporated in Eqs. \eqref{FEDE1ch3} and \eqref{FEDE2ch3} for which we consider some of the well-known forms of $G(T)$ in the following section.

\section{Dynamical system analysis with power law model }\label{dynamicalsystemanalysis}
%********************************
%********************************
We consider $G(T)= \alpha \left(-T\right)^{\beta}$, where $\alpha$ and $\beta$ are model parameters \cite{Mirza_2017,briffa2023,Bengochea:2008gz}. Subsequently,  Eqs. \eqref{FEDE1ch3} and \eqref{FEDE2ch3} reduce to
 
 \begin{align}
 \rho_{DE} &= X+V(\phi)+ F(\phi)\left[\alpha\left(1-2\beta\right)\right]\left(-T\right)^{\beta},\label{FEDErho}\\ 
 p_{DE} &= X-V(\phi)- F(\phi) \left[\alpha\left(1-2\beta\right)\right](-T)^{\beta}+8\dot{H} F(\phi) T \alpha \beta (\beta-1)(-T)^{\beta-2}\nonumber\\ 
  & \,\,\, +4 (-\alpha \beta) (-T)^{\beta-1} F(\phi) \dot{H}+4H(-\alpha \beta (-T)^{\beta-1}) F_{,\phi} \dot{\phi}\,\label{FEDEp}.
 \end{align}
 
Also, the Klein-Gordon equation for a scalar field in Eq. \eqref{KleinGordon2} becomes, 
 \begin{equation}
 \ddot{\phi}+3H\dot{\phi}+\alpha(-T)^{\beta} F_{,\phi}+V_{,\phi}(\phi)=0.\label{KleinGordonM1}
 \end{equation}
In order to analyse the model, we wish to form an autonomous dynamical system with the following dynamical phase space variables, 
\begin{align} 
x &= \frac{\kappa\dot{\phi}}{\sqrt{6}H}, \quad
y = \frac{\kappa \sqrt{V}}{\sqrt{3}H}, \quad
u = \frac{\kappa^2 F(\phi)\left[\alpha (1-2\beta)(-T)^{\beta}\right]}{3H^2},\ \quad \rho = \frac{\kappa\sqrt{\rho_r}}{\sqrt{3}H},\nonumber\\
\lambda &= -\frac{V_{,\phi}(\phi)}{\kappa V(\phi)} \quad,
\Gamma = \frac{V(\phi)V_{,\phi\phi}}{V_{,\phi}^2(\phi)},\quad
\sigma = -\frac{F_{,\phi}(\phi)}{\kappa F(\phi)}  \quad,  \Theta = \frac{F(\phi)F_{,\phi\phi}}{F_{,\phi}^2(\phi)}.\label{dynamicalvariables}
\end{align}
The standard density parameters in Eq. \eqref{densityparameters} in terms of dynamical variables are obtained as,
\begin{eqnarray} \label{densityeqs}
\Omega_r & = & \rho^2, \nonumber\\
\Omega_m & = & 1-x^2-y^2-u-\rho^2,\nonumber\\
\Omega_{DE} & = & x^2+y^2+u.
\end{eqnarray}
In terms of the dynamical variables, we may have 
\begin{equation} \label{eq25}
\frac{\dot{H}}{H^2} = \frac{u \left(2 \beta  \left(\sqrt{6} \lambda  x-3\right)+3\right)+(2 \beta -1) \left(\rho ^2+3 x^2-3 y^2+3\right)}{2 (2 \beta -1) (\beta  u-1)}.
\end{equation}
The deceleration parameter $q=-1-\frac{\dot{H}}{H^2}$ is a measure of the accelerating or decelerating behaviour of the Universe and  may be expressed as  
\begin{equation}\label{decelerationEq}
q =-1-\frac{u \left(2 \beta  \left(\sqrt{6} \lambda  x-3\right)+3\right)+(2 \beta -1) \left(\rho ^2+3 x^2-3 y^2+3\right)}{2 (2 \beta -1) (\beta  u-1)}. 
\end{equation}
The EoS parameters for the total and the DE sector, respectively become
\begin{eqnarray}
\omega_{tot} &=&-1-\frac{u \left(2 \beta  \left(\sqrt{6} \lambda  x-3\right)+3\right)+(2 \beta -1) \left(\rho ^2+3 x^2-3 y^2+3\right)}{3 (2 \beta -1) (\beta  u-1)}, \label{eq26}\\
\omega_{DE} &=& \frac{u \left(\beta  \left(-2 \beta  \left(\rho ^2+3\right)+\rho ^2-2 \sqrt{6} \lambda  x+9\right)-3\right)-3 (2 \beta -1) (x-y) (x+y)}{3 (2 \beta -1) (\beta  u-1) \left(u+x^2+y^2\right)}.\label{eq27}
\end{eqnarray}
The attractor points are the critical points that fall in the stability condition for stable nodes or stable spirals, which can be reached through cosmological evolution. Another type of critical point with zero eigenvalues are the non-hyperbolic critical point \cite{coley2003dynamical}. In this case, the linear stability theory fails to determine the stability of the critical point. Hence, we applied the specific condition in which if the number of vanishing eigenvalues is equal to the dimension of the set of critical points, it is normally hyperbolic. The stability can be analyzed by deriving the conditions for which the non-vanishing eigenvalues are negative.  While studying the cosmic scenarios, it is imperative to find solutions where the energy density of the scalar field matches with the background energy density $(\frac{\rho_{DE}}{\rho_{m}}=\tau_1)$, where $\tau_1$ is a non--zero constant. The cosmological solutions satisfying this property are known as the scaling solutions \cite{Otalora:2013dsa,Otalora:2013tba}. In this study, we also analyse the scaling solutions with two different non-minimally coupled scalar field functions $F(\phi)$.

For $N$, as defined in chapter \ref{Chapter2}, $a(t)$ being the scale factor; we obtain the autonomous dynamical system for the power law model as, 
\begin{eqnarray}
\frac{dx}{dN} &=& -3 x+\frac{\sqrt{6} \lambda  y^2}{2}+\frac{\sqrt{6} \sigma  u}{2-4 \beta}\nonumber\\
   &+& \left(-\frac{x \left(u \left(2 \beta  \left(\sqrt{6} \lambda  x-3\right)+3\right)+(2 \beta -1) \left(\rho ^2+3 x^2-3 y^2+3\right)\right)}{2 (2 \beta -1) (\beta  u-1)}\right),\label{dynamicaleq1}\\
\frac{dy}{dN} &=& \frac{\left(-\sqrt{6}\right) \lambda  x}{2}\left(-\frac{yu \left(2 \beta  \left(\sqrt{6} \lambda  x-3\right)+3\right)+(2 \beta -1) \left(\rho ^2+3 x^2-3 y^2+3\right)}{2(2 \beta -1) (\beta  u-1)}\right), \label{dynamicaleq2}\\
\frac{du}{dN} &=&
\frac{(\beta -1) u \left(u \left(2 \beta  \left(\sqrt{6} \lambda  x-3\right)+3\right)+(2 \beta -1) \left(\rho ^2+3 x^2-3 y^2+3\right)\right)}{(2 \beta -1) (\beta  u-1)}\nonumber\\
&-&\sqrt{6} \sigma  u x, \label{dynamicaleq3}\\
\frac{d\rho}{dN} &=& -\rho  \left(\frac{u \left(2 \beta  \left(\sqrt{6} \lambda  x-3\right)+3\right)+(2 \beta -1) \left(\rho ^2+3 x^2-3 y^2+3\right)}{2 (2 \beta -1) (\beta  u-1)}+2\right), \label{dynamicaleq4}\\
\frac{d\lambda}{dN} &=& -\sqrt{6}\left(\Gamma-1\right)\lambda^2 x, \label{dynamicaleq5}\\
\frac{d\sigma}{dN} &=& -\sqrt{6}\left(\Theta-1\right)\lambda^2 x.\label{dynamicaleq6}
\end{eqnarray}
In this study, we will consider the scalar field potential as \cite{samaddar2023qualitative,Gonzalez-Espinoza:2020jss,duchaniya2023dynamical},
\begin{align}
    V(\phi)=V_{0}e^{-\kappa\tau\phi}\,,
\end{align}
where $\tau$ is a constant; $V_0$ provides the strength of the potential. This will result in $\Gamma=1$, hence the system Eq. \eqref{dynamicaleq5}, will not contribute in the following cases.

\subsection{\texorpdfstring{$ F(\phi)=F_0e^{-\eta  \kappa \phi} $}{}}\label{Exponentialcouplingfun}
Let us choose the non-minimally coupled scalar field function as $F(\phi)=F_0e^{-\eta  \kappa \phi} $, where $F_0$ and $\eta$ are constants. This exponential form of the coupling function is widely studied in the literature to study the dynamical system analysis in scalar field models \cite{Gonzalez-Espinoza:2020jss,roy2018dynamical}. In this case, we have  
 $\Gamma=1$ and $\Theta=1$ and the autonomous dynamical system presented in  Eqs. \eqref{dynamicaleq1} to \eqref{dynamicaleq6} reduces to four equations with four independent variables $x,\, y,\, u,\, \rho$. The critical points for this system are presented in Table \ref{CH3Table1} along with the values of $\omega_{tot}$ and the standard density parameters $\Omega_{m}$, $\Omega_{r}$, $\Omega_{DE}$. The stability of the critical point is analyzed using the eigenvalues of the Jacobian matrix at each critical point. For this case, the eigenvalues, along with the stability conditions, are presented in Table \ref{CH3Table2}. 

\begin{figure}[H]
    \centering
\includegraphics[width=53mm]{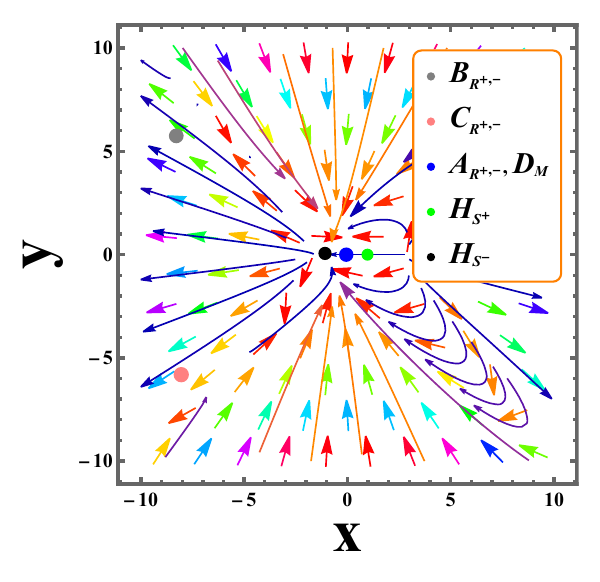}
\includegraphics[width=53mm]{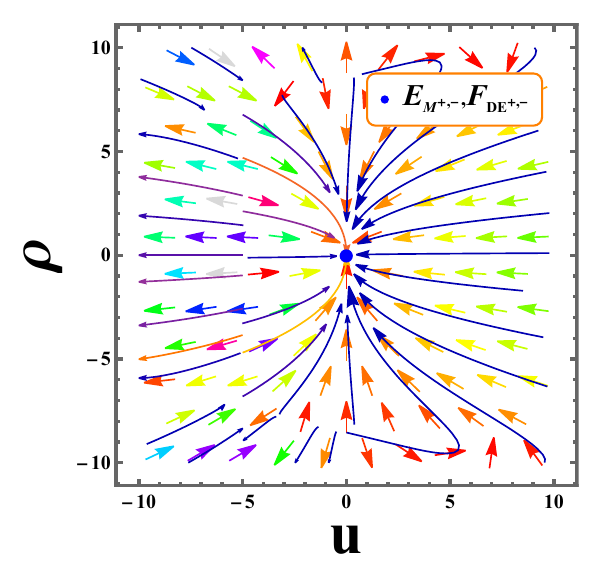}
\includegraphics[width=53mm]{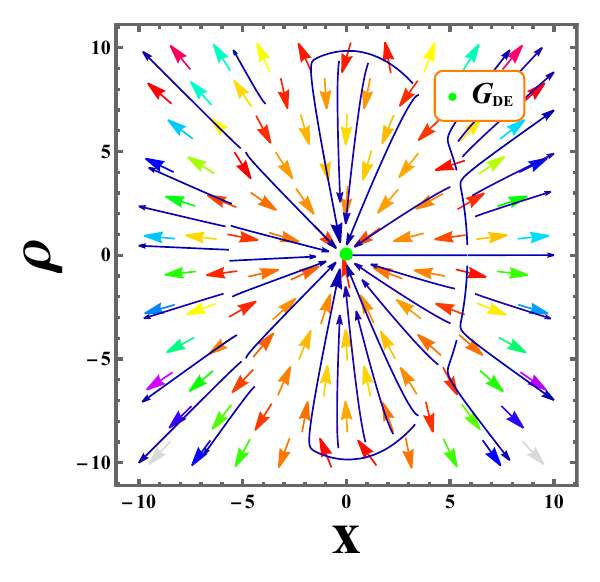}
    \caption{ 2D phase portraits for model \ref{Exponentialcouplingfun}. }
    \label{CH3phasespace2dm1}
\end{figure}

\begin{figure}[H]
\centering
\includegraphics[width=60mm]{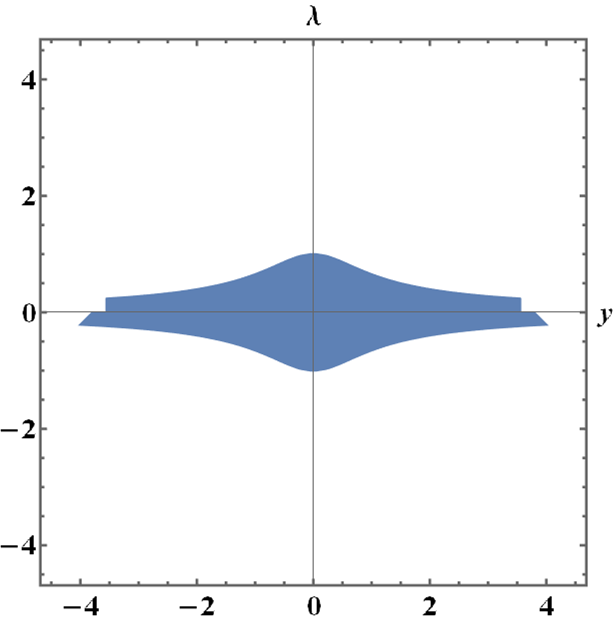}
\includegraphics[width=60mm]{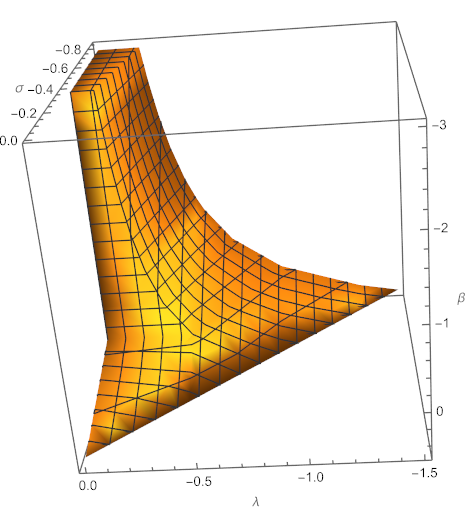}
\caption{2D and 3D region plots for model \ref{Exponentialcouplingfun} .} \label{CH3regionplotm1}
\end{figure}

 \begin{itemize}
\item \textbf{Radiation-dominated critical points:} The critical points $A_{R_{\pm}},B_{R_{\pm}}$, and $C_{R_{\pm}}$ are the points in the radiation-dominated phase. The critical point $A_{R_{\pm}}$ with $\Omega_{r}=1$ represents the standard radiation-dominated era of the Universe evolution. The eigenvalues of the Jacobian matrix at this critical point contain both positive and negative signs hence it is an unstable saddle critical point. The value of $\omega_{tot}=\frac{1}{3}$ at this critical point. The other two critical points $B_{R_{\pm}}$, and $C_{R_{\pm}}$ are the scaling radiation-dominated solutions with $\Omega_{DE}=\frac{4}{\lambda^2}$. These critical points represent the non-standard radiation-dominated era with $\Omega_{r}=1-\frac{4}{\lambda^2}$. The eigenvalues at these critical points can be observed from Table \ref{CH3Table2}, which contains one positive eigenvalue. The other eigenvalues may take negative values in the range $\beta>\frac{1}{2} \left(2-\sigma\right)\land \left(-\frac{8}{\sqrt{15}}\leq \lambda<-2\lor 2<\lambda\leq \frac{8}{\sqrt{15}}\right)$ leading to saddle points in this range and therefore, they become unstable. The behaviour of phase space trajectories at these critical points can be analysed from Fig. \ref{CH3phasespace2dm1}. The plot is for  $u=8,\, \rho=4.5,\, \beta=-0.2, \,  \sigma=-0.30, \, \lambda=-0.2$. One may observe that the trajectories move away from these critical points, ascertaining the unstable behaviour. In this case the value of $\Omega_{DE}$ for critical points $A_{R}, B_{R}$ is $\frac{4}{\lambda^2}$, hence these critical points are lying into the physically viable region for $\lambda<-2\lor \lambda>2$ at which we have $0<\Omega_{DE}<1$.
\end{itemize}
\begin{table}[H]
     % title of Table
    \centering % used for centering table
     \scalebox{0.8}{
    \begin{tabular}{|c |c |c |c|c| c| c| c|} % centered columns (5 columns)
    \hline\hline %inserts double horizontal lines
    \parbox[c][0.9cm]{1.3cm}{{Name}
    }& $ \{ x_{c}, \, y_{c}, \, u_{c}, \, \rho_{c} \} $ & {Existence Condition} &  {$\omega_{tot}$}&{$\omega_{DE}$}& $\Omega_{r}$& $\Omega_{m}$& $\Omega_{DE}$\\ [0.5ex] % inserts table %headin$g$
    \hline\hline % inserts single horizontal line
    \parbox[c][1.3cm]{1.3cm}{$A_{R^{\pm}}$ } &$\{ 0, 0, 0, \pm 1 \}$ & $2 \beta -1\neq 0$ &  $\frac{1}{3}$&$1$& $1$& $0$ & $0$ \\
    \hline
    \parbox[c][1.3cm]{1.3cm}{$B_{R^{\pm}}$ } & $\{\frac{2 \sqrt{\frac{2}{3}}}{\lambda}, \,  \frac{2}{\sqrt{3}\lambda}, \, 0 ,  \, \pm \sqrt{1-\frac{4}{\lambda^2}}\}$ & $\lambda\ne 0\, ,2 \beta -1\neq 0 $&  $\frac{1}{3}$&  $\frac{1}{3}$&$1-\frac{4}{\lambda^2}$ & $0$ & $\frac{4}{\lambda^2}$ \\
    \hline
    \parbox[c][1.3cm]{1.3cm}{$C_{R^{\pm}}$ } & $\{\frac{2 \sqrt{\frac{2}{3}}}{\lambda}, \,  -\frac{2}{\sqrt{3}\lambda}, \, 0 ,  \, \pm \sqrt{1-\frac{4}{\lambda^2}}\}$ & $\lambda\ne 0\, ,2 \beta -1\neq 0 $&  $\frac{1}{3}$&  $\frac{1}{3}$&$1-\frac{4}{\lambda^2}$ & $0$ & $\frac{4}{\lambda^2}$\\
    \hline
   \parbox[c][1.3cm]{1.3cm}{$D_{M}$ } &  $\{0, \, 0, \, 0, \, 0 \}$ & $2 \beta -1\neq 0$ &  $0$& $1$&$0$ & $1$ & $0$\\
   \hline
   \parbox[c][1.3cm]{1.3cm}{$E_{M^{\pm}}$} &   $\{\frac{\sqrt{\frac{3}{2}}}{\lambda},\pm\frac{\sqrt{\frac{3}{2}}}{\lambda},0,0\}$ & $2 \beta -1\neq 0$ &  $0$&$0$&$0$ & $1-\frac{3}{\lambda^2}$ & $\frac{3}{\lambda^2}$\\
   \hline
   \parbox[c][1.3cm]{1.3cm}{$F_{DE^{\pm}}$} & $\{\frac{\lambda}{\sqrt{6}},\pm\sqrt{1-\frac{\lambda^2}{6}},0,0 \}$ & $2 \beta -1\neq 0$ &  $-1+\frac{\lambda^2}{3} $&  $-1+\frac{\lambda^2}{3} $& $0$ & $0$ & $1$\\
   \hline
 \parbox[c][1.3cm]{1.3cm}{$G_{DE}$} &  $\{0, y,1-y^2,0 \}$ & $\begin{tabular}{@{}c@{}}$\beta=0\,, \sigma =\frac{\lambda  y^2}{y^2-1}\,,$\\  $y^2-1\neq 0$\end{tabular}$& $-1$ &$-1$&$0$ & $0$ & $1$\\
 \hline
 \parbox[c][1.3cm]{1.3cm}{$H_{S^{\pm}}$} &  $\{\pm1, 0,0,0 \}$ & $\begin{tabular}{@{}c@{}}$2 \beta -1\neq 0$\end{tabular}$& $1$ &$1$&$0$ & $0$ & $1$\\
 \hline
 \end{tabular}}
\caption{Critical points with the existence condition for model \ref{Exponentialcouplingfun}.}
% is used to refer to this table in the text
\label{CH3Table1}
\end{table}
\begin{itemize}
\item{\textbf{Matter-dominated critical points:}} The critical points $ D_{M}, E_{M^{\pm}}$ are the matter-dominated critical points. As is mentioned in Table \ref{CH3modelIIeigenvalues}, the critical point $D_{M}$ is a standard matter-dominated critical point with $\Omega_{m}=1$. Also, there is a saddle point at $D_{M}$ showing unstable behaviour. In Fig. \ref{CH3phasespace2dm1}, the phase space trajectories move away from this critical point, ensuring the unstable behaviour. The other critical point $E_{M^{\pm}}$ represents non-standard matter-dominated critical point with $\Omega_{M}=1-\frac{3}{\lambda^2}$, and is lying in the physical viable region for $\lambda<-\sqrt{3}\lor \lambda>\sqrt{3}$, where we have $0<\Omega_{DE}<1$. Moreover, the critical point $E_{M^{\pm}}$ can alleviate the coincidence problem \cite{Kofinas:2014aka}. This critical point is a matter-dominated scaling solution. At this critical point, we have  $\Omega_{DE}=\frac{3}{\lambda^2}$. From the eigenvalues presented in Table \ref{CH3Table2}, it may be observed that this critical point is stable at $2 \sigma<\lambda\leq 2 \sqrt{\frac{6}{7}}\land \beta >\frac{\lambda-\sigma}{\lambda}$. The attracting phase space trajectories at this critical point can be observed from Fig. \ref{CH3phasespace2dm1}. Since, in this case, we have $\omega_{tot}=0$,  both the critical points are unable to describe the accelerating phase of the Universe.
\end{itemize}
In the following Table \ref{CH3Table2}, $\chi_1=\frac{\sqrt{-\left(1-2 \beta \right){}^2 \lambda^2 \left(15 \lambda^2-64\right)}}{2 \left(2 \beta-1\right) \lambda^2}$, the detailed analysis of each of the critical points are given below.
 From now onward, we use the symbol $\lor$ to denote ``or", and the symbol $\land$ will be used to denote ``and".
\begin{table}[H]
%\small\addtolength{\tabcolsep}{-7pt}
     % title of Table
    \centering % used for centering table
    \scalebox{0.685}{
    \begin{tabular}{|c |c |c |c| c|} % centered columns (5 columns)
    \hline\hline %inserts double horizontal lines
    \parbox[c][0.9cm]{2.5cm}{{Critical points}
    }& Eigenvalues & Stability Conditions \\ [0.5ex] % inserts table %headin$g$
    \hline\hline % inserts single horizontal line
    \parbox[c][1.3cm]{1.3cm}{$A_{R^{\pm}}$ } & $\Bigl\{-1,1,2,-4 (\beta -1)\Bigl\}$ & Saddle at $\beta >1$\\
    \hline
    \parbox[c][1.3cm]{1.3cm}{$B_{R^{\pm}}$ } & $\Bigl\{1,-\chi_{1}-\frac{1}{2},\frac{1}{2} \left(\chi_{1}-1\right),-4 \beta-\frac{4 \sigma}{\lambda}+4\Bigl\}$ & Saddle at $\left(\beta>\frac{1}{2} \left(2-\sigma\right)\land \left(-\frac{8}{\sqrt{15}}\leq \lambda <-2\lor 2<\lambda\leq \frac{8}{\sqrt{15}}\right)\right)$\\
    \hline
   \parbox[c][2cm]{1.3cm}{$C_{R^{\pm}}$ } & $\Bigl\{1,-\chi_{1}-\frac{1}{2},\frac{1}{2} \left(\chi_{1}-1\right),-4 \beta-\frac{4 \sigma}{\lambda}+4\Bigl\}$  & Saddle at $\left(\beta>\frac{1}{2} \left(2-\sigma\right)\land \left(-\frac{8}{\sqrt{15}}\leq \lambda <-2\lor 2<\lambda\leq \frac{8}{\sqrt{15}}\right)\right)$\\
   \hline
   \parbox[c][1.3cm]{1.3cm}{$D_{M}$} &  $\Bigl\{\frac{1-2 \beta}{2 \left(2 \beta-1\right)},-\frac{3}{2},\frac{3}{2},-3 \left(\beta-1\right)\Bigl\}$ &  Saddle at $\beta >1$  \\
   \hline
    \parbox[c][1.3cm]{1.3cm}{$E_{M^{\pm}}$} &  $\begin{tabular}{@{}c@{}} $\Bigl\{-\frac{1}{2},-\frac{3 \sqrt{-\left(1-2 \beta\right){}^2 \lambda^2 \left(7 \lambda^2-24\right)}}{4 \left(2 \beta-1\right) \lambda^2}-\frac{3}{4},$\\$\frac{3}{4} \left(\frac{\sqrt{-\left(1-2 \beta\right){}^2 \lambda^2 \left(7 \lambda^2-24\right)}}{\left(2 \beta-1\right) \lambda^2}-1\right),-3 \beta-\frac{3 \sigma}{\lambda}+3\Bigl\}$\end{tabular}$ &$\begin{tabular}{@{}c@{}} Stable at  \\ $2 \sigma<\lambda\leq 2 \sqrt{\frac{6}{7}}\land \beta >\frac{\lambda-\sigma}{\lambda}$ \end{tabular}$ \\
 \hline
 \parbox[c][1.3cm]{1.3cm}{$F_{DE^{\pm}}$} &  $\Bigl\{\frac{1}{2} \left(\lambda^2-6\right),\frac{1}{2} \left(\lambda^2-4\right),\lambda^2-3,-\beta \lambda^2-\lambda \sigma +\lambda^2\Bigl\}$ & $\begin{tabular}{@{}c@{}}  \\Stable at \\$\beta\in \mathbb{R}\land \Big(\left(-\sqrt{3}<\lambda<0\land \sigma<\lambda-\beta \lambda \right)$\\$
 \lor \left(0<\lambda<\sqrt{3}\land \sigma>\lambda-\beta \lambda \right)\Big)$\end{tabular}$  \\
 \hline
 \parbox[c][1.3cm]{1.3cm}{$G_{DE}$} & $\begin{tabular}{@{}c@{}} $\Bigl\{-3,-2,-\frac{\sqrt{3} \sqrt{\left(y^2-1\right) \left(\left(4 \lambda^2+3\right) y^2-3\right)}}{2 \left(y^2-1\right)}-\frac{3}{2},$\\$\frac{1}{2} \left(\frac{\sqrt{3} \sqrt{\left(y^2-1\right) \left(\left(4 \lambda^2+3\right) y^2-3\right)}}{y^2-1}-3\right)\Bigl\}$\end{tabular}$ & $\begin{tabular}{@{}c@{}} Stable at\\ $\lambda\in \mathbb{R}\land \lambda\neq 0 $\\$\land \left(-\sqrt{3} \sqrt{\frac{1}{4 \lambda^2+3}}\leq y<0\lor 0<y\leq \sqrt{3} \sqrt{\frac{1}{4 \lambda^2+3}}\right)$\end{tabular}$ \\
 \hline
\parbox[c][1.3cm]{1.3cm}{$H_{S^{\pm}}$} & $\begin{tabular}{@{}c@{}} $\Bigl\{1,3,\frac{1}{2} \left(6-\sqrt{6} \lambda \right),-6 \beta-\sqrt{6} \sigma+6\Bigl\}$\end{tabular}$ & $\begin{tabular}{@{}c@{}} Saddle at\\ $\sigma\in \mathbb{R}\land \beta>\frac{1}{6} \left(6-\sqrt{6} \sigma \right)\land |\lambda|<\sqrt{6}$\end{tabular}$ \\
 \hline
\end{tabular}}
    \caption{Eigenvalues and the stability conditions for model \ref{Exponentialcouplingfun}.}
    % is used to refer to this table in the text
    \label{CH3Table2}
\end{table}
%%%%%
\begin{itemize}
\item{\textbf{DE-dominated critical points:}} The critical points $F_{DE^{\pm}}$ and $G_{DE}$ are representing the standard DE-dominated era with $\Omega_{DE}=1$. The critical point $F_{D}$ is a stable DE-dominated solution with the stability range $\beta\in \mathbb{R}\land \Big(\left(-\sqrt{3}<\lambda<0\land \sigma<\lambda-\beta \lambda \right)
 \lor \left(0<\lambda<\sqrt{3}\land \sigma>\lambda-\beta \lambda \right)\Big)$. This critical point is a stable attractor, and the attracting behaviour of the phase space trajectories can be studied from Fig. \ref{CH3phasespace2dm1}. The value of $\omega_{tot}=-1+\frac{\lambda^2}{3}$ and can explain accelerated expansion of the Universe at $-\sqrt{2}<\lambda<\sqrt{2}$. This critical point will exist at the condition $2\beta-1\ne0$. The critical point $G_{DE}$ is a de Sitter solution with $\omega_{tot}=-1$. This solution exists at $\beta=0$. From Table \ref{CH3Table2} this critical point is stable and show stability at $\lambda\in \mathbb{R}\land \lambda\neq 0 \land \left(-\sqrt{3} \sqrt{\frac{1}{4 \lambda^2+3}}\leq y<0\lor 0<y\leq \sqrt{3} \sqrt{\frac{1}{4 \lambda^2+3}}\right)$. To get better clarity, the 2D plot for this range in terms of the dynamical variable $y$ and $\lambda$ is plotted in Fig. \ref{CH3regionplotm1}.  The phase space trajectories at this critical point show the attracting behaviour, and therefore, this critical point is a stable attractor. One interesting result is that the power law form with exponential scalar field potential is quite capable of describing different evolutionary phases of the  Universe. The common parametric range for which we have stable critical points $E_{M^{\pm}}, \, F_{DE^{\pm}},\, G_{DE}$ is  $-\frac{\sqrt{3}}{2}<\sigma<0\&\&\left(\beta \leq \frac{1}{2}\land -\frac{\sigma }{\beta _{10}-1}<\lambda<0\right)$. For a better visualisation of stability conditions pertaining to the critical points, the same has been shown graphically in Fig. \ref{CH3regionplotm1}.
  \item{\textbf{Stiff matter-dominated critical points:}}
 In the era of stiff matter, when $\omega=\frac{p}{\rho}=1$, the energy density \(\rho\) changes as \(\rho \propto a(t)^{-6}\). This decrease is much faster than that for radiation (\(\rho \propto a^{-4}\)) or matter (\(\rho \propto a^{-3}\)). The critical points $H_{s}^{\pm}$ correspond to a non-accelerating, DE-dominated Universe with stiff DE. In this scenario, the equation of state parameter $\omega_{tot}=1$. This critical point shows saddle point nature at $\sigma\in \mathbb{R}\land \beta>\frac{1}{6} \left(6-\sqrt{6} \sigma \right)\land |\lambda|<\sqrt{6}$, and same can be observed from Fig. \ref{CH3phasespace2dm1}.
\end{itemize} 
In above Fig. \ref{CH3phasespace2dm1} we have $x=0,\, y=-4.9,\, u=8,\, \rho=4.5,\, \beta=-0.2, \beta=0$ for ($G_{DE}$),  \, $ \sigma=-0.30, \, \lambda=-0.2$. In Fig. \ref{CH3regionplotm1} stability condition of critical point $G_{DE}$ and 3D Region plot for the combined range of stability conditions for critical points for model parameters for $-\frac{\sqrt{3}}{2}<\sigma<0$ and $\left(\beta \leq \frac{1}{2}\land -\frac{\sigma }{\beta _{10}-1}<\lambda<0\right)$ for model \ref{Exponentialcouplingfun}.
It is to note that the combined stability range for critical points $E_{M^{\pm}}, F_{DE^{\pm}}, G_{DE}$ is $-\frac{\sqrt{3}}{2}<\sigma<0\&\&\left(\beta \leq \frac{1}{2}\land -\frac{\sigma }{\beta _{10}-1}<\lambda<0\right)$.

\begin{figure}[H]
    \centering
\includegraphics[width=60mm]{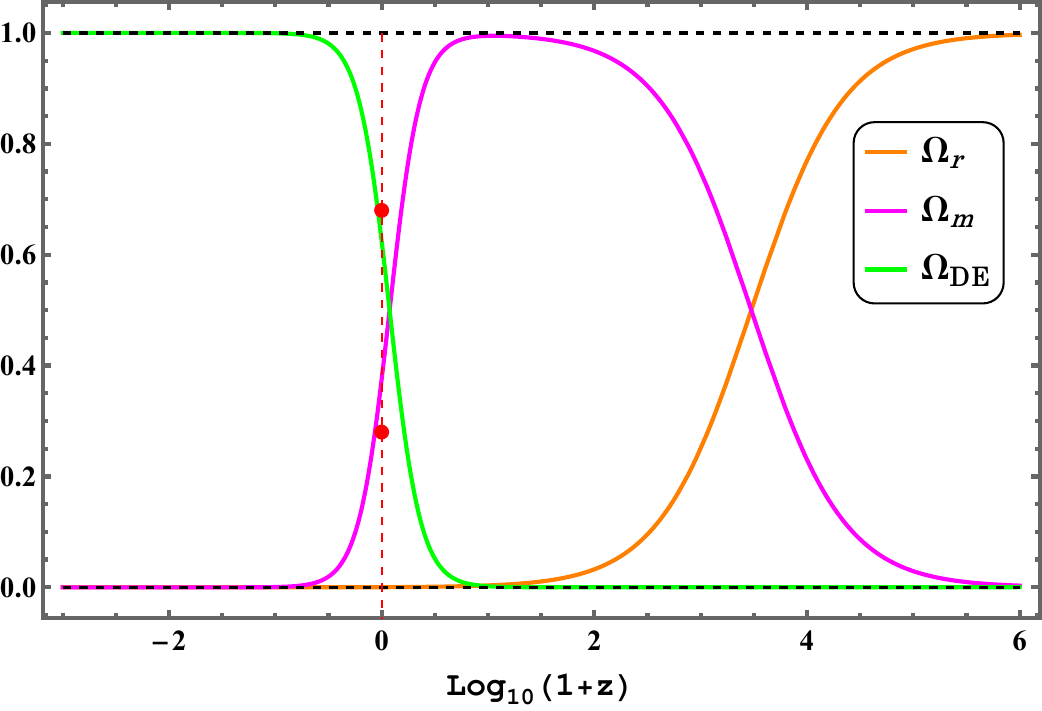}    \includegraphics[width=60mm]{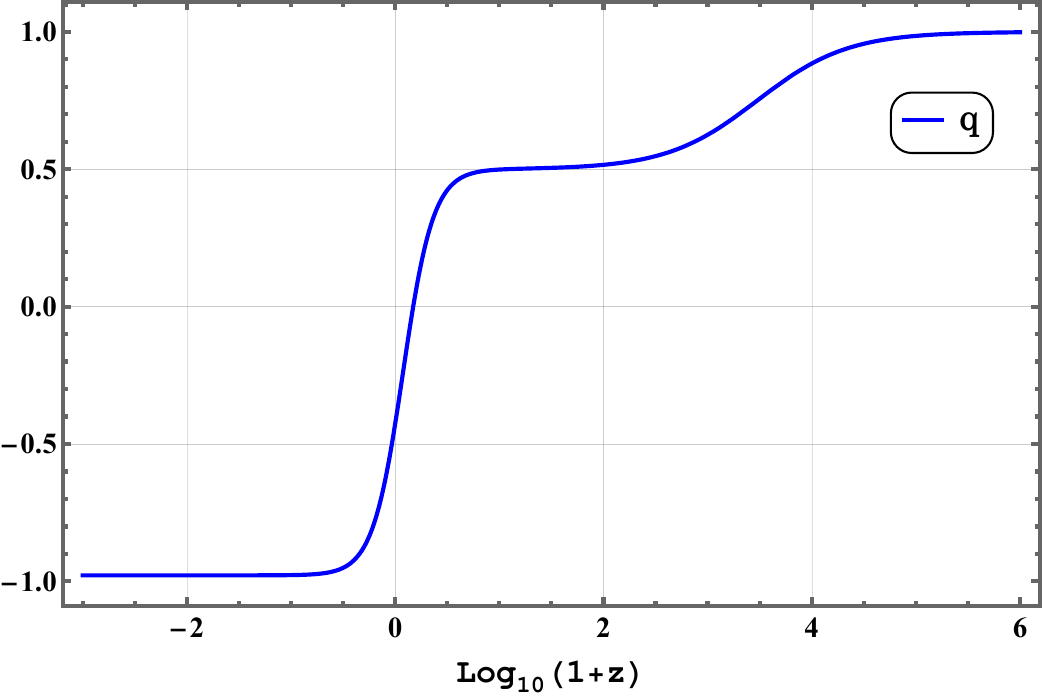}
    \caption{Density parameters and deceleration parameter for model \ref{Exponentialcouplingfun}. } \label{CH3evolutionm1}
\end{figure}

\begin{figure}[H]
    \centering
\includegraphics[width=70mm]{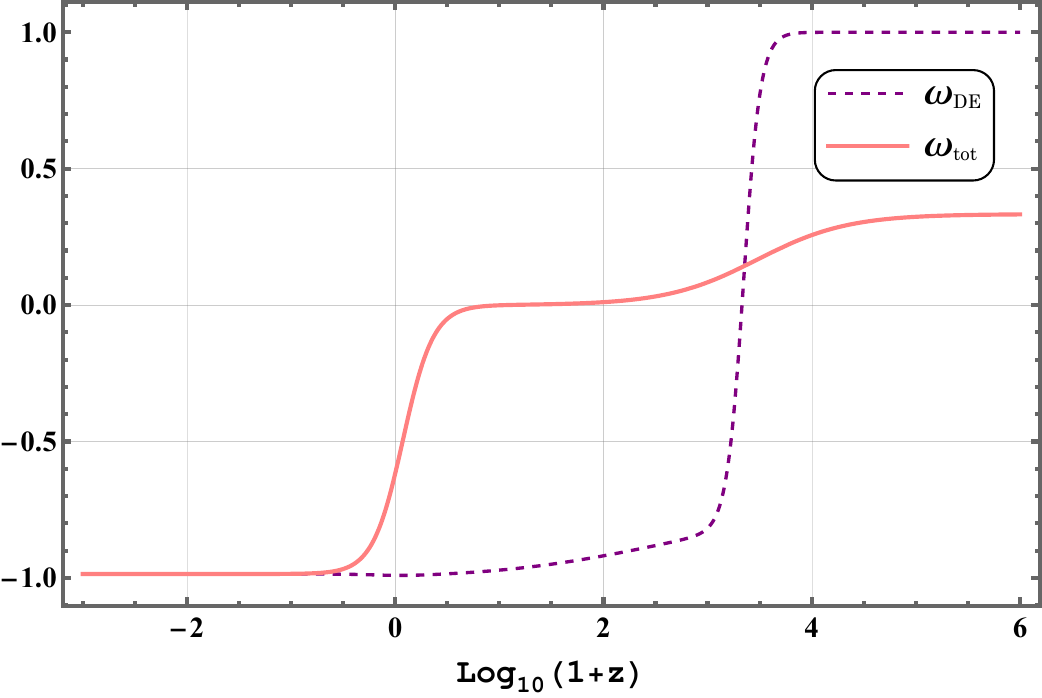}
    \caption{ The EoS for DE $(\omega_{DE})$ and total EoS $(\omega_{tot})$ parameter for model \ref{Exponentialcouplingfun}. } \label{CH3Eosplotm1}
\end{figure}

In Fig. \ref{CH3evolutionm1}, we have shown the evolutionary behaviour of the energy density pertaining to the matter, radiation, DE phase, and the deceleration parameter. It can be observed that the redshift of radiation
matter equality is around $z \approx 3387$, and the transition to the  accelerated phase at $z \approx 0.62$ (Ref. Fig. \ref{CH3evolutionm1}), which is compatible to the
 $\Lambda$CDM value. From the evolution plots of standard density parameters, we observe that at present, $\Omega_{m}\approx 0.3$, which agrees with the Planck observation results \cite{Planck:2018vyg}. The dominant presence of DE at the present epoch is quite visible from the derived value of the DE density parameter $\Omega_{DE} \approx 0.7$ \cite{Kowalski_2008}. The deceleration parameter shows a decreasing behaviour from early epoch to late phase. It lies in the negative domain in the present epoch as well as at the late times and hence describes an accelerating behaviour of the Universe at the present epoch and late epoch. The present value of $q \approx -0.614^{+0.002}_{-0.002}$ and is compatible with the observational study made in \cite{Capozziellomnras}. In Fig. \ref{CH3Eosplotm1}, the total equation of state $\omega_{tot}$ and the equation of state of DE $\omega_{DE}$ are shown. We observe that at late times, both the curves approach to $\Lambda$CDM behaviour, though $\omega_{DE}$ shows a phantom behaviour in the recent past epochs.

In Figs. \ref{CH3evolutionm1}, \ref{CH3Eosplotm1} we have for radiation $(\Omega_r)$, matter $(\Omega_m)$, DE $(\Omega_{DE})$ in redshift.  Right Panel--Deceleration parameter $(q)$ in redshift for $\beta=-0.2, \,  \sigma=-0.30, \, \lambda=-0.2$ for the initial conditions $x_0=10^{-8.89}, \, y_0=10^{-2.89}, \, u_0=10^{-5.96}, \, \rho_0= 10^{-0.9}$ for model \ref{Exponentialcouplingfun}.
\subsection{\texorpdfstring{$ F(\phi)=F_{0} \phi^n$}{}}\label{Powerlawcouplingfun}
\begin{figure}[H]
    \centering
\includegraphics[width=60mm]{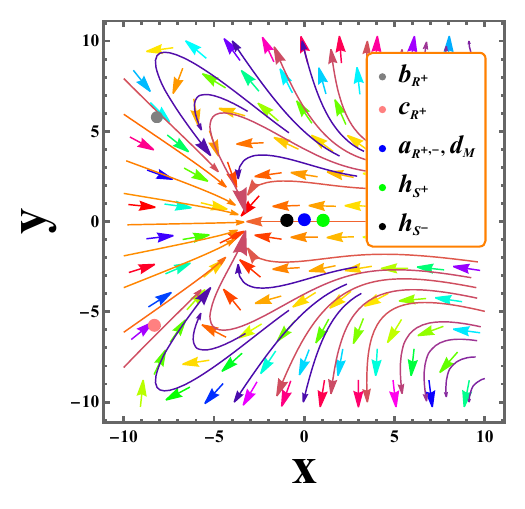}
\includegraphics[width=60mm]{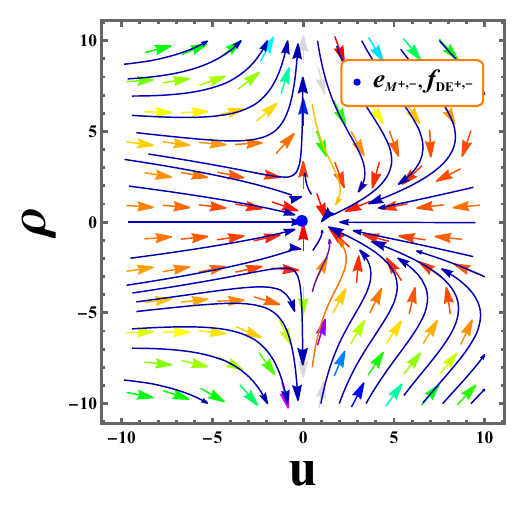}
\includegraphics[width=60mm]{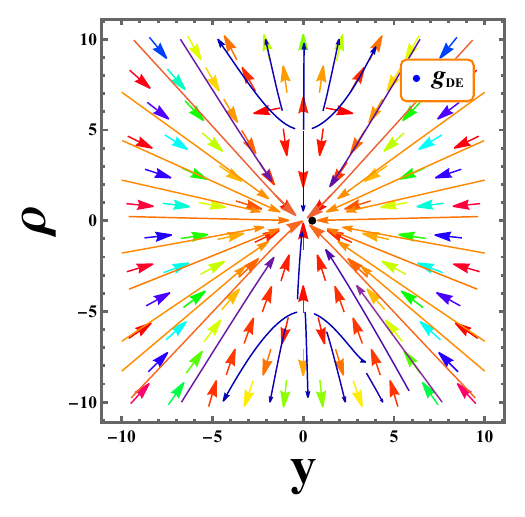}
    \caption{2D phase portrait for model \ref{Powerlawcouplingfun}.}
    \label{CH3phasespacem1.2}
\end{figure}

\begin{figure}[H]
    \centering
\includegraphics[width=60mm]{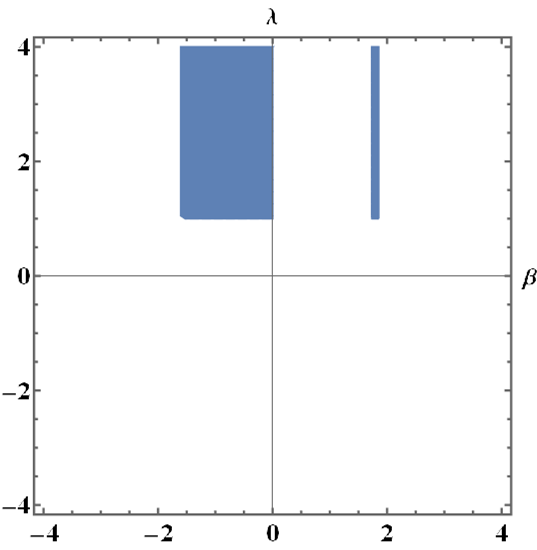}
    \caption{2D region plot for model \ref{Powerlawcouplingfun}.}
    \label{CH32dregionplotm1.2}
\end{figure}

We may consider a power law scalar function $F(\phi)=F_{0} \phi^n$ for the dynamical system analysis within the $f(T,\phi)$ gravity model. Here, $F_0$ and $n$ are constant model parameters. Using Eq. \eqref{dynamicalvariables}, we get $\Gamma=1$ and $\Theta=\frac{n-1}{n}$ so that the dynamical system will have five independent variables $\left(x,\, y,\, u,\, \rho,\, \sigma\right)$ variables. Apart from the system of equations Eqs. \eqref{dynamicaleq1} to \eqref{dynamicaleq4}, we have an additional equation in the system,
\begin{eqnarray}
\frac{d\sigma}{dN} &=& \frac{\sqrt{6} \sigma ^2 x}{n}.\label{DynamicalEq7}
\end{eqnarray}
The critical points and the value of $\omega_{tot}, \Omega_{r}, \Omega_{m}$ and $\Omega_{DE}$ for this case as obtained from solving the autonomous dynamical system are presented in Table \ref{CH3modelIIcriticalpoints}. The stability of these critical points can be analysed using the eigenvalues obtained for the Jacobian. The eigenvalues with the stability conditions at these critical points are presented in Table \ref{CH3modelIIeigenvalues}. 
Below, we present a detailed analysis of these critical points obtained from solving the autonomous dynamical system while considering different phases of evolution.
 \begin{itemize}
\item \textbf{Radiation-dominated critical points:} In the radiation-dominated phase of the Universe, there appear the critical points $a_{R^{\pm}}, b_{R^{\pm}}, c_{R^{\pm}}$. The critical point $a_{R^{\pm}}$ is the standard radiation-dominated critical point with $\Omega_{r}=1$, which has the eigenvalues encompassing zero, positive, and negative values Ref. Table \ref{CH3modelIIeigenvalues}.  Because of this, the critical point becomes non-hyperbolic and appears as a saddle point. The phase space trajectories move away from this critical point. The saddle point behaviour can be observed from the 2D phase portrait in Fig. \ref{CH3phasespacem1.2}. The other two critical points $b_{R^{\pm}}, c_{R^{\pm}}$ are the non-standard radiation-dominated critical points with $\Omega_{r}=1-\frac{4}{\lambda^2}$. These critical points are the scaling radiation-dominated solutions with $\Omega_{DE}=\frac{4}{\lambda^2}$. The eigenvalues at these critical points contain zero, positive, and negative values and become non-hyperbolic, saddle, and unstable. The phase space trajectories move away from these critical points and can be observed in Fig. \ref{CH3phasespacem1.2}. Since the predicted value of the $\omega_{tot}$ in the present case is $\frac{1}{3}$, these critical points may not describe the late-time accelerated expansion phenomena.
\item \textbf{Matter-dominated critical points:} The critical points $d_{M}$ and $ e_{M^{\pm}}$ describe the matter-dominated phase of the Universe. The critical point $d_{M}$ has existence condition $2\beta n-n \ne 0$ which is same for $e_{M^{\pm}}$ but $e_{M^{\pm}}$ are described at $\beta=2$. The critical point $d_{M}$ describes a standard matter-dominated epoch with $\Omega_{m}=1$. The eigenvalues at $d_{M}$ are presented in Table \ref{CH3modelIIeigenvalues}, which show this critical point's non-hyperbolic saddle point behaviour. From the phase space diagram presented in Fig. \ref{CH3phasespacem1.2}, the trajectories at this critical point move away and show a saddle point behaviour. The critical point $e_{M^{\pm}}$ is a non-standard matter-dominated critical point with $\Omega_{m}=1-\frac{3}{\lambda^2}$. The eigenvalues at this critical point show that this is a normally hyperbolic critical point and show stability at $-2 \sqrt{\frac{6}{7}}\leq \lambda<-\sqrt{3}\lor \sqrt{3}<\lambda\leq 2 \sqrt{\frac{6}{7}}$. The phase space trajectories at this critical point show attractor behaviour and can be observed from Fig. \ref{CH3phasespacem1.2}. Moreover $e_{M^{\pm}}$ is a scaling matter-dominated solution with $\Omega_{DE}=\frac{3}{\lambda^2}$, and is lying in the physical viable region for $\lambda<-\sqrt{3}\lor \lambda>\sqrt{3}$. Moreover, the critical point $e_{M^{\pm}}$ can alleviate the coincidence problem \cite{Kofinas:2014aka}.
\end{itemize}
\begin{table}[H]
     % title of Table
    \centering % used for centering table
    \scalebox{0.78}{\begin{tabular}{|c |c |c |c| c| c|c| c|} % centered columns (5 columns)
    \hline\hline %inserts double horizontal lines
    \parbox[c][0.9cm]{1.3cm}{{Name}
    }& $ \{ x_{c}, \, y_{c}, \, u_{c}, \, \rho_{c}, \, \sigma_{c} \} $ & {Existence Condition} &  \textbf{$\omega_{tot}$}&$\omega_{DE}$ &$\Omega_{r}$& $\Omega_{m}$& $\Omega_{DE}$\\ [0.5ex] % inserts table %headin$g$
    \hline\hline % inserts single horizontal line
    \parbox[c][1.3cm]{1.3cm}{$a_{R^{\pm}}$ } &$\{ 0, 0, 0, \pm 1, \, \sigma \}$ & $2 \beta  n-n\neq 0$ &  $\frac{1}{3}$& $1$ &$1$& $0$ & $0$ \\
    \hline
    \parbox[c][1.3cm]{1.3cm}{$b_{R^{\pm}}$ } & $\{\frac{2 \sqrt{\frac{2}{3}}}{\lambda}, \,  \frac{2}{\sqrt{3}\lambda}, \, 0 ,  \, \pm \sqrt{1-\frac{4}{\lambda^2}}\, , 0\}$ & $ \lambda\ne 0\, ,2 \beta  n-n\neq 0 $&  $\frac{1}{3}$&$\frac{1}{3}$&$1-\frac{4}{\lambda^2}$ & $0$ & $\frac{4}{\lambda^2}$ \\
    \hline
    \parbox[c][1.3cm]{1.3cm}{$c_{R^{\pm}}$ } & $\{\frac{2 \sqrt{\frac{2}{3}}}{\lambda}, \,  -\frac{2}{\sqrt{3}\lambda}, \, 0 ,  \, \pm \sqrt{1-\frac{4}{\lambda^2}}\, , 0\}$ & $\lambda\ne 0\, , 2 \beta  n-n\neq 0$&  $\frac{1}{3}$& $\frac{1}{3}$& $1-\frac{4}{\lambda^2}$ & $0$ & $\frac{4}{\lambda^2}$ \\
    \hline
   \parbox[c][1.3cm]{1.3cm}{$d_{M}$ } &  $\{0, \, 0, \, 0, \, 0, \, \sigma \}$ & $2 \beta  n-n\neq 0$ &  $0$& $1$ &$0$ & $1$ & $0$\\
   \hline
   \parbox[c][1.3cm]{1.3cm}{$e_{M^\pm}$} &   $\{\frac{\sqrt{\frac{3}{2}}}{\lambda }, \, {\pm} \frac{\sqrt{\frac{3}{2}}}{\lambda }, \, 0, \, 0, \, 0\}$ & $2 \beta  n-n\neq 0, \, \beta=2$ &  $0$&$0$&$0$ & $1-\frac{3}{\lambda^2}$ & $\frac{3}{\lambda^2}$\\
   \hline
   \parbox[c][1.3cm]{1.3cm}{$f_{DE^{\pm}}$} & $\{\frac{\lambda}{\sqrt{6}}, \,\pm \sqrt{1-\frac{\lambda^2}{6}} \, 0, \, 0,\, 0\}$ & $2 \beta  n-n\neq 0, \, \beta=2$ &  $-1+\frac{\lambda^2}{3}$&$-1+\frac{\lambda^2}{3}$ &$0$ & $0$ & $1$\\
 \hline
 \parbox[c][1.3cm]{1.3cm}{$g_{DE}$} &  $\{ 0, \, \frac{1}{2}, \, \frac{3}{4}, \, 0, \, \frac{-\lambda}{3} \}$ & $\beta=0, \, n=2$ &  $-1$ &$-1$&$0$ & $0$ & $1$\\
 \hline
 \parbox[c][1.3cm]{1.3cm}{$h_{S^{\pm}}$} &  $\{\pm1, 0,0,0 ,0 \}$ & $\begin{tabular}{@{}c@{}}$2 \beta  n-n\neq 0$\end{tabular}$& $1$ &$1$&$0$ & $0$ & $1$\\
 \hline
 \end{tabular}}
\caption{Critical points with the existence condition for model \ref{Powerlawcouplingfun}.}
% is used to refer to this table in the text
\label{CH3modelIIcriticalpoints}
\end{table}
%%%%%%
\begin{itemize}
\item{\textbf{DE-dominated critical points:}} The critical points $f_{DE^{\pm}}$ and $ g_{DE}$ describe the DE-dominated phase. The critical points $f_{DE^{\pm}}$ exists for $2\beta n -n \ne 0, \beta=2$ and is a standard DE-dominated critical point with $\Omega_{DE}=1$. This critical point is normally hyperbolic and shows stability at $\sqrt{3}<\lambda <0\lor 0<\lambda <\sqrt{3}$. It is interesting to note that, this critical point can describe the late-time cosmic acceleration phenomena at $-\sqrt{2}<\lambda<\sqrt{2}$. The phase space trajectories are attracting at this critical point and can be observed from Fig. \ref{CH3phasespacem1.2}. The critical point $ g_{DE}$ is a de Sitter solution and can be obtained at $\beta=0, n=2$. The phase space plot is plotted for $x=0,\, y=-4.9,\, u=8,\, \rho=4.5,\, \beta=0, \,  \sigma=-0.30, \, \lambda=-0.2, n=2$ where the model parameters are constrained using the common stability range of the critical points $\left(-3 \sqrt{\frac{2}{7}}\leq \lambda<0\lor \sqrt{3}<\lambda\leq 2 \sqrt{\frac{6}{7}}\right) \land \beta>1.$ For ample clarity, we have plotted a 2D plot for this range in Fig. \ref{CH32dregionplotm1.2}. The critical point $g_{DE}$ becomes normally hyperbolic and shows stable behaviour within the range $-3 \sqrt{\frac{2}{7}}\leq \lambda <0\lor 0<\lambda\leq 3 \sqrt{\frac{2}{7}}$.
\item{\textbf{Stiff matter-dominated critical points:}}
  The critical point $h_{S^{\pm}}$ shows the value of $\omega_{tot}=1$ hence describing the stiff matter-dominated era. Here, the value of $\Omega_{DE}=1$. This critical point shows saddle point nature at $\beta>1\land |\lambda|<\sqrt{6}$. The phase space trajectories can be analysed from Fig. \ref{CH3phasespacem1.2}.
\end{itemize}
%\begin{landscape}
\begin{table}[H]
     % title of Table
    \centering % used for centering table
    \scalebox{0.72}{
    \begin{tabular}{|c |c |c |c| c|} % centered columns (5 columns)
    \hline\hline %inserts double horizontal lines
    \parbox[c][1.3cm]{2.4cm}{\textbf{Critical points}
    }& Eigenvalues & Stability conditions \\ [0.5ex] % inserts table %headin$g$
    \hline\hline % inserts single horizontal line
    \parbox[c][1.3cm]{1.3cm}{$a_{R^{\pm}}$ } & $\Bigl\{0,-1,1,2,-4 (\beta -1)\Bigl\}$ & Saddle at $\beta >1$ \\
    \hline
    \parbox[c][2.0cm]{1.3cm}{$b_{R^{\pm}}$ } & $\begin{tabular}{@{}c@{}}$\Bigl\{0,1,-4 \left(\beta-1\right),-\frac{\sqrt{-\left(1-2 \beta\right){}^2 \lambda^2 \left(15 \lambda^2-64\right)}}{2 \left(2 \beta-1\right) \lambda^2}-\frac{1}{2},$\\$\frac{1}{2} \left(\frac{\sqrt{-\left(1-2 \beta\right){}^2 \lambda^2 \left(15 \lambda^2-64\right)}}{\left(2 \beta -1\right) \lambda^2}-1\right)\Bigl\}$\end{tabular}$ & Saddle at $\beta>1\land \left(-\frac{8}{\sqrt{15}}\leq \lambda<-2\lor 2<\lambda\leq \frac{8}{\sqrt{15}}\right)$\\
    \hline
   \parbox[c][1.3cm]{1.3cm}{$c_{R^{\pm}}$ } & $\begin{tabular}{@{}c@{}}$\Bigl\{0,1,-4 \left(\beta-1\right),-\frac{\sqrt{-\left(1-2 \beta\right){}^2 \lambda^2 \left(15 \lambda^2-64\right)}}{2 \left(2 \beta -1\right) \lambda^2}-\frac{1}{2},$\\$\frac{1}{2} \left(\frac{\sqrt{-\left(1-2 \beta\right){}^2 \lambda^2 \left(15 \lambda^2-64\right)}}{\left(2 \beta-1\right) \lambda ^2}-1\right)\Bigl\}$\end{tabular}$ & Saddle at $\beta>1\land \left(-\frac{8}{\sqrt{15}}\leq \lambda<-2\lor 2<\lambda\leq \frac{8}{\sqrt{15}}\right)$\\
   \hline
   \parbox[c][1.3cm]{1.3cm}{$d_{M}$} &  $\Bigl\{0,-\frac{3}{2},-\frac{1}{2},\frac{3}{2},-3 (\beta -1)\Bigl\}$ &  Saddle at $\beta >1$  \\
   \hline
 \parbox[c][1.3cm]{1.3cm}{$e_{M^\pm}$} &  $\Bigl\{0,-3,-\frac{1}{2},\frac{3 \left(-\lambda^2-\sqrt{24 \lambda^2-7 \lambda ^4}\right)}{4 \lambda^2},\frac{3 \left(\sqrt{24 \lambda^2-7 \lambda^4}-\lambda^2\right)}{4 \lambda^2}\Bigl\}$ &$\begin{tabular}{@{}c@{}} Stable at \\ $-2 \sqrt{\frac{6}{7}}\leq \lambda<-\sqrt{3}\lor \sqrt{3}<\lambda\leq 2 \sqrt{\frac{6}{7}}$ \end{tabular}$\\
 \hline
 \parbox[c][1.3cm]{1.3cm}{$f_{DE^\pm}$} &  $\Bigl\{0,-\lambda^2,\frac{1}{2} \left(\lambda^2-6\right),\frac{1}{2} \left(\lambda^2-4\right),\lambda^2-3\Bigl\}$ &  $\begin{tabular}{@{}c@{}} \\Stable at \\ $\sqrt{3}<\lambda <0\lor 0<\lambda <\sqrt{3}$\end{tabular}$  \\
 \hline
 \parbox[c][1.3cm]{1.3cm}{$g_{DE}$} & $\begin{tabular}{@{}c@{}} $\Bigl\{0,-3,-2,\frac{1}{4} \left(-\sqrt{2} \sqrt{18-7 \lambda^2}-6\right),\frac{1}{4} \left(\sqrt{2} \sqrt{18-7 \lambda^2}-6\right)\Bigl\}$ \end{tabular}$& $\begin{tabular}{@{}c@{}} Stable at\\ $-3 \sqrt{\frac{2}{7}}\leq \lambda <0\lor 0<\lambda\leq 3 \sqrt{\frac{2}{7}}$\end{tabular} $\\
 \hline
 \parbox[c][1.3cm]{1.3cm}{$h_{S^{\pm}}$} & $\begin{tabular}{@{}c@{}} $\Bigl\{0,1,3,-6 \left(\beta -1\right),\frac{1}{2} \left(\sqrt{6} \lambda\pm6\right)\Bigl\}$\end{tabular}$ & $\begin{tabular}{@{}c@{}} Saddle at\\ $\beta>1\land |\lambda|<\sqrt{6}$\end{tabular}$ \\
 \hline
\end{tabular}}
    \caption{Eigenvalues and the stability for model \ref{Powerlawcouplingfun}.}
    % is used to refer to this table in the text
    \label{CH3modelIIeigenvalues}
\end{table}
%\end{landscape}
The common stability range for critical points $e_{M^{\pm}},f_{DE^{\pm}},g_{DE}$ as obtained from the analysis of the autonomous dynamical system for the present power law behaviour of the functional $F(\phi)$ is $\left(-3 \sqrt{\frac{2}{7}}\leq \lambda<0\lor \sqrt{3}<\lambda\leq 2 \sqrt{\frac{6}{7}}\right) \land \beta>1 $.
The combined range of stability conditions of critical points is $\left(-3 \sqrt{\frac{2}{7}}\leq \lambda<0\lor \sqrt{3}<\lambda\leq 2 \sqrt{\frac{6}{7}}\right) \land \beta>1$, and is plotted in Fig. \ref{CH32dregionplotm1.2}.

\begin{figure}[H]
    \centering
\includegraphics[width=60mm]{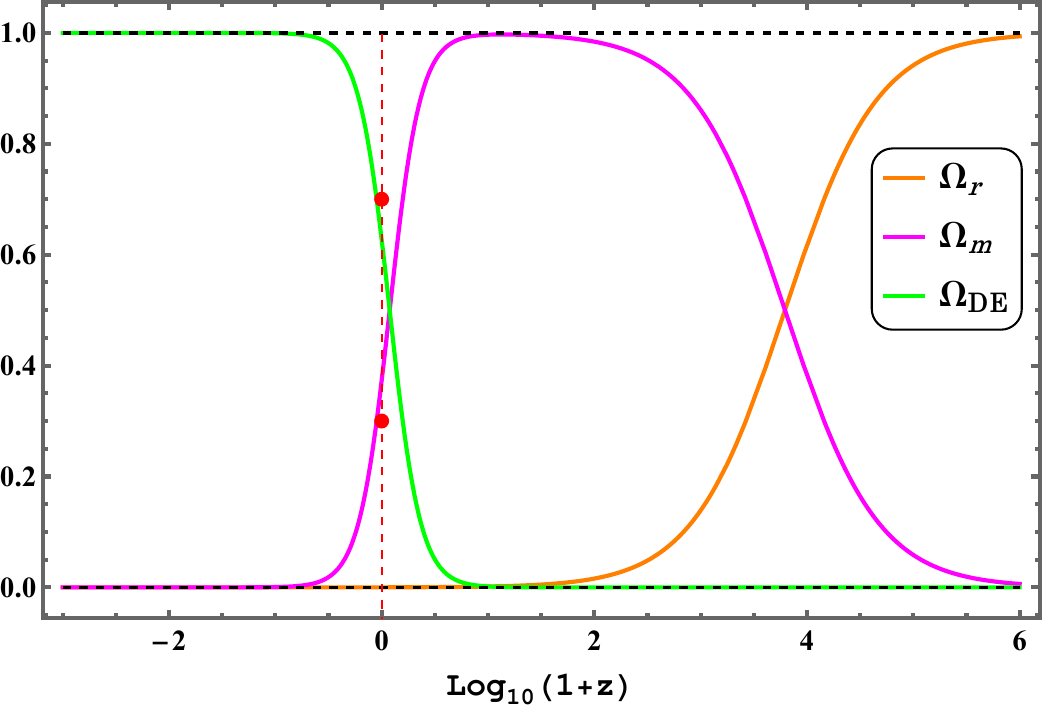}    \includegraphics[width=60mm]{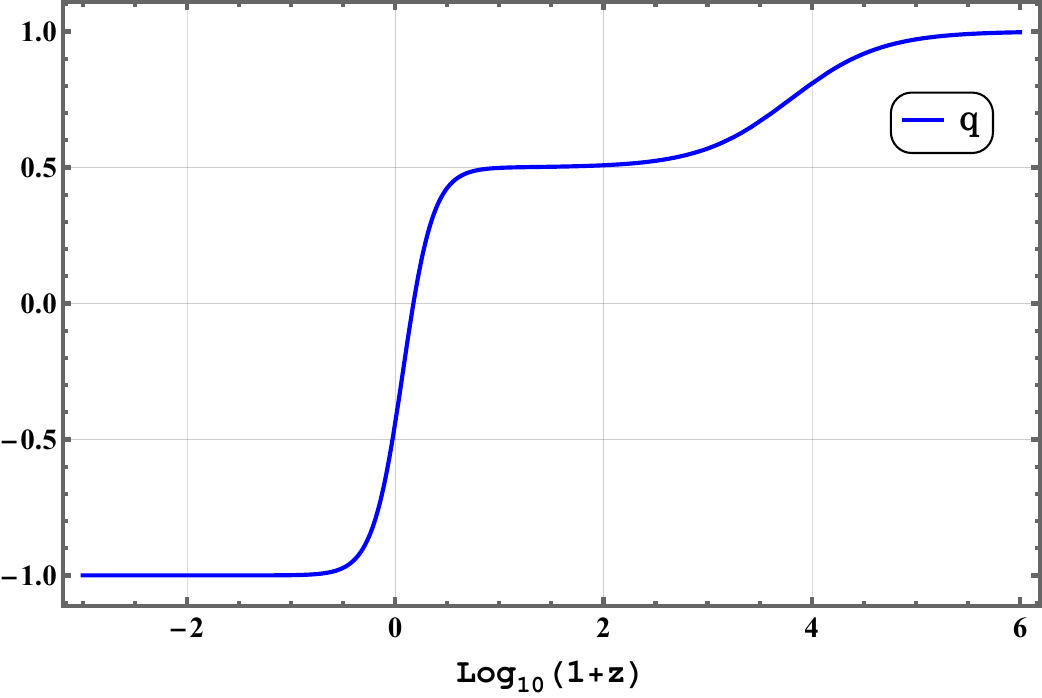}
    \caption{ Density parameter and deceleration parameter for model \ref{Powerlawcouplingfun}. } \label{CH3evolutionm1.2}
\end{figure}

\begin{figure}[H]
    \centering
\includegraphics[width=60mm]{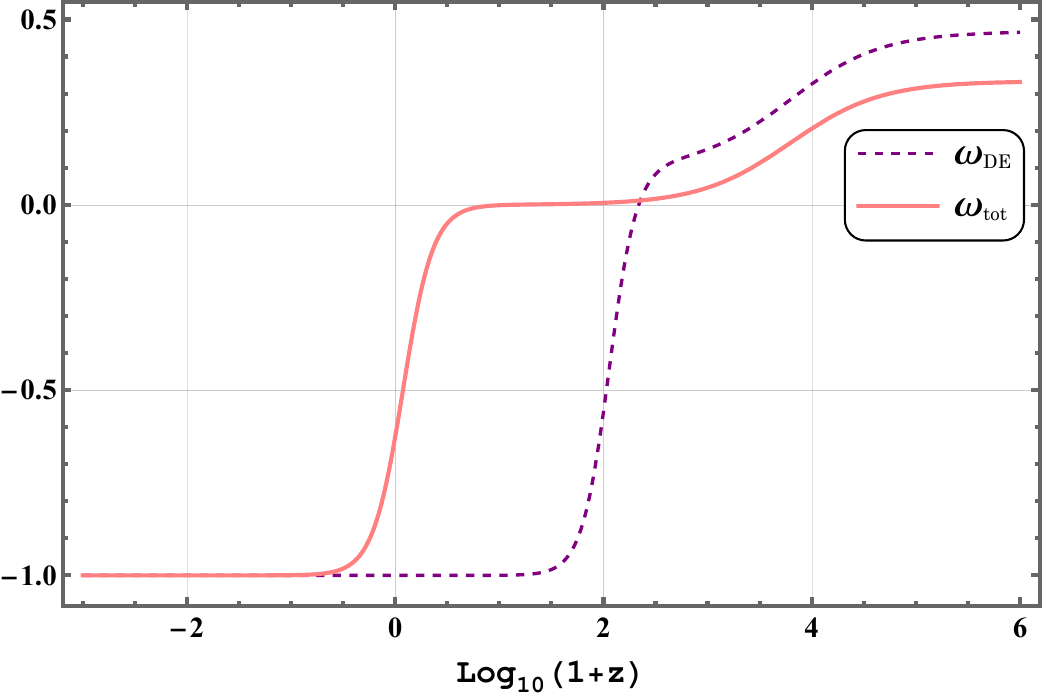}
    \caption{The EoS for DE $(\omega_{DE})$ and total EoS $(\omega_{tot})$ parameter for model \ref{Powerlawcouplingfun}.} \label{CH3Eosplotplotm1.2}
\end{figure}

In Fig. \ref{CH3evolutionm1.2}, we have shown the evolutionary behaviour of $\Omega_{r}, \, \Omega_{m} \,, \Omega_{DE}$ from which one can conclude that at the early phase, radiation dominated the DE and thereafter upto the late epochs, the DE dominates both radiation and matter. The matter-radiation equality occurred at the redshift around $z \approx 3387$ and the value of $\Omega_{DE}\approx 0.7$ \cite{Kowalski_2008} and $\Omega_{m}\approx 0.3$ \cite{Planck:2018vyg}. The transition from deceleration to acceleration occurs at the low redshift $z \approx 0.65$, which can be observed from the plot of the deceleration parameter presented in Fig. \ref{CH3evolutionm1.2}. The transition redshift, as obtained from this dynamical system analysis, is quite compatible with that in the $\Lambda$CDM model. The value of the deceleration parameter at the present time is $q_{0}=-0.4944^{+0.036}_{-0.036}$ and is comparable to that constrained in \cite{Capozziellomnras}. From Fig.~\ref{CH3Eosplotplotm1.2}, the behaviour of $\omega_{DE}$ shows deceleration to acceleration from early to late times, and at the present epoch, we obtain $\omega_{DE}\approx -0.993^{+0.05}_{-0.02}$ which is well within the constrained values \cite{Planck:2015bue,Planck:2018vyg}.
 The Fig. \ref{CH3phasespacem1.2} plotted for $x=0,\, y=-4.9,\, u=8,\, \rho=4.5,\, \beta=-0.2, \, \beta=2$ \, (for $e_{M^{\pm}}, f_{DE^{\pm}}$), \, $\beta=0$ (for $g_{DE}$),$ \,   \sigma=-0.30, \, \lambda=-0.2, \, n= -1.2, \, n= 2$ (for $g_{DE}$). In Figs. \ref{CH3evolutionm1.2}, \ref{CH3Eosplotplotm1.2}  $\beta=1.1, \,  n=0.30, \, \lambda=-0.2$ for the initial conditions $x_0=10^{-8.89}, \, y_0=10^{-2.89}, \, u_0=10^{-5.96}, \, \rho_0= 10^{-0.9}, \, \sigma_{0}=-10 \times 10^{-19}$.

\section{Conclusion}\label{conclusionch3}
In this work, we have studied the dynamical system analysis for the recently developed scalar-torsion $f(T, \phi)$ gravity \cite{Gonzalez-Espinoza:2021mwr,Gonzalezreconstruction2021,Gonzalez-Espinoza:2020jss}  formalism. Two well-motivated non-minimally coupling functions of the scalar field $F(\phi)$, one is exponential \cite{Gonzalez-Espinoza:2020jss,roy2018dynamical} and the other one is power law form \cite{roy2018dynamical} of the scalar field are considered in the analysis. These forms of the coupling functions can be obtained through the reconstruction \cite{Gonzalezreconstruction2021} using the power law form of the torsion scalar $T$ in $f(T,\phi)$ gravity formalism.   

Model \ref{Exponentialcouplingfun} produces seven critical points. The critical points $A_{R^{\pm}}, B_{R^{\pm}}, C_{R^{\pm}}$  are in the radiation-dominated epoch of the Universe. These critical points show unstable behaviour, which can be seen from the 2D phase space plots Fig. \ref{CH3phasespace2dm1}. The critical points $D_{M}, E_{M^{\pm}}$ are the matter-dominated critical points. The critical point $D_{M}$ shows saddle point behaviour and represents a standard matter-dominated era with $\Omega_{m}=1$. The critical point $E_{m^{\pm}}$ is a stable scaling matter-dominated solution and shows stability within the range $2 \sigma<\lambda\leq 2 \sqrt{\frac{6}{7}}\land \beta >\frac{\lambda-\sigma}{\lambda}$.  Moreover, the critical points $F_{DE^{\pm}}, G_{DE}$ represent the DE-dominated era of the evolutionary Universe. The DE-dominated critical points show stable behaviour and are the late-time attractors that can be visualised from the $2D$ phase space plots presented in Fig. \ref{CH3phasespace2dm1}. The critical point $f_{DE^{\pm}}$ describes the accelerated expansion of the Universe evolution with the range $-\sqrt{2}<\lambda<\sqrt{2}$. The exponential coupling function as envisaged in the present model is capable of producing the critical points representing matter, radiation, and the DE-dominated era with a common stability range of the model parameters $-\frac{\sqrt{3}}{2}<\sigma<0\&\&\left(\beta \leq \frac{1}{2}\land -\frac{\sigma }{\beta _{10}-1}<\lambda<0\right)$. The same range can be visualised from the 3D plot presented in Fig. \ref{CH3regionplotm1}.   

In model \ref{Powerlawcouplingfun},  we have analysed the power law coupling scalar field coupling function \cite{roy2018dynamical} with the exponential scalar field potential. The additional dynamical variable $\sigma$ needs to be considered in addition to the four independent dynamical variables $x$, $y$, $u$, $\rho$. In this case, we obtain three critical points $a_{R^{\pm}}, b_{R^{\pm}}, c_{R^{\pm}}$, which represents the radiation-dominated epoch of the Universe evolution and are the unstable critical points. The critical points $d_{m}$ and $e_{M}^{\pm}$ represent the matter-dominated era with $\omega_{tot}=0$. Critical point $d_{M}$ shows saddle point behaviour, whereas the critical point $e_{M^{\pm}}$ is a stable scaling matter-dominated solution. In this case, the critical points $f_{DE^{\pm}}$ and $g_{DE}$ represent the DE-dominated epoch of Universe evolution.  The critical point $f_{DE^{\pm}}$ explain the accelerated expansion within the range $-\sqrt{2}<\lambda<\sqrt{2}$ and the critical point $g_{DE}$ is the de Sitter solution. We obtain the common range of the model parameters where these critical points show stability as $\left(-3 \sqrt{\frac{2}{7}}\leq \lambda<0\lor \sqrt{3}<\lambda\leq 2 \sqrt{\frac{6}{7}}\right) \land \beta>1$ and the same range can be visualised from Fig. \ref{CH32dregionplotm1.2}. 

As a final say, we conclude that both the models, as described through the choices of the scalar field functional $F(\phi)$, are viable cosmological models and are capable of describing different important phases of the evolution of the Universe, including the possible description of the late-time cosmic speed up issue. This work can be extended, and these models can be validated to constrain the viable ranges of the model parameters using different recent cosmological observations.

%Chapter 4
\chapter{Dynamical system analysis in teleparallel gravity with boundary terms} 

\label{Chapter4} % For referencing the chapter elsewhere, use \ref{Chapter1} 

\lhead{Chapter 4. \emph{Dynamical system analysis in teleparallel gravity with boundary terms}} % This is for the header on each page - perhaps a shortened title
\vspace{7 cm}
*The work in this chapter is covered by the following publications: \\

\textbf{S.A. Kadam}, Ninaad P. Thakkar, and B. Mishra, ``Dynamical system analysis in teleparallel gravity with boundary term", \textit{European Physical Journal C}, \textbf{83}, 809 (2023).\\

\textbf{S. A. Kadam}, Santosh V. Lohakare and B. Mishra, ``Dynamical complexity in teleparallel Gauss–Bonnet gravity", \textit{Annals of Physics}, \textbf{460}, 169563 (2024).

%----------------------------------------------------------------------------------------
%\section{Summary of Results}
\clearpage
 \section{Introduction}
 In the present thesis, the preceding two chapters, chapter \ref{Chapter2}, \ref{Chapter3}, are focused on the examination of dynamical system analysis and the assessment of the impact of different scalar field potentials. This section will concentrate on comprehending the role of the teleparallel Gauss-Bonnet and the teleparallel boundary term. This chapter provides a summary of two distinct studies. An autonomous dynamical system is formulated for both higher-order gravity theories. Dealing with these teleparallel higher-order gravity formalisms poses a significant challenge in obtaining analytical solutions to the system of equations. Therefore, the approach we have adopted is crucial in examining the qualitative behavior of the solution without looking into the exact cosmological solutions of the system. Several interesting aspects emerge during these investigations, one of which is the role of the variable $\lambda=\frac{\ddot{H}}{H^2}$ when constructing an autonomous dynamical system. This variable is essential for defining the dynamical system in its autonomous form. Depending on its values, we will analyze the various phases of the evolution of the Universe. A comprehensive analysis is obtained in this chapter.
\section{\texorpdfstring{$f(T, B)$}{} gravity field equations}\label{f(TB)fieldequations}
In flat space-time, the formalism in Sec  \ref{ftbformulasec} imply that the torsion scalar $T$ and the boundary term $B$ respectively reduce to,
\begin{equation} \label{TandB}
T=6H^{2}, \hspace{2cm} B=6(3H^{2}+\dot{H}).
\end{equation}
Now, the field equations of $f(T,B)$ gravity can be derived on varying the action Eq. \eqref{f(TBaction)} for the metric \eqref{FLATFLRW} and tetrad \eqref{FLRWTETRAD} as, 
\begin{eqnarray}
   3H \dot{f_B}-3H^2(3f_B+2f_T)-3\dot{H}f_B+\frac{1}{2}f(T,B)&=&\kappa^2\rho \,,\nonumber\label{eq:6}\\
  \ddot{f_B} -3H^2(3f_B+2f_T)-\dot{H}(3f_B+2f_T)-2H\dot{f_T} +\frac{1}{2}f(T,B)&=&-\kappa^2 p\,. \label{eq:7}
\end{eqnarray}
One of the most important properties of this theory is that it satisfies the continuity equation $(\dot{\rho}_i+3H(\rho_i+p_i)=0)$ for $i= m, r, DE$, that is matter, radiation, and the DE  respectively. To better understand the contributions of the modified Lagrangian, we consider the $f (T, B)$ gravity Lagrangian mapping, $f(T, B) \rightarrow - T + \Tilde{f}(T, B)$, then the Friedmann equations in Eq. \eqref{FriedmanEQ} can be used to obtain the expression for energy density and pressure for the DE phase can be obtained as,
\begin{eqnarray}
3H^2(3\Tilde{f}_B+2\Tilde{f}_T)-3H \dot{\Tilde{f}}_B+3\dot{H}\Tilde{f}_B-\frac{1}{2}\Tilde{f} (T,B)&=&\kappa^2\rho_{DE} \,, \label{eq:12}\\
-3H^2(3\Tilde{f}_B+2\Tilde{f}_T)-\dot{H}(3\Tilde{f}_B+2\Tilde{f}_T)-2H\dot{\Tilde{f}}_T+\ddot{\Tilde{f}}_B+\frac{1}{2}\Tilde{f}(T,B)&=&\kappa^2 p_{DE}\,.\label{eq:13}
\end{eqnarray}
The expression EoS parameter for the DE phase can be written as,
\begin{equation}
\omega_{DE}=-1+\frac{\ddot{\Tilde{f}}_{B}-3H\dot{\Tilde{f}}_{B}-2\dot{H}\Tilde{f}_{T}-2H\dot{\Tilde{f}}_{T}}{3H^{2}(3\Tilde{f}_{B}+2\Tilde{f}_{T})-3H\dot{\Tilde{f}}_{B}+3\dot{H}\Tilde{f}_{B}-\frac{1}{2}\Tilde{f}(T,B)}\,.\label{eq:14}
\end{equation}
Next, we shall define the dynamical variables and express the cosmological parameters in terms of dynamical variables. The cosmological behaviour of the models will be studied through some functional forms of $\Tilde{f}(T, B)$.

\section{Dynamical system analysis in \texorpdfstring{$f(T, B)$}{} 
gravity}\label{sec:dynamicalsystemanalysis}
From the theoretical point of view, any proposed cosmological model should contain at least part that explains ``Inflation $\rightarrow$ Radiation $\rightarrow$ Matter $\rightarrow$ DE " \cite{bohmer2017dynamicalsystem}. To attain the aforementioned suggested cosmological model, inflation must be an unstable point in order for the Universe to have an inflation exit, whereas radiation and matter points must be saddle points in order for these eras to be long enough. The final phase of the DE era should be a stable period of accelerated expansion.
In order to analyse this, we consider the Universe filled with two fluids such that $\rho=\rho_{m}+\rho_{r}$, where $\rho_m$ and $\rho_{r}$ respectively be the energy density for matter and radiation. In the matter dominated phase $p_{m}=0$ and hence $\omega_m$ vanishes, whereas in the radiation phase, $\omega_{r}=\frac{1}{3}$. With these, we define the dynamical variables as,
\begin{eqnarray}
& & X=\tilde{f}_{B} \,,\quad Y=\tilde{f}_{B} \frac{\dot{H}}{H^2}\,,\quad Z=\frac{\dot{\tilde{f}}_B}{H} \,,\quad 
V=\frac{\kappa^2 \rho_r}{3H^2}\,,\quad  W=-\frac{\tilde{f}}{6H^2}\,.\label{dynamicalvariablech4}
\end{eqnarray}
The standard density parameters expressions for matter $(\Omega_m)$, radiation $(\Omega_r)$ and DE $(\Omega_{DE})$ as presented in Eq. \eqref{densityparameters} can be constrained in terms of a dynamical variable can be written as,
\begin{align}
\Omega_{m}+\Omega_{r}+W+2\tilde{f}_{T}+Y+3X-Z=1\,,
\end{align}
where
\begin{align}
    \Omega_{DE}=W+2\tilde{f}_{T}+Y+3X-Z\,.
\end{align}
The total EoS parameter and DE EoS parameter are respectively obtained in dynamical variables as,
\begin{align}
\omega_{tot}&=-1-\frac{2 Y}{3 X}\,,\nonumber\\
\omega_{DE}&=-\frac{(V+3) X+2 Y}{3 X (2 \tilde{f}_{T} +W+3 X+Y-Z)}\,.
    \label{expressionomegadeomegatot}
\end{align}
To express the autonomous dynamical system, we define the parameter $\lambda=\frac{\ddot{H}}{H^3}$ \cite{Odintsov:2018,Franco:2020lxx} and is treated as a constant throughout the analysis. To note the value of the parameter $\lambda=8, \frac{9}{2},$ connects with the radiation, matter dominated phase, respectively, whereas for DE, it depends on the dynamical variables $X$ and $Y$. The values of this parameter is not dependent on choice of $f(T,B)$, moreover these values can be retraced using the power law solution. Subsequently, we have obtained the autonomous dynamical system as follows,
\begin{eqnarray}
\frac{dX}{dN}&=&Z\,,\nonumber\\
\frac{dY}{dN}&=&\lambda  X+\frac{Y (Z-2 Y)}{X}\,,\nonumber\\
\frac{dZ}{dN}&=&6 \tilde{f}_{T} -V+3 W+\frac{2 \tilde{f}_{T}  Y}{X}-\frac{Y Z}{X}-\frac{2 Y}{X}+ 9 X+3 Y+2\tilde{f}^{\,'}_{T}-3\,,\nonumber\\
\frac{dV}{dN}&=&-\frac{2 V (2 X+Y)}{X}\,,\nonumber\\
\frac{dW}{dN}&=&-\frac{2 W Y}{X}-\lambda  X-\frac{2 \tilde{f}_{T}  Y}{X}-6 Y\,.\label{GDS}
\end{eqnarray}
Now, to study the stability analysis, we need some form of $\Tilde{f}(T, B)$, and hence we have considered two forms of $\Tilde{f}(T, B)$ that lead to two models.
\subsection{Logarithmic boundary term coupling model} \label{logbmodel}
\begin{figure}[H]
    \centering
    \includegraphics[width=60mm]{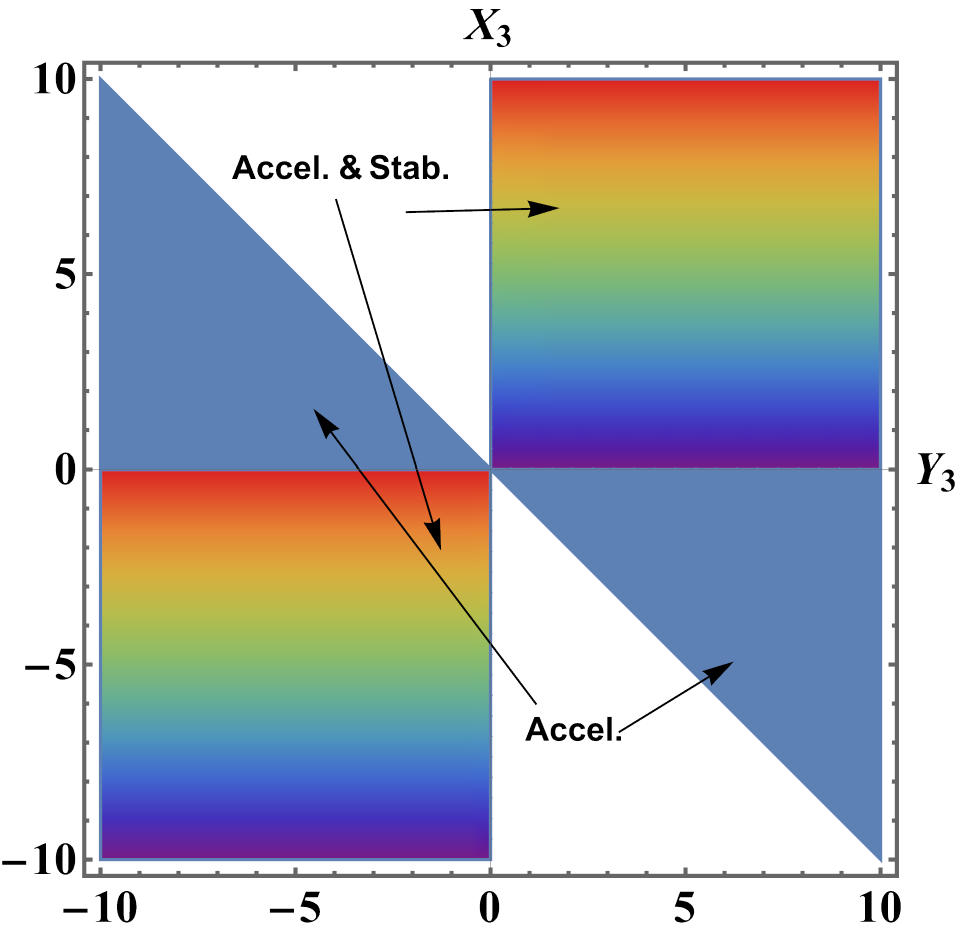}
    \caption{2D plot for stability and acceleration region for critical point $C_{3}$ for model \ref{logbmodel}. } \label{modelIregionplot}
\end{figure}

\begin{figure}[H]
    \centering
    \includegraphics[width=60mm]{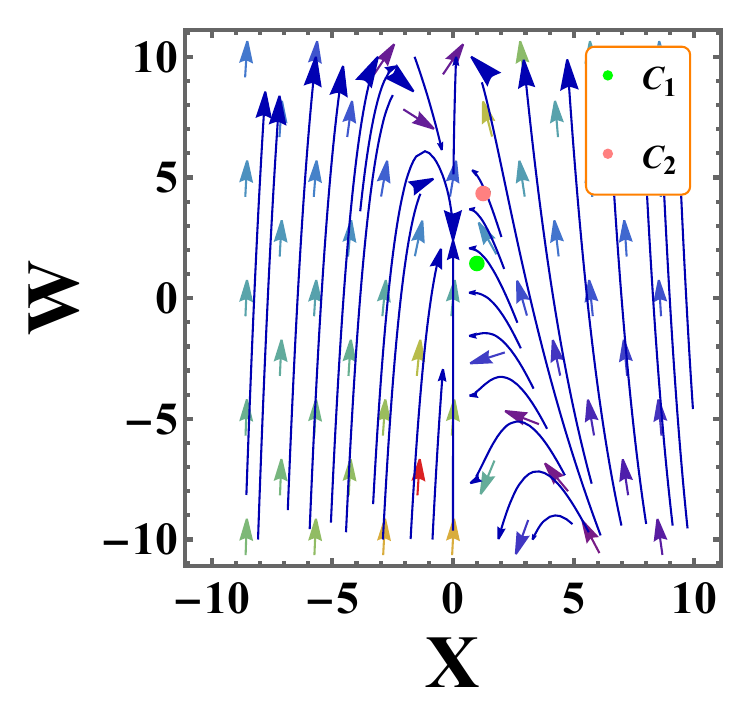}
    \includegraphics[width=60mm]{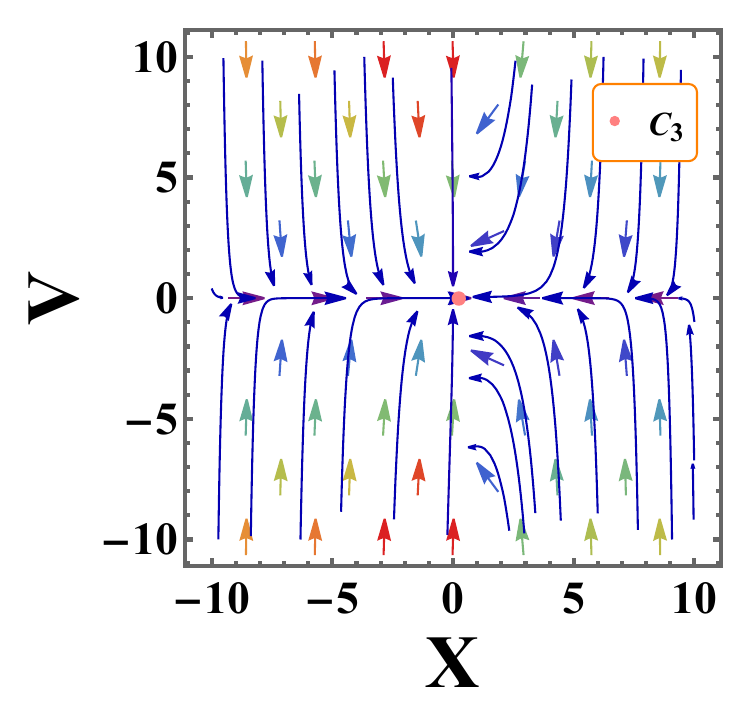}
    \caption{2D phase portrait for model \ref{logbmodel} .} \label{modelI2dphaseportrait}
\end{figure}

\begin{figure}[H]
    \centering
    \includegraphics[width=60mm]{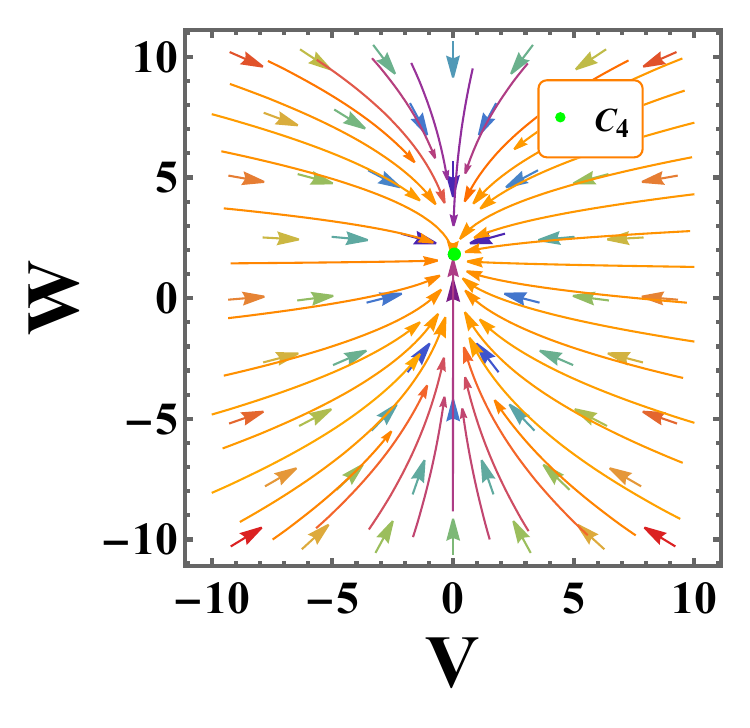}
    \caption{2D phase portrait for model \ref{logbmodel} .} \label{modelI2dphaseportrait2}
\end{figure}

\begin{figure}[H]
    \centering
    \includegraphics[width=60mm]{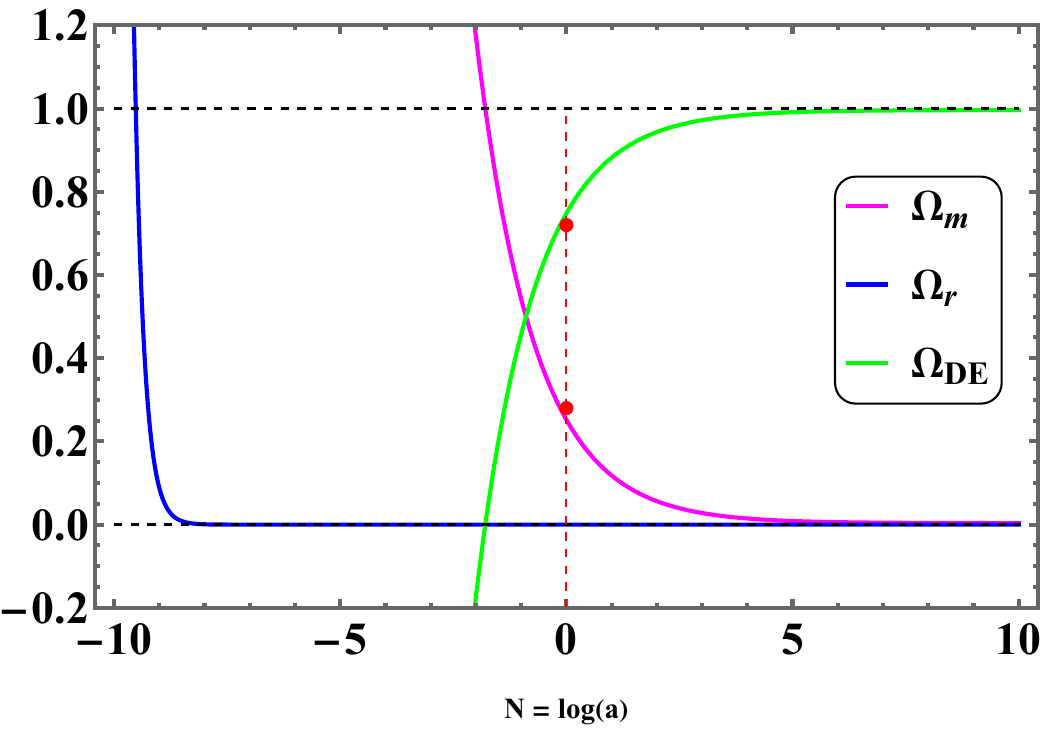}
    \caption{Density parameters for model \ref{logbmodel}.}\label{model1evolutionch4}
\end{figure}

\begin{figure}[H]
    \centering
    \includegraphics[width=65mm]{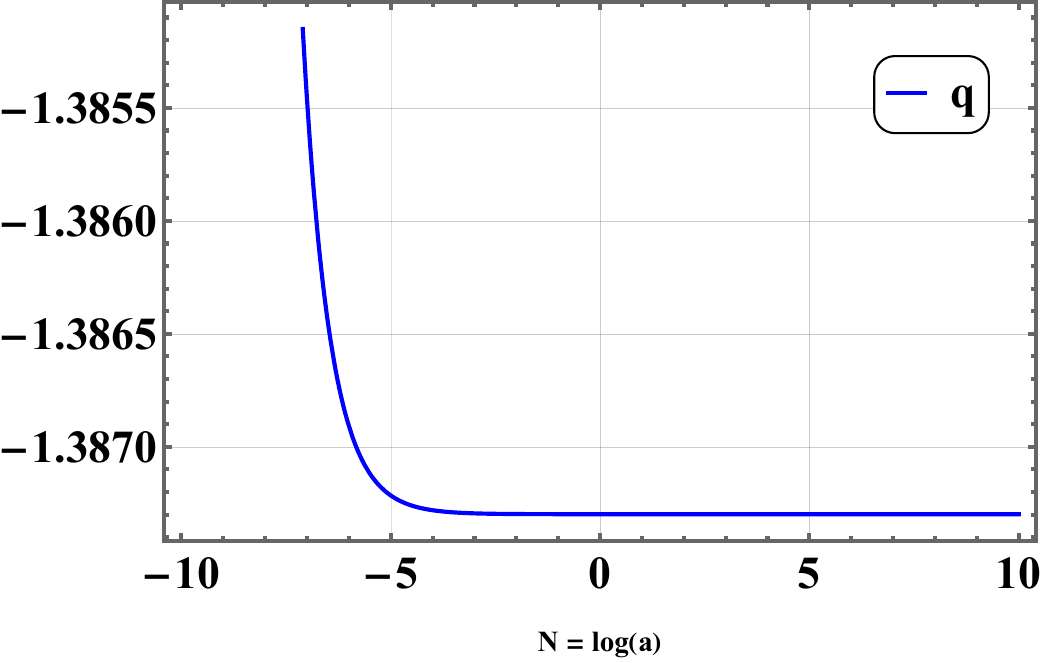}
     \includegraphics[width=60mm]{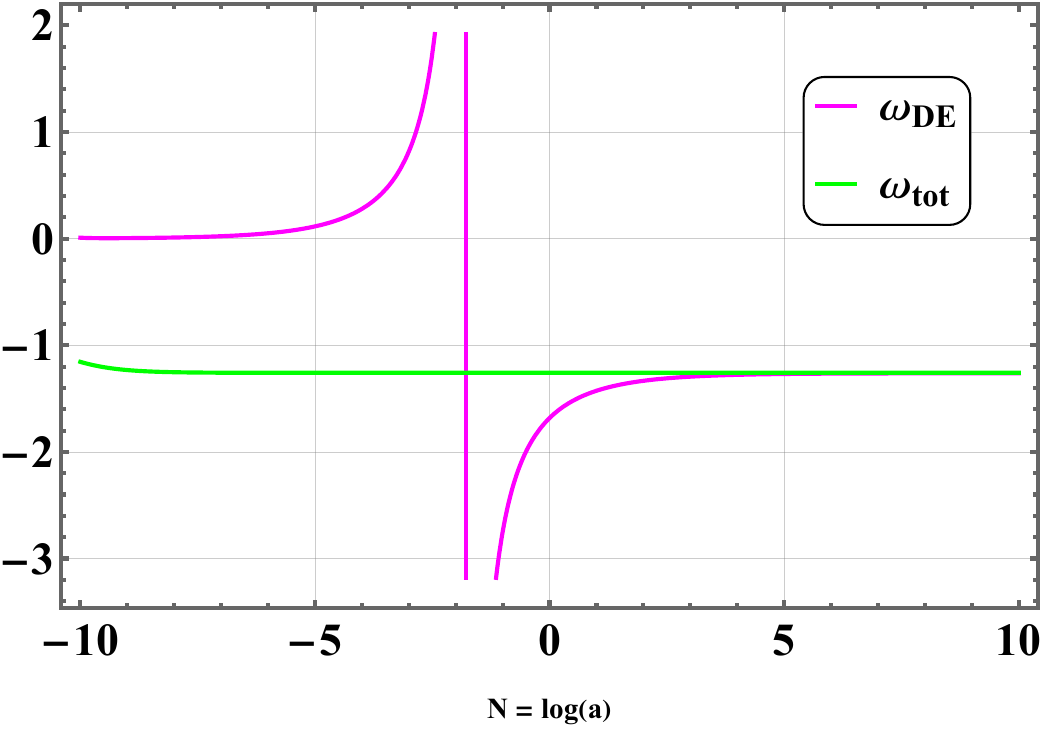}
    \caption{Deceleration parameter for Model \ref{logbmodel}.} \label{modelIomegadedecelech4}
\end{figure}

We consider,
\begin{equation*}
\Tilde{f}(T,B) =\xi T + \alpha B log (B) \,.
\end{equation*}
This specific form of $f(T, B)$ has been successful in addressing the late-time cosmic phenomena issue \cite{Escamilla-Rivera:2019ulu,bamba2011eos}, Noether symmetry \cite{Capozziello:2016eaz}. Also, the critical points can be analysed in the presence of a non-canonical scalar field and the exponential potential function in $f(T, B)$ gravity framework \cite{paliathanasis2021epjp}. In the absence of a scalar field, the cosmological aspects through the behavior of critical points analysis may provide some deeper insight into the evolution of the Universe in different evolution phases. The dynamical variable $Z$ from Eq. \eqref{dynamicalvariablech4} can be written as, $Z=\alpha\left[\frac{6Y+X\lambda}{3X+Y}\right]$ and treated as the dependent variable. The autonomous dynamical system for this setup can be obtained as,
\begin{align}
\frac{dX}{dN}&=\frac{\alpha  (\lambda  X+6 Y)}{3 X+Y}\,,\nonumber\\
\frac{dY}{dN}&=X \left(\lambda -\frac{2 Y^2}{X^2}\right)+\frac{\alpha  Y (\lambda  X+6 Y)}{X (3 X+Y)}\,,\nonumber\\
\frac{dV}{dN}&=-\frac{2 V (2 X+Y)}{X}\,,\nonumber\\
\frac{dW}{dN}&=-\frac{2 W Y+\lambda  X^2+6 X Y+2 \xi  Y}{X}\,.
    \label{Dynamicalsystemmodel-I}
\end{align}
The standard density parameters for DE and matter can be calculated as
\begin{align}
    \Omega_{DE}&=2 \xi +W+3 X+Y-\frac{\alpha  (\lambda  X+6 Y)}{3 X+Y}\nonumber\,,\\ 
    \Omega_{m}&=-2 \xi -V-W-3 X-Y+1+\frac{\alpha  (\lambda  X+6 Y)}{3 X+Y}\,.
\end{align}
and the EoS parameter for DE is given as.
\begin{align}
   \omega_{DE} =-\frac{(V+3) X+2 Y}{3 X \left(2 \xi +W-\frac{\alpha  (\lambda  X+6 Y)}{3 X+Y}+3 X+Y\right)}\nonumber\,.
\end{align}
 We shall find the critical points of the dynamical system with their existence conditions are presented in Table \ref{modelIcriticalpoint},
\begin{table}[H]
    \centering
    \begin{tabular}{|c|c|c|c|c|c|}
        \hline\hline
        \parbox[c][0.8cm]{3.8cm}{\centering Critical points} & 
        \parbox[c][0.8cm]{0.8cm}{\centering X} & 
        \parbox[c][0.8cm]{0.8cm}{\centering Y} & 
        \parbox[c][0.8cm]{0.8cm}{\centering V} & 
        \parbox[c][0.8cm]{1.5cm}{\centering W} & 
        \parbox[c][0.8cm]{4.2cm}{\centering Exist for} \\ [0.5ex]
        \hline\hline
        \parbox[c][1cm]{3.8cm}{\centering\textbf{$C_1= (X_1,Y_1,V_1,W_1)$}} & 
        \parbox[c][1cm]{0.8cm}{\centering$X_1$} & 
        \parbox[c][1cm]{0.8cm}{\centering$-2 X_1$} & 
        \parbox[c][1cm]{0.8cm}{\centering$V_1$} & 
        \parbox[c][1cm]{1.4cm}{\centering$-\xi - X_1$} & 
        \parbox[c][1cm]{4.2cm}{\centering$X_1 \neq 0, \alpha = 0, \lambda = 8$} \\
        \hline
        \parbox[c][1.3cm]{3.8cm}{\centering$C_2= (X_2,Y_2,V_2,W_2)$} & 
        \parbox[c][1.3cm]{0.8cm}{\centering$X_2$} & 
        \parbox[c][1.3cm]{0.8cm}{\centering$-\frac{3}{2} X_2$} & 
        \parbox[c][1.3cm]{0.8cm}{\centering$0$} & 
        \parbox[c][1.3cm]{1.4cm}{\centering$W_2$} & 
        \parbox[c][1.3cm]{4.2cm}{\centering$X_2 \neq 0, \xi = -W_2 - 3 X_2 - Y_2, \alpha = 0, \lambda = \frac{9}{2}$} \\
        \hline
        \parbox[c][1.3cm]{3.8cm}{\centering$C_3= (X_3,Y_3,V_3,W_3)$} & 
        \parbox[c][1.3cm]{0.8cm}{\centering$X_3$} & 
        \parbox[c][1.3cm]{0.8cm}{\centering$Y_3$} & 
        \parbox[c][1.3cm]{0.8cm}{\centering$0$} & 
        \parbox[c][1.3cm]{3.1cm}{\centering$-3 X_3 - Y_3 - \xi$} & 
        \parbox[c][1.3cm]{4.2cm}{\centering$X_3 \neq 0, Y_3(3 X_3 + 2 Y_3) \neq 0, \alpha = 0, \lambda = \frac{2 Y_3^2}{X_3^2}$} \\
        \hline
        \parbox[c][1.3cm]{3.8cm}{\centering$C_4= (X_4,Y_4,V_4,W_4)$} & 
        \parbox[c][1.3cm]{0.8cm}{\centering$X_4$} & 
        \parbox[c][1.3cm]{0.8cm}{\centering$0$} & 
        \parbox[c][1.3cm]{0.8cm}{\centering$0$} & 
        \parbox[c][1.3cm]{0.8cm}{\centering$W_4$} & 
        \parbox[c][1.3cm]{4.2cm}{\centering$X_4 \neq 0, W_4, \alpha, \xi = \text{arbitrary}, \lambda = 0$} \\
        \hline
    \end{tabular}
    \caption{The critical points for model \ref{logbmodel}}
    \label{modelIcriticalpoint}
\end{table}
 We have given eigenvalues of each criticalpoints for the analysis below:
\begin{table}[H]
     % title of Table
    \centering % used for centering table
    \begin{tabular}{|c|c|} % centered columns (5 columns)
    \hline\hline %inserts double horizontal lines
    \parbox[c][1.3cm]{2.3cm}{Critical points}& Eigenvalues \\ [0.5ex] % inserts table %heading
    \hline\hline % inserts single horizontal line
    \parbox[c][0.3cm]{1.3cm}{$C_1$ } & $\{0,0,4,8\}$   \\
    \hline
    \parbox[c][0.3cm]{1.3cm}{$C_2$ } & $\left\{6,3,-1,0\right\}$  \\
    \hline
   \parbox[c][0.3cm]{1.3cm}{$C_3$ } &  $\left\{0,-4\frac{Y_3}{X_3}, -2\frac{Y_3}{X_3}, -2\frac{(2X_3+Y_3)}{X_3}\right\}$ \\
   \hline
   \parbox[c][0.3cm]{1.3cm}{$C_4$ } &  $\{0,0,0,-4\}$   \\
    \hline
    \end{tabular}
    \caption{Eigenvalues corresponding to the critical point for model \ref{logbmodel}}
    % is used to refer to this table in the text
    \label{modelIeigenvaluesch4}
\end{table}
\begin{center}
\begin{table}[H]
    % title of Table
    \centering % used for centering table
    \scalebox{0.8}{
    \begin{tabular}{|c|c|c|c|c|} % centered columns (5 columns)
    \hline\hline %inserts double horizontal lines
    \parbox[c][1.3cm]{2.2cm}{Critical points} & Stability conditions & $q$ & $\omega_{tot}$ & $\omega_{DE}$ \\ [0.5ex] % inserts table %heading
    \hline\hline % inserts single horizontal line
    \parbox[c][0.8cm]{2.2cm}{\textbf{$C_1$}} & Unstable & $1$ & $\frac{1}{3}$ & $-\frac{V_1-1}{3 \xi }$ \\
    \hline
    \parbox[c][0.8cm]{2.2cm}{$C_2$}  &  Unstable& $\frac{1}{2}$ & $0$& 0\\
    \hline
    \parbox[c][0.8cm]{2.2cm}{$C_3$}  &  $\begin{tabular}{@{}c@{}}     Stable for\\$\left(X_3<0\land Y_3<0\right)\lor$\\$ \left(X_3>0\land Y_3>0\right)$\end{tabular}$ & $-\frac{X_3+Y_3}{X_3}$  & $-1-\frac{2 Y_3}{3 X_3}$ & $\frac{3 X_3+2 Y_3}{6 Y_3-3 \xi X_3}$ \\
     \hline
    \parbox[c][0.8cm]{2.2cm}{$C_4$}  &    Nonhyperbolic  & $-1$  & $-1$ & $-\frac{1}{2 \xi+W_4+3 X_4}$ \\
    [1ex] % [1ex] adds vertical space
    \hline %inserts a single line
    \end{tabular}}
     \caption{Stability condition, EoS parameter and deceleration parameter for model \ref{logbmodel}}
    % is used to refer to this table in the text
    \label{modelIstabilitycondich4}
\end{table}
\end{center}
\begin{center}
\begin{table}[H]
    % title of Table
    \centering % used for centering table
    \scalebox{0.8}{
    \begin{tabular}{|c|c|c|c|c|c|} % centered columns (5 columns)
    \hline\hline %inserts double horizontal lines
   \parbox[c][1.3cm]{1.3cm}{Critical Points} & Evolution Eqs.& Universe phase & $\Omega_{m}$& $\Omega_{r}$ & $\Omega_{DE}$\\ [0.5ex] % inserts table %heading
    \hline\hline % inserts single horizontal line
   \parbox[c][0.8cm]{1.3cm}{ $C_1$ }& $\dot{H}=-2 H^{2}$ & $a(t)= t_{0} (2 t+c_{2})^\frac{1}{2}$ & $-\xi-V_1+1$& $V_1$ & $\xi$\\
    \hline
    \parbox[c][0.8cm]{1.3cm}{$C_2$} & $\dot{H}=-\frac{3}{2}H^{2}$ & $a(t)= t_{0} (\frac{3}{2}t+c_{2})^\frac{2}{3}$ & $W_2+\frac{3 X_2}{2}+1$& $0$ & $-W_2-\frac{3 X_2}{2}$\\
    \hline
    \parbox[c][0.8cm]{1.3cm}{$C_3$ }&  $\dot{H}=-\frac{Y_3}{ X_3}H^2$ & $a(t)= t_{0} (\frac{Y_3}{X_3}t+c_{2})^\frac{X_3}{Y_3}$ & $-\xi +\frac{2 Y_3}{X_3}+1$& $0$ & $\xi -\frac{2 Y_3}{X_3}$\\
    \hline
    \parbox[c][0.8cm]{1.3cm}{$C_4$ }&  $\dot{H}=0$ & $a(t)= t_{0} e^{c_2 t}$ & $-2 \xi -W_4-3 X_4+1$& $0$ & $2 \xi +W_4+3 X_4$\\
    [1ex] % [1ex] adds vertical space
    \hline %inserts single line
    \end{tabular}}
     \caption{Evolution Eqs., phase of the Universe, density parameters for model \ref{logbmodel}}
     % is used to refer this table in the text
    \label{modelIdensityparameterch4}
\end{table}
\end{center}
\begin{itemize}
\item{\textbf{Radiation-dominated critical points:}}
The critical point $C_1$ with $\lambda=8$ describes the radiation-dominated era. The value of the parameter $\omega_{tot}=\frac{1}{3}$, $q=1$. This critical point will describe the standard radiation-dominated era for $V_{1}=1,\xi=0$ at which the contribution of the standard density parameter for DE will vanish. The eigenvalues at this critical point are presented in Table \ref{modelIeigenvaluesch4}, which shows this critical point is unstable in nature and the corresponding value of the deceleration and EoS parameter is that of radiation-dominated Table \ref{modelIstabilitycondich4}. The evolution equation, along with the exact solution, is obtained and presented in Table \ref{modelIdensityparameterch4}. The exact solution obtained at this critical point is in the power law $a(t)=t_0 (t)^h$ form with $h=\frac{1}{2}$ which explains the radiation-dominated era. The behavior of phase space trajectories at this critical point shows that this critical point is the saddle point and hence unstable, as can be seen in Fig.  \ref{modelI2dphaseportrait}.

\item{\textbf{Matter-dominated critical points:}}
The value of the EoS parameter ($\omega_{tot}$) vanishes at critical point $C_2$ critical point, hence, this critical point represents the CDM (Cold DM) phase of the evolution of Universe. The value of dynamical variable $\lambda$ is $\frac{9}{2}.$ This critical point describes a non-standard CDM-dominated era with the small contribution of the DE density parameter refer Table \ref{modelIdensityparameterch4}. The evolution equation with the exact solution at this critical point is presented in Table \ref{modelIdensityparameterch4}, the power law solution with index $\frac{2}{3}$ indicates that this will describe the CDM-dominated era. The existence of positive and negative eigenvalues at this critical point, as presented in Table \ref{modelIeigenvaluesch4} shows that this critical point is unstable. The phase space trajectories at this critical point show that this critical point is a saddle point, which can be analyzed in Fig. \ref{modelI2dphaseportrait}, which will support the stability condition obtained from the sign of the eigenvalues.

\item{\textbf{DE-dominated critical points:}} 
At critical point $C_3$, the value of $q, \omega_{DE}$ and $\omega_{tot}$ is dependent on the dynamical variables $X, Y$ hence, this critical point can describe the early as well as the late phases of the Universe evolution. The critical point will describe the de Sitter solution at $Y_3=0$ and describe accelerating expansion of the Universe at parametric range $\left(X_3<0\land Y_3<-X_3\right)\lor \left(X_3>0\land Y_3>-X_3\right)$. To get better clarity, the range of parameters where it describes accelerating expansion and stability is plotted in the region plot in Fig. \ref{modelIregionplot}. From this, we can analyze that the value of $Y_3$ is near 0 in stability, existence, and in the parametric range where parameters are capable to describe the accelerated expansion of the Universe at critical point $C_3$, hence this critical point is capable of describing a DE dominated era of Universe evolution. The phase space trajectories at this critical point are attractors and can be analyzed in Fig. \ref{modelI2dphaseportrait}. This critical point is describing the standard DE-dominated era with $\Omega_{DE}=1$ at $\xi=1, Y_3=0$ refer Table \ref{modelIdensityparameterch4}. The exact solution obtained at this critical point is in the power law $a(t)=t_0 (t)^h$ form with $h=\frac{X_3}{Y_3}$ depending upon the value of dynamical variables $X_3$ and $Y_3$, the corresponding phase of the Universe evolution can be analysed. The eigenvalues at this critical point are normally hyperbolic \cite{coley2003dynamical} and are stable in the parameter range as described in Table \ref{modelIstabilitycondich4}.

\item{\textbf{de Sitter solution:}} The value of $q=-1$, $\omega_{tot}=-1$ at the critical point $C_4$, hence this critical point explains the de Sitter solution. The exact cosmological solution at this critical point is described in Table \ref{modelIdensityparameterch4} which takes the de Sitter solution form. From this one can observe that this critical point should explain a standard DE-dominated era at $\xi=\frac{1}{2}, W_4 =0, X_4 =0$. However, due to the presence of zero eigenvalues, this critical point is non-hyperbolic in nature. The linear stability theory fails to provide information regarding the stability of the critical point if the critical point contains zero eigenvalues. Moreover, since the system equations do not satisfy the central manifold condition (after separating from linear and non-linear parts, the non-linear part at zero is not vanishing), therefore it fails to describe the stability of this critical point. The 2D phase space diagram to analyse the behaviour of the phase space trajectories at this critical point in Fig. \ref{modelI2dphaseportrait2} has been given, where it is observed that the phase space trajectories are attracted towards this critical point. Hence this critical point is an attractor.
\end{itemize}
The above 2D phase plots are for $\lambda=0.3, \xi=-2.4, \alpha=1.3$.
In Fig. \ref{modelIregionplot}, the stability and acceleration region of the critical point $C_3$ has been shown. Though the light blue shaded area in the figure shows the acceleration, the stability could not be established. The evolution plot of standard density parameters in terms of the $N$  as defined in chapter \ref{Chapter2}, in Fig. \ref{model1evolutionch4}, we can find the value of $\Omega_{r} \approx 0.019, \Omega_{m} \approx 0.3$ and $\Omega_{DE} \approx 0.7$. The vertical dashed line represents the present time. The blue curve represents the evolution of the standard density parameter for radiation, and it can be observed that this curve dominates the other two curves at the early evolution, decreasing gradually from early to late time, and tends to zero at late times. The behavior of $q$ and EoS for $\omega_{tot}, \omega_{DE}$ can be analyzed in Fig. \ref{modelIomegadedecelech4}. Since the plot of $q$ lies in the negative region, hence it is capable of describing the current accelerated expansion of the Universe. The value of $q$ at the present time is $-1.387$ which is approximately the same as the current observation study \cite{Feeney:2018}. The plot of $\omega_{tot}$ at present value takes the value $-1.233$ which agrees with $\omega_0=-1.29 ^{+0.15}_{-0.12}$ \cite{DIVALENTINO:2016}.
The initial conditions used to plot Figs. \ref{model1evolutionch4}, \ref{modelIomegadedecelech4} are $X=-1.2 \times 10^{2.2},\, Y=2.2 \times 10^{-3.4},\,  V=1.02 \times 10^{2.6},\, W=4.5\times 10^{-8.1},\,  \lambda=0.3,\, \xi=-2.4, \alpha=1.3$.
\subsection{Power law model}\label{ModelIIch4}
\begin{figure}[H]
    \centering
    \includegraphics[width=60mm]{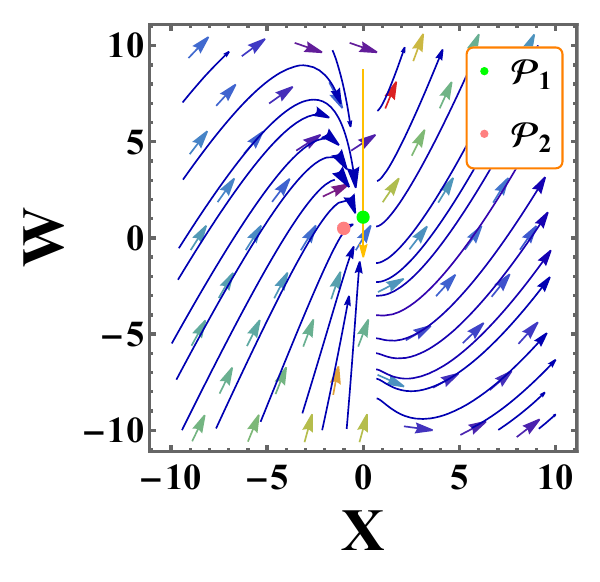}
    \includegraphics[width=60mm]{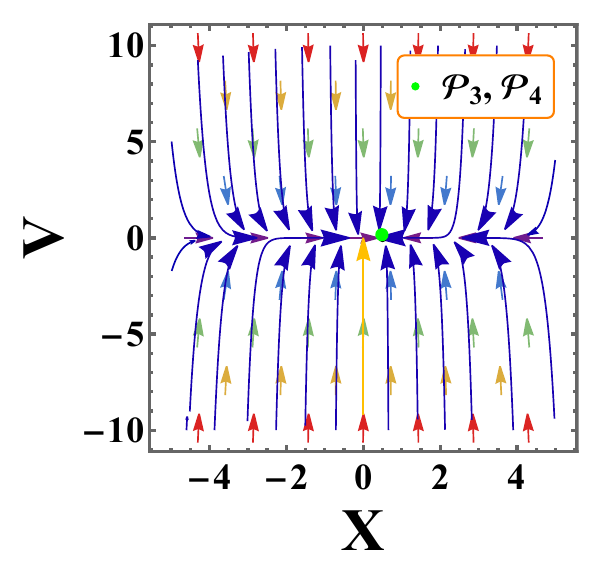}    
    \caption{2D phase portrait for the dynamical system for model \ref{ModelIIch4}.} \label{modelII2dphaseportraitch4}
\end{figure}

We consider,
\begin{equation}
\Tilde{f}(T,B) = \zeta T + \beta (-B)^p \,,
\end{equation}
where $\Tilde{f_2}(B)= \beta (-B)^p$, is nonlinear and capable of studying observational tests for the theory by referring to the most recent SN-Ia data \cite{Bengochea:2008gz}. This is a prominent form of $\tilde{f}(T, B)$ gravity due to its usefulness in explaining the present cosmic expansion and $H_0$ tension \cite{briffa202}. Similar to the first model, we get a relationship for the dynamical variable $Z=\left[\frac{(p-1) X (\lambda  X+6 Y)}{3 X+Y}\right],$ hence the dynamical variable $Z$ is treated as a dependent and the others are independent. The terms $\tilde{f}_{T}=\zeta $ and $\tilde{f}^{\,'}_{T}=0\,$ will convert the general dynamical system in Eq. \eqref{GDS} into an autonomous form as follows,
 \begin{align}
\frac{dX}{dN}&=\frac{(p-1) X (\lambda  X+6 Y)}{3 X+Y}\,,\nonumber\\
\frac{dY}{dN}&=\frac{(p-1) Y (\lambda  X+6 Y)}{3 X+Y}+X \left(\lambda -\frac{2 Y^2}{X^2}\right)\,,\nonumber\\
\frac{dV}{dN}&=-\frac{2 V (2 X+Y)}{X}\,,\nonumber\\
\frac{dW}{dN}&=-\frac{2 W Y+\lambda  X^2+6 X Y+2 \zeta  Y}{X}\,.
    \label{Dynamicalsystemmodel-II}
\end{align}
The standard density parameter for DE and matter can be given as,
\begin{align}
    \Omega_{DE}&=2 \zeta -\frac{(p-1) X (\lambda  X+6 Y)}{3 X+Y}+W+3 X+Y\nonumber\,,\\ 
    \Omega_{m}&=-2 \zeta +\frac{(p-1) X (\lambda  X+6 Y)}{3 X+Y}-V-W-3 X-Y+1\,.
\end{align}
The EoS parameter for DE can be written as.
\begin{align}
   \omega_{DE} =-\frac{(V+3) X+2 Y}{3 X \left(2 \zeta -\frac{(p-1) X (\lambda  X+6 Y)}{3 X+Y}+W+3 X+Y\right)}\nonumber\,.
\end{align}
Next, we have calculated and presented the critical points for this dynamical system in Table \ref{model2criticalpoint}
\begin{table}[H]
 % title of Table
\centering % used for centering table
\scalebox{0.9}{\begin{tabular}{|c|c|c|c|c|c|} % centered columns (5 columns)
\hline\hline %inserts double horizontal lines
 \parbox[c][0.8cm]{3.3cm}{Critical points} & $x$ & $y$ &$v$ & $w$ & Exists for\\ [0.5ex] % inserts table %heading
\hline\hline % inserts a single horizontal line
 \parbox[c][1cm]{3.3cm}{\textbf{$\mathcal{P}_1= (x_1,y_1,v_1,w_1)$}} & $x_{1}$ & $-2 x_{1}$  & $v_1$ &$-\zeta -x_1$ & $x_1\neq 0, p=1, \lambda =8.$ \\
\hline
\parbox[c][1cm]{3.3cm}{$\mathcal{P}_2= (x_2,y_2,v_2,w_2)$} & $x_{2}$ & $-\frac{3x_{2}}{2}$  & $0$ & $w_2$ & $ x_2 \ne 0, \zeta =-w_2-\frac{3 x_2}{2}, p=1, \lambda=\frac{9}{2}.$\\
\hline
\parbox[c][1cm]{3.3cm}{$\mathcal{P}_3= (x_3,y_3,v_3,w_3)$} & $x_3$ & $\sqrt{\frac{\lambda}{2}} x_3$  & $0$ &$-\zeta-\frac{1}{2} \left(\sqrt{2\lambda} +6\right) x_3$&$3 x_3+y_3\neq 0, p=1, \lambda=\text{arbitrary}.$ \\
\hline
\parbox[c][1cm]{3.3cm}{$\mathcal{P}_4= (x_4,y_4,v_4,w_4)$} & $x_4$ & $0$  & $0$ &$0$& $ p=\text{arbitrary}, x_4 \ne 0, \lambda=0.$\\
\hline
\end{tabular}}
\caption{The critical points for model \ref{ModelIIch4}. }
 % is used to refer this table in the text
\label{model2criticalpoint}
\end{table}
To study the stability of each critical point, the eigenvalues at each critical point of the Jacobian matrix are calculated and presented in Table \ref{modelIIeigenvaluesch4}. Depending upon the sign of the eigenvalues, the stability of the critical point can be concluded. The stability conditions for each critical point, along with the values of $q, \,\omega_{tot}$ and $\omega_{DE}$ are presented in Table \ref{modelIIstabilitycondi} along with the detailed descriptions.
\begin{itemize}
     \item{\textbf{Radiaiton-dominated Critical points:}} The value of parameter $\lambda$ is $8$ at $\mathcal{P}_{1}$, hence this critical point represents radiation-dominated era with $\omega_{tot}=\frac{1}{3}$. This critical point is in the standard radiation-dominated era at $V_1=1, \zeta=0$, where $\Omega_{r}=1$ and $\Omega_{m}=\Omega_{DE}=0$ can be observed from Table \ref{modelIIdensityparameter}. The exact solution retraced at this critical point is in the power law $a(t)=t_0 (t)^h$ form with $h=\frac{1}{2}$ which identifies the radiation-dominated era. According to the sign of the eigenvalues presented in Table \ref{modelIIeigenvaluesch4}, this critical point remains as a saddle point and hence is unstable. The same behaviour can be confirmed from the phase space trajectories Fig. \ref{modelII2dphaseportraitch4}. The exact cosmological solution at this critical point is shown in Table \ref{modelIIdensityparameter}. 
\end{itemize}
\begin{table}[H]
     % title of Table
    \centering % used for centering table
    \begin{tabular}{|c|c|} % centered columns (5 columns)
    \hline\hline %inserts double horizontal lines
    \parbox[c][0.3cm]{2.5cm}{Critical points}& Eigenvalues \\ [0.5ex] % inserts table %heading
    \hline\hline % inserts single horizontal line
    \parbox[c][1cm]{1.3cm}{$\mathcal{P}_1$ } & $\{0,0,4,8\}$ \\
    \hline
    \parbox[c][0.3cm]{1.3cm}{$\mathcal{P}_2$ } & $\{0,-1,3,6\}$  \\
    \hline
   \parbox[c][0.3cm]{1.3cm}{$\mathcal{P}_3$ } &  $\left\{0,-4-\sqrt{2 \lambda},-2 \sqrt{2 \lambda},-\sqrt{2 \lambda}\right\}$   \\
   \hline
   \parbox[c][0.3cm]{1.3cm}{$\mathcal{P}_4$ } &  $\{0,0,0,-4\}$   \\
    \hline
    \end{tabular}
    \caption{Eigenvalues corresponding to each critical point for model \ref{ModelIIch4}.}
    % is used to refer to this table in the text
    \label{modelIIeigenvaluesch4}
\end{table}
\begin{center}
\begin{table}[H]
    % title of Table
    \centering % used for centering table
    \begin{tabular}{|c|c|c|c|c|} % centered columns (5 columns)
    \hline\hline %inserts double horizontal lines
    \parbox[c][0.5cm]{2.5cm}{Critical points} & Stability conditions & $q$ & $\omega_{tot}$ & $\omega_{DE}$ \\ [0.5ex] % inserts table %heading
    \hline\hline % inserts single horizontal line
    \parbox[c][0.8cm]{1.3cm}{\textbf{$\mathcal{P}_1$}} & Unstable & $1$ & $\frac{1}{3}$ & $-\frac{v_1-1}{3 \zeta }$ \\
    \hline
    \parbox[c][0.8cm]{1.3cm}{$\mathcal{P}_2$}  &  Unstable& $\frac{1}{2}$ & $0$& 0\\
    \hline
    \parbox[c][0.8cm]{1.3cm}{$\mathcal{P}_3$}  &  \begin{tabular}{@{}c@{}}     Stable for\\ $x_3\in \mathbb{R}\land \lambda >0$ \end{tabular} & $-1-\sqrt{\frac{\lambda}{2}}$  & $-1-\frac{\sqrt{2 \lambda}}{3} $ & $-\frac{1}{\zeta}+\frac{\sqrt{2 \lambda}}{3 \zeta}$ \\
    \hline
    \parbox[c][0.8cm]{1.3cm}{$\mathcal{P}_4$}  &  $\text{Nonhyperbolic}$ & $-1$  & $-1$ & $-\frac{1}{2 \zeta+w_4+3 x_4}$ \\
    [1ex] % [1ex] adds vertical space
    \hline %inserts a single line
    \end{tabular}
     \caption{Stability condition, EoS and deceleration parameters for model \ref{ModelIIch4}.}
    % is used to refer to this table in the text
    \label{modelIIstabilitycondi}
\end{table}
\end{center}
To identify the phase of the evolution of the Universe the exact cosmological solutions at the critical points are calculated and presented in Table \ref{modelIIdensityparameter}.

\begin{table}[H]
     % title of Table
    \centering % used for centering table
    \scalebox{0.8}{
    \begin{tabular}{|c|c|c|c|c|c|} % centered columns (5 columns)
    \hline\hline %inserts double horizontal lines
   \parbox[c][0.6cm]{2.5cm}{Critical points} & Evolution Eqs. & Universe phase & $\Omega_{m}$& $\Omega_{r}$ & $\Omega_{DE}$\\ [0.5ex] % inserts table %heading
    \hline\hline % inserts single horizontal line
   \parbox[c][0.8cm]{1.3cm}{ $\mathcal{P}_1$ }& $\dot{H}=-2 H^{2}$ & $a(t)= t_{0} (2 t+c_{2})^\frac{1}{2}$ & $1-\zeta-v_1$& $v_1$ & $\zeta$\\
    \hline
    \parbox[c][0.8cm]{1.3cm}{$\mathcal{P}_2$} & $\dot{H}=-\frac{3}{2}H^{2}$ & $a(t)= t_{0} (\frac{3}{2}t+c_{2})^\frac{2}{3}$ & $1-\zeta$& $0$ & $\zeta$\\
    \hline
    \parbox[c][0.8cm]{1.3cm}{$\mathcal{P}_3$ }&  $\dot{H}=\sqrt{\frac{\lambda}{2}}H^2$ & $a(t)= t_{0} (-\sqrt{\frac{\lambda}{2}}t+c_{2})^{\sqrt{\frac{2}{\lambda}}}$ & $0$& $0$ & $1$\\
    \hline
    \parbox[c][0.8cm]{1.3cm}{$\mathcal{P}_4$ }&  $\dot{H}=0$ & $a(t)= t_{0} e^{c_2 t}$ & $1-2 \zeta-w_4-3 x_4$& $0$ & $2 \zeta+w_4+3 x_4$\\
    [1ex] % [1ex] adds vertical space
    \hline %inserts single line
    \end{tabular}}
    \caption{Phase of the Universe, density parameters for model \ref{ModelIIch4}. }
     % is used to refer this table in the text
    \label{modelIIdensityparameter}
\end{table}

\begin{figure}[H]
    \centering
    \includegraphics[width=60mm]{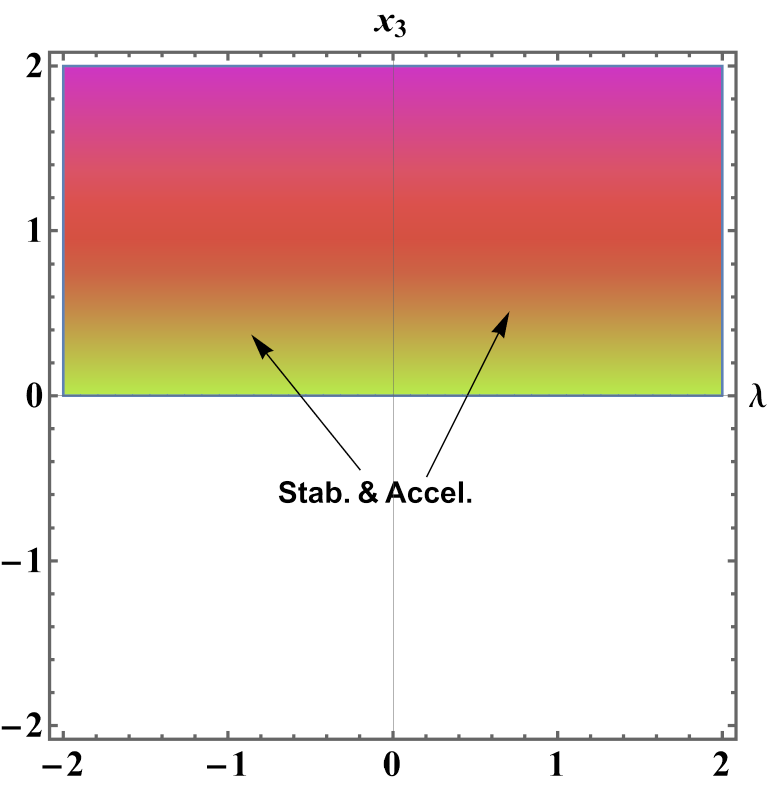}
    \caption{ Stability and acceleration of the Universe for model \ref{ModelIIch4}.} \label{modelIIregionplotch4}
\end{figure}

\begin{figure}[H]
    \centering
    \includegraphics[width=60mm]{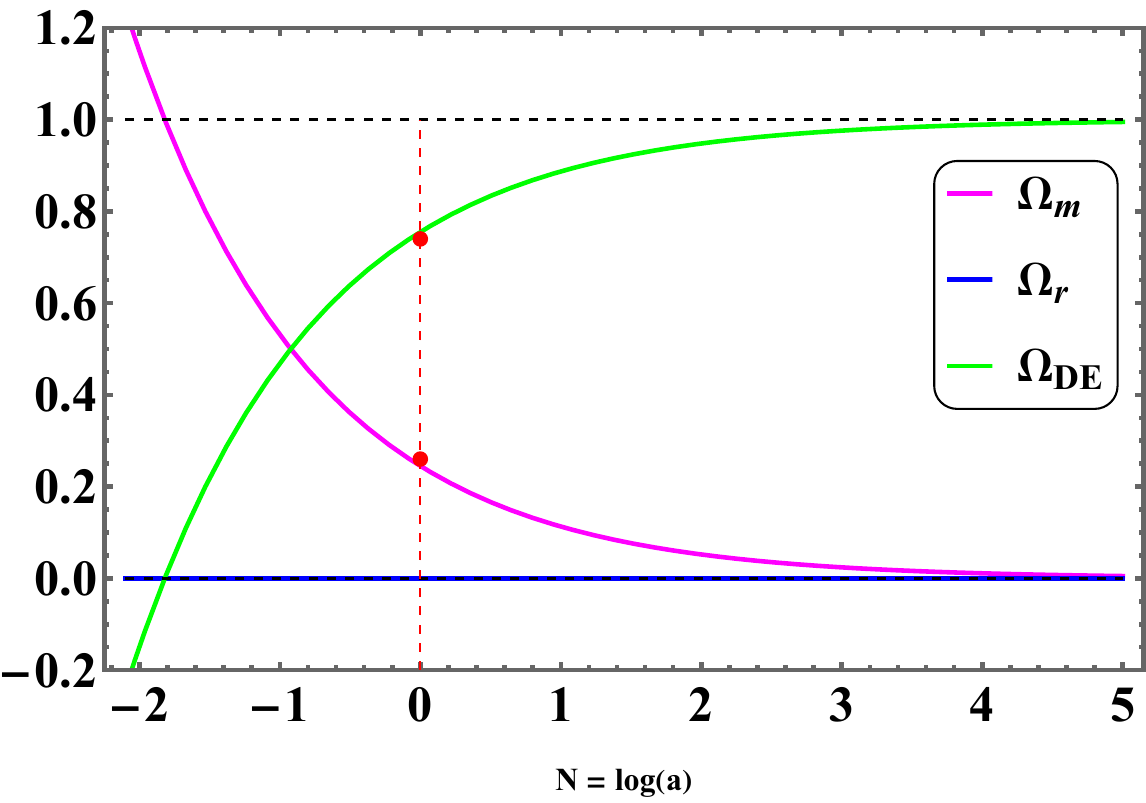}
    \caption{Evolution of the density parameters for model \ref{ModelIIch4}.} \label{model2evolutionch4}
\end{figure}

\begin{figure}[H]
    \centering
    \includegraphics[width=60mm]{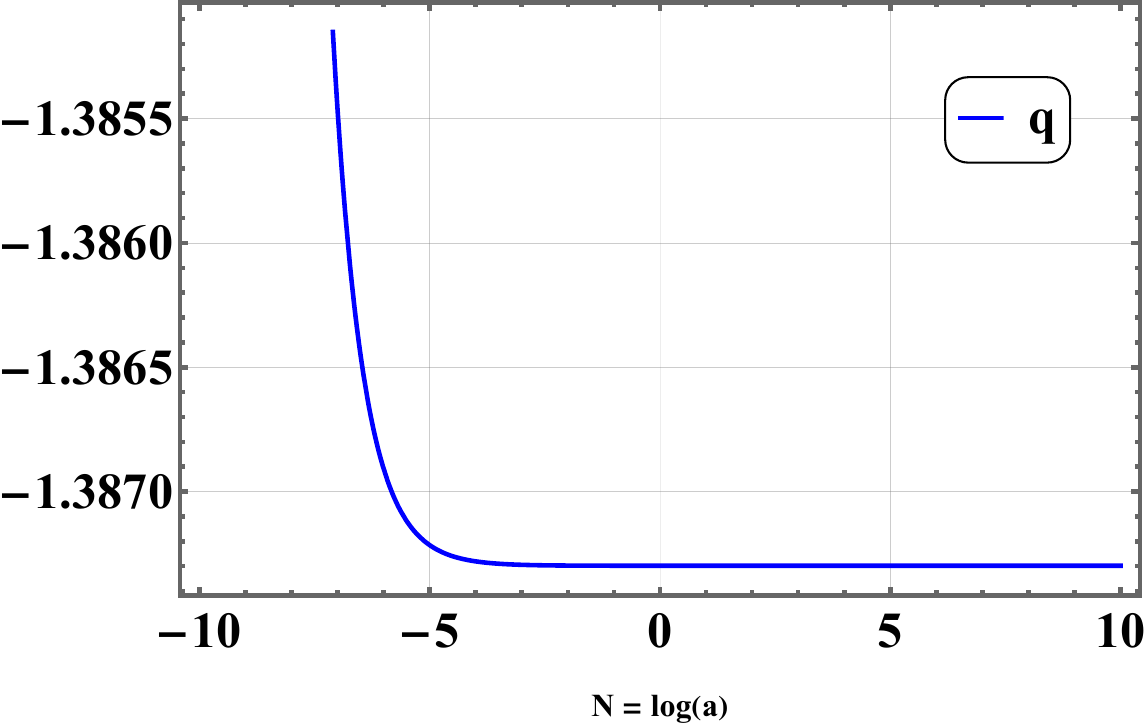}
    \includegraphics[width=56mm]{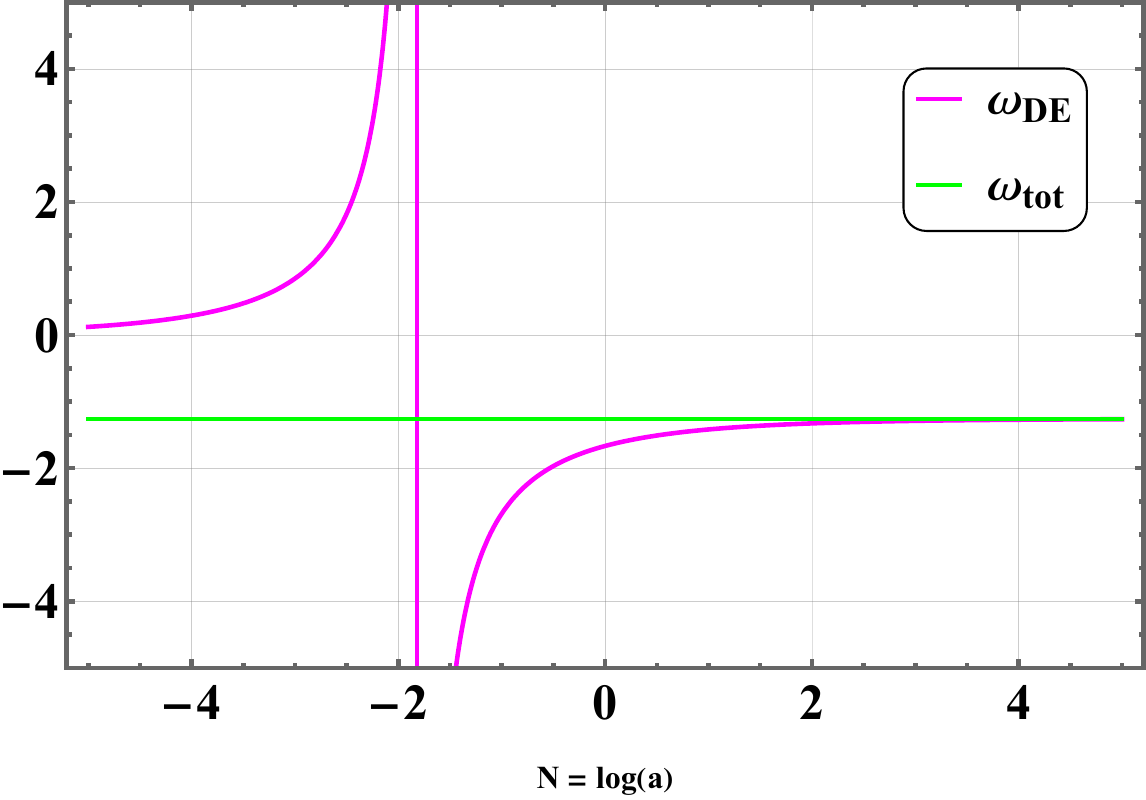}
    \caption{Deceleration $(q)$ and EoS parameters $\omega_{DE}$, $\omega_{tot}$ for model \ref{ModelIIch4}.} \label{model2omegadedecelech4}
\end{figure}

\begin{itemize}
\item{\textbf{Matter-dominated critical points:}} This critical point is representing the CDM-dominated era, where $\omega_{tot}=0$. This can be observed from the exact cosmological solution obtained at this critical point with $a(t)=t_0 (t)^h$, where $h=\frac{2}{3}$  as presented in Table \ref{modelIIdensityparameter}. The value of $\Omega_{m}$ at this critical point is $1$ for $\zeta=0$ and hence represents a standard CDM-dominated era. Since there is an eigenvalue at the Jacobian matrix with a positive sign, this critical point is a saddle point, Ref. Table \ref{modelIIeigenvaluesch4}. From Fig. \ref{modelII2dphaseportraitch4}, the phase space trajectories at this critical point are moving away from it, hence unstable in behaviour. The parameter $\lambda$ will take the value $\frac{9}{2}$.

  \item{\textbf{DE-dominated critical points:}} The value of $q$ and $\omega_{tot}$ at critical point $\mathcal{P}_{3}$ are $\lambda$ dependent.  Since the value of $\Omega_{DE}=1$, this critical point will represent a standard DE-dominated era and will describe accelerated expansion within the range $x_3\in \mathbb{R}\land \lambda\geq 0$. The stability conditions at this critical point are presented in Table \ref{modelIIstabilitycondi}. For clear visualisation of the parametric range where critical point $\mathcal{P}_3$ is stable and describes accelerating behaviour, we have plotted region plot for parameters $\lambda$ and $x_3$ and is presented in Fig. \ref{modelIIregionplotch4}. This plot lies in the upper half plane and with the inclusion of $x_3$-axis due to the acceleration range of parameter $\lambda$. We observed that this critical point can describe the de Sitter solution at $\lambda=0$ and shows stability at the points near $\lambda=0$. The eigenvalues at these critical points are normally hyperbolic and show stability as described in Table \ref{modelIIstabilitycondi}. This critical point is a late time attractor and the same can be visualized from the behavior of phase space trajectories from Fig. \ref{modelII2dphaseportraitch4}. The exact solution obtained at this critical point is in the power law $a(t)=t_0 (t)^h$ form with $h=\sqrt{\frac{2}{\lambda}}$ which can explain the different epochs of the Universe evolution depending upon the value of $\lambda$.

   \item{\textbf{de Sitter solution :}} This critical point is the de Sitter solution with $q=\omega_{tot}=-1$. The de Sitter solution at this critical point is obtained and presented in Table \ref{modelIIdensityparameter}. The standard DE-dominated era can be described by this critical point at $\zeta=\frac{1}{2}, w_4=0, x_4=0$. From the eigenvalues one can see that there are three zeros and one negative eigenvalue, hence linear stability theory will fail to confirm the stability at this point. Therefore, we have moved forward to obtain stability using central manifold theory (CMT). But in this case, while applying CMT, we have observed that after co-ordinates shift to the center, this system will not satisfy the CMT condition (after we separated the linear and non-linear parts of the system equations, the nonlinear part does not vanish as same in first model critical point $C_3$). We have plotted and analysed the 2D phase portrait at this critical point presented in Fig. \ref{modelII2dphaseportraitch4}. The phase space trajectories are attracting towards this critical point, hence this critical point is a late time attractor.  
\end{itemize}
Fig. \ref{modelII2dphaseportraitch4} is plotted for $\lambda=0.3,\, \zeta=1.0001,\,p=-1$. Graphically we have presented $\Omega_{DE}, \Omega_{m}, \Omega_{r}$ in Fig. \ref{model2evolutionch4} with the initial condition: $X=1.2 \times 10^{2.2}, Y=2.2 \times 10^{-3.4}, V=1.02 \times 10^{2.6}, W=4.5 \times10^{-8.1}$. It has been observed that the value of $\Omega_{m}\approx 0.3$ and $\Omega_{DE} \approx 0.7$ at the present time. The plot for $\Omega_{r}$ vanishes throughout the evolution. The plot for $\Omega_{DE}$ dominates both $\Omega_{r},\,  \Omega_{m}$ at the late phase of cosmic evolution. The plots for $q,\, \omega_{DE},\, \omega_{tot}$ is presented in Fig. \ref{model2omegadedecelech4}.  The value of $q$ at the present time is $-1.387$ which is approximately the same as the current observation study \cite{Feeney:2018}. The plot of $\omega_{tot}$ at present takes the value $-1.255$ which agrees with \cite{Hinshaw:2013}.
The initial conditions for Fig. \ref{model2evolutionch4} and Fig. \ref{model2omegadedecelech4} are $X=1.2 \times 10^{2.2}, Y=2.2 \times 10^{-3.4}, V=1.02 \times 10^{2.6}, W=4.5 \times10^{-8.1},\, \lambda=0.3,\, \zeta=1.0001,\, p=-1$. 

\section{\texorpdfstring{$f(T, T_G)$}{} gravity field equations}\label{f(TTG)fieldequations}
 Using the definitions in Eq. \eqref{TORSIONSCALAR} and in Eq. \eqref{eq:T_G_def}, the torsion scalar and the teleparallel equivalent of the Gauss-Bonnet term can be calculated as,
\begin{equation}\label{T,T_G}
    T =6H^2\,,\quad T_G = 24H^2\left(\dot{H}+H^2\right),\,
\end{equation}
where the Gauss-Bonnet term turns out to have the same value for this background as its curvature analog. The Friedmann equations for this set-up obtained to be \cite{Kofinas:2014daa, Bahamonde:2016kba}
\begin{subequations}
\begin{align}
   F - 12 H^2 F_T - T_G F_{T_G} + 24 H^3 \dot{F}_{T_G} = 2 \kappa^2 \rho, \label{1stFEch4}\\ 
   F - 4 (\dot{H} +3 H^2) F_{T} - 4 H \dot{F}_{T}-T_{G} F_{T_{G}}
   + \frac{2}{3 H} T_G \dot{F}_{T_G} + 8 H^2 \ddot{F}_{T_{G}} = -2\kappa^2 p\label{2ndFEch4}.
\end{align}
\end{subequations}
An over-dot refers to derivatives with respect to cosmic time $t$. To establish the deviation of this theory from GR and to better probe the role from the modified Lagrangian, we consider $F(T, T_G)=-T+f(T, T_G)$. Subsequently, Eqs. \eqref{1stFEch4} and \eqref{2ndFEch4} reduce respectively as,
\begin{subequations}
\begin{align}
   6H^2+f-12H^2 f_{T}-T_{G} f_{T_G}+24H^3 \dot{f}_{T_G}=2\kappa^2 \rho \,,\label{1stfe} \\ 
     2(2\dot{H}+3H^2)+f-4(\dot{H}+3H^2)f_{T}-4H\dot{f}_{T} - T_{G} f_{T_G}+\frac{2}{3H} T_G \dot{f}_{T_G}+8H^2 \ddot{f}_{T_G}=-2\kappa^2 p\,.\label{2ndfe}
\end{align}
\end{subequations}
 These equations will reduce to GR when $f=0$ or $f=T_G$; in the latter case, as expected, the $T_G$ terms will get canceled. Now, using the Friedmann equation Eq. \eqref{FriedmanEQ}, the expressions for energy density and pressure for the DE sector can be retrieved as
\begin{subequations}
\begin{align}
    \frac{-1}{2\kappa^2}\left(f-12H^2 f_{T}-T_G f_{T_G}+24H^3 \dot{f}_{T_G}\right)&=\rho_{DE}\,,\label{fede1ch4}\\
    \frac{1}{2\kappa^2}\left(f-4(\dot{H}+3H^2) f_{T}-4H\dot{f}_T-T_G f_{T_G}+\frac{2}{3H} T_G \dot{f}_{T_{G}}+8H^2 \ddot{f}_{T_G}\right)&=p_{DE}\,.\label{fede2CH4}
\end{align}
\end{subequations}
The expression for the EoS parameter for DE can be written as, 
\begin{align}
    \omega_{DE}=-1+\frac{8H^2 \ddot{f}_{T_G}-4\dot{H}f_{T}-4H\dot{f}_T + \frac{2}{3H} T_G \dot{f}_{T_G}  -24 H^3 \dot{f}_{T_G}}{12H^2 f_T +T_G f_{T_G} -24 H^3 \dot{f}_{T_G}-f}\,. \label{omegade}
\end{align}
\section{Dynamical system analysis in \texorpdfstring{$f(T,T_G)$}{} 
gravity}\label{DynamicalanalysisCH4}
We shall first define the appropriate dynamical variables to analyze the dynamical system of higher-order teleparallel gravity \cite{Franco:2020lxx, Franco:2021}. By differentiating the dynamical variables with respect to $N$, as defined in chapter \ref{Chapter2}, the expressions for the autonomous dynamical system can be obtained and, subsequently, the critical points. We consider the Universe to be filled with two fluids, $\rho=\rho_{m}+\rho_{r}$, where $\rho_m$ and $\rho_{r}$ respectively denote the energy density for matter and radiation phase. In the matter phase, the matter pressure $p_m=0$ and therefore $\omega_m$ also vanishes. In the radiation phase,  the EoS parameter, $\omega_{r}=\frac{1}{3}$. Therefore, with these considerations, we define the dynamical variables as follows
\begin{align}
X=f_{T_G} H^2 \,,\quad Y=\dot{f}_{T_G}H \,,\quad Z=\frac{\dot{H}}{H^2} \,,\quad 
V=\frac{\kappa^2 \rho_r}{3H^2}\,,\quad
W=-\frac{f}{6H^2}\,.\label{dynamical variables}
\end{align}
The standard density parameters in Eq. \eqref{densityparameters} can be expressions for matter $(\Omega_m)$, radiation $(\Omega_r)$ and DE $(\Omega_{DE})$ phase are respectively satisfy the constrain equation with,
\begin{align}
\Omega_{m}+\Omega_{r}+W+2f_{T}+4XZ+4X-4Y=1,
\end{align}
and
\begin{align}
    \Omega_{DE}=W+2f_{T}+4XZ+4X-4Y.
\end{align}
To express the autonomous dynamical system, we define the parameter $\lambda=\frac{\ddot{H}}{H^3}$ \cite{Odintsov:2018}, and so the general form of the dynamical system can be obtained as follows,
\begin{align}
\frac{dX}{dN}&=2XZ+Y\,,\nonumber\\
\frac{dY}{dN}&=-\frac{3}{4}-\left(Z+2\right)Y-\frac{Z}{2}+\frac{3W}{4}+\left(Z+3\right)\frac{f_{T}}{2}+\frac{\dot{f}_{T}}{2H}\nonumber +3X\left(Z+1\right)-\frac{V}{4}\,,\nonumber\\
\frac{dZ}{dN}&=\lambda-2Z^2\,,\nonumber\\
\frac{dW}{dN}&=-2f_{T}Z-8XZ^2-4X\lambda-16XZ-2WZ\,,\nonumber\\
\frac{dV}{dN}&=-4V-2VZ\,.\label{generaldynamicalsystem4_2}
\end{align}
To form the autonomous dynamical system, we need a form for $f(T, T_G)$ in which the terms $f_{T},\frac{\dot{f}_{T}}{H}$ can be written in terms of the dynamical variables, which we shall discuss by considering two forms of $f(T, T_G)$ as two models.

\subsection{Mixed power law model}\label{Model_Ich4_2}
We consider the mixed power law form of $f(T, T_G)$ \cite{Capozziello:2016eaz} as
\begin{align}
f(T,T_G)=f_{0} T_{G}^{k} T^{m},\label{firstmodel4_2}
\end{align}
where $f_{0}, m, k$ are the arbitrary constants, the GR limit can be recovered for vanishing index powers \cite{Escamilla-Rivera:2019ulu}. The motivation to consider this form is to study the role of parameter $\lambda$ in its most general form. From Eq. \eqref{dynamical variables}  and Eq. \eqref{firstmodel4_2}, we can write $f_{T}=-m W$, and this will guarantee the autonomous dynamical system and the dynamical variable $X$ becomes, \begin{align}
X&=f_{T_G} H^2=f_{0} k G^{k-1} T^{m} H^{2}\,,\nonumber\\
&=\frac{kfH^2}{G}=\frac{kf}{24(\dot{H}+H^2)}=\frac{kf}{6H^2}\left(\frac{1}{4\left(\frac{\dot{H}}{H^2}+1\right)}\right)=\frac{-Wk}{4(Z+1)}.
\end{align}
The parameter $X$ depends on $W$ with the condition on $Z\ne -1$. This condition also guarantees $T_G=24H^4(Z+1)$, i.e. the non-vanishing teleparallel Gauss-Bonnet term. The dynamical variable $W$ can be rewritten as
\begin{align}
W=-\frac{4X(Z+1)}{k} \label{eqforz4_2}\,,
\end{align}
from Eq. (\ref{eqforz4_2}), an expression for $\lambda$ can be obtained,
\begin{align}
\lambda&=\frac{1}{1-k}\left[2(Z+1) Z (m-2)+2Z^2(k+1)+4kZ- \frac{(Z+1)Y}{X}\right]\label{eqforlambda4_2}\,.
\end{align}
From Eqs. \eqref{eqforz4_2} and \eqref{eqforlambda4_2}, one can note that the system equations become singular at $k=0$ or $k=1$, i.e. we can not use it in the limit to $f(T)$ gravity ($m$ arbitrary) and in the limit to GR $(m=0, k=0)$. This is a shortcoming of the chosen variables since the original Eqs. \eqref{1stfe} and \eqref{2ndfe} are regular in these limits. Although the parameter $f_0$ does not play any role in the further analysis, the system equations will converge to GR spontaneously for the case $k=0, m=0$, but since due to the singularity of system equations at $k=0$, we will drop this value in our analysis. We will consider $X$, $Y$, $Z$, $V$ as independent variables, $W$ and $\lambda$ as dependent variables and then inserting $f_{T}$ and $\frac{\dot{f}_{T}}{H}$ in to the system Eq. \eqref{generaldynamicalsystem4_2}, the system equation we obtain,
\begin{align}
\frac{dX}{dN}&=2XZ+Y\,,\nonumber\\
\frac{dY}{dN}&=-\frac{k^2 (V-12 X (Z+1)+4 Y (Z+2)+2 Z+3)}{4 (k-1) k}+\frac{-8 X (Z+1) (m (Z-3)+3)+V+2 Z+3}{4 (k-1)}\nonumber\\& \,\,\,\,\,\,\,\, -\frac{4 (2 m-1) X (Z+1) (2 m Z+3)}{4 (k-1) k}+\frac{4 Y (2 m (Z+1)+Z+2)}{4 (k-1)},\nonumber\\
\frac{dZ}{dN}&=\frac{(Z+1) (Y-2 X Z (2 k+m-2))}{(k-1) X},\nonumber\\
\frac{dV}{dN}&=-4V-2VZ\,.\label{DynamicalSystem1stmodel}
\end{align}
We get the density parameter for DE and matter, respectively, in dynamical variables as,
\begin{subequations}
\begin{align}
\Omega_{DE}&=-4 Y+\frac{4 X (Z+1) (k+2 m-1)}{k},  \\
\Omega_{m}&=1 - V + 4 Y -[\frac{4}{k} (-1 + k + 2 m) X (1 + Z)]\,.
\end{align}
\end{subequations}
The critical points of the dynamical system can be obtained by considering $\frac{dX}{dN}=0, \frac{dY}{dN}=0, \frac{dZ}{dN}=0, \frac{dV}{dN}=0$, and are obtained as in Table \ref{TABLE-I}. 
\begin{table*}[!htb]
 % title of Table
\centering % used for centering table
\scalebox{0.95}{
\begin{tabular}{|*{6}{c|}}\hline 
    \parbox[c][0.8cm]{3.3cm}{Critical points} & $X$ & $Y$ & $Z$ & $V$ & Exist for\\ [0.5ex]\hline \hline % inserts table %heading \hline\hline % inserts a single horizontal line
    \parbox[c][1.2cm]{3cm}{$A_1= (x_1,y_1,z_1,v_1)$} & $x_1$ & $4 x_1$ & $-2$ & $v_1$ & \begin{tabular}{@{}c@{}}$v_{1}-4 x_{1}-1\neq 0$, $m=\frac{v_{1}-12 x_{1}-1}{v_{1}-4 x_{1}-1},$ $k=\frac{1-m}{2},$ \\ $-v_{1} x_{1}+8 x_{1}^2+x_{1}\neq 0$ \end{tabular}\\
    \hline
    \parbox[c][1cm]{3cm}{$A_2= (x_2,y_2,z_2,v_2)$} & $x_{2}$ & $3 x_{2}$ & $-\frac{3}{2 }$ & $0$ &  \begin{tabular}{@{}c@{}}$8 x_{2}+1\neq 0, x_{2}\ne 0,$\\ $k=\frac{1-m}{2},$ $m^2-1\neq 0 $
    \end{tabular}\\
    \hline
    \parbox[c][0.8cm]{3cm}{$A_3= (x_3,y_3,z_3,v_3)$} & $\frac{1}{4}$ & $0$ & $0$ & $0$ & $k^2-k\neq 0,$ $m=\frac{1}{2}$ \\
    \hline
    \parbox[c][1.2cm]{3cm}{$A_4= (x_4,y_4,z_4,v_4)$} & $x_4$ & $0$ & $0$ & $0$ &  \begin{tabular}{@{}c@{}}$4 x_4-1\neq 0,$ $\left(2 m-1\right) x_4 \left(8 x_4+1\right) \left(8 m x_4-1\right)\neq 0,$\\ $k=-\frac{4 (2 m x_4-x_4)}{4 x_4-1}$.\end{tabular}\\
    \hline
    \parbox[c][1.5cm]{3cm}{$A_5= (x_5,y_5,z_5,v_5)$} & $x_5$ & $y_5$ & $-\frac{y_5}{2 x_5}$ & $0$ & \begin{tabular}{@{}c@{}}$y_{5} (12 x_{5}-2 y_{5}+1) (3 x_{5}-y_{5})\neq 0$\\$m=\frac{4 x_{5}+2 y_{5}+1}{12 x_{5}-2 y_{5}+1}$ \\$k=\frac{1-m}{2},$ $m^2-1\neq 0$ \end{tabular}\\
    \hline
\end{tabular}}
\caption{The critical points for model \ref{Model_Ich4_2}.}
\label{TABLE-I}
\end{table*}
 We have summarized the stability of all the critical points for first model in Table \ref{TABLE-II}. To identify the phase of evolution, the value of the deceleration parameter $q$ and the EoS parameters ($\omega_{tot}$, $\omega_{DE}$) are presented, corresponding to each critical point.
\begin{table*}[!htb]
    \centering % used for centering table
    \scalebox{0.95}{\begin{tabular}{|*{5}{c|}}\hline
    \parbox[c][0.8cm]{2.5cm}{Critical points} & Stability Conditions & $q$ & $\omega_{tot}$ & $\omega_{DE}$ \\ [0.5ex] % inserts table %heading
    \hline\hline % inserts single horizontal line
    \parbox[c][0.7cm]{1cm}{$A_1$} & Unstable & $1$ & $\frac{1}{3}$ & $\frac{1}{3}$ \\
    \hline
    \parbox[c][1cm]{1cm}{$A_2$} & $\begin{tabular}{@{}c@{}} Stable for\\$m<-1\land \frac{1-m}{6 m-10}<x_2\leq \frac{8 m-8}{9 m^2-48 m+71}$\\ Otherwise Unstable\end{tabular}$ & $\frac{1}{2}$ & $0$ & $0$ \\
    \hline
    \parbox[c][1cm]{1cm}{$A_3$} & $\begin{tabular}{@{}c@{}} Stable for\\$\frac{1}{4}<k\leq \frac{13}{25}$\end{tabular}$  & $-1$ & $-1$ & $-1$\\
    \hline
    \parbox[c][1cm]{1cm}{$A_4$} & $\begin{tabular}{@{}c@{}} Stable for\\$x_4<-\frac{1}{6}\land \left(m\leq \frac{17 x_4+2}{72 x_4^2+24 x_4+2}\lor m>\frac{4 x_4+1}{12x_4+1}\right)$\end{tabular}$  & $-1$ & $-1$ & $-1$\\
    \hline
    \parbox[c][1cm]{1cm}{$A_5$} & $\begin{tabular}{@{}c@{}} Stable for\\$\left(x_5<0\land y_5>3 x_5\right)\lor \left(x_5>0\land y_5<3 x_5\right)$\end{tabular}$  & $-1+\frac{y_5}{2 x_5}$ & $-1+\frac{y_5}{3 x_5}$ & $-1+\frac{y_5}{3 x_5}$\\
    [1ex] % [1ex] adds vertical space
    \hline %inserts a single line
    \end{tabular}}
    \caption{Stability condition, deceleration and EoS parameter for model \ref{Model_Ich4_2}.}
    \label{TABLE-II}
\end{table*}

\begin{figure}[H]
    \centering
    \includegraphics[width=55mm]{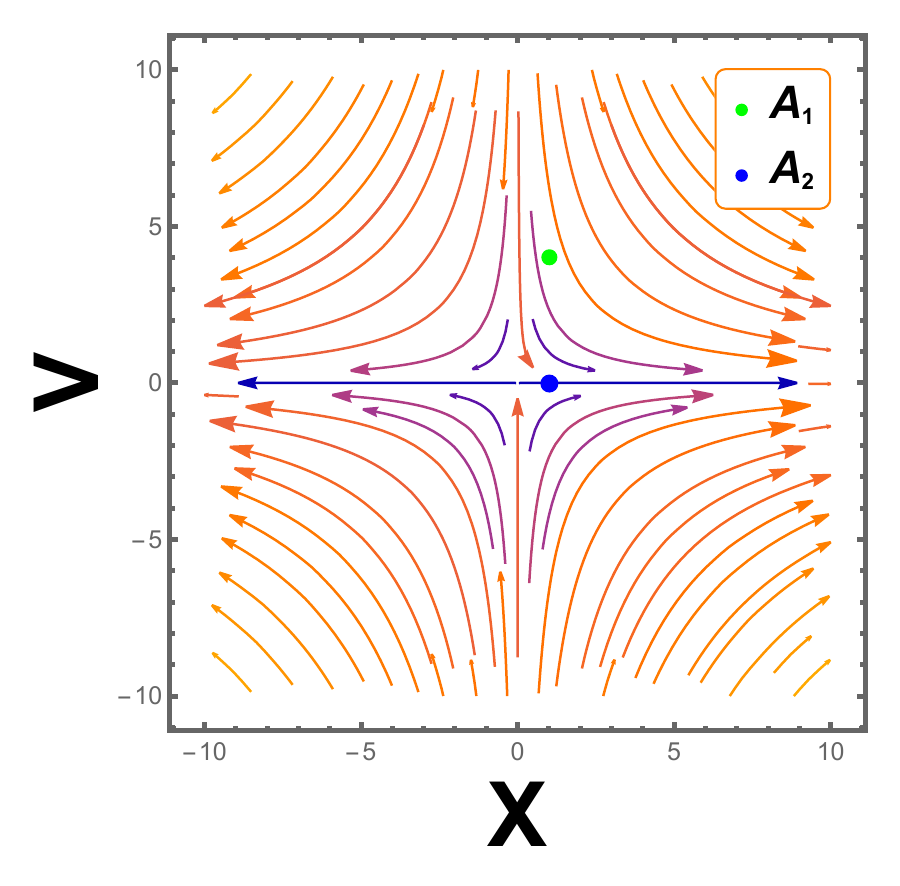}
    \includegraphics[width=55mm]{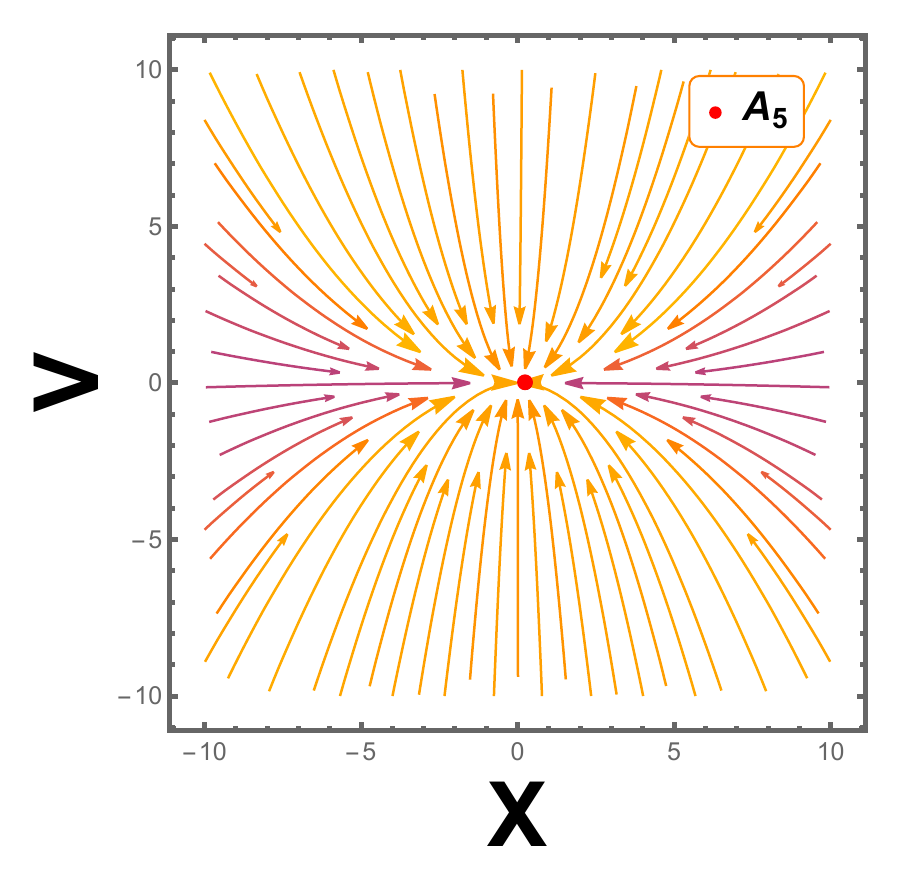}
    \includegraphics[width=55mm]{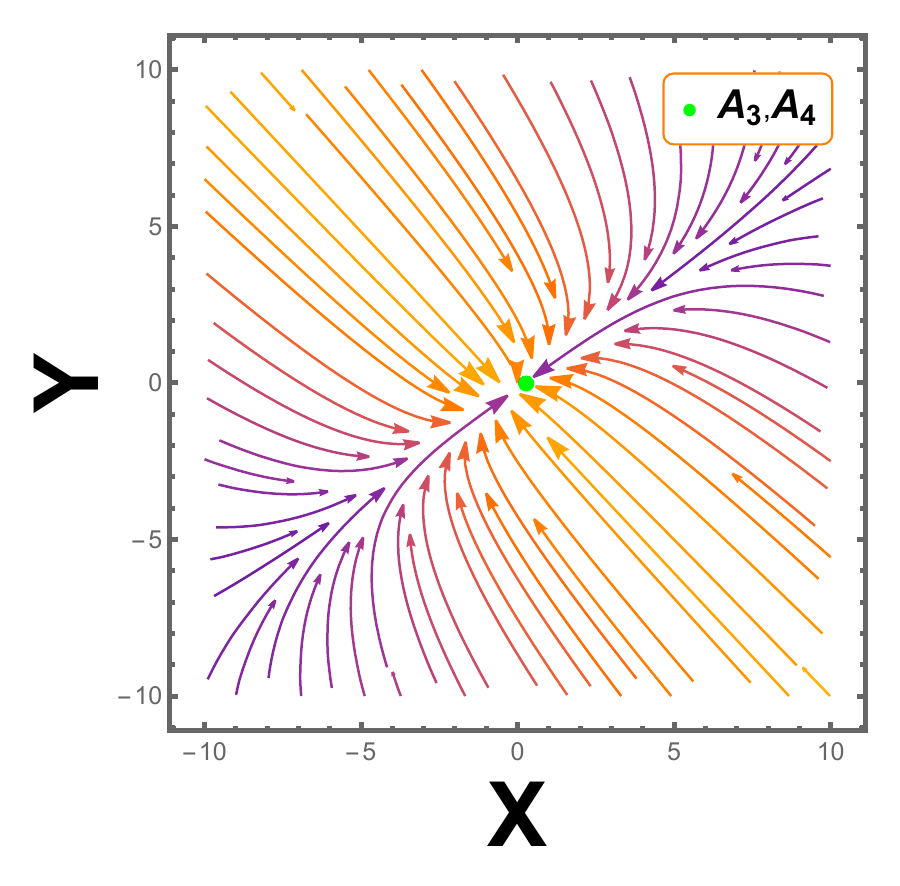}
    \caption{2D phase portrait with $k=0.029$, $m=0.5$ for model \ref{Model_Ich4_2}. } \label{Fig1} 
\end{figure}

The cosmological solutions for the corresponding evolution equation at each critical point, along with the standard density parameters ($\Omega_{m}$, $\Omega_{DE}$, $\Omega_{r}$) for first model are presented in Table \ref{TABLE-III}.
\begin{table*}[!htb]
     % title of Table
    \centering % used for centering table
    \begin{tabular}{|*{6}{c|}}\hline
    \parbox[c][0.8cm]{2.5cm}{Critical points} & Evolution Eqs. & Universe phase & $\Omega_{m}$& $\Omega_{r}$ & $\Omega_{DE}$\\ [0.5ex] % inserts table %heading
    \hline\hline % inserts single horizontal line
    \parbox[c][0.7cm]{1cm}{$A_1$} & $\dot{H}=-2 H^{2}$ & $a(t)= t_{0} (2 t+c_{2})^\frac{1}{2}$ & $0$& $v_1$ & $1-v_1$\\
    \hline
    \parbox[c][0.7cm]{1cm}{$A_2$} & $\dot{H}=-\frac{3}{2}H^{2}$ & $a(t)= t_{0} (\frac{3}{2}t+c_{2})^\frac{2}{3}$ & $\frac{2 \left(3 m-5\right) x_2}{m-1}+1$& $0$ & $\frac{2 \left(5-3 m\right) x_2}{m-1}$\\
    \hline
    \parbox[c][0.7cm]{1cm}{$A_3$} & $\dot{H}=0$ & $a(t)= t_{0} e ^{c_{1}t} $ & $0$& $0$ & $1$\\
    \hline
    \parbox[c][0.7cm]{1cm}{$A_4$} &$\dot{H}=0$ & $a(t)= t_{0} e ^{c_{1}t} $ &  $0$& $0$ & $1$\\
    \hline
    \parbox[c][0.7cm]{1cm}{$A_5$} &  $\dot{H}=-\frac{y_5}{2 x_5}H^2$ & $a(t)= t_{0} (\frac{y_5}{2 x_5}t+c_{2})^\frac{2x_5}{y_5}$ & $0$& $0$ & $1$\\
    [1ex] % [1ex] adds vertical space
    \hline %inserts single line
    \end{tabular}
    \caption{Phase of the Universe, density parameters for model \ref{Model_Ich4_2}.}
     % is used to refer this table in the text
    \label{TABLE-III}
\end{table*}
The eigenvalues for the Jacobian matrix of the dynamical system in Eq. (\ref{DynamicalSystem1stmodel}) at each critical point are presented in Table \ref{TABLE-IV}.
\begin{table*} [!htb]
\scalebox{0.88}{
\begin{tabular}{|*{2}{c|}}\hline \hline
\parbox[c][0.5cm]{2.5cm}{Critical points} & \parbox[c][1cm]{15cm}{Eigenvalues}\\ \hline
\hline
\parbox[c][0.5cm]{0.5cm}{$A_1$} & \parbox[c][1cm]{15cm}{$\left\{0,1,\frac{x_1 \left(v_1-8 x_1-1\right)-r}{2 x_1 \left(-v_1+8 x_1+1\right)},\frac{r+x_1 \left(v_1-8 x_1-1\right)}{2 x_1 \left(-v_1+8 x_1+1\right)}\right\}$}\\ \hline
\parbox[c][0.5cm]{0.5cm}{$A_2$} & \parbox[c][1.5cm]{15cm}{$\left\{0,-1,-\frac{\sqrt{\left(m^2-1\right) x_2 \left(m \left(\left(9 m-48\right) x_2-8\right)+71 x_2+8\right)}}{4 \left(m^2-1\right) x_2}-\frac{3}{4},\frac{\sqrt{\left(m^2-1\right) x_2 \left(m \left(\left(9 m-48\right) x_2-8\right)+71 x_2+8\right)}}{4 \left(m^2-1\right) x_2}-\frac{3}{4}\right\}$}\\ \hline
\parbox[c][0.5cm]{0.5cm}{$A_3$} & \parbox[c][1cm]{15cm}{$\left\{-4,-3,\frac{-\sqrt{25 k^2-38 k+13}-3 k+3}{2 \left(k-1\right)},\frac{\sqrt{25 k^2-38 k+13}-3 k+3}{2 \left(k-1\right)}\right\}$}\\ \hline
\parbox[c][0.5cm]{0.5cm}{$A_4$} & \parbox[c][1cm]{15cm}{$\left\{-4,-3,-\frac{s}{2 x_4 \left(8 m x_4-1\right)}-\frac{3}{2},\frac{3 x_4 \left(1-8 m x_4\right)+s}{2 x_4 \left(8 m x_4-1\right)}\right\}$}\\ \hline
\parbox[c][0.5cm]{0.5cm}{$A_5$} & \parbox[c][1cm]{15cm}{$\left\{0,-\frac{4 x_5-y_5}{x_5},-\frac{-8 x_5 y_5+12 x_5^2+y_5^2}{2 x_5 \left(2 x_5-y_5\right)},-\frac{-5 x_5 y_5+6 x_5^2+y_5^2}{x_5 \left(2 x_5-y_5\right)}\right\}$}\\ 
\hline
\end{tabular}}
\caption{Eigenvalues corresponding to each critical point for model \ref{Model_Ich4_2}.}
\label{TABLE-IV}
\end{table*}
where, 
\begin{align}
r&=\sqrt{-\left(x_1 \left(v_1-8 x_1-1\right) \left(-v_1 \left(9 x_1+2\right)+2 v_1^2+8 x_1^2+x_1\right)\right)}\,,\nonumber\\
s&=\sqrt{x_4 \left(8 m x_4-1\right) \left(2 m \left(6 x_4+1\right){}^2-17 x_4-2\right)}\,.\nonumber
\end{align}
\begin{itemize}
    \item{\textbf{Radiation-dominated critical points:}} The critical point $A_{1}$ represents radiation dominated era with $\omega_{tot}=\omega_{DE}=\frac{1}{3}$ and  $q=1$, it describes the standard radiation dominated era for $v_{1}=1$, where $\Omega_{r}=1$ and $\Omega_{m}=0,\Omega_{DE}=0$. From Fig. \ref{Fig1}, it can be analysed that critical point $A_1$ representing radiation-dominated era exists at a parametric range $m=\frac{v_{1}-12 x_{1}-1}{v_{1}-4 x_{1}-1},$ $k=\frac{1-m}{2}$, and is unstable (saddle) due to presence of a positive eigenvalue. From the 2D phase space in Fig. \ref{Fig1}, we can observe that phase space trajectories move away from the critical point, further confirming the saddle point behavior.

\item{\textbf{Matter-dominated critical points:}} The critical point $A_{2}$ describes a non-standard cold DM-dominated era with the negligible contribution of DE density $\Omega_{DE}=\frac{2 \left(5-3 m\right) x_2}{m-1}$. This critical point will represent a standard cold DM-dominated era for $x_2=0$ or $m=\frac{3}{5}$. The CDM-dominated era can also be described at the critical point $A_2$ in the parametric range $k=\frac{1-m}{2},$ $m^2-1\neq 0$. The eigenvalues for the Jacobian matrix at this critical point are non-hyperbolic in nature, as presented in Table \ref{TABLE-IV} and show stability at the condition described in Table \ref{TABLE-II}. This critical point will either be a saddle or unstable node for the parameter range lying outside the stability condition described in Table \ref{TABLE-II}. We have plotted 2D and 3D phase portraits for $m=0.5$, for which stability condition on $x_2$ is $-0.0714286<x_2\le -0.0812183$. The coordinate of the phase portraits for 2D and 3D is in the unstable range for a critical point $A_2$, hence describing the saddle point nature of this critical point. The power law solution corresponding to the evolution equation at this critical point and the standard density parameters corresponding to different phases of the Universe evolution are presented in Table \ref{TABLE-III}. In this case, the values of $\omega_{tot}=\omega_{DE}=0$, and the deceleration parameter will take the value $q=\frac{1}{2}$ which is positive; hence this critical point can not describe the current cosmic acceleration.  

\item{\textbf{de Sitter solutions :}} The critical points $A_3$ and $A_4$ both are the de Sitter solutions with $\Omega_{DE}=1, \Omega_{m}=0, \Omega_{r}=0$. The value of $\omega_{tot}=\omega_{DE}=q=-1$; hence these two critical points describe the current accelerated expansion of the Universe. The de Sitter solution can be explained at the critical point $A_3$, and it exists in the parametric range $k^2-k\neq 0,$ $m=\frac{1}{2}$. The critical point $A_4$ is valid in the parametric range $\left(2 m-1\right) x_4 \left(8 x_4+1\right) \left(8 m x_4-1\right)\neq 0,$ $k=-\frac{4 (2 m x_4-x_4)}{4 x_4-1}$. The eigenvalues of the Jacobian matrix at both the critical points are presented in Table \ref{TABLE-IV} and are hyperbolic in nature. The stability conditions are described in Table \ref{TABLE-II}. The phase space trajectories show attractor behavior, which can be observed from Fig. \ref{Fig1}.

\item{\textbf{DE-dominated critical points:}} The value of the deceleration parameter, EoS parameters for a critical point $A_5$ is dependent on coordinates $x$ and $y$ as the value of $q=-1+\frac{y_5}{2 x_5},\omega_{tot}=\omega_{DE}=-1+\frac{y_5}{3 x_5}$. This critical point represents a  DE-dominated era with $\Omega_{DE}=1.$ This can explain the current cosmic acceleration at $\left(x_5<0\land y_5>2 x_5\right)\lor \left(x_5>0\land y_5<2 x_5\right)$. From the eigenvalues presented in Table \ref{TABLE-IV}, we can infer that the critical point is non-hyperbolic in nature and is stable at the stability condition presented in Table \ref{TABLE-II}.  Since the phase space trajectories behavior can be analyzed from Fig. \ref{Fig1} are attracting towards the critical point $A_{5}$, this critical point is an attractor.
\end{itemize}
\begin{figure}[H]
    \centering
    \includegraphics[width=60mm]{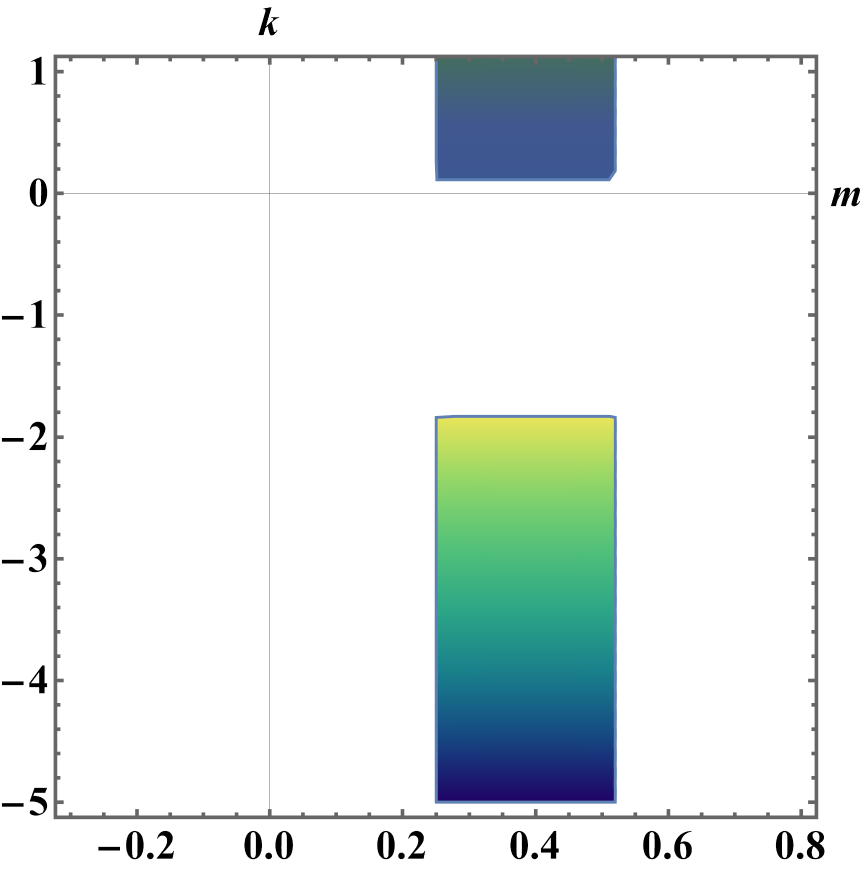}
    \caption{Region plot for critical points $A_3$ and $A_4$ at $x_4=-\frac{1}{3}$ for model \ref{Model_Ich4_2}. } \label{regionplotm1}
\end{figure}
In Fig. \ref{regionplotm1}, we have given the region plot, which will be helpful to visualize the ranges of the model parameters $m$ and $k$ at which both the critical points ($A_3, A_4$) are stable and confirms the accelerating behavior. Mathematically, one can obtain the range as, $\left((\frac{1}{4} < k \leq \frac{13}{25}) \wedge \left(m > \frac{1}{6}\right)\right) \vee \left((\frac{1}{4} < k \leq \frac{13}{25}) \wedge \left(m \leq -\frac{11}{6}\right)\right)$.
\begin{figure}[H]
    \centering
    \includegraphics[width=60mm]{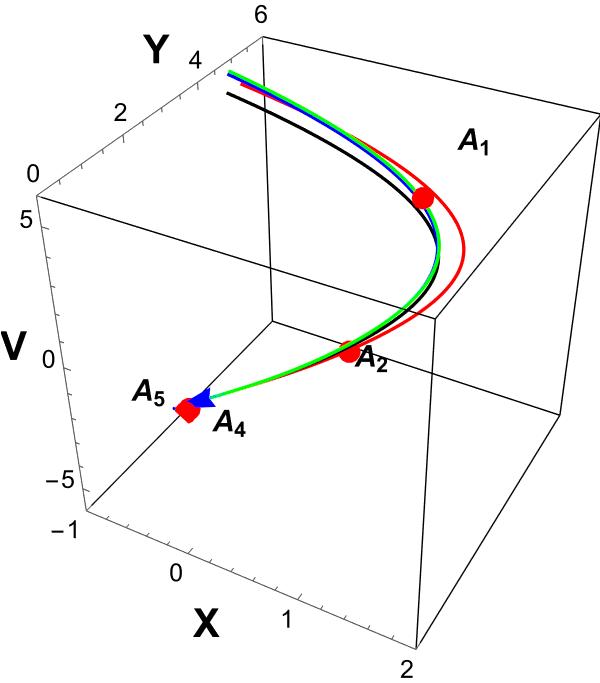}
    \caption{3D phase portrait with $k=0.029, m=0.5$ for model \ref{Model_Ich4_2}. } \label{3-D Model-I}
\end{figure}
The 3D phase portrait presented in Fig. \ref{3-D Model-I} allows us to analyze the behavior of trajectories at the critical points representing different phases of Universe evolution. Phase space trajectory passes through the critical point $A_1 \rightarrow A_2 \rightarrow A_4, A_5.$  We can see from the figure that the chosen trajectory evolves from a radiation-dominant solution corresponding to critical point $A_1$ to an accelerating solution corresponding to critical points $A_4$ and $A_5$. There may be first red dots in the trajectory transition between $A_1$ (radiation) and $A_2$ (matter) and last dots in the trajectory transition between $A_4$ and $A_5$ (de Sitter). Our final attractors' points, $A_4$ and $A_5$ represent the de Sitter epoch with cosmic acceleration.
\begin{figure}[H]
    \centering
    \includegraphics[width=60mm]{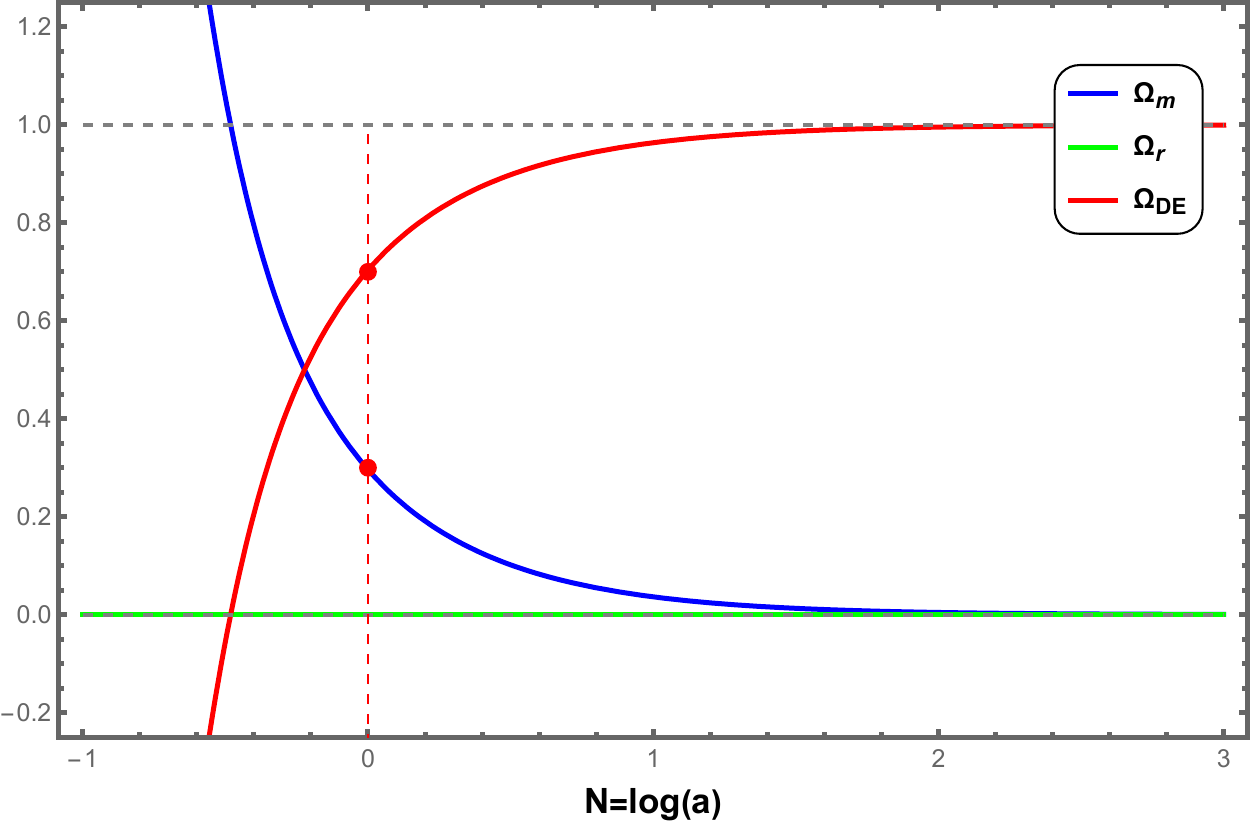}
    \caption{Evolution of the density parameters for model \ref{Model_Ich4_2}. } \label{fig:evolution I}
\end{figure}
\begin{figure}[H]
    \centering
    \includegraphics[width=60mm]{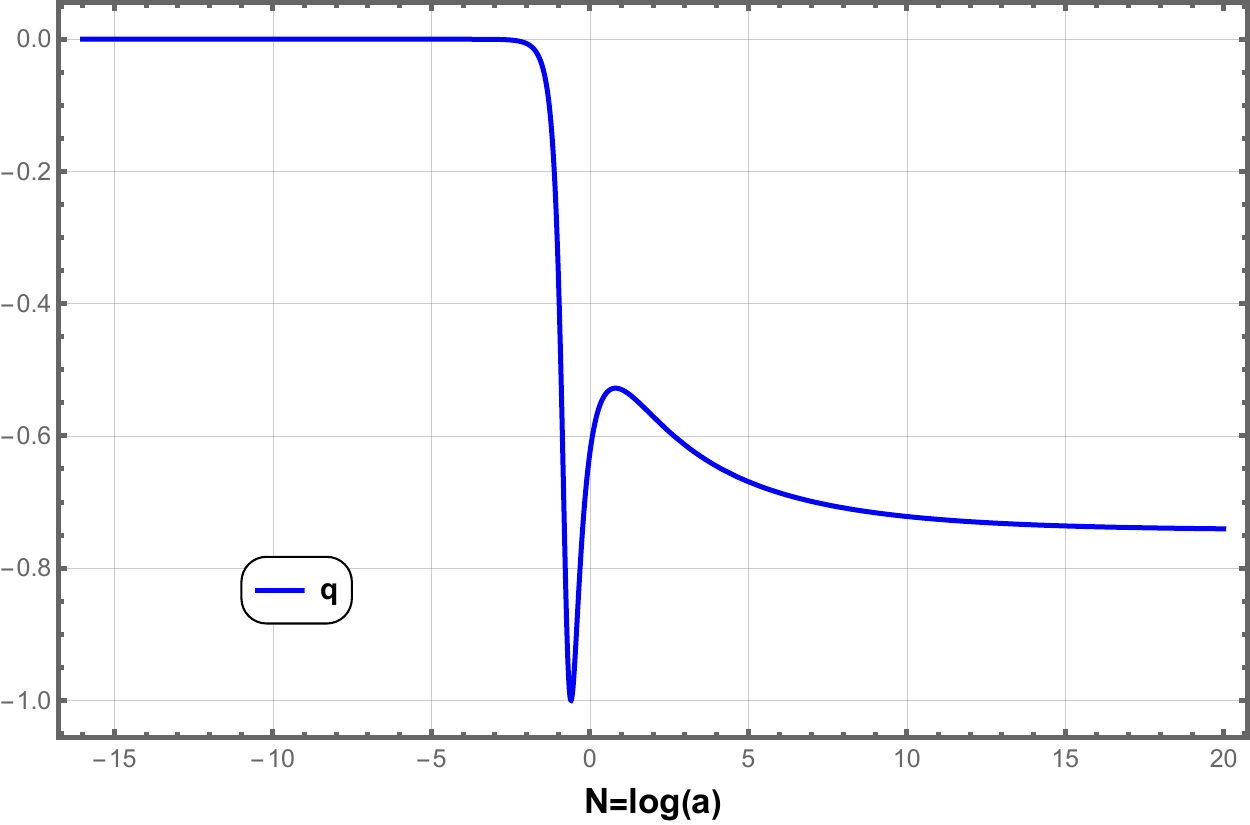}
    \includegraphics[width=60mm]{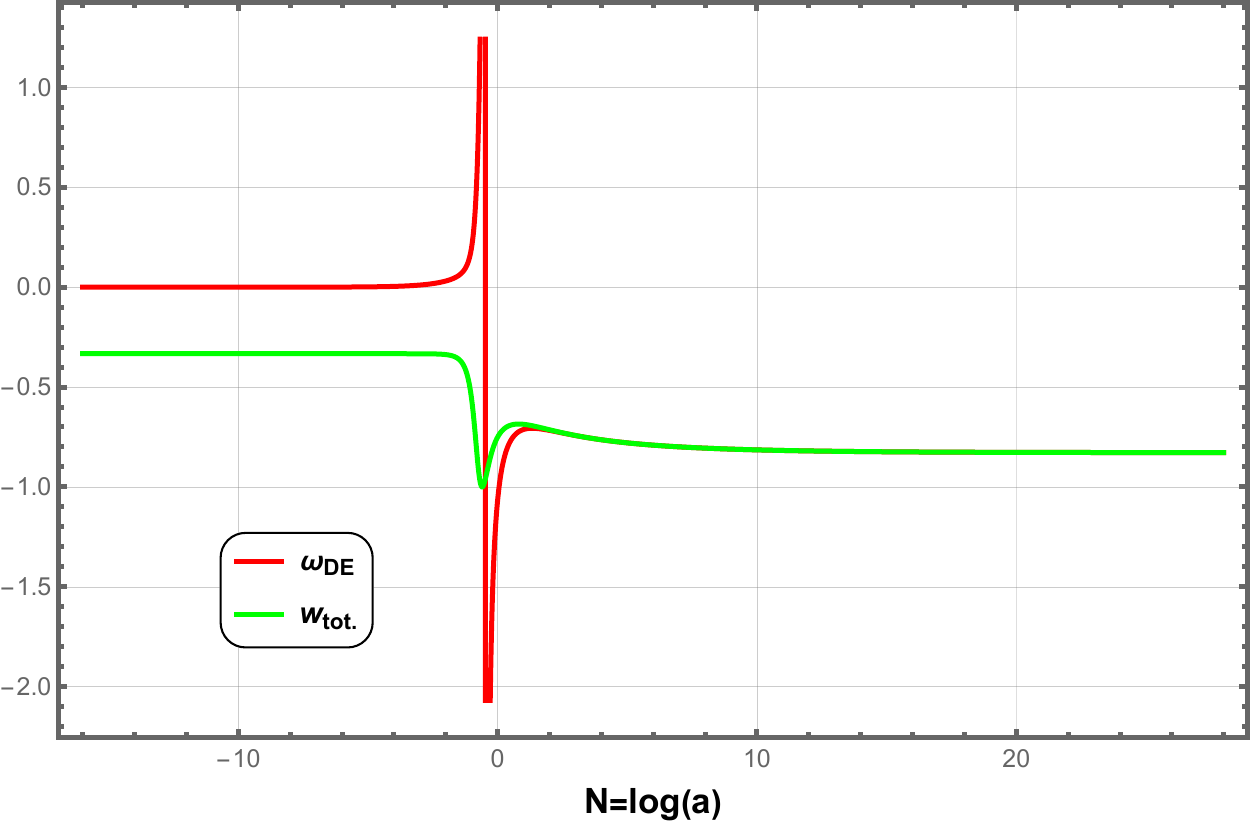}
    \caption{Deceleration and EoS parameter for model \ref{Model_Ich4_2}.} \label{Fig2}
\end{figure}
The initial conditions for above plots are $X=-10^{-1},Y={10^{-5}},Z=10^{-10},V=10^{-11}$, $k=0.029$, $m=0.5$.
The evolution plot for the standard density parameter is plotted in Fig. \ref{fig:evolution I}. The vertical dashed line represents a present time of cosmic evolution, at which the standard density parameter for matter and DE shows values approximately equal to $0.3$ and $0.7$, respectively. The value of the deceleration parameter at present time is $q_{0}=-0.663$ which is compatible with the current observation study $q_0=-0.528^{+0.092}_{-0.088}$ \cite{Gruber:2014}. The plots for EoS parameters are presented in Fig. \ref{Fig2}, from which we can study the early and late time evolution of the Universe. The present value of $\omega_{DE}=-1.05$ which is compatible with the WMAP+CMB result [$\omega_0=-1.073 ^{+0.090}_{-0.089}$] \cite{Hinshaw:2013}. The best way to constrain these value is to use the current observational data sets which is studied in Chapter \ref{Chapter6}.
\subsection{Sum of separated power law model}\label{Model-IICH4_2}
We consider the sum of the separated power law \cite{Capozziello:2016eaz} form of $f(T, T_G)$ as
\begin{align}
f(T,T_G)=g_{0} T_{G}^{k} + t_{0}T^{m},
\label{secondtmodel4_2}\end{align}
where $m$ is arbitrary and $k\neq 1$. In this case, from the dynamical system variables defined in Eq. \eqref{dynamical variables}, we find $f_{T}=-m W -\frac{4m}{k}\left(1+Z\right)X$. In this case, similar to the first model, the dynamical variable $W$ can not be written as dependent on variables $X$ and $Z$; therefore, $W$ has to be treated as an independent variable, whereas the dynamical variable $Y$ shows the dependency on the variable $X$ as $Y=\frac{X \left(k-1\right) \left(2 Z^2 +\lambda +4Z \right)}{\left(Z+1\right)}$. In this case, the only term containing an explicit time-dependence form is the parameter $\lambda$, which plays an important role in identifying the particular epoch of the evolution of the Universe. The particular value of the parameter $\lambda$ can be reproduced by exact cosmological solutions representing a particular epoch of evolution. For $\lambda=0$, it can produce the exact de Sitter scale factor, whereas, for $\lambda=\frac{9}{2}$, it is the matter-dominated exact solution. Further, the radiation-dominated scale factor reproduces for $\lambda=8$. The physical significance can be determined by the behavior of the EoS parameter ($\omega_{tot},\omega_{DE}$) and also the attracting solutions of this theory \cite{Odintsov:2018}. Following methods from the past literature, we consider the dynamical variable constant ($\lambda$) in the further analysis of this model \cite{Franco:2020lxx,Odintsov:2018,odintsov2017autonomous}. By referring to the general dynamical system defined in Eq. ($\ref{generaldynamicalsystem4_2}$), we can write the dynamical system in autonomous form as,
\begin{align}
\frac{dX}{dN}&=\frac{(k-1) X (\lambda +2 Z (Z+2))}{Z+1}+2 X Z\,,\nonumber\\
\frac{dZ}{dN}&=\lambda-2Z^2\,,\nonumber\\
\frac{dV}{dN}&=-4V-2VZ \,,\nonumber\\
\frac{dW}{dN}&=\frac{8 m X Z (Z+1)}{k}+2 (m-1) W  Z-4 X (\lambda +2 Z (Z+2))\,.\label{DynamicalSystem2ndstmodel}
\end{align}
The density parameter for the DE and matter can be written as,
\begin{subequations}
\begin{align}
    \Omega_{DE}&=-2 m W +4 X Z+4 X+W-\frac{8 m X (Z+1)}{k}-\frac{4 (k-1) X (\lambda +2 Z (Z+2))}{Z+1}\,,\\
    \Omega_{m}&=1-V+2 m W -4 X Z-4 X-W +\frac{8 m X (Z+1)}{k}+\frac{4 (k-1) X (\lambda +2 Z (Z+2))}{Z+1}\,.
\end{align}
\end{subequations}
The critical points for the dynamical system in Eq. (\ref{DynamicalSystem2ndstmodel}) with the existing condition are presented in Table \ref{TABLE-V}.
\begin{table*} [!htb]
\small\addtolength{\tabcolsep}{-3pt}
\begin{tabular}{|*{7}{c|}}\hline
\parbox[c][1cm]{2.5cm}{Critical points} & \parbox[c][1cm]{0.5cm}{X} & \parbox[c][1cm]{0.5cm}{Z} & \parbox[c][1cm]{0.5cm}{V} & \parbox[c][1cm]{0.5cm}{W} & \parbox[c][1.5cm]{1.5cm}{Exist for} \\ 
\hline
\hline
\parbox[c][1cm]{4cm} {$B_1=(X_1,Z_1,V_1,W_1)$} & 0 & -2 & $V_1$ & 0 & $k\neq 0$, $\lambda=8$, arbitrary m, $V_1$. \\\hline
\parbox[c][1cm]{4cm} {$B_2=(X_2,Z_2,V_2,W_2)$} & $0$ & $-\frac{3}{2}$ & $0$ & $0$  & $\lambda=\frac{9}{2}, k \neq 0$. \\
\hline
\parbox[c][2cm]{4cm} {$B_3=(X_3,Z_3,V_3,W_3)$} & $X_3$ & $\epsilon_1$ & $0$ & $W_3$  & \begin{tabular}{@{}c@{}} $\epsilon_1 \ne 0, \epsilon_1+1\neq 0, k=\frac{1}{2},$\\ $W_3 =-8 (X_3 \epsilon_1+X_3), \lambda= 2 \epsilon_1^2$ \end{tabular}\\
\hline
\parbox[c][1cm]{4cm} {$B_4=(X_4,Z_4,V_4,W_4)$} & $0$ & $\epsilon_2$ & $0$ & $W_4$ & \begin{tabular}{@{}c@{}}$W_4  \epsilon_2\neq 0$,\\
$k \epsilon_2+k\neq 0, m=1$, $\lambda =2 \epsilon_2^2$.\end{tabular} \\
\hline
\end{tabular}
\caption{The critical points for model \ref{Model-IICH4_2}}
\label{TABLE-V}
\end{table*}
In this case, it can be noted that there are less number of critical points than in  first model; also, since $\lambda$ is independent, we can categorize the critical points for different phases of evolution on the basis of the value of $\lambda$. The system will describe radiation, matter, and the DE-dominated era for $\lambda=8,\frac{9}{2}, \text{value depending on coordinates X and Y}$ respectively. The stability conditions, the values of the deceleration parameter, $\omega_{DE}, \omega_{tot}$ at each critical point are presented in Table \ref{TABLE-VI}.
\begin{table*}[!htb]
    % title of Table
    \centering % used for centering table
    \begin{tabular}{|*{5}{c|}}\hline
    \parbox[c][0.8cm]{2.5cm}{Critical points} & Stability conditions & $q$ & $\omega_{tot}$ & $\omega_{DE}$ \\ [0.5ex] % inserts table %heading
    \hline\hline % inserts single horizontal line
    \parbox[c][0.6cm]{1cm}{$B_1$} & Unstable & $1$ & $\frac{1}{3}$ & - - \\
    \hline
    \parbox[c][0.6cm]{1cm}{$B_2$} & Unstable & $\frac{1}{2}$ & $0$ & - - \\
    \hline
    \parbox[c][0.6cm]{1cm}{$B_3$} &  \begin{tabular}{@{}c@{}} Stable for \\ $m<1\land \epsilon_1>0$ \end{tabular} & $-1-\epsilon_1$ & $-1-\frac{2 \epsilon_1}{3}$ & $\frac{2 \epsilon_1+3}{12 X_3-12 X_3 \epsilon_1}$ \\
    \hline
    \parbox[c][0.6cm]{1cm}{$B_4$} & \begin{tabular}{@{}c@{}}Stable for\\ $k<\frac{1}{2}\land \epsilon_2>0$ \end{tabular}& $-1-\epsilon_2$ & $-1-\frac{2 \epsilon_2}{3}$ & $\frac{-W_4 \epsilon_2+2 \epsilon_2+3}{3 W_4}$ \\
    \hline
    \end{tabular}
    \caption{Stability condition, deceleration and EoS parameter for model \ref{Model-IICH4_2}.} 
    % is used to refer to this table in the text
    \label{TABLE-VI}
\end{table*}
The evolution equations and the standard density parameters for radiation, matter, and DE at each critical point are presented in Table \ref{TABLE-VII}.
\begin{table}
   % title of Table
    \centering % used for centering table
    \begin{tabular}{|*{6}{c|}}\hline
    \parbox[c][0.8cm]{2.4cm}{Critical points} & Evolution Eqs. & Universe phase & $\Omega_{r}$& $\Omega_{m}$ & $\Omega_{DE}$\\ [0.5ex] % inserts table %heading
    \hline\hline % inserts single horizontal line
    \parbox[c][0.7cm]{1cm}{$B_1$} & $\dot{H}=-2 H^{2}$ & $a(t)= t_{0} (2 t+c_{2})^\frac{1}{2}$ & $V_1$& $1-V_1$ & $0$\\
    \hline
    \parbox[c][0.7cm]{1cm}{$B_2$} & $\dot{H}=-\frac{3}{2}H^{2}$ & $a(t)= t_{0} (\frac{3}{2}t+c_{2})^\frac{2}{3}$ & $0$& $1$ & $0$\\
    \hline
    \parbox[c][0.7cm]{1cm}{$B_3$} & $\dot{H}=\epsilon_1 H^{2}$ & $a(t)= t_{0} (-\epsilon_1 t+c_{2})^\frac{-1}{\epsilon_1}$ & $0$& $1-4 X_3 \left(\epsilon_1-1\right)$ & $4 X_3 \left(\epsilon_1-1\right)$\\
    \hline
    \parbox[c][0.7cm]{1cm}{$B_4$} & $\dot{H}=\epsilon_2 H^{2}$ & $a(t)= t_{0} (-\epsilon_2 t+c_{2})^\frac{-1}{\epsilon_2}$ & $0$ & $W_4+1$ & $-W_4$\\
    
    \hline %inserts single line
    \end{tabular}
     \caption{Phase of the Universe, density parameters for model \ref{Model-IICH4_2}.} 
     % is used to refer this table in the text
    \label{TABLE-VII}
\end{table}
The stability of the critical points is obtained from the signature of the eigenvalues and is presented in Table \ref{TABLE-VIII}. 
\begin{table}
\begin{tabular}{|*{2}{c|}}\hline
\parbox[c][0.5cm]{2.5cm}{Critical points} & \parbox[c][1cm]{8cm}{Eigenvalues}\\ \hline
\hline
\parbox[c][0.5cm]{1cm}{$B_1$} & \parbox[c][1cm]{8cm}{$\left\{0,8,-4 \left(m-1\right),-4 \left(2 k-1\right)\right\}$}\\ 
\hline
\parbox[c][0.5cm]{1cm}{$B_2$} & \parbox[c][1cm]{8cm}{$\left\{-1,6,-3 \left(2 k-1\right),-3 \left(m-1\right)\right\}$}\\
\hline
\parbox[c][0.5cm]{1cm}{$B_3$} & \parbox[c][1cm]{8cm}{$\left\{0,-4 \epsilon_1,2 \left(m-1\right) \epsilon_1,-2 \left(\epsilon_1+2\right)\right\}$}\\ 
\hline
\parbox[c][0.5cm]{1cm}{$B_4$} & \parbox[c][1cm]{8cm}{$\left\{0,-4 \epsilon_2,2 \left(2 k-1\right) \epsilon_2,-2 \left(\epsilon_2+2\right)\right\}$}\\ 
\hline
\end{tabular}
\caption{Eigenvalues corresponding to each critical point for model \ref{Model-IICH4_2}.} % title of Table
\label{TABLE-VIII}
\end{table}

\begin{figure}[H]
    \centering
    \includegraphics[width=60mm]{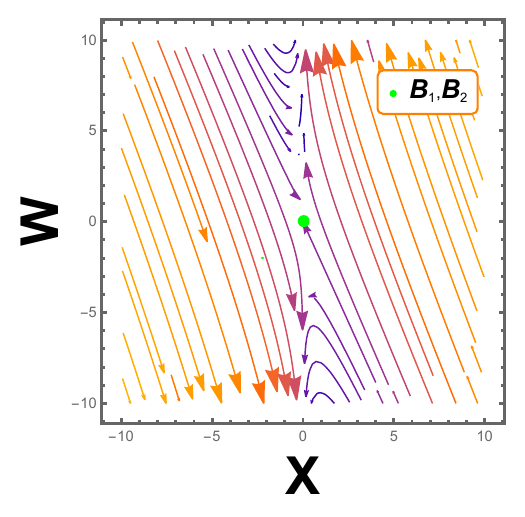}
    \includegraphics[width=60mm]{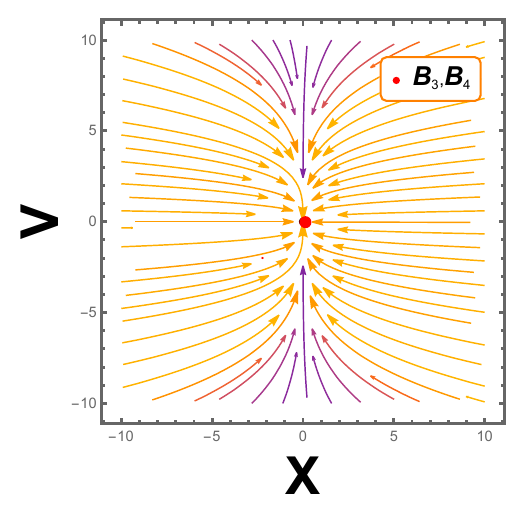}
    \caption{2D phase portrait for  model \ref{Model-IICH4_2}. } \label{Fig3}
\end{figure}

The description for each critical point for the dynamical system in Eq. (\ref{DynamicalSystem2ndstmodel}) are as follows:
\begin{itemize}
    \item{\textbf{Radiation-dominated critical points:}} The critical point $B_1$ is describing the radiation-dominated era with the parametric value of $Z=-2$, and $\lambda=8$. The value of deceleration parameter $q=1, \omega_{tot}=\frac{1}{3}$. The value of $\omega_{DE}$ is undetermined at $B_1$. From Table \ref{TABLE-VI}, one can observe that the critical points describing the radiation-dominated era are unstable. From Table \ref{TABLE-VIII}, in this case, the eigenvalues have zero as one of the eigenvalues; hence, these critical points are non-hyperbolic in nature. This critical point has at least one positive eigenvalue; hence, this critical point is unstable. The exact cosmological solution and the corresponding evolution equation are described in Table \ref{TABLE-VII}. The values of the standard density parameter imply that the critical point $B_1$ defines a standard radiation-dominated era at $V_1=1$. This critical point represents the non-standard radiation-dominated era in which a small amount of matter density parameter contributes. The phase space trajectories near this critical point can be analyzed in Fig. \ref{Fig3}. The critical point representing the radiation-dominated era is plotted in a single plot. The phase space trajectories are moving away from all these critical points; hence, we can analyze the saddle point behavior, which is unstable.
\item\textbf{Matter-dominated critical point: }  The critical point $B_2$ is describing the cold-DM-dominated era. From Table \ref{TABLE-VII}, we can observe that the critical point $B_{2}$ represents the standard cold DM-dominated era. Again from Table \ref{TABLE-VI}, we see the value of $q=\frac{1}{2}$ and $\omega_{tot}=0$ at this critical point. From Table \ref{TABLE-VIII}, we observe that the eigenvalues are unstable, and the existence of one zero eigenvalue at this critical point implies that this critical point is non-hyperbolic. The behavior of phase space trajectories can be analyzed from Fig. \ref{Fig3}. The trajectories at these critical points move away from the critical points; hence, critical points show saddle point behavior.
\end{itemize}
\begin{itemize}
\item{\textbf{DE-dominated critical points:}} In critical points $B_3, B_4$, the value of parameter $\lambda$ and the parameter $Z$ is depend on $\epsilon_1$ and $\epsilon_2$. The critical point  $B_3$ and at $B_{4}$ may describe current cosmic acceleration, respectively, at $\epsilon_1>-1$ and $\epsilon_2>-1$. The de Sitter solution can be analyzed at $\epsilon_1, \epsilon_2=0$ respectively for  critical points $B_{3}$ and $B_{4}$. The hyperbolic or non-hyperbolic nature of these critical points depends on the coordinate values and shows stability at the conditions described in Table \ref{TABLE-VI}. The exact cosmological solution and the evolution equations at these critical points are described in Table \ref{TABLE-VII}. From the behavior of phase space trajectories in Fig. \ref{Fig3}, it can be observed that phase space trajectories are attracting towards both the critical points; hence, these critical points are attractors in behavior.
\end{itemize}
\begin{figure}[H]
    \centering
    \includegraphics[width=60mm]{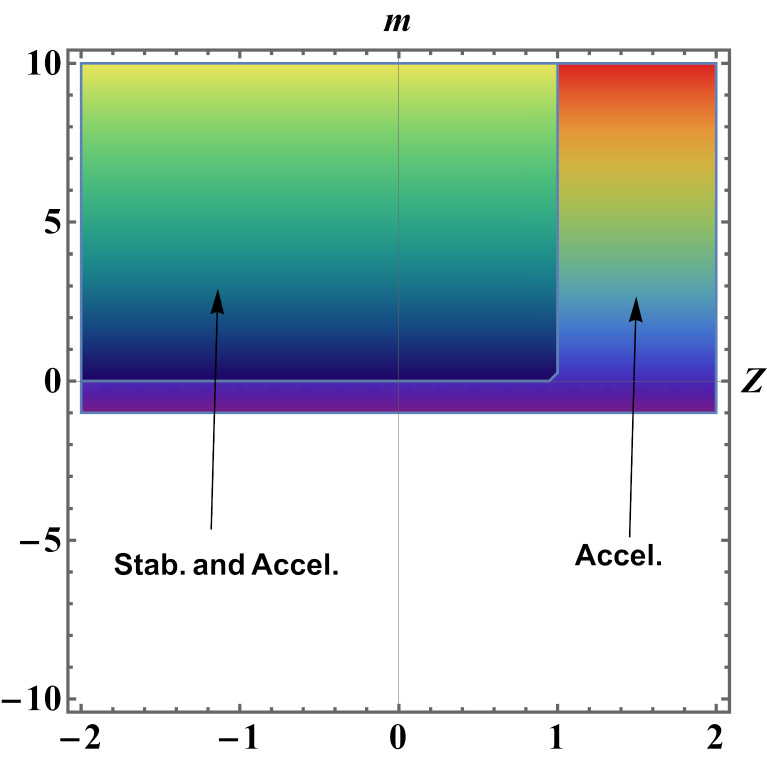}
    \includegraphics[width=60mm]{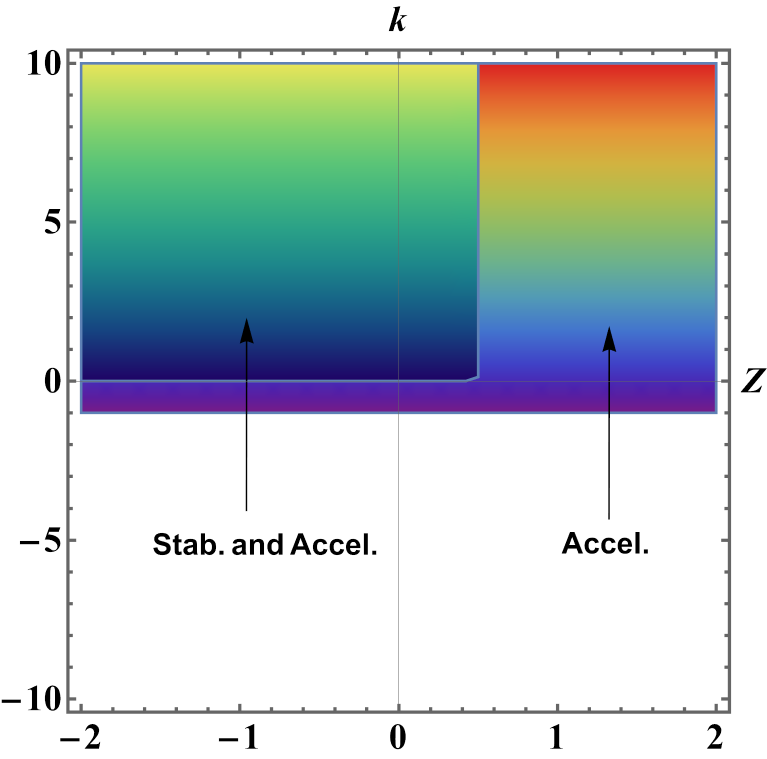}
    \caption{Region plots for critical points $B_3$ and $B_4$ showing stability and acceleration for model \ref{Model-IICH4_2}. } \label{regionplotm2}
\end{figure}
We wish to mention here that the fixed points $B_1$ and $B_2$ exist only for the value of $\lambda$ as $8$ and $\frac{9}{2}$, respectively. For the choice of the variables of the model, all three different epochs of the evolution of the Universe could not be obtained in a single phase space for fixed $\lambda$. Fig. \ref{regionplotm2} describes the critical points $B_3$ and $B_4$ are stable and show accelerating behavior for the range of the model parameters $\left((m < 1) \wedge (z > 0)\right) \wedge \left((k<\frac{1}{2}) \wedge (z > 0)\right)$.
\begin{figure}[H]
    \centering
    \includegraphics[width=60mm]{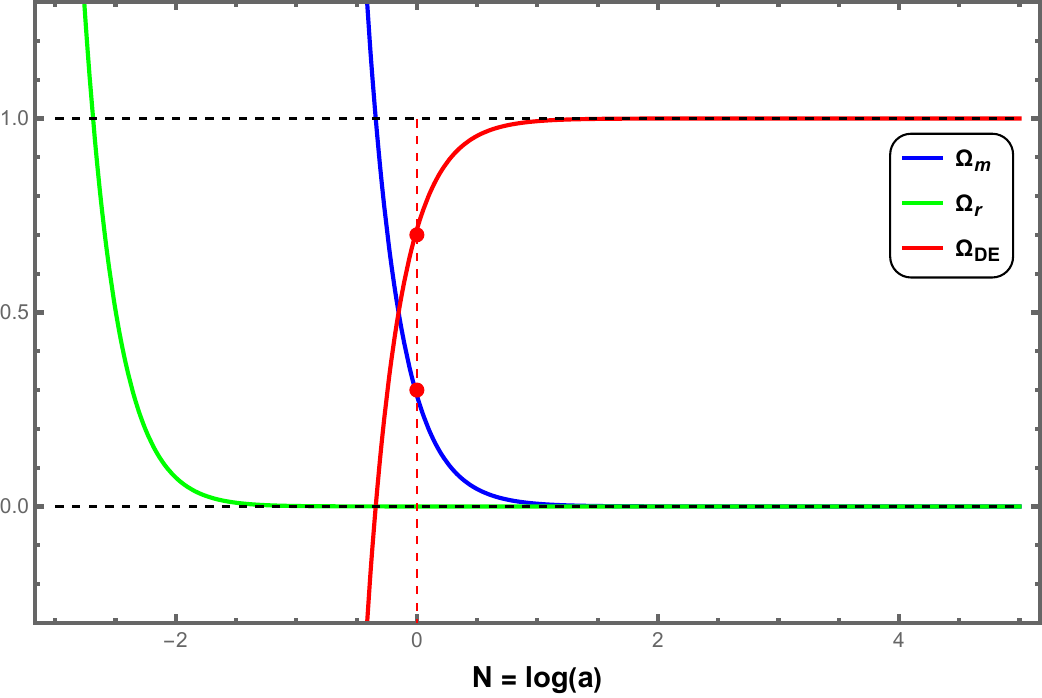}
    \caption{Evolution of the density parameters for model \ref{Model-IICH4_2}.} \label{fig:evolution II}
\end{figure}
The evolution of density parameters $\Omega_{r}$, $\Omega_{m}$, $\Omega_{DE}$ have been presented in Fig. \ref{fig:evolution II}. The vertical dashed line represents the present time at which the values of $\Omega_{DE} \approx 0.7$ and $\Omega_{m} \approx 0.3.$ At the early epoch, we can see that the evolution curve for $\Omega_{r}$ is dominating the other two curves, but it will go on decreasing from the early to the late time of cosmic evolution. The deceleration parameter $q$ and EoS parameters in redshift $N = log(a)$ have been given in Fig. \ref{Fig4}. Currently, the value of the deceleration parameter is obtained as $q_{0}=-1.345$, which agrees with the range provided in Ref. \cite{Feeney:2018}. The present value of the DE EoS parameter has been obtained as $\omega_{DE}=-1.08$ and is approximately the same as in Ref. \cite{Hinshaw:2013}.
\begin{figure}[H]
    \centering
    \includegraphics[width=60mm]{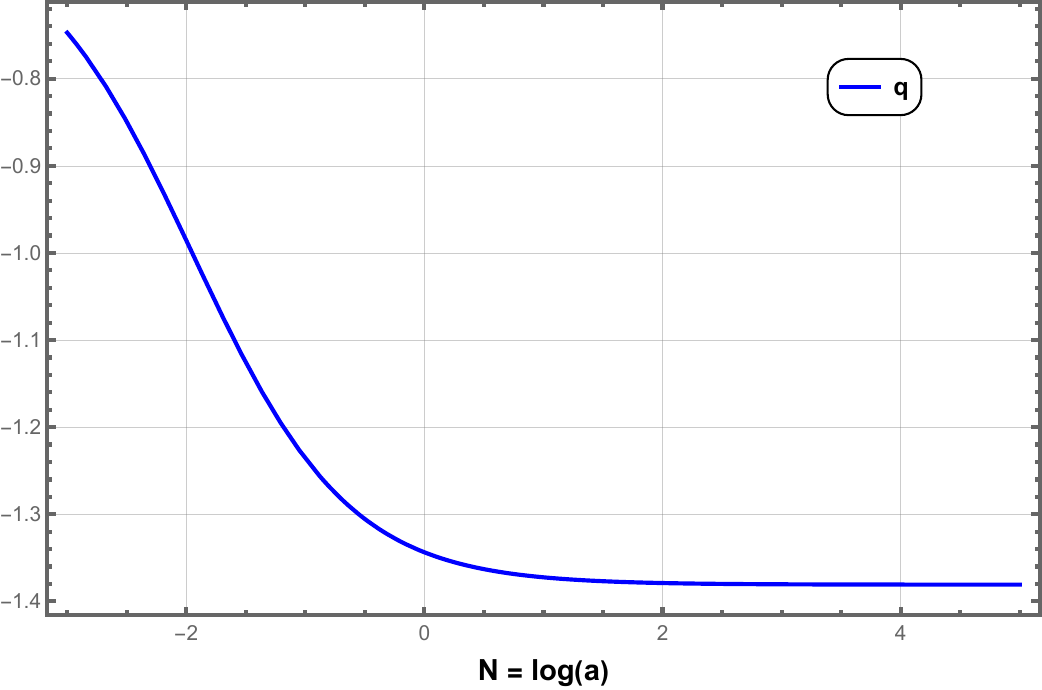}
    \includegraphics[width=60mm]{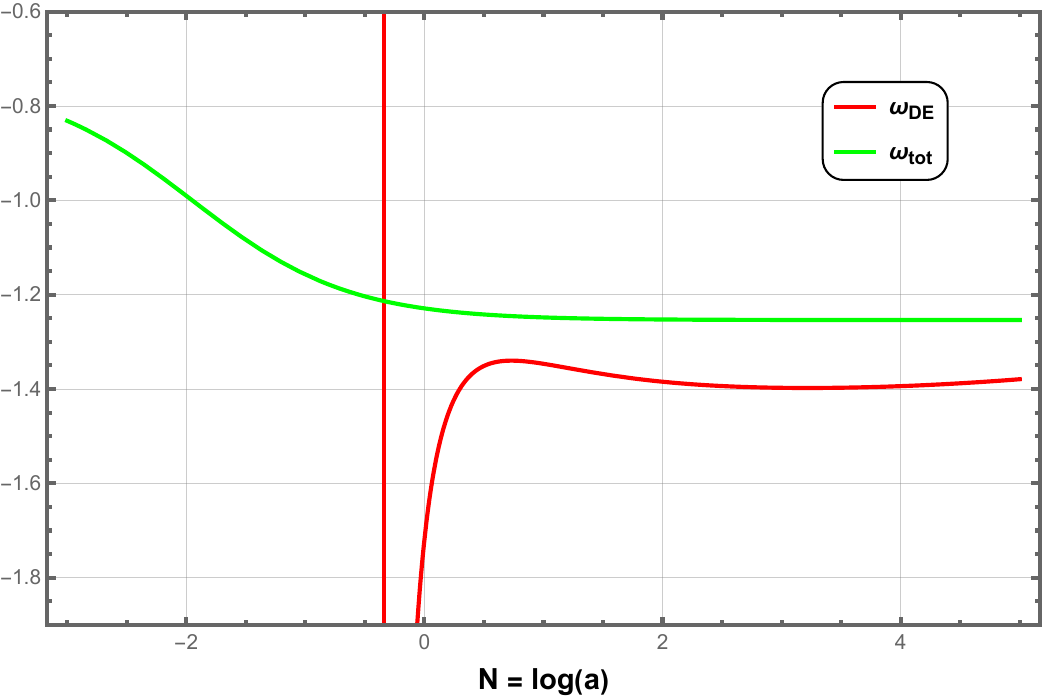}
    \caption{Deceleration and EoS parameter   for model \ref{Model-IICH4_2}.} \label{Fig4}
\end{figure}
The initial conditions in this case are $X=10^{-1.3},\,  Z=0.02 \times 10^{1.2},\, V=0.02 \times 10^{-2.5},\, W=0.0021 \times 10^{-3.7}$,\, $ m=0.67$,\, $k=0.785$.

\section{Conclusion}\label{conclusionch4}
 The general dynamical system, which is dependent on the form of $\Tilde{f}(T, B)$, is presented in Eq. (\ref{GDS}). In Subsec \ref{logbmodel}, the form of $\Tilde{f}$ is the linear term of the torsion scalar $T$ along with the logarithmic form of the boundary term $B$. This model can convert a general dynamical system into an autonomous form and is presented in Eq. \eqref{Dynamicalsystemmodel-I}. The critical points, along with the existing condition, are presented in Table \ref{modelIcriticalpoint}. The stability of each critical point is estimated based on the sign of the eigenvalues of the Jacobian matrix at each critical point. The stability conditions for model \ref{logbmodel} are presented in Table \ref{modelIstabilitycondich4}. To identify the evolutionary epoch at each critical point, the values of standard density parameters are also calculated and presented in Table \ref{modelIdensityparameterch4}. From this, it has been observed that for the first model, the critical point $C_{1}$ representing the radiation-dominated era and the critical point $C_{2}$ representing the matter-dominated era show saddle points and hence are unstable. The same can be verified by analyzing the behavior of phase space trajectories presented in Figs. \ref{modelI2dphaseportrait} and \ref{modelI2dphaseportrait2}. At the critical point $C_{3}$, the value of $\omega_{DE}$ and $\omega_{tot}$ which are dependent on the coordinate $X, Y$, we can study the DE phase of the Universe evolution. For better visibility, the existence and stability region for a critical point $C_3$ is plotted and presented in Fig. \ref{modelIregionplot}. To analyze the behavior of standard density parameters, the plot for $\Omega_{DE}, \Omega_{r}$ and $\Omega_{m}$ in terms of the redshift is presented in Fig. \ref{model1evolutionch4}. From this plot, we can conclude that the density parameter for radiation and matter goes on decreasing and vanishes at a late time, whereas the plot for DE is increasing from an early to a late time of cosmic expansion. The plot for $q$ and $\omega_{DE}$, $\omega_{tot}$  are also presented in Fig. \ref{modelIomegadedecelech4}, which shows compatibility with the observation study made in \cite{Feeney:2018,DIVALENTINO:2016}.  One of the important findings of the addition of boundary term is that the logarithmic form of boundary term is capable to describe a critical point in all the evolutionary phases of the Universe, which might not be possible in the context of $f(T)$ gravity formalism.

Further, we have another form of $\Tilde{f}$ with general index $p$ of the negative boundary term $B$. This case enables us to analyse the role of boundary term more clearly in the extended teleparallel gravity and we present the analysis in Subsec \ref{ModelIIch4}. This form has been widely studied in the literature \cite{Bengochea:2008gz,briffa202}. Hence, it is interesting to investigate in the higher order gravity. The autonomous dynamical system is presented in Eq. \eqref{Dynamicalsystemmodel-II} and to analyse the different phases of the evolution of the Universe, we obtained the critical points, which are presented in Table \ref{model2criticalpoint}. The critical point $\mathcal{P}_3$, the deceleration parameter, and $\omega_{tot}$ show dependence on parameter $\lambda$ whose value contributes to the identification of different phases of the evolution of the Universe. This critical point made a difference in the description of both models in terms of the DE-dominated era. The stability conditions are obtained from the eigenvalues and presented in Table \ref{modelIIstabilitycondi} along with the values of $q$, $\omega_{tot}$, $\omega_{DE}$. The stability behaviour has been also confirmed with the behaviour of phase space trajectories presented in Fig. \ref{modelII2dphaseportraitch4}. In this case, the exact cosmological solutions and the value of $\Omega_{DE}$, $\Omega_{r}$, $\Omega_{m}$ at each critical point have been derived and are presented in Table \ref{modelIIdensityparameter}. From this study, we have concluded that this form is capable of representing and describing all three important phases of the evolution of the Universe.

We have also performed the dynamical system analysis in a modified $f(T, T_G)$ gravity framework. Two well-motivated forms of $f(T, T_G)$ are considered, such as (i) mixed power law and (ii) sum of the separated power law. It is possible to select functions that assign the models according to the Noether symmetry approach, and then those functions (and then the models) should be physically applicable. The critical points of each model have been obtained, and its stability behavior has been examined. In the first model, the critical points, $A_1,\, A_2,\, A_3,\, A_4,$ and $A_5$, are obtained for some specific ranges of the model parameters. The critical points $A_3$, $A_4$, and $A_5$ are showing stable behavior. The range of model parameters at which the stability and accelerating behavior of $A_3$ and $A_4$ can be achieved as $\left((\frac{1}{4} < k \leq \frac{13}{25}) \wedge \left(m > \frac{1}{6}\right)\right) \vee \left((\frac{1}{4} < k \leq \frac{13}{25}) \wedge \left(m \leq -\frac{11}{6}\right)\right)$. Whereas the stability of critical point $A_5$ is dynamically variable dependent, it has a significant role in analyzing the early and future fate of the Universe. From the evolution plot, the present value of the density parameter was obtained to be $\Omega_{DE} \approx 0.7$ and $\Omega_{m} \approx 0.3$. The present value of deceleration and DE EoS parameters has been obtained respectively as $q=-0.663$ and $\omega_{DE}=-1.05$. In the second model, we have obtained four critical points such as    $B_1,\, B_2,\, B_3,\,$ and $B_4$. The two stable points $B_3$ and $B_4$ are showing accelerating and stable behavior in the range of the model parameters  $\left((m < 1) \wedge (z > 0)\right) \wedge \left((k<\frac{1}{2}) \wedge (z>0)\right)$. For this model, density parameters' present value is $\Omega_{DE} \approx 0.7$ and $\Omega_{m} \approx 0.3$. Further, the present value of deceleration and DE EoS parameters are $q=-1.345$ and $\omega_{DE}=-1.08$.

% Chapter 5
\chapter{Constraining \texorpdfstring{$f(T,B,T_G,B_G)$}{} gravity by dynamical system analysis} % Main chapter title

\label{Chapter5} % For referencing the chapter elsewhere, use \ref{Chapter1} 

\lhead{Chapter 5. \emph{Constraining $f(T, B, T_G, B_G)$ gravity by dynamical system analysis}} % This is for the header on each page - perhaps a shortened title

\vspace{10 cm}
* The work in this chapter is covered by the following publication: \\

\textbf{S. A. Kadam} and B. Mishra, ``Constraining $f(T, B, T_G, B_G)$ gravity by dynamical system analysis", \textit{Physics of the Dark Universe},  \textbf{46}, 101693 (2024).

%----------------------------------------------------------------------------------------
%\section{Summary of Results}
 \clearpage
 \section{Introduction}\label{ch5introduction}
 In the previous chapter, we analysed the modification in TEGR by considering the contribution from the teleparallel boundary terms like $B$ and $T_G$; here, both of these modifications have separate motives to consider in the study of Universe dynamics. The teleparallel Gauss-Bonnet term includes whole degrees of freedom related to the teleparallel torsion scalar $T$ \cite{Kofinas:2014aka,bahamonde2019noether}. Whether the inclusion of boundary term preserves the Lorentz invariance and acts as a bridge between the TEGR and GR with the relation $R=-T+B$ \cite{Bahamonde:2016grb,Bahamonde:2015zma}. Moving forward and considering these teleparallel boundary terms in one framework may produce interesting results in studying different cosmic phenomena, including the study of different phases of the evolution of the Universe. With this motive, researchers have recently published one of the generalized frameworks of TEGR $f(T, B, T_G, B_G)$ gravity \cite{Bahamonde:2016kba}, the cosmologically viable models in this formalism are presented using the Noether symmetry approach in \cite{bahamonde2019noether}. We have also taken these interests forward to frame the autonomous dynamical system in this general teleparallel formalism, and in this work, we test the behavior of these models, which are cosmologically viable.

\section{\texorpdfstring{$f(T,B,T_G,B_G)$}{}  gravity field equations}
The detailed general formalism of this gravity is presented in chapter \ref{Chapter1} in Sec \ref{1.6.4}.
 The field equations of $f=-T+\mathcal{F}(T, B, T_G, B_G)$ gravity \cite{Bahamonde:2016kba,bahamonde2019noether,bahamonde:2021teleparallel} can be obtained as,
\begin{eqnarray} 
   f +6 H \dot{f}_B - 2T f_{T} -4 T H \dot{f}_{T_G} -B f_{B} -T_{G} f_{T_G}  = 2 \kappa^2 \rho, \label{1stFEch5}\\ 
   f-4H\dot{f}_{T}-8H^2 \ddot{f}_{T_G} -B f_{B} - \frac{2B f_{T}}{3} +\frac{2 T_G \dot{f}_{T_G}}{3H}-T_G f_{T_G} +2 \ddot{f}_{B}= -2\kappa^2 p,\label{2ndFEch5}
\end{eqnarray} 
where $\rho$ and $p$ respectively denote energy density and pressure for matter and radiation. To find the deviation from TEGR, we use the transformation $f=-T+\mathcal{F}(T, B, T_G, B_G)$, which in turn provides the pressure and energy density expression for the DE phase.
One can retrieve the field equations required to analyse the dynamics of DE as,
\begin{eqnarray} 
    \frac{-1}{2\kappa^2}\left(\mathcal{F}+6H\dot{\mathcal{F}}_B-2T \mathcal{F}_{T}-4TH\dot{\mathcal{F}}_{T_G} -B \mathcal{F}_B-T_G \mathcal{F}_{T_G}\right)&=\rho_{DE}\,,\label{fede1ch5}\\
    \frac{1}{2\kappa^2}\Big(\mathcal{F}-4H\dot{\mathcal{F}}_T-8H^2\ddot{\mathcal{F}}_{T_G} -B \mathcal{F}_B -\frac{2B} {3} \mathcal{F}_{T} +\frac{2T_G}{3H} \dot{\mathcal{F}_{T_G}} -T_G\mathcal{F}_{T_G}+2\ddot{\mathcal{F}_{B}}\Big)&=p_{DE}\,,\label{FEDE2ch5}
\end{eqnarray} 
where $\rho_{DE}$ and $p_{DE}$ respectively denotes pressure and energy density for DE phase. Subsequently, it satisfies the continuity equation, as presented in Eq. \eqref{inConservationEq}, which also holds for the expressions of pressure and energy density of matter and radiation phases. 

\section{Dynamical system in \texorpdfstring{$f(T,B,T_G,B_G)$}{} 
gravity}\label{DynamicalanalysisCh5}
The field equations obtained in Eqs. \eqref{1stFEch5} and \eqref{2ndFEch5} are highly non-linear, and to analyse this, the dynamical system approach can be applied, which can be applied to a wide range of space-time \cite{paliathanasis2021epjp}. Applying the dynamical system approach, the evolution equations are reduced to an ordinary differential equation, which describes a self-consistent phase space.
This allows for an initial analysis of these theories and suggests what kinds of models should be examined or identifying possible observational evidence \cite{Franco:2021,Escamilla-Rivera:2019ulu,briffa202}. Two boundary terms, such as $B$ and $T_G$, are included in the formalism, and the other boundary term vanishes because of the space-time chosen.
Now, we have defined the dynamical variables as \cite{Franco:2020lxx,Escamilla-Rivera:2019ulu,KADAM2024Aop},
\begin{align}
X=\mathcal{F}_{T_G} H^2 \,,\quad Y=\dot{\mathcal{F}}_{T_G}H \,,\quad Z=\frac{\dot{H}}{H^2} \,,\quad 
V=\frac{\kappa^2 \rho_r}{3H^2}\,,\quad
W=-\frac{\mathcal{F}}{6H^2}\,,\quad  \Psi=\mathcal{F}_B \,,\quad \Theta=\mathcal{F}^{'}_{B}.\label{generaldynamicalvariablesch5}
\end{align}
The system is not closed at this point, and the formulation is incomplete without further assumptions. For the dynamical system to be autonomous, we define the additional variable $\lambda=\frac{\ddot{H}}{H^3}$ \cite{Odintsov:2018,Franco:2021,KADAM2024Aop}. The parameter $\lambda$ is the sole term in the dynamical system that exhibits explicit time-dependence, or correspondingly depends on $N$ for a general form of the Hubble rate. Scenarios like $\lambda$ is constant were examined with, the situation where $\lambda$ equals 0 can be derived from an exact de Sitter scale factor, namely $a(t) = e^{\Lambda t}$, with $\Lambda$ being constant, or from a quasi de Sitter scale factor given by $a(t) = e^{H_0 t−H_i t^2}$, where $H_0$ and $H_i$ are constants. In principle, there could be additional scale factors that yield a constant parameter $\lambda$, but the physical importance of each scenario is defined by the behavior of the equation of state parameter $\omega$. It places limitations on the study. The choices we can carry forward for variable $\lambda $ are either assumed to be constant or require models that can be rewritten in terms of the other variables.
The density parameters satisfies the constrained equation,
\begin{align}
\Omega_{m}+V+\Omega_{DE}=1.
\end{align}
We have denoted the standard radiation density parameter $\Omega_{r}=V$. Subsequently, $\Omega_{m}$ show dependency on the other dynamical variables as,
\begin{align}
\Omega_{m}&=1-V-W+\Theta-2\mathcal{F}_{T}-4Y-\left(3+Z\right)\Psi+4ZX+4X,\nonumber\\
    \Omega_{DE}&=W-\Theta+2\mathcal{F}_{T}+4Y+(3+Z)\Psi-4ZX-4X.
\end{align}
Now, the general form of the autonomous dynamical system in this formalism can be formed by, with  $N$ as defined in chapter \ref{Chapter2},
\begin{eqnarray}
\frac{dX}{dN}&=&2XZ+Y\,,\nonumber\\
\frac{dY}{dN}&=&YZ+\Gamma\,,\nonumber\\
\frac{dZ}{dN}&=&\lambda-2Z^2\,,\nonumber\\
\frac{dV}{dN}&=&-4V-2VZ\,,\nonumber\\
\frac{dW}{dN}&=& -2Z\mathcal{F}_T+4\lambda x +16ZX+8Z^2X-6Z\Psi-\lambda \Psi -2ZW\,,\nonumber\\
\frac{d\Psi}{dN}&=&\Theta \,,\nonumber\\
\frac{d\Theta}{dN}&=&\Xi-\Theta Z\,.
\label{generaldynamicalsystem}
\end{eqnarray}
We denote $\Gamma=\ddot{\mathcal{F}}_{T_G}$ and $ \Xi=\frac{\ddot{\mathcal{F}}_B}{H^2}$. One can relate $\Gamma$ and $ \Xi$ in terms of the dynamical variables as,
\begin{align}
    \Gamma=\frac{1}{4} \left[ 3+2 Z +V-3W-2\mathcal{F}^{'}_T-9 \Psi-3 Z \Psi-\left(6+2Z\right)\mathcal{F}_T - 8 Z Y -8Y+12ZX+12X+\Xi\right]\,.\nonumber
\end{align}
In the above equation, the terms $\mathcal{F}_T$ and $\Xi$ must have to convert into the dynamical variable to form the autonomous dynamical system. To do this, we need to have some form of $\mathcal{F}(T, B, T_G, B_G)$. Hence, in the following subsections, we have considered two forms of $\mathcal{F}(T, B, T_G, B_G)$ to frame the cosmological models and understand the evolutionary behaviour of the Universe.

\subsection{Mixed power law}\label{Model-I}
\begin{figure}[H]
    \centering
    \includegraphics[width=60mm]{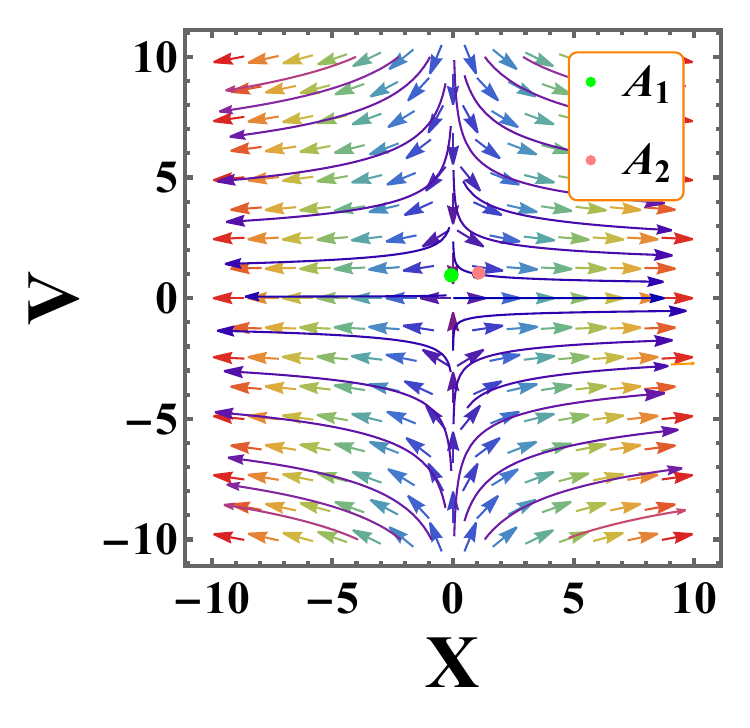}
    \includegraphics[width=60mm]{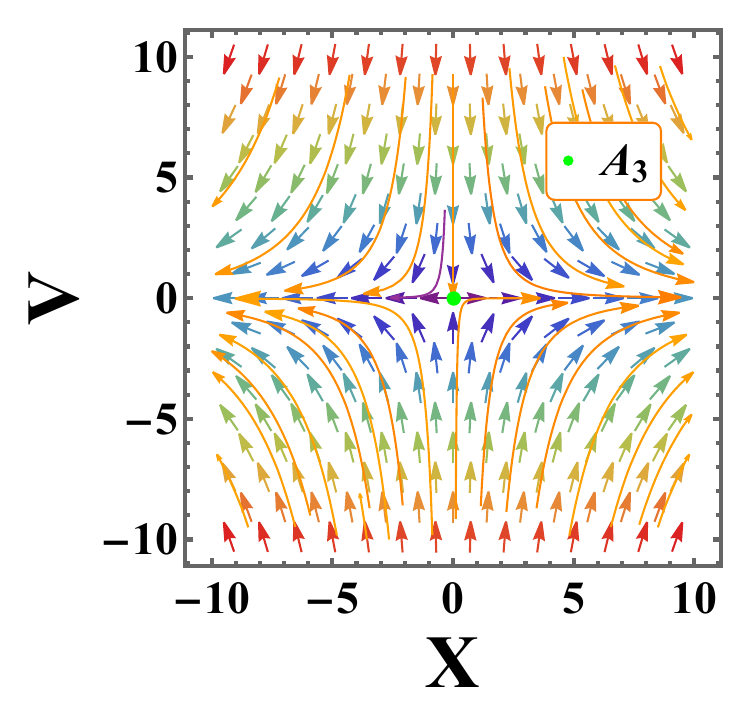}
    \includegraphics[width=60mm]{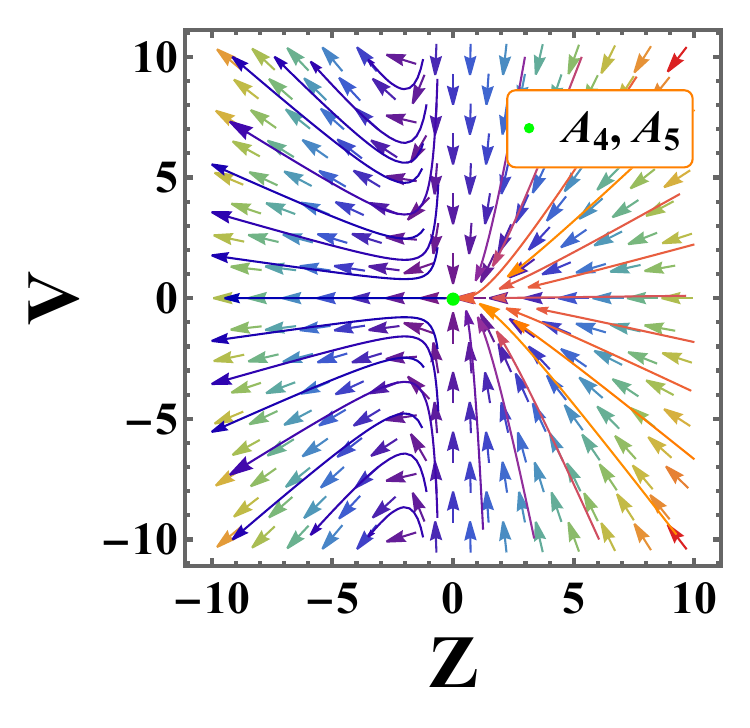}
    \caption{2D phase portrait for the dynamical system, for model \ref{Model-I}. } \label{phasespacem1}
\end{figure}

\begin{figure}[H]
 \centering
  \includegraphics[width=60mm]{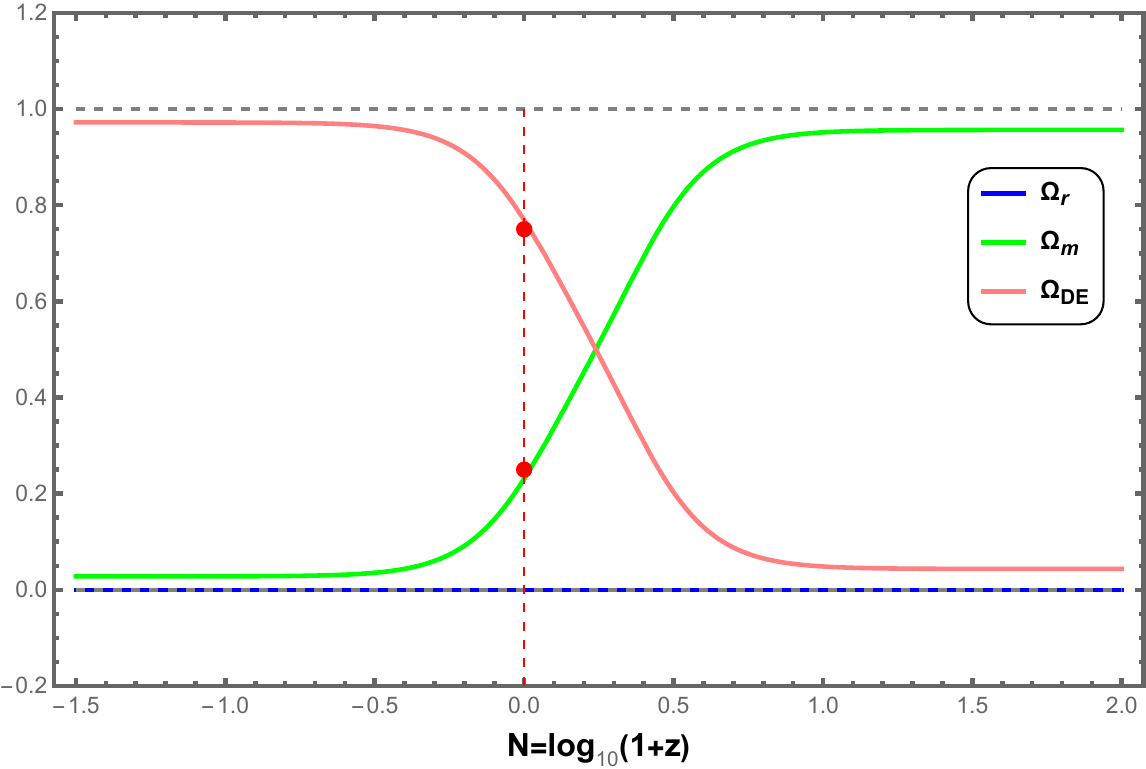}
    \includegraphics[width=60mm]{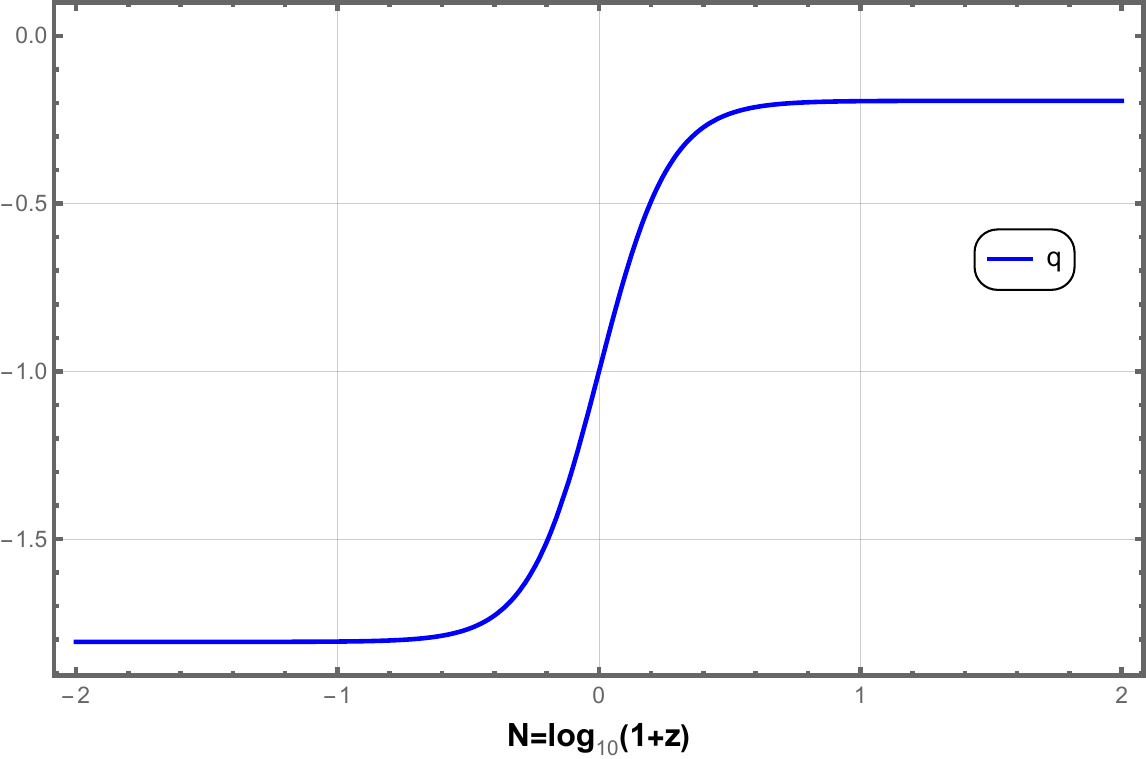}
    \caption{Behaviour of density parameters for matter, radiation, and DE, $q$ for  Model \ref{Model-I}. } \label{evolutionm1}
\end{figure}

First, we consider the mixed power law form of $\mathcal{F}(T, B, T_G, B_G)$ \cite{bahamonde2019noether} as
\begin{align}
\mathcal{F}(T, B, T_G, B_G)=f_{0} T^{m} B^{n}T_{G}^{k} \,,\label{firstmodel}
\end{align}
where $f_{0}, m, n, k$ are arbitrary constants. This form is capable of converting $\mathcal{F}_{T}$ into the dynamical variables as $\mathcal{F}_{T}=-m W$, and this will guarantee the autonomous dynamical system. The variables from Eq. \eqref{generaldynamical variables} can show the  dependency relations as \begin{align}
X&=\mathcal{F}_{T_G} H^2=f_{0} k T_G^{k-1} T^{m} B^{n}H^{2} \,,\nonumber\\
&=\frac{k\mathcal{F}H^2}{T_G}=\frac{-k\mathcal{F}}{24(\dot{H}+H^2)}=\frac{-k\mathcal{F}}{6H^2}\left(\frac{1}{4(\frac{\dot{H}}{H^2}+1)}\right)=\frac{Wk}{4(Z+1)}\,.
\end{align}
which implies,
\begin{align}
W=\frac{4X(Z+1)}{k}\,. \label{eqforz}
\end{align}
Some other dynamical variables from Eq. \eqref{generaldynamicalvariablesch5} can have a direct relationship as,
\begin{align}
\Psi&=\frac{-4n}{k}\frac{(Z+1)X}{ (Z+3)}\,,\nonumber\\
\Theta&=-\frac{4 (n-1) n X (Z+1) (\lambda +6 Z)}{k (Z+3)^2}\,,\nonumber\\
Y&= X\left[2mZ+n\frac{(6Z+\lambda)}{(Z+3)}+\frac{(k-1)(\lambda+4Z+2Z^2)}{(1+Z)}\right]\,.\label{depedency relations}
\end{align}
The variables $W,\Psi, \Theta, Y$ can be written in terms of the variables $X, Z, \lambda$, and hence are treated as the dependent variables. The general autonomous system in Eq. \eqref{generaldynamicalsystem} reduces into three independent variables: $X, Z, V$. In this case, the variable $\lambda$ is treated as a constant that plays a crucial role in identifying the evolutionary epochs of the Universe. The motivation behind considering $\lambda$ as a constant is that cosmological solutions can be retraced for their constant value, such as when  $\lambda=8$, it refers to the study of radiation-dominated epochs, $\lambda=0$ leads to de Sitter Universe and $\lambda$ = $\frac{9}{2}$ refers to matter-domination era. With this setup, the dynamical system in Eq. (\ref{generaldynamicalsystem}) can be written as,
\begin{align}
\frac{dX}{dN}&=X \left(2 m Z+\frac{n (\lambda +6 Z)}{Z+3}+\frac{(k-1) (\lambda +2 Z (Z+2))}{Z+1}+2 Z\right)\,,\nonumber\\
\frac{dZ}{dN}&=\lambda -2 Z^2,\nonumber\\
\frac{dV}{dN}&=-2 V(Z+2).\label{DynamicalSystem1stmodelch5}
\end{align}
The density parameters can be presented as,
\begin{align}
\Omega_{DE}&=-\Theta -2 m W+3 \Psi +W-4 X Z-4 X+4 Y+\Psi  Z\,,  \\
\Omega_{m}&=1+\Theta +2 m W-3 \Psi -V-W+4 X Z+4 X-4 Y-\Psi  Z\,.
\end{align}
Using the dependency relations in Eq. \eqref{depedency relations}, the density parameters become,
\begin{eqnarray} 
    \Omega_{DE}&=&\frac{4 X}{k}\left[-\frac{n (\lambda +\lambda  (Z-k (Z+3))+Z (-6 k (Z+3)+Z (Z+13)+21)+9)}{(Z+3)^2}\right]\nonumber\\
    &+&\frac{4X}{k}\left[\frac{n^2 (Z+1) (\lambda +6 Z)}{(Z+3)^2}+\frac{(k-1) (\lambda  k+(2 k-1) Z (Z+2)-1)}{Z+1}\right]\nonumber\\
 &+&\frac{4X}{k}\left[2 m ((k-1) Z-1)\right]\,,\\
 \Omega_{m}&=&-V+\left[\frac{-4 X \left(n^2 (Z+1)^2 (\lambda +6 Z)-n \sigma_1 (Z+1)+(k-1) \sigma_3 (Z+3)^2+\sigma_2\right)}{(Z+1) \left(k (Z+3)^2\right)}\right]\nonumber\\
 &+&\left[\frac{(Z+1) \left(k (Z+3)^2\right)}{(Z+1) \left(k (Z+3)^2\right)}\right]\,,
\end{eqnarray} 
where,
\begin{eqnarray} \nonumber
    \sigma_1&=&\lambda +\lambda  (Z-k (Z+3))+Z (-6 k (Z+3)+Z (Z+13)+21)
    +9\,,\nonumber\\
    \sigma_{2}&=&2 m (Z+1) (Z+3)^2 ((k-1) Z-1)\,,\nonumber\\
   \sigma_{3}&=&\lambda  k+(2 k-1) Z (Z+2)-1 \,.\nonumber\\
\end{eqnarray} 
Further as the dynamical variable $Z=\frac{\dot{H}}{H^2},$ the total EoS parameter $(\omega_{tot})$ and the deceleration parameter ($q$) can be presented in the form of dynamical variables as,
\begin{align}
    \omega_{tot}&=-1-\frac{2 Z}{3} , \nonumber\\
    q&=-1-Z\,.\label{EosdeceleM1}
\end{align}
To obtain the critical points at different epochs, we consider $\frac{dX}{dN}=0, \, \frac{dZ}{dN}=0, \, \frac{dV}{dN}=0$. Here, we discuss the stability criteria of the critical points based on the nature of eigenvalues. In an $n$-dimensional system, if $n$ number of eigenvalues exist for each critical point, then the classification is to be followed from Sec \ref{dmsoverview}. Now, for Model \ref{Model-I}, the coordinates and the values of $\omega_{tot}$, $q$ at each critical point are presented in Table \ref{modelIcriticalpoints}.  The eigenvalues and the stability conditions associated with each critical point are presented, and the conditions for which these critical points describe the accelerating phase of the Universe are presented in Table \ref{modelIeigenvalues}. The standard density parameters for each critical point are given in Table \ref{modelIdensityparametersm-1}.
\begin{table}[H]
     % title of Table
    \centering % used for centering table
    \begin{tabular}{|c |c |c |c| c|} % centered columns (5 columns)
    \hline\hline %inserts double horizontal lines
    \parbox[c][1.3cm]{2.4cm}{Critical points} & Co-ordinates (X,\, Z,\, V) &  Existence condition&  $\omega_{tot}$&  \textbf{$q$}\\ [0.5ex] % inserts table %heading
    \hline\hline % inserts single horizontal line
    \parbox[c][1.3cm]{1.3cm}{$A_1$ } &$\left[0, \, -2, \, V\right]$& $\lambda =8$ & $\frac{1}{3}$ &   $1$\\
    \hline
    \parbox[c][1.3cm]{1.3cm}{$A_2$ } & $\left[ X, \, -2, \,  V \right]$ & $\lambda =8,\, k=\frac{-m-n+1}{2}$ &$\frac{1}{3}$&  $1$ \\
    \hline
   \parbox[c][1.3cm]{1.3cm}{$A_3$ } &  $\left[ 0, \, -\frac{3}{2}, \, 0 \right]$& $\lambda =\frac{9}{2} $& $0$ &  $\frac{1}{2}$\\
   \hline
   \parbox[c][1.3cm]{1.3cm}{$A_4$} &  $\left[ X, \,\sqrt{\frac{\lambda}{2}}, \, 0\right]$& $ k=\frac{-m-n+1}{2}$ & $-1-\frac{1}{3} \sqrt{2} \sqrt{\lambda }$ &  $-1-\sqrt{\frac{\lambda}{2}}$ \\
   \hline
 \parbox[c][1.3cm]{1.3cm}{$A_5$} &  $\left[ X, \, 0, \, 0 \right]$ & $\lambda =0$& $-1$ &  $-1$ \\
 \hline
    \end{tabular}
    \caption{Critical points and corresponding values of $\omega_{tot}$, $q$ for model \ref{Model-I}.}
    % is used to refer to this table in the text
    \label{modelIcriticalpoints}
\end{table}
\begin{table}[H]
     % title of Table
    \centering % used for centering table
   \scalebox{0.8}{ \begin{tabular}{|c |c |c |c|} % centered columns (5 columns)
    \hline\hline %inserts double horizontal lines
    \parbox[c][1.3cm]{2.4cm}{Critical points}
    &  Eigenvalues &  Stability&  Acceleration \\ [0.5ex] % inserts table %heading
    \hline\hline % inserts single horizontal line
    \parbox[c][1.3cm]{1.3cm}{$A_1$ } & $\left[8,0,-4 \left(m+n+2 k-1\right)\right]$ & Saddle at $\left(k|m\right)\in \mathbb{R}\land n>-2 k-m+1$ & Never \\
    \hline
    \parbox[c][1.3cm]{1.3cm}{$A_2$ } & $\left[8,0,0\right]$ & Unstable& Never \\
    \hline
   \parbox[c][1.3cm]{1.3cm}{$A_3$ } &  $\left[6,-1,-3 \left(m+n+2 k-1\right)\right]$ & Saddle at $\left(k|m\right)\in \mathbb{R}\land n>-2 k-m+1$& Never\\
   \hline
   \parbox[c][1.3cm]{1.3cm}{$A_4$ } &  $\left[0,-\sqrt{2} \sqrt{\lambda}-4,-2 \sqrt{2} \sqrt{\lambda}\right]$ &Stable for $\lambda>0$ &  $\lambda\geq 0$ \\
   \hline
   
   \parbox[c][1.3cm]{1.3cm}{$A_5$ } &  $\left[0,0 -4\right]$ & Nonhyperbolic & Always \\
 \hline
    \end{tabular}}
    \caption{Eigenvalues with stability and acceleration conditions for model \ref{Model-I}.}
    % is used to refer to this table in the text
    \label{modelIeigenvalues}
\end{table}
Where in Table \ref{modelIdensityparametersm-1},
\begin{align}
    \chi _1&=\left(\sqrt{2} \sqrt{\lambda}+2\right) \left(\sqrt{2} \sqrt{\lambda}+6\right){}^2 \left(m+n-1\right)\,,\nonumber\\
    \chi _2&=2 \sqrt{2} \lambda^{3/2} \left(10 n+3\right)+\lambda^2 \left(4 n+1\right)+8 \lambda\left(7 n+2\right)+12 \sqrt{2} \sqrt{\lambda} \left(2 n-1\right)-36\,.
\end{align}
The common range of the model parameters, where this model can describe the unstable radiation, saddle matter, and the stable DE-dominated era, is $n\geq-2 k-m+1$.  
A detailed discussion of each critical point is given below, according to the history of the Universe evolution.
\begin{table}[H] 
     % title of Table
    \centering % used for centering table
    \scalebox{0.8}{\begin{tabular}{|c |c |c |c|} % centered columns (5 columns)
    \hline\hline %inserts double horizontal lines
    \parbox[c][1.3cm]{2.4cm}{ Critical points 
    }& $\Omega_r$ & $\Omega_m$&$\Omega_{DE}$\\ [0.5ex] % inserts table %heading
    \hline\hline % inserts single horizontal line
    \parbox[c][1.3cm]{1.3cm}{$A_1$ } & $V$ & $1-V$& $0$ \\
    \hline
    \parbox[c][1.3cm]{1.3cm}{$A_2$ } & $V$ & $1-\frac{4 X \left(m-8 n^2+11 n-3\right)}{m+n-1}-V$ &$\frac{4 X \left(m-8 n^2+11 n-3\right)}{m+n-1}$\\
    \hline
   \parbox[c][1.3cm]{1.3cm}{$A_3$ } &  $0$ & $1$ & $0$ \\
   \hline
   \parbox[c][1.3cm]{1.3cm}{$A_4$ } &  $0$ & $1-\frac{8 X \left(\left(10 \sqrt{2} \lambda^{3/2}+\lambda^2+72 \lambda +108 \sqrt{2} \sqrt{\lambda}+108\right) m-\left(n-1\right) \chi _2\right)}{\chi _1}$ &$\frac{8 X \left(\left(10 \sqrt{2} \lambda^{3/2}+\lambda^2+72 \lambda+108 \sqrt{2} \sqrt{\lambda}+108\right) m-\left(n-1\right) \chi _2\right)}{\chi _1}$ \\
   \hline
   \parbox[c][1.3cm]{1.3cm}{$A_5$ } &  $0$ &  $\frac{4 X (k+2 m+n-1)}{k}+1$ &$-\frac{4 X (k+2 m+n-1)}{k}$ \\
 \hline
    \end{tabular}}
    \caption{Standard density parameters for model \ref{Model-I}.}
    % is used to refer to this table in the text
    \label{modelIdensityparametersm-1}
\end{table}
\begin{itemize}
\item \textbf{Radiation-dominated critical points:} Both critical points  $A_1$, $A_2$ are representing the early Universe radiation dominated era with $\omega_{tot}=\frac{1}{3}$ and $q=1$. Both have eigenvalue eight as presented in Table \ref{modelIeigenvalues} with the positive signature are unstable. The value of $\Omega_{r}$ is dependent on the dynamical variable $V$ refer Table \ref{modelIdensityparametersm-1} will represent a standard radiation-dominated era at $V=1$ for $A_1$, and for $A_2$ it is $V=1, X=0$. Since the value of $\omega_{tot}=\frac{1}{3}$, these critical points will not describe the current accelerating phase, and the phase space is plotted for the model parameter values $m=0.499, \, n=0.4408, k=0.0301, \, \lambda=8$. These values satisfy the existence condition for a critical point $A_2$ and the condition at which the critical point $A_1$ is saddle. On analyzing the phase space trajectories presented in Fig. \ref{phasespacem1}, one can observe that the phase space trajectories move away at these critical points, confirming the saddle point nature. Moreover, in this case the value of the parameter $\lambda$ is 8.

\item \textbf{Matter-dominated critical point:} This critical point $A_3$ represents standard matter-dominated era with $\Omega_{m}=1$ refer Table \ref{modelIdensityparametersm-1}. The values of $\omega_{tot}=0$ and $q=\frac{1}{2}$ at this critical point as presented in Table \ref{modelIcriticalpoints}. The value of the parameter $\lambda$ at this critical point is $\frac{9}{2}$. The phase space trajectory behavior can be analysed using Fig. \ref{phasespacem1}. The phase space trajectories are moving away from this critical point, and hence, this critical point is showing saddle point behavior. This critical point can be described in the important condition on model parameters $k=\frac{-m-n+1}{2}$, which is a similar condition that arises in \cite{KADAM2024Aop} for this particular model. This condition also plays a crucial role in obtaining a nontrivial Noether vector, as discussed in \cite{Capozziello:2016eaz,Bahamonde:2016grb} where particular forms of the model containing both $B$ and $ T_G$ have been discussed.

\item \textbf{DE-dominated critical points:} The critical point $A_4$ at which we have $\omega_{tot}=-1-\frac{1}{3} \sqrt{2} \sqrt{\lambda }$ and $q=-1- \sqrt{\lambda }$ as presented in Table \ref{modelIcriticalpoints}. This critical point will describe the acceleration of the Universe at $\lambda \ge 0$ and show stability at $\lambda > 0$ refer Table \ref{modelIeigenvalues}. The eigenvalues at this critical point contain zero, and the dimension of the set of eigenvalues are equal to the number of vanishing eigenvalues; hence, it is normally hyperbolic and shows stability at $\lambda > 0$ \cite{coley2003dynamical}. As the variable $\lambda$ is constant, the normally hyperbolic condition is applicable to show stability for $A_4$. For critical point $A_5$, the central manifold theory fails to show stability, as the approximation coefficients vanish. This critical point will not represent a standard DE-dominated solution as the standard density parameter for matter contributes a small amount. The phase space trajectories attract at this critical point. The phase space plot is plotted $V$ vs $Z$ for this critical point. The trajectories from the region $Z>0$ show an attractor nature at this critical point and are moving away in the region $Z<0$, and the same can be analysed from Fig. \ref{phasespacem1}.
The critical point $A_5$ represents a de Sitter solution with $\omega_{tot}=q=-1$. This critical point is nonhyperbolic due to the presence of two zero eigenvalues. This critical point can describe the acceleration of the Universe and will describe the standard DE-dominated era at $X=-\frac{k}{4 (k+2 m+n-1)}$. The phase space trajectories at this critical point show attracting behavior. The value of $\lambda$ at this critical point is 0, as expected. 
\end{itemize}
In Fig. \ref{evolutionm1} $q$ in terms of the redshift for the dynamical system with $m=0.499, \, n=0.4408, \, k=0.0301, \, \lambda=1.301 $ with the initial condition $X_{0}=-1.022\times 10^{-1.988},\, Z_{0}=10\times 10^{-2.5},\, V_{0}=0.2\times 10^{-50}\,. $
From the evolution plots of standard density parameters presented in Fig. \ref{evolutionm1}, we observe that the matter density parameter was dominating the DE density parameter at early times and decreasing over time. Its values have been observed as $\Omega_{m}\approx 0.3$, which is in agreement with \cite{Planck:2018vyg}. The DE density parameter is increasing from early to late, and at the present time, its value is observed to be $\Omega_{DE} \approx 0.7$ \cite{Kowalski_2008}. The radiation density plot vanishes throughout the evolution. The value of $\Omega_{r}\approx 0$ currently agrees with the \cite{Arbey_2021,Planck:2018vyg}. The deceleration parameter lies in the negative region and is capable of describing the accelerating behaviour of the Universe; the present value of $q_{0} \approx -1^{+0.1}_{-0.1}$ and is compatible with the result as in \cite{Capozziellomnras}. The 2D phase space plots in Fig. \ref{phasespacem1} are for $A_1, A_2$, ($\lambda=8, Z=-1.99$), for $A_3$ ($\lambda=\frac{9}{2}, Z=-1.499$) and for $A_4, \, A_5$($\lambda=0, \, X=1$), the model parameters values are $m=0.499, \, n=0.4408, \, k=0.0301$ 
\subsection{Sum of separated power law (particular case)}\label{Model-II}
\begin{figure}[H]
    \centering
    \includegraphics[width=60mm]{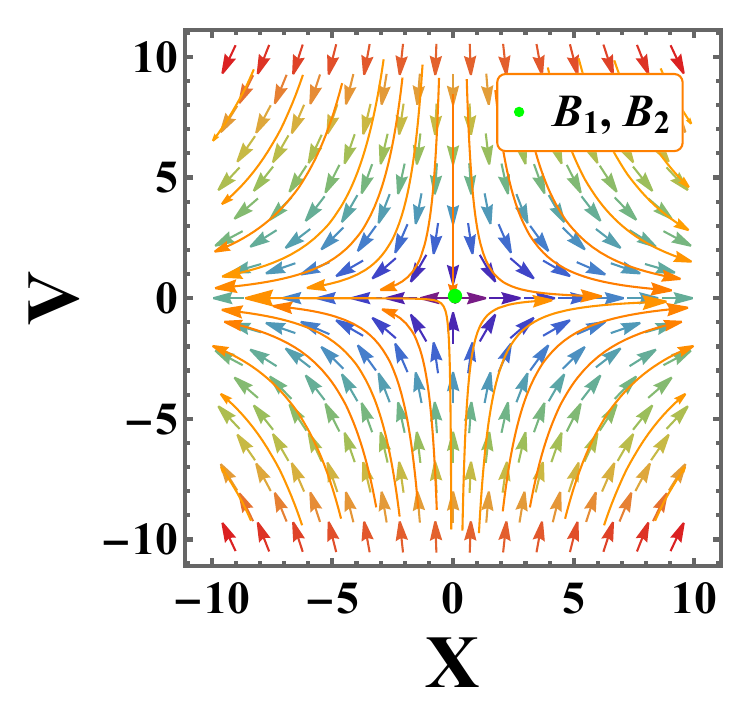}
    \includegraphics[width=60mm]{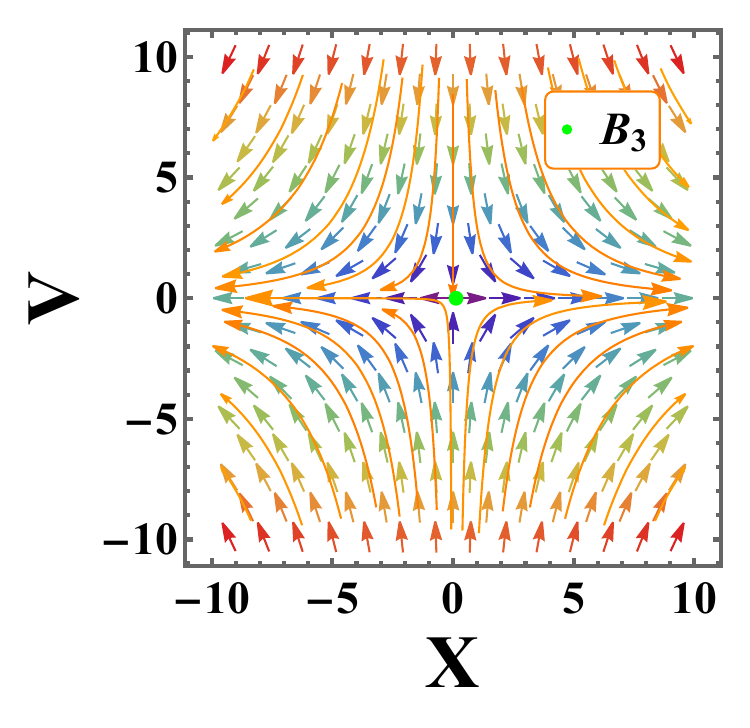}
    \includegraphics[width=60mm]{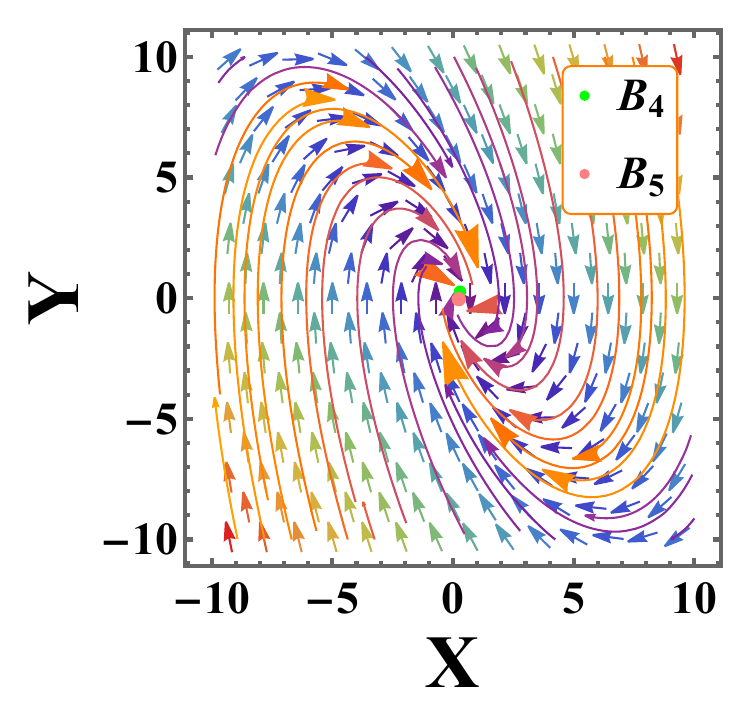}
    \caption{2D phase portrait for model \ref{Model-II}.}  \label{phasespacem2}
\end{figure}

\begin{figure}[H]
    \centering
    \includegraphics[width=60mm]{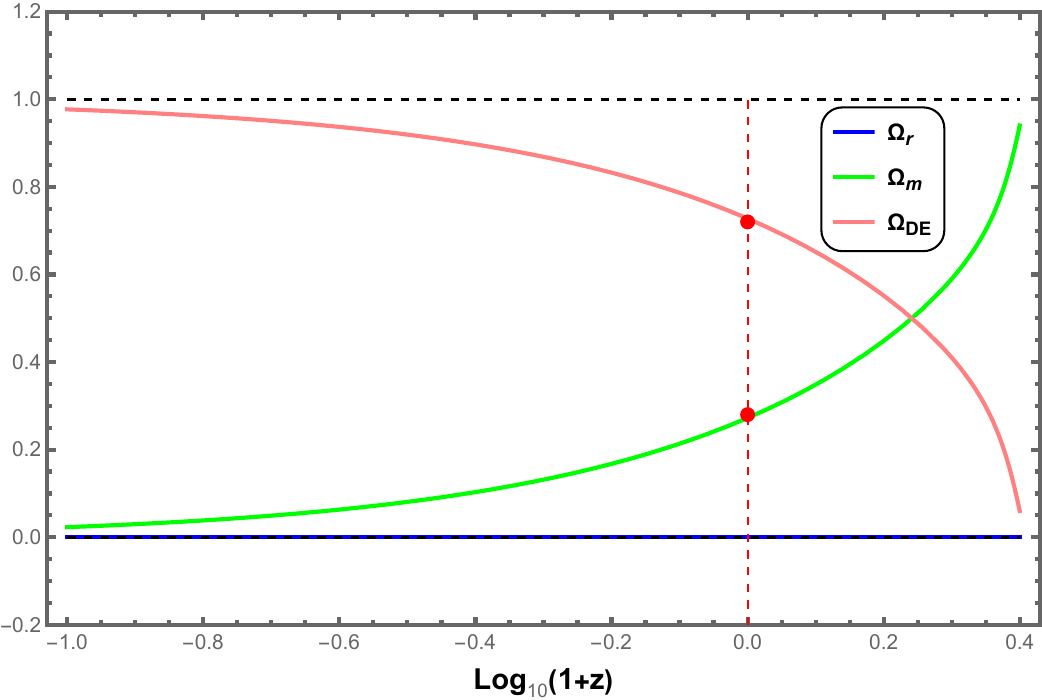}
    \includegraphics[width=60mm]{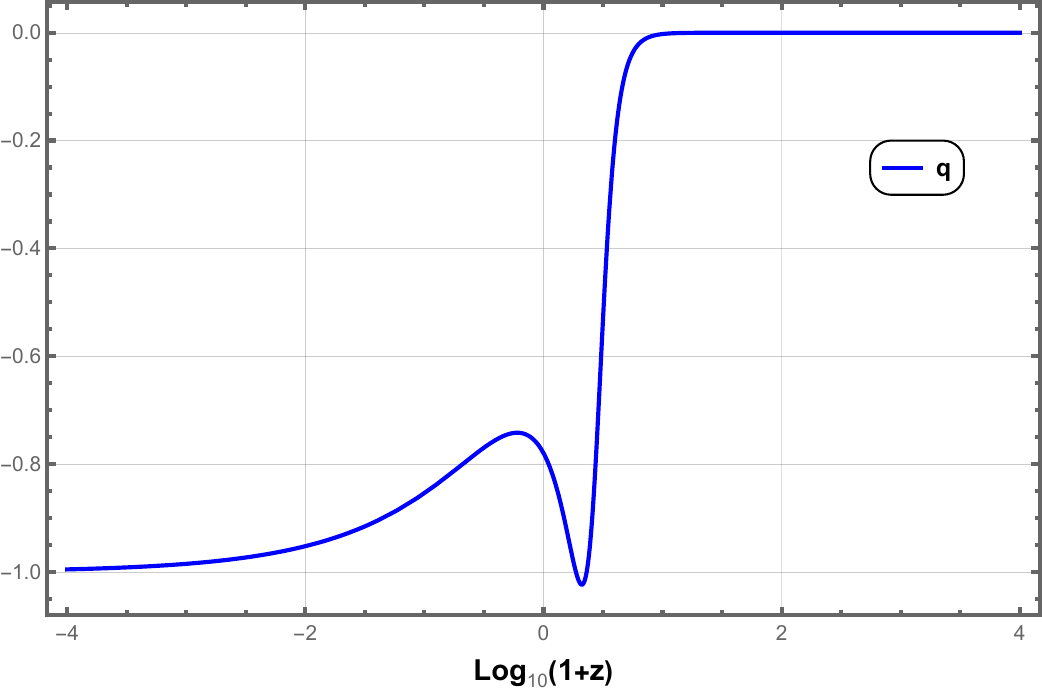}
    \caption{Behaviour of density parameters for matter and DE, $q$ for model \ref{Model-II}. } \label{evolutionm2}
\end{figure}

The second form considered for $\mathcal{F}(T,B,T_G,B_G)$ is a particular form of $\mathcal{F}(T, B, T_G, B_G)= t_{0} T^m + b_{0}B^{n}+g_{0}T_{G}^{k}$, which has been successful in providing a viable cosmological model in Noether symmetry approach \cite{bahamonde2019noether}. For $k=0$ and $n=0$, the dynamical system analysis has been studied respectively in \cite{Franco:2020lxx} and \cite{KADAM2024Aop}. However, we are interested in analyzing a novel particular form in which the teleparallel boundary term $B$ and the Gauus-Bonnet invariant $T_G$ contribute. We define the second form as,
\begin{align}
\mathcal{F}(T, B, T_G, B_G)= b_{0}B+g_{0}T_{G}^{k} \,.\label{secondmodel}
\end{align}
In this case, we have demonstrated that either one of $Y$ and $\lambda$ is dependent, not both simultaneously. Here, we follow the approach in which the dynamical variable $Y$ is treated as independent and obtain the dependency relation for the variable $\lambda$; this will allow us to analyse all different phases of Universe evolution in a single phase space. To obtain an autonomous dynamical system, we have $\mathcal{F}_T =0, \Psi=\mathcal{F}_B =b_0$, and $\Theta=\mathcal{F^{'}}_B =0$. The other dependency relations for the dynamical variables can be presented as,
\begin{align}
W&=\frac{4 X}{k} \left( Z+1\right)-b_{0}  \left(3+Z\right) \,,\nonumber\\
\lambda&=\frac{Y (Z+1)}{(k-1) X}-2 Z (Z+2) \,.\label{dependencyrelationsmodel2}
\end{align}
 Hence, the dynamical variable $W$ is a dependent variable, and the variable $\lambda$ can be written in terms of the model parameters and the other dynamical variables. The general autonomous dynamical system presented in Eq. \eqref{generaldynamicalsystem} can be reduced to the four independent variables as follows,
\begin{align}
\frac{dX}{dN}&=2XZ+Y\,,\nonumber\\
\frac{dY}{dN}&=\frac{1}{4} \left(\frac{12 (k-1) X (Z+1)}{k}+V-4 Y (Z+2)+2 Z+3\right)\,,\nonumber\\
\frac{dZ}{dN}&=\frac{(Z+1) (Y-4 (k-1) X Z)}{(k-1) X}\,,\nonumber\\
\frac{dV}{dN}&=-4V-2VZ\,.
\label{dynamicalsystemmodel2}
\end{align}
Also, with respect to the dynamical variables, the density parameters for DE and matter can be obtained respectively as,
\begin{align}
\Omega_{DE}&=4 Y-\frac{4 (k-1) X (Z+1)}{k}\,,  \\
\Omega_{m}&=1+\frac{4 (k-1) X (Z+1)}{k}-V-4 Y\,.
\end{align}
Since, in both the models, the dynamical parameter $Z$ is an independent variable, hence the equations for the total EoS parameter($\omega_{tot}$) and the deceleration parameter ($q$) 
 are as same as in model \ref{Model-I}, Ref. Eq.\eqref{EosdeceleM1}. The critical points for the system in Eq. (\ref{dynamicalsystemmodel2}) are presented in Table \ref{modelIIcriticalpoints}. The eigenvalues and the stability conditions associated with each critical point are given in Table \ref{modelIIeigenvalues} with the conditions where these points can describe the current and the late-
 time acceleration of the Universe. The standard density parameters for each critical point are given in Table \ref{densityparametersm-2}.
\begin{table}[H]
     % title of Table
    \centering % used for centering table
    \scalebox{0.8}{\begin{tabular}{|c |c |c| c| c|} % centered columns (5 columns)
    \hline %inserts double horizontal lines
    \parbox[c][1.3cm]{2.4cm}{ Critical points
    }&  Co-ordinates  \,  (X,\, Y,\, Z, \,V) & Existence condition&  $\omega_{tot}$ & $q$\\ [0.5ex] % inserts table %heading
    \hline 
    \parbox[c][1.3cm]{1.3cm}{$B_1$ } &$\Bigl\{ X, \, 4X, \, -2, \, 1-12 X\Bigl\}$ & $ k=\frac{1}{2}$ & $\frac{1}{3}$& $1$ \\
    \hline  
    \parbox[c][1.3cm]{1.3cm}{$B_2$ } & $\Bigl\{X, \, 3 X, \, -\frac{3}{2},  \, 0 \Bigl\}$ &$k=\frac{1}{2}$& $0$& $\frac{1}{2}$ \\
 \hline   
      \parbox[c][1.3cm]{1.3cm}{$B_3$ } &  $\Bigl\{\frac{1}{8}, \, \frac{1}{4}, \, -1, \, 0 \Bigl\}$ & $k^2-k\neq 0$ & $-\frac{1}{3}$& $0$ \\
   \hline
   \parbox[c][1.3cm]{1.3cm}{$B_4$} &  $\Bigl\{ X, \,
   \frac{1}{2} (1-4 X), \, 1-\frac{1}{4 X}, \, 0 \Bigl\}$ & $k=\frac{1}{2}$ & $-1-\frac{2}{3} \left(1-\frac{1}{4 X}\right)$& $-2+\frac{1}{4 X}$ \\
\hline   
 \parbox[c][1.3cm]{1.3cm}{$B_5$} &  $\Bigl\{\frac{-k}{4 \left(k-1\right)}, \, 0,  \, 0, \, 0  \Bigl\}$ & $\frac{1}{1-k}\neq 0$ & $-1$& $-1$ \\
 \hline
    \end{tabular}}
    \caption{Critical points and corresponding values of $\omega_{tot}$, $q$ for model \ref{Model-II}.}
    % is used to refer to this table in the text
    \label{modelIIcriticalpoints}
\end{table}
\begin{table}[H]
     % title of Table
    \centering % used for centering table
    \scalebox{0.8}{
    \begin{tabular}{|c |c |c |c|} % centered columns (5 columns)
    \hline\hline %inserts double horizontal lines
    \parbox[c][0.8cm]{2.4cm}{ Critical points 
    }&  Eigenvalues  &  Stability &  Acceleration \\ [0.5ex] % inserts table %heading
    \hline\hline % inserts single horizontal line
    \parbox[c][1.3cm]{1.3cm}{$B_1$ } & $\left[0,1,\frac{-\sqrt{4 X-47 X^2}-X}{2 X},\frac{\sqrt{4 X-47 X^2}-X}{2 X}\right]$ & Saddle at $\frac{1}{12}<X\leq \frac{4}{47}$ & Never \\
    \hline
    \parbox[c][1.3cm]{1.3cm}{$B_2$ } & $\left[0,-1,\frac{-\sqrt{8 X-71 X^2}-3 X}{4 X},\frac{\sqrt{8 X-71 X^2}-3 X}{4 X}\right]$ & $\begin{tabular}{@{}c@{}}Saddle for $0<X<\frac{1}{10}\,,$\\Stable for $\frac{1}{10}<X \leq \frac{8}{71} $\end{tabular}$ & Never \\
    \hline
   \parbox[c][1.3cm]{1.3cm}{$B_3$ } &  $\left[-2,-2,-1,\frac{2 \left(2 k-1\right)}{k-1}\right]$ & $\begin{tabular}{@{}c@{}}Saddle for $k<\frac{1}{2}\lor k>1\,,$\\Stable for $\frac{1}{2}<k<1 $\end{tabular}$ & Never\\
   \hline
   \parbox[c][1.3cm]{1.3cm}{$B_4$ } &  $\left[0,-\frac{12 X-1}{2 X},-\frac{10 X^6-X^5}{2 X^6},-\frac{16 X^6-X^5}{4 X^6}\right]$ &$\begin{tabular}{@{}c@{}}Saddle for $\frac{1}{16}<X<\frac{1}{12}\,,$\\Stable for $X<0\lor X>\frac{1}{10}$\end{tabular}$ &  $X<0\lor X>\frac{1}{8}$ \\
   \hline
   \parbox[c][1.3cm]{1.3cm}{$B_5$ } &  $\left[-4,-\frac{3 \left(k^2-k\right)}{\left(k-1\right) k},\frac{-3 k^2+3 k-\mu_1}{2 \left(k-1\right) k},\frac{-3 k^2+3 k+\mu_1}{2 \left(k-1\right) k}\right]$ & $\begin{tabular}{@{}c@{}}Saddle for $k<0\lor k>1\,,$\\Stable for $\frac{8}{25}\leq k<\frac{1}{2}$\end{tabular}$ & Always \\
 \hline
    \end{tabular}}
    \caption{Eigenvalues for model \ref{Model-II}.\\}
    % is used to refer to this table in the text
    \label{modelIIeigenvalues}
\end{table}
 here, $\mu_1=\sqrt{25 k^4-58 k^3+41 k^2-8 k}$ .
\begin{table}[H]
     % title of Table
    \centering % used for centering table
    \begin{tabular}{|c |c |c |c|} % centered columns (5 columns)
    \hline\hline %inserts double horizontal lines
    \parbox[c][0.8cm]{2.4cm}{ Critical points 
    }& $\Omega_r$ & $\Omega_m$&$\Omega_{DE}$\\ [0.5ex] % inserts table %heading
    \hline\hline % inserts single horizontal line
    \parbox[c][0.8cm]{0.8cm}{$B_1$ } & $1-12 X$ & $0$& $12 X$ \\
    \hline
    \parbox[c][0.8cm]{0.8cm}{$B_2$ } & $0$ & $1-10 X$ &$10 X$\\
    \hline
   \parbox[c][0.8cm]{0.8cm}{$B_3$ } &  $0$ & 0 & 1 \\
   \hline
   \parbox[c][0.8cm]{0.8cm}{$B_4$ } &  $0$ & $0$ & $1$ \\
   \hline
   \parbox[c][0.8cm]{0.8cm}{$B_5$ } &  $0$ &  $0$ &$1$ \\
 \hline
    \end{tabular}
    \caption{Standard density parameters for model \ref{Model-II}.}
    % is used to refer to this table in the text
    \label{densityparametersm-2}
\end{table}
From the above tables, we can conclude that the critical point describing radiation, matter, and the DE-dominated epochs, i.e., $B_1, \, B_2, \, B_4$ respectively, exists at $k=\frac{1}{2}$, and the remaining two critical points $B_3$ and $B_5$ show stability at $\frac{1}{2}<k<1, \frac{8}{25} \le k <\frac{1}{2}$. Hence, these critical points show the existence and stability in the combined range of parameters $\frac{8}{25} \le k < 1$.  Next, we shall provide detailed discussions on the behaviour of the critical points obtained:
\begin{itemize}
\item \textbf{Radiation-dominated critical point:} The critical point  $B_1$ represents the non-standard radiation-dominated era of the evolution of the Universe. This critical point will represent the standard radiation-dominated era at $X=0$, which can be analysed from Table \ref{densityparametersm-2}. The presence of eigenvalues $1, 0$ confirms this critical point is saddle within the range $\frac{1}{12} < X \le \frac{4}{47}$ and non-hyperbolic in nature. The phase space trajectories can be analysed through Fig. \ref{phasespacem2}. The trajectories are moving away from this critical point hence the saddle point nature of the critical point can be observed. As presented in Table \ref{modelIIcriticalpoints}, the existance condition for critical point $B_{1}$ is $k=\frac{1}{2}$.

\item \textbf{Matter-dominated critical point:} At the critical point $B_2$, the value of $\omega_{tot}=0$ and $q=\frac{1}{2}$, hence it describe matter-dominated era. This critical point represents a standard matter-dominated era at $X=0$ where $\Omega_{m}=1$ can be visualized from Table \ref{densityparametersm-2}. The presence of eigenvalue 0 implies that this critical point is normally hyperbolic and is showing stable behavior at $\frac{1}{10} < X \le \frac{8}{71}$ where the other eigenvalues lie in the negative region and are saddle within the range $0 < X \le \frac{1}{10}$. The phase space trajectories in Fig. \ref{phasespacem2} explain the saddle point nature of this critical point.

\item \textbf{Transition critical points:}
As we are aware, accelerating behaviour can be obtained at $q<0$ and $\omega <-\frac{1}{3}$. At the critical point $B_3$, we have $q=0$ and $\omega =-\frac{1}{3}$ Ref. Table \ref{modelIIcriticalpoints}. Hence, this critical point can not describe the accelerating expansion. This critical point shows its existence at $k^2 \ne k$ and is saddle for $k<\frac{1}{2}\lor k>1\,,$ and stable for $\frac{1}{2}<k<1$. The nature of the phase space trajectories can be studied from Fig. \ref{phasespacem2}; it represents the critical point $B_3$ as a saddle point. But due to the choice of dynamical variable $Z=\frac{\dot{H}}{H^2}$, the deceleration parameter will only show dependency on the variable $Z$. It may be this is the cause of the deceleration parameter failing to produce the transition phase and the same can be observed in Fig. \ref{evolutionm2}.

\item \textbf{DE-dominated critical points:}
The critical point $B_4$ in which the $\omega_{tot}, q$ depends on the dynamical variable $X$ and is capable of explaining both the early phases and the late time of cosmic expansion depending upon the choice of the value for variable $X$. From Table \ref{modelIIeigenvalues}, one can analyse the eigenvalues of the Jacobian matrix at this critical point for the system presented in Eq. (\ref{dynamicalsystemmodel2}). At this critical point, eigenvalues contain zero and hence are non-hyperbolic. Since the number of vanishing eigenvalues are equal to the dimension of the set of eigenvalues, it is normally hyperbolic and is stable at $X<0$ or $X>\frac{1}{10}$, and the acceleration of the Universe can be described at $X<0$ or $X>\frac{1}{8}$, the same can be seen in Table \ref{modelIIeigenvalues}. This critical point is an attractor, and the attracting nature of the phase space trajectories can be studied in Fig. \ref{phasespacem2}.

The critical point $B_5$ is a stable de Sitter solution with $\omega_{tot}=q=-1$ showing stability within the range $\frac{8}{25} \le k \le \frac{1}{2}$. This is also an attractor solution, and the phase space diagram is presented in Fig. \ref{phasespacem2}. One thing to be noted is that these critical points represent the standard DE-dominated era of Universe evolution with $\Omega_{DE}=1$ Ref. Table \ref{densityparametersm-2}.
\end{itemize}

In Fig. \ref{evolutionm2} $k=0.38$ and for the initial conditions $X_{0}=0.4 \times 10^{-1.9}, \, Y_{0}=-10.1 \times 10^{-2.5}, \, Z_{0}=10^{-2.5}, \, V_{0}=5\times 10^{-10}$. The behavior of standard density parameters for matter and DE have been plotted in Fig. \ref{evolutionm2}. These parameters take the value $\Omega_m \approx 0.3$ and $\Omega_{DE} \approx 0.7$ at present, the density parameter for radiation is vanishing, and these results show compatibility to the present observational studies \cite{Arbey_2021,Planck:2018vyg,Kowalski_2008}. The deceleration parameter lies in the negative region at present, and late-times shows agreement in describing the late-time accelerating behavior. The observed value of the deceleration parameter at present time is $q_{0}=-0.7931^{+0.02}_{-0.02}$ and is in agreement with \cite{Capozziellomnras}.  In this scenario, the matter density exhibits divergent behavior at early times. This could be attributed to the presence of the teleparallel Gauss-Bonnet term $T_G$. Furthermore, higher-order terms such as the teleparallel Gauss-Bonnet term $T_G$ and the boundary term $B$ contribute to the early divergence in matter density as described in \cite{Franco:2021}.
In Fig. \ref{phasespacem2}, for $B_1 , B_2 (Y=0, \, Z=2, \, k=0.5) $, $B_3  (Y=0, \, Z=2, \, k=1.5) $ and $B_4 , B_5 (V=0, \, Z=0, \, k=0.5).$

\section{Conclusion}\label{conclusionch5}
This chapter presents a systematic dynamical system analysis for two well-motivated cosmological viable models \cite{bahamonde2019noether} in the higher order teleparallel $f(T, B, T_G, B_G)$ gravity formalism. The model \ref{Model-I} can describe the early radiation, matter-dominated epochs, and the DE-dominated epochs at a particular value of $\lambda$. The saddle nature of the critical points at the early radiation and matter-dominated epochs, along with the stable DE-dominated solution have been analysed in both cases.

The model \ref{Model-I} describes a radiation-dominated era at saddle critical points $A_1$ and unstable critical point $A_2$ followed by the critical point $A_3$, which represents a matter-dominated era can be analysed using Tables \ref{modelIcriticalpoints} and \ref{modelIeigenvalues}. The saddle point nature of critical point $A_3$ can be observed from Fig. \ref{phasespacem1}. The stable behavior of DE solutions $A_4$ and the attracting nature of critical point $A_5$ have been obtained. Moreover, the critical point $A_5$ represents the de Sitter solution; both are attractors, and the same can be visualized using Fig. \ref{phasespacem1}. The phase space trajectories show attracting nature from the positive side of the dynamical variable $Z$, as the critical point $A_4$ shows stability at $\lambda >0$, which defines the $Y, Z$ coordinate of this critical point. Since the value of $\omega_{tot}, q$ at $A_4$ depends upon free parameter $\lambda$,  it shows its ability to describe both the early and the late time epochs of the evolution of the Universe. For model \ref{Model-I}, the 2D phase space, the plots of standard density parameters for matter, radiation and DE, and the deceleration parameter have been presented in Figs. \ref{phasespacem1}-\ref{evolutionm1}. The results obtained from these figures are compatible with the current observational studies \cite{Planck:2018vyg,Kowalski_2008}.

Similar to the model \ref{Model-I}, model \ref{Model-II} describes the radiation, matter, and DE attractor solutions. The critical point $B_1$ is the only point representing the radiation-dominated era in model \ref{Model-II}. The key difference in both the models is an additional critical point describing the transition phase (critical point $B_3$) is present in model \ref{Model-II}.  In the matter-dominated solution, the critical point $B_4$ is normally hyperbolic and is showing stability at $X< 0$ or $X>\frac{1}{10}$. The de Sitter solution in model \ref{Model-II} represents standard DE era and is stable at $\frac{8}{25} \le k \le \frac{1}{2}$. The model parameter $k$ takes the value $\frac{1}{2}$ at which critical points represent the radiation, matter, and DE eras of Universe evolution. The plots, in this case, are presented in Figs. \ref{phasespacem2} and \ref{evolutionm2} are compatible with the observational studies \cite{Planck:2018vyg,Kowalski_2008,Capozziellomnras} at the present time.  The common range of the model parameters, where model \ref{Model-I} can describe the saddle radiation and matter-dominated critical points ($A_1, A_3$), is $n\geq-2 k-m+1$. However, the critical points describe radiation, DE ($A_2, A_4$) show existence at $k=\frac{-m-n+1}{2}$. For model \ref{Model-II}, the critical points show the existence and stability in the combined range of parameters $\frac{8}{25} \le k < 1$.

This study extends the analysis made in \cite{Franco:2020lxx,
KADAM2024Aop} and combine the results in a more general way to test the cosmological applications of the presented models. Moreover, this guarantees to move forward to check the observational outcomes of these models. 

% Chapter 6
\chapter{Teleparallel gravity and quintessence the role of boundary term couplings} % Main chapter title

\label{Chapter6} % For referencing the chapter elsewhere, use \ref{Chapter1} 

\lhead{Chapter 6. \emph{Teleparallel gravity and quintessence the role of boundary term couplings}} % This is for the header on each page - perhaps a shortened title

\vspace{10 cm}
The work in this chapter is covered by the following publication: \\

\textbf{S. A. Kadam}, L. K. Duchaniya and B. Mishra, ``Teleparallel gravity and quintessence: the role of nonminimal boundary couplings", \textit{Annals of Physics}, \textbf{470}, 169808 (2024).

%----------------------------------------------------------------------------------------
%\section{Summary of Results}
 \clearpage
\section{Introduction}\label{ch6intro} 
Initially, to study the late time cosmic acceleration, the works are done in the quintessence scalar field where the power law coupling potentials are considered \cite{Wetterich1988,Ratra1988}, for review in curvature formalism one can refer to \cite{Copeland:2006wr}. The coupling coefficient $\xi \phi^2$ of $T$ was considered in the teleparallel modifications and was analysed in \cite{WEI2012430,Geng:2011}. To analyse the scaling attractors in the teleparallel gravity formalism, the $\xi \phi^2$ coupling coefficient is generalized to $\xi f(\phi)$ in \cite{Otalora:2013tba}. Moreover, in \cite{Bahamonde_2015}, this formalism has further modified the teleparallel approach by considering the inclusion of the nonminimal coupling $\xi \phi^2$, $\chi \phi^2$ to the $T$, and $B$ respectively. The more general form where the general scalar field functions $f(\phi), g(\phi)$ are incorporated to study the generalised second law of thermodynamics and the Noether symmetry approach in \cite{Zubair_2017,Gecim_2018}. Moreover, the generalized non-minimal coupling of a tachyonic
 scalar field with the teleparallel boundary term is studied in \cite{Bahamonde_2019}. The  Lorentzian wormholes are constructed in this formalism by Noether symmetries \cite{Bahamonde_2016}. During the literature study, we came across the scope of the study of dynamical system analysis, which has not been studied previously in this general scalar-tensor formalism. Hence in this study, we aim to construct an autonomous dynamical system to analyse the viability of the different well-motivated scalar field potentials to analyse the different evolution epochs of the Universe. We consider the action formula in which the scalar field is non-minimally coupled to both $T$ and $B$ as follows,
\begin{equation}\label{Scalar_Torsion_Lagrangian}
    \mathcal{S} =\int \Bigl\{ \frac{-T}{2 \kappa^2} - \frac{1}{2}\left(f(\phi) T+g(\phi)B - \partial_{\mu} \phi \partial^{\mu} \phi\right) -V(\phi)+ \mathcal{S}_{\text{m}}+\mathcal{S}_{\text{r}}\Bigl\}e\, \mathrm{d}^4 x \,.
\end{equation}
The gravitational field equations and their solutions rely on the spin connection in modified teleparallel gravity. Therefore, it is essential to establish a procedure for determining the specific spin connection corresponding to each tetrad field to solve the field equations effectively. In the realm of FLRW cosmologies, it has been shown that the diagonal tetrad serves as a suitable representation and is expressed in Eq. \eqref{FLRWTETRAD}. The appropriate spin connection associated with this tetrad is the vanishing spin connection, leading to physically meaningful outcomes \cite{Krssak:2015oua, Hohmann:2018rwf, Gonzalez_Espinoza_2020}. This tetrad choice leads to the flat FLRW metric as presented in \eqref{FLATFLRW}
The tetrad field allows for expressing the $T$ and $B$ in terms of the scale factor and its time derivatives as, in Eq. \eqref{TandB}
 where $H \equiv \frac{\dot{a}}{a}$ is the Hubble parameter, and a dot represents the derivative
with respect to time. 
Varying the above action Eq. \eqref{Scalar_Torsion_Lagrangian}, the field equations can be obtained as,
\begin{align}
   3H^2 =& \kappa^2\Bigl\{\rho_{m}+\rho_{r}-3H^2 f(\phi)+3Hg^{'}(\phi)\dot{\phi}+V(\phi)+\frac{\dot{\phi}^2}{2}\Bigl\}, \label{1stFE}\\ 
   3H^2 + 2\dot{H} =&-\kappa^2 \Bigl\{ p_{r} -V(\phi)+\left(3H^2 +2\dot{H}\right)f(\phi) +2H f^{'}(\phi) \dot{\phi} +\frac{\dot{\phi}^2}{2}-\dot{\phi}^2 g^{''}(\phi) -g^{'}(\phi) \ddot{\phi}\Bigl\} \label{2ndFE}\,.
\end{align}
Where prime denotes the differentiation with respect to the scalar field $\phi$. With the same setting, the Klein-Gordon equation can be obtained as,
\begin{align}
\ddot{\phi}+3H\dot{\phi}+\left(\frac{B}{2} g^{'} (\phi) +\frac{T}{2} f^{'} (\phi) \right)+V^{'} (\phi)=0\,.
\end{align}
Comparing Eqs. \eqref{1stFE} and \eqref{2ndFE} to the Friedmann equations presented in Eq. \eqref{FriedmanEQ}, one can retrace the evolution equations required to analyse the dynamics of the DE as,
\begin{align}
    -3H^{2} f(\phi)+3Hg^{'}(\phi) \dot{\phi} +V(\phi) +\frac{\dot{\phi}^2}{2} &=\kappa^2 \rho_{DE}\,,\label{fede1ch6}\\
   -V(\phi)+ \left(3H^2+2\dot{H}\right) f(\phi) +2H f^{'} (\phi) \dot{\phi} +\frac{\dot{\phi}^2}{2} -\dot{\phi}^2 g^{''}(\phi) -g^{'} (\phi) \ddot{\phi} &=\kappa^2 p_{DE}\,.\label{FEDE2ch6}
\end{align}
One can easily verify that the equations Eqs. \eqref{fede1ch6} and \eqref{FEDE2ch6} are satisfying the standard conservation equation as stated in Eq. \eqref{inConservationEq}. The standard density parameters for matter, radiation, and the DE as presented in Eq. \eqref{densityparameters} can be written as which satisfies the constrained equation as presented in Eq. \eqref{CONSTRAINEQ}.

\section{Analysis in nonminimal coupling purely to the
 boundary term}\label{exponentialcoupling}
 We will analyse the scalar field model non-minimally coupled to the boundary term for the case $f(\phi)=0$ and 
 $g(\phi)\ne 0$. For the case  $f(\phi)\ne 0$ and $g(\phi)= 0$, one may see the Ref. \cite{Otalora:2013tba}, where as for the particular case $f(\phi)=\xi$, the detailed analysis is in Ref. \cite{WEI2012430,Geng:2011}. We define the dynamical variables as follows,
\begin{align}
x=\frac{\kappa \dot{\phi}}{\sqrt{6}H}\,, \quad y=\frac{\kappa \sqrt{V}}{\sqrt{3}H}\,,\quad u=\kappa g^{'}(\phi)\,, \quad \rho=\frac{\kappa \sqrt{\rho_{r}}}{\sqrt{3}H}\,, \quad \lambda=\frac{-V^{'}(\phi)}{\kappa V(\phi)}\,.\label{generaldynamical variables}
\end{align}
 The first Friedmann Eq. \eqref{1stFE} can be obtained in the form of dynamical variables as,
\begin{align}
1=\Omega_{m}+\rho^{2}+\sqrt{6} x u +y^2 +x^2,
\end{align}
where,
\begin{align}
    \Omega_{\phi}=\sqrt{6} x u +y^2 +x^2.
\end{align}
 To frame an autonomous dynamical system, we have to consider the exponential coupling function to the boundary term. In this case, so we consider the coupling function as, $g(\phi)=g_{0} e^{-\alpha \phi \kappa}$ \cite{Zlatev_1999,Bahamonde_2019}. Now the dynamical system can be obtained by taking the differentiation of the dynamical variables with  $N$ as defined in chapter \ref{Chapter2},
\begin{align}
\frac{dx}{dN}&=-3 x-\frac{9}{\sqrt{6}} u+ \frac{3 \lambda y^2}{\sqrt{6}}-\left(x +\frac{3 u}{\sqrt{6}}\right)\frac{\dot{H}}{H^2} \,,\nonumber\\
\frac{dy}{dN}&=-\sqrt{\frac{3}{2}} x \lambda y -y \frac{\dot{H}}{H^2},\nonumber\\
\frac{du}{dN}&=-\sqrt{6} \alpha  u x\,,\nonumber\\
\frac{d\rho}{dN}&=-2 \rho -\rho \frac{\dot{H}}{H^2}\,,\nonumber\\
\frac{d\lambda}{dN}&=-\sqrt{6} x \lambda ^2 \left(\Gamma -1\right)\,,\nonumber\\
\label{general_dynamicalsystemcase_I}
\end{align}
where
\begin{align}
    \frac{\dot{H}}{H^2}&=-\frac{\rho ^2+9 u^2+6 \alpha  u x^2+3 \sqrt{6} u x-3 y^2 (\lambda  u+1)+3 x^2+3}{3 u^2+2}\,.\label{HdotbyH^2}
\end{align}
Using Eq. \eqref{HdotbyH^2} into Eq. \eqref{general_dynamicalsystemcase_I}, the autonomous dynamical system can be generated as follows,  
\begin{align}
\frac{dx}{dN}&=\frac{2 x \left(\rho ^2+9 u^2-3 \lambda  u y^2-3 y^2-3\right)+6 x^3 (2 \alpha  u+1)+3 \sqrt{6} u x^2 (2 \alpha  u+3)}{6 u^2+4}\nonumber\\
           &+\frac{\sqrt{6} \left(u \left(\rho ^2-3 y^2-3\right)+2 \lambda  y^2\right)}{6 u^2+4}\,,\nonumber\\
\frac{dy}{dN}&=-\sqrt{\frac{3}{2}} \lambda  x y+\frac{y \left(\rho ^2+9 u^2+6 \alpha  u x^2+3 \sqrt{6} u x-3 y^2 (\lambda  u+1)+3 x^2+3\right)}{3 u^2+2}\,,\nonumber\\
\frac{du}{dN}&=-\sqrt{6} \alpha  u x\,,\nonumber\\
\frac{d\rho}{dN}&=\frac{\rho  \left(\rho ^2+3 \left(u^2+u x \left(2 \alpha  x+\sqrt{6}\right)-\lambda  u y^2+x^2\right)-3 y^2-1\right)}{3 u^2+2}\,,\nonumber\\
\frac{d\lambda}{dN}&=-\sqrt{6} x f\,,
\label{dynamicalsystem}
\end{align}
where $f=\lambda^2 (\Gamma-1)$ and can be written in a generalise way as, $f=\alpha_{1}\lambda^2+\alpha_{2}\lambda+\alpha_{3}$ \cite{Roy_2018}. The potentials in Table \ref{Potentialfunctions} will be used for further study.
\begin{table}[H]
     % title of Table
    \centering % used for centering table
    \scalebox{0.8}{
    \begin{tabular}{|c |c |c |c| c| c|} % centered columns (5 columns)
    \hline
 \multicolumn{6}{|c|}{\textbf{List of Potential Functions}} \\
    \hline %inserts double horizontal lines
    \parbox[c][1.5cm]{1.5cm}{\textbf{Name / Refs.}
    }&\textbf{Potential function} $V(\phi)$ & \textbf{$f$} &  \textbf{$\alpha_{1}$}& $\alpha_{2}$& $\alpha_{3}$\\ [0.5ex] % inserts table %headin$g$
    \hline % inserts single horizontal line
    \parbox[c][1.2cm]{1.2cm}{$P_{1}$ \cite{Zlatev_1999}} & $V_{0}e^{-\kappa \phi}$ & $0$&  $0$& $0$ & 
    $0$  \\
    \hline
   \parbox[c][1.2cm]{1.2cm}{$P_{2}$ \cite{Sahni_2000}} &  $Cosh(\xi \phi)-1$ & $-\frac{\lambda^2}{2}+\frac{\xi^2}{2}$ &  $-\frac{1}{2}$& $0$ & $\frac{\xi^2}{2}$ \\
   \hline
   \parbox[c][1.2cm]{1.2cm}{$P_{3}$ \cite{SAHNI2000}} &   $V_{0} Sinh^{-\eta}(\beta \phi)$ & $\frac{-\lambda^2}{\eta}-\eta \beta^2$ &  $\frac{1}{\eta}$& $0$ & $-\eta \beta^2$ \\
   \hline
    \end{tabular}}
    \caption{List of potentials functions with corresponding $f$ and values of $\alpha_{1},\alpha_{2}, \alpha_{3}.$}
    % is used to refer to this table in the text
    \label{Potentialfunctions}
\end{table}
In Table \ref{Potentialfunctions}, $V_{0}, \xi, \eta, \beta$ are the constants. We have demonstrated the dynamical analysis for each of the above cases in detail in the following sections.

\subsection{Potential \texorpdfstring{$P_1: V_{0}e^{-\kappa \phi}$}{}}\label{P_1}
The exponential potential is considered in the literature to study the teleparallel DE scalar field models \cite{Roy_2018,
Gonzalez-Espinoza:2020jss,duchaniya2023dynamical}. This potential can also be considered to study the generalized teleparallel non-minimally coupled tachyonic models in the presence of the boundary term $B$ \cite{Bahamonde_2019}. 

The critical points throughout this study are the points at which the autonomous dynamical system vanishes and can be obtained through $\frac{dx}{dN}=0,\,\frac{dy}{dN}=0,\,\frac{du}{dN}=0,\,\frac{d\rho}{dN}=0.$ In this case, the value of $f=0$; hence, the above autonomous dynamical system reduces to the four dimensions. The critical points, along with the value of $q, \omega_{tot}$ and the standard density parameters $\Omega_{r}, \, \Omega_{m},\, \Omega_{DE}$ are presented in the following Table \ref{P1criticalpoints}.
The detailed analysis of the critical points is presented as follows,
\begin{itemize}
    \item{\textbf{Radiation-dominated critical points} $A_R, B_R$}: \\
The critical point $A_R$ represents a standard radiation-dominated era with $\Omega_r=1$. The critical point $B_R$ represents a non-standard radiation-dominated era with $\Omega_r=1-\frac{4}{\lambda^2}$. This critical point is a scaling solution with $\Omega_{DE}=\frac{4}{\lambda^2}$. The physical condition $0<\Omega_{DE}<1$, imposes the condition on $|\lambda|>2$. The value of the deceleration parameter $q=1$, the total EoS parameter $\omega_{tot}=\frac{1}{3}$ at both of these critical points; hence, these critical points describe the early time radiation-dominated era of the Universe.
\item{\textbf{Matter-dominated critical points} $C_M, D_M$}: \\
The critical points $C_M$ and $D_M$ both are the critical points describing the matter-dominated epochs of the evolution of the Universe. The value of the parameters $q=\frac{1}{2}, \, \omega_{tot}=0$ at both of these critical points and hence describing the early matter-dominated epoch. The critical point $C_M$ is the standard matter-dominated epoch with $\Omega_{m}=1$. At $D_M$, the small amount of standard density parameter for $DE$ $ \Omega_{DE}=\frac{3}{\lambda^2}$ contributes. This critical point is a non-standard matter-dominated critical point with $\Omega_{m}=1-\frac{3}{\lambda^2}$. Moreover, critical point $D_M$ is the scaling solution, the physical condition on $\Omega_{DE}$ obtains the condition on $\lambda$ as, $|\lambda| >\sqrt{3}$. This critical point also appears in the study of the standard quintessence model \cite{copelandLiddle} and the study of the teleparallel DE model \cite{Xu_2012}. 
\item{\textbf{DE-dominated critical points} $E_{DE}, F_{DE}, G_{DE}$}: \\
The critical point $E_{DE}$ describes a DE-dominated epoch of the Universe evolution with $\Omega_{DE}=1$. The value of $\omega_{tot}=-1+\frac{\lambda^2}{3}$, which falls in the quintessence regime. This critical point is a late-time attractor solution, and the values of the $q, \omega_{tot}$ show compatibility with the current observation studies. This critical point also exists in the study of standard quintessence \cite{copelandLiddle} and in the teleparallel DE models \cite{Xu_2012}. The critical point $F_{DE}$ corresponds to an accelerating Universe with complete DE domination ($\omega_{tot}=-1$). The DE behaves like a cosmological constant. This critical point is a novel critical point that is not present in standard quintessence \cite{copelandLiddle}, and also, in terms of the coordinates, it varies from \cite{Xu_2012}. Similar to $F_{DE}$, the critical point $G_{DE}$ is also behaving like a cosmological constant. This solution describes a standard DE-dominated era with $\Omega_{DE}=1$.
\end{itemize}

\begin{table}[H]
     % title of Table
    \centering % used for centering table
    \begin{tabular}{|c |c |c |c| c| c| c|} % centered columns (5 columns)
    \hline\hline %inserts double horizontal lines
    \parbox[c][0.6cm]{2.4cm}{Critical points
    }& $ \{ x, \, y, \, u, \, \rho \} $ & $q$ &  \textbf{$\omega_{tot}$}& $\Omega_{r}$& $\Omega_{m}$& $\Omega_{DE}$\\ [0.5ex] % inserts table %headin$g$
    \hline\hline % inserts single horizontal line
    \parbox[c][1.3cm]{1.3cm}{$A_{R}$ } &$\{ 0, 0, 0,  1 \}$ & $1$ &  $\frac{1}{3}$& $1$& $0$ & $0$ \\
    \hline
    \parbox[c][1.3cm]{1.3cm}{$B_{R}$ } & $\left\{\frac{2 \sqrt{\frac{2}{3}}}{\lambda }, \, \, \frac{2}{\sqrt{3} \lambda }, \, \, 0, \, \, \sqrt{1-\frac{4}{\lambda^2}}\right\}$ & $1$&  $\frac{1}{3}$& $1-\frac{4}{\lambda ^2}$ & $0$ & $\frac{4}{\lambda ^2}$ \\
    \hline
   \parbox[c][1.3cm]{1.3cm}{$C_{M}$ } &  $\{0,\, \, 0,\, \, 0, \, \, 0\}$ & $\frac{1}{2}$ &  $0$& $0$ & $1$ & $0$\\
   \hline
   \parbox[c][1.3cm]{1.3cm}{$D_{M}$} &   $\left\{\frac{\sqrt{\frac{3}{2}}}{\lambda }, \, \, \frac{\sqrt{\frac{3}{2}}}{\lambda },\, \, 0, \, \, 0\right\}$ & $\frac{1}{2}$ &  $0$& $0$ & $1-\frac{3}{\lambda ^2}$ & $\frac{3}{\lambda ^2}$\\
   \hline
 \parbox[c][1.3cm]{1.3cm}{$E_{DE}$} &  $\left\{\frac{\lambda }{\sqrt{6}}, \, \, \sqrt{1-\frac{\lambda^2}{6}},\, \, 0,\, \, 0\right\}$ & $-1+\frac{\lambda ^2}{2}$ & $-1+\frac{\lambda ^2}{3}$ & $0$ & $0$ & $1$\\
 \hline
 \parbox[c][1.3cm]{1.3cm}{$F_{DE}$} &  $\left\{ 0,\, \, 1, \, \, \frac{\lambda }{3}, \, \, 0\right\}$ & $-1$ &  $-1$ & $0$ & $0$ & $1$\\
 \hline
 \parbox[c][1.3cm]{1.3cm}{$G_{DE}$} &  $\left\{ 0,\, \, 1, \, \, 0, \, \, 0\right\}$ & $-1$ &  $-1$ & $0$ & $0$ & $1$\\
 \hline
 \parbox[c][1.3cm]{1.3cm}{$H_{S}$} &  $\left\{ 1,\, \,0, \, \, 0, \, \, 0\right\}$ & $2$ &  $1$ & $0$ & $0$ & $1$\\
 \hline
    \end{tabular}
    \caption{Critical points analysis: values of $q$,\, $\omega_{tot}$,\, $\Omega_{r}$,\, $\Omega_{m}$ 
 and $\Omega_{DE}$ for potential \ref{P_1} $(P_1)$.}
    % is used to refer to this table in the text
    \label{P1criticalpoints}
\end{table}
\begin{itemize}
 \item{\textbf{Critical point representing the stiff DE $H_{S}$:}}\\
  In the stiff matter era, $\omega=\frac{p}{\rho}=1$ and the energy density \(\rho\) evolves as \(\rho \propto a(t)^{-6}\). This is a much more rapid decrease than for radiation (\(\rho \propto a^{-4}\)) or matter (\(\rho \propto a^{-3}\)) \cite{Chavanis2015}. The critical point $H_{S}$ is corresponding to a non-accelerating, DE-dominated Universe with
 a stiff DE. In this case, the EoS parameter $\omega_{tot}=1$. This critical point exists in studying the standard quintessence model and the teleparallel DE model \cite{copelandLiddle, Xu_2012}.
\end{itemize}
\textbf{The Eigenvalues and the stability conditions:}
\begin{itemize}
\item \textbf{Stability of critical points} $A_{R}, B_{R}:$ \\
The eigenvalues of the critical points $A_R$ are as 
$\Bigl\{\nu_{1}=2, \, \nu_{2}=-1,\, \nu_{3}=1, \, \nu_{4}=0\Bigl\}.$ According to the sign of these eigenvalues, one can conclude that this critical point is a saddle point. The eigenvalues of the critical points $B_R$ are as $\Bigl\{\nu_{1}=-\frac{4 \alpha }{\lambda }, \, \nu_{2}=1, \, \nu_{3}=-\frac{\sqrt{64 \lambda ^4-15 \lambda ^6}}{2 \lambda ^3}-\frac{1}{2}, \, \nu_{4}=\frac{\sqrt{64 \lambda ^4-15 \lambda ^6}}{2 \lambda ^3}-\frac{1}{2}\Bigl\}.$ This critical point is not an unstable critical point but is showing the saddle behaviour at $\Bigl\{0<\lambda <2\land \alpha <0\Bigl\}.$ Both of these critical points represent the radiation-dominated epoch of the evolution of the Universe.
\item \textbf{Stability of critical points} $C_{M}, D_{M}:$ \\
The eigenvalues of the critical point $C_{M}$ are $\Bigl\{\nu_{1}=-\frac{3}{2} ,\, \nu_{2}=\frac{3}{2} ,\, \nu_{3}=-\frac{1}{2} ,\, \nu_{4}=0 ,\, \Bigl\}$. The existence of the positive and the negative eigenvalue implies that this critical point is a saddle critical point. The eigenvalues at the critical point $D_{M}$ are $\Bigl\{\nu_{1}=-\frac{3 \alpha }{\lambda } ,\, \nu_{2}=-\frac{1}{2} ,\, \nu_{3}=-\frac{3 \left(\lambda ^3+\sqrt{24 \lambda ^4-7 \lambda ^6}\right)}{4 \lambda ^3} ,\, \nu_{4}=\frac{3 \sqrt{24 \lambda ^4-7 \lambda ^6}}{4 \lambda ^3}-\frac{3}{4} \Bigl\}$. It is stable at $\Bigl\{\left(\alpha <0\land -2 \sqrt{\frac{6}{7}}\leq \lambda <-\sqrt{3}\right)\lor \left(\alpha >0\land \sqrt{3}<\lambda \leq 2 \sqrt{\frac{6}{7}}\right)\Bigl\}$, and saddle at \\
$\Bigl\{\left(\alpha <0\land \sqrt{3}<\lambda \leq 2 \sqrt{\frac{6}{7}}\right)\lor \left(\alpha >0\land -2 \sqrt{\frac{6}{7}}\leq \lambda <-\sqrt{3}\right)\Bigl\}$.\\
The saddle point nature of these critical points represents early-time matter-dominated epochs of the evolution of the Universe as expected.
\item \textbf{Stability of critical points} $E_{DE}, F_{DE}, G_{DE}:$ \\
The eigenvalues at the Jacobian matrix for a critical point $E_{DE}$ are $\Bigl\{\nu_{1}=-\alpha  \lambda,\, \nu_{2}=\frac{1}{2} \left(\lambda ^2-6\right),\, \nu_{3}=\frac{1}{2} \left(\lambda ^2-4\right),\, \nu_{4}=\lambda ^2-3 \Bigl\}$. This critical point is a stable late-time attractor solution with stability at $\Big(-\sqrt{3}<\lambda <0\land \alpha <0\Big)\lor \left(0<\lambda <\sqrt{3}\land \alpha >0\right)$ and is saddle at $\left(-\sqrt{3}<\lambda <0\land \alpha >0\right)\lor \left(0<\lambda <\sqrt{3}\land \alpha <0\right)$. This critical point is unstable for $\Big(\lambda <-\sqrt{6}\land \alpha >0\Big)\lor \Big(\lambda >\sqrt{6}\land \alpha <0\Big)$. The critical point $F_{DE}$ represents the cosmological constant (de Sitter) solution with $\omega_{tot}=-1$. The eigenvalues at this critical point are $\Bigl\{\nu_{1}=-3,\, \nu_{2}=-2,\, \nu_{3}=-\frac{3 \left(\sqrt{\left(\lambda ^2+6\right) \left(8 \alpha  \lambda +\lambda ^2+6\right)}+\lambda ^2+6\right)}{2 \left(\lambda ^2+6\right)},\, \nu_{4}=\frac{3}{2} \left(\frac{\sqrt{\left(\lambda ^2+6\right) \left(8 \alpha  \lambda +\lambda ^2+6\right)}}{\lambda ^2+6}-1\right)\Bigl\}$. It shows stability at $\Bigl\{8 \alpha  \lambda +\lambda ^2+6\geq 0\land ((\lambda >0\land \alpha <0)\lor (\alpha >0\land \lambda <0))\Bigl\}$ and saddle at $\Bigl\{ (\lambda <0\land \alpha <0)\lor (\lambda >0\land \alpha >0)\Bigl\}$. The eigenvalues at the critical point $G_{DE}$ are $\Bigl\{\nu_{1}=-2,\, \nu_{2}=0,\, \nu_{3}=-3,\,\nu_{4}=-3\Bigl\}$. Note that the critical
point with zero eigenvalues are termed a non-hyperbolic critical point. This is the critical point containing a zero eigenvalue, and the other three eigenvalues are negative; hence, to check its stability, the linear stability theory fails to obtain the stability of such a critical point. Hence, We have obtained stability at this critical point using the Central Manifold Theory (CMT), and the detailed application of CMT is presented in the appendix. From CMT, it has been concluded that this critical point is stable at $\alpha>0$. 

\item \textbf{Stability of critical point} $H_{S}:$ \\
The eigenvalues at this critical point are $\Bigl\{\nu_{1}=3,\,\nu_{2}=1,\,\nu_{3}=-\sqrt{6} \alpha,\,\nu_{4}=3-\sqrt{\frac{3}{2}} \lambda\Bigl\}$. This critical point is saddle at $\left(\alpha >0\land \lambda >\sqrt{6}\right)$ and is unstable at $\left(\alpha <0\land \lambda <\sqrt{6}\right)$.
\end{itemize}
\textbf{Numerical results:}\\
In this study, we have analysed critical points representing different epochs of the evolution of the Universe. Among these critical points, $A_{R}, B_{R}$ are the critical points describing the radiation-dominated epoch of the evolution of the Universe. Moreover, these critical points show saddle point behaviour. The critical points $C_{M}, D_{M}$ represent the matter-dominated epoch of the Universe, and both show saddle point behaviour. The critical points $E_{DE}, F_{DE}$ and $G_{DE}$ are the DE- solutions. From the stability analysis, one can confirm that these critical points are the late time stable attractors within the particular range of the model parameters. Now, we will analyse the numerical solution using the autonomous system presented in Eq. \eqref{dynamicalsystem}. The numerical solutions were calculated using the ND-solve command in Mathematica. Our analysis was based on the Hubble and Supernovae Ia (SNe Ia) observational data sets, presented in detail in the appendix. 

In Fig. \ref{Eosdensitym1}, the EoS parameters for DE, total, and the EoS parameter of $\Lambda$CDM are compared. The plots demonstrate the evolution of $\omega_{tot}$, $\omega_{DE}$, and $\omega_{\Lambda CDM}$ towards $-1$ at late times. The present value of $\omega_{DE}(z=0)=-1$, consistent with the observational result of Planck collaboration \cite{Planck:2018vyg}. The evolution of energy densities of radiation, DE, and DM is illustrated in Fig. \ref{Eosdensitym1}. From these plots, one can observe that radiation prevails in the early times of the Universe. Moreover, the plots describe the sequence by showing that the DM-dominated era can be described as a small time period and, finally, the emergence of the cosmological constant at the late time is observed. 
\begin{figure}[H]
 \centering
  \includegraphics[width=57mm]{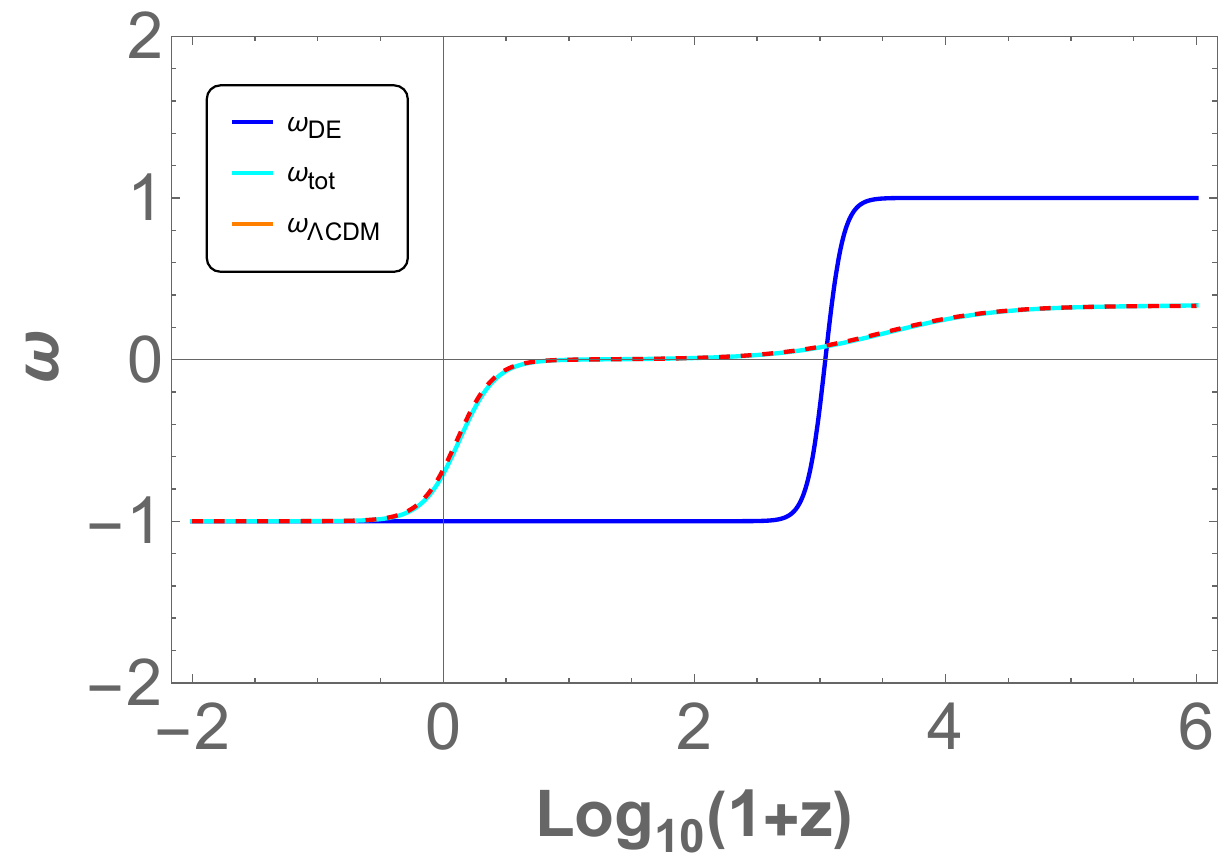}
    \includegraphics[width=60mm]{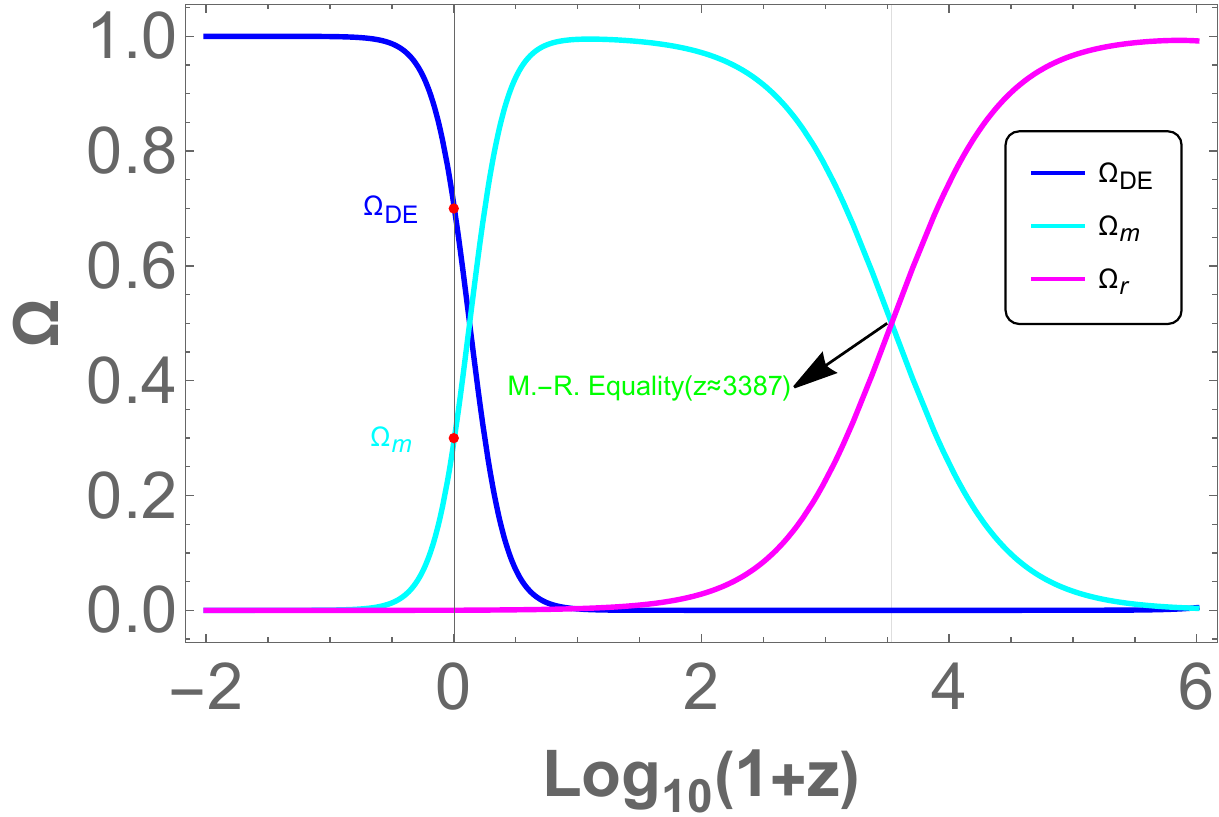}
    \caption{Evolution of EoS and standard density parameters for Model \ref{P_1}. } \label{Eosdensitym1}
\end{figure}    
In this case, the plots are plotted for the initial conditions are $x_C=10^{-8.89},\,y_C=10^{-2.89},\,u_C=10^{-5.96},\,\rho_C=10^{-0.75},\, \lambda=-0.01, \alpha=-5.2$. 

\begin{figure}[H]
 \centering
  \includegraphics[width=57mm]{Figures/Chapter-6/Eosm1.pdf}
    \includegraphics[width=60mm]{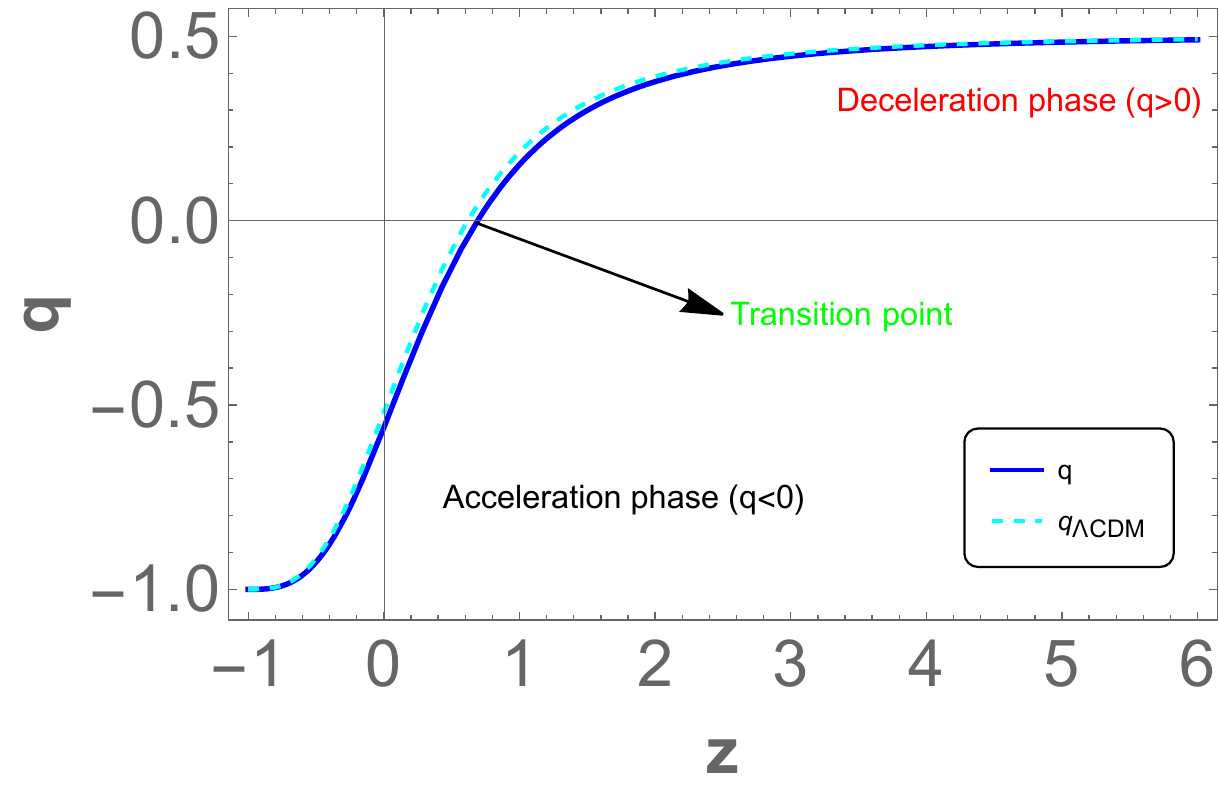}
    \caption{Evolution of Hubble and deceleration parameters for model \ref{P_1}.} \label{h(z)q(z)m1}
\end{figure}
\begin{figure}[H]
    \centering
\includegraphics[width=60mm]{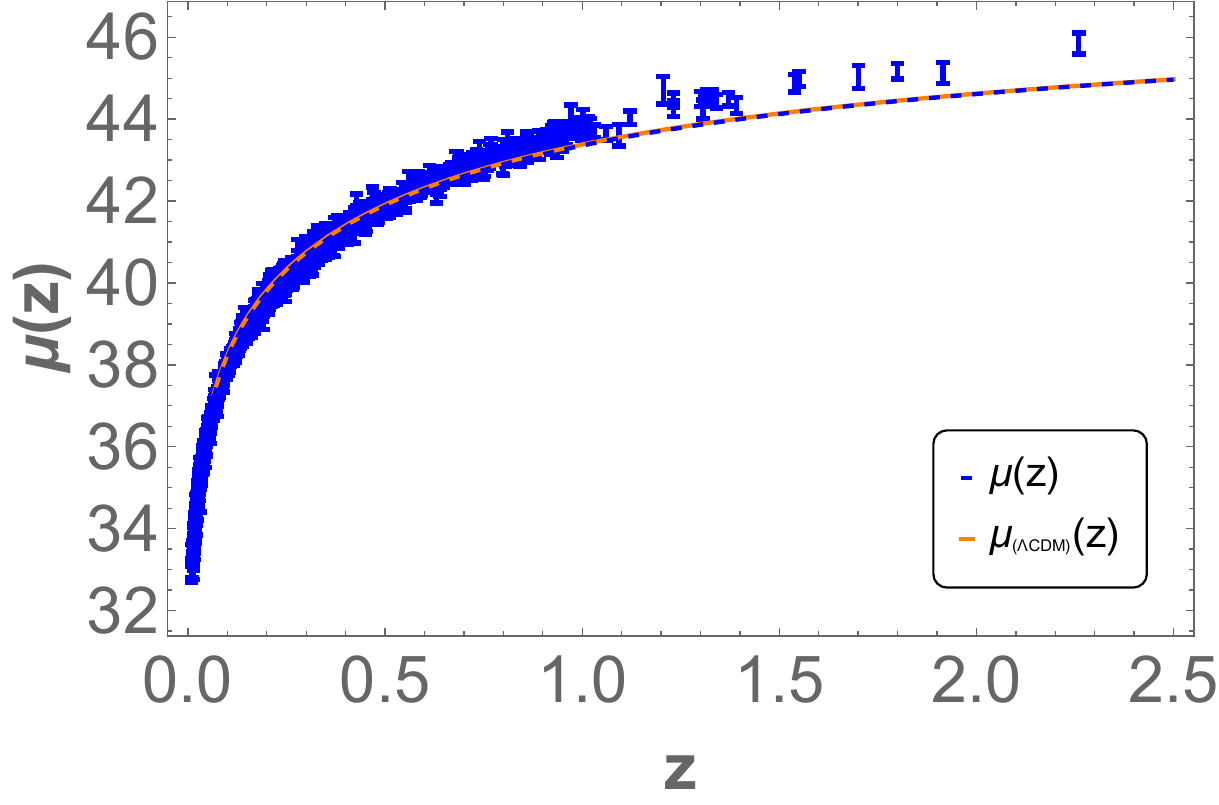}\caption{Plot of the observed distance modulus function $\mu(z)$ and the predicted $\Lambda$CDM model distance modulus function $\mu_{\Lambda CDM}(z)$ for model \ref{P_1}.} \label{mu(z)m1}
\end{figure}

The plot shows the contribution amount of DM $\Omega_{m}\approx 0.3$ and DE $\Omega_{DE}\approx 0.7$ density parameters. The time of matter-radiation equality is around $z\approx 3387$ and is denoted with a pointed arrow in Fig.  \ref{Eosdensitym1}. Fig.  \ref{h(z)q(z)m1} illustrates the Hubble rate evolution and the Hubble data points \cite{Moresco_2022_25}, with $H_{0}=70$ Km/(Mpc sec) \cite{Planck:2018vyg}, showing that the model closely aligns with the standard $\Lambda$CDM model. In Fig.  \ref{h(z)q(z)m1}, the behavior of the deceleration parameter is examined; it shows transient behavior at $z\approx 0.66$ and is consistent \cite{PhysRevD.90.044016a}. At the present time, the value of the deceleration parameter is $q\approx -0.53$ \cite{PhysRevResearch.2.013028}. Fig.  \ref{mu(z)m1} illustrates the evolution of the modulus function $\mu(z)$, showing that the model curve aligns well with the $\Lambda$CDM model modulus function $\mu_{\Lambda CDM}$ including data from 1048 Supernovae Ia (SNe Ia).

\begin{figure}[H]
    \centering
\includegraphics[width=60mm]{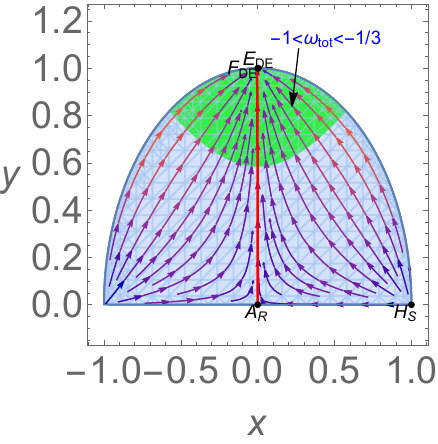}\caption{2D phase space for model \ref{P_1}.} \label{2dm1}
\end{figure}

In the 2D phase space portrait shown in Figure \ref{2dm1}, where the region is plotted by using the conditions $0<\Omega_{m}\leq1$ and $y>0$. The green-colored shaded region corresponds to the accelerating expansion of the Universe. Specifically it describes the quintessence behavior ($-1<\omega_{tot}<-\frac{1}{3}$), with the stable critical points $E_{DE}$ and $F_{DE}$ lying within this region. These critical points characterize the late-time cosmic acceleration phase of the Universe.
The dynamical variables $x$ and $y$ are used to plot the phase portrait. The critical points $A_{R}$ and $E_{DE}$ are connected by the trajectory showing a red line. The phase space trajectories indicate that the solution transitions from a saddle critical point ($A_{R}$) to a stable critical point ($E_{DE}, F_{DE}$). The behavior of the trajectories suggests that the critical points $A_{R}$, $H_{S}$, $D_{M}$, and $B_{R}$ exhibit saddle behavior, while critical points $E_{DE}$ and $F_{DE}$ demonstrate stable behavior.  The initial conditions and the values of the model parameters are consistent with those of Figure \ref{Eosdensitym1}.  In Fig. \ref{2dm1}, the accelerating (quintessence) phase of the Universe, $-1 < \omega_{tot} < -\frac{1}{3}$, is highlighted in the green/shaded region.

\subsection{Potential \texorpdfstring{$P_2: Cosh(\xi \phi)-1$}{}}\label{P_2}
This potential plays an important role in studying the dynamics of the phantom and the quintessence DE, DM models \cite{Roy_2018,Sahni_2000}. The critical points are presented in Table \ref{P2criticalpoints}. In this case, from Table \ref{Potentialfunctions}, the value of $\alpha_1=-\frac{1}{2}$ and $\alpha_3=\frac{\xi^2}{2}$, hence we will analyse this potential with the autonomous dynamical system having five independent variables. 
\begin{table}[H]
     % title of Table
    \centering % used for centering table
    \scalebox{0.85}{\begin{tabular}{|c |c |c |c| c| c| c|} % centered columns (5 columns)
    \hline\hline %inserts double horizontal lines
    \parbox[c][0.9cm]{2.4cm}{Critical points}& $ \{ x, \, y, \, u, \, \rho, \, \lambda \} $ & $q$ &  \textbf{$\omega_{tot}$}& $\Omega_{r}$& $\Omega_{m}$& $\Omega_{DE}$\\ [0.5ex] % inserts table %headin$g$
    \hline\hline % inserts single horizontal line
    \parbox[c][1.3cm]{1.3cm}{$a_{R}$ } &$\{ 0, 0, 0,  1,  \xi \}$ & $1$ &  $\frac{1}{3}$& $1$& $0$ & $0$ \\
    \hline
    \parbox[c][1.3cm]{1.3cm}{$b_{R}$ } & $\left\{\frac{2 \sqrt{\frac{2}{3}}}{\xi },\frac{2}{\sqrt{3} \xi },0,\frac{\sqrt{\xi ^2-4}}{\xi },\xi \right\}$ & $1$&  $\frac{1}{3}$& $1-\frac{4}{\xi ^2}$ & $0$ & $\frac{4}{\xi ^2}$ \\
    \hline
   \parbox[c][1.3cm]{1.3cm}{$c_{M}$ } &  $\{0,\, \, 0,\, \, 0, \, \, 0, \, \, \xi\}$ & $\frac{1}{2}$ &  $0$& $0$ & $1$ & $0$\\
   \hline
   \parbox[c][1.3cm]{1.3cm}{$d_{M}$} &   $\left\{\frac{\sqrt{\frac{3}{2}}}{\xi },\frac{\sqrt{\frac{3}{2}}}{\xi },0,0,\xi\right\}$ & $\frac{1}{2}$ &  $0$& $0$ & $1-\frac{3}{\xi ^2}$ & $\frac{3}{\xi ^2}$\\
   \hline
 \parbox[c][1.3cm]{1.3cm}{$e_{DE}$} &  $\left\{\frac{\xi }{\sqrt{6}},\sqrt{1-\frac{\xi ^2}{6}},0,0,\xi\right\}$ & $-1+\frac{\xi ^2}{2}$ & $-1+\frac{\xi ^2}{3}$ & $0$ & $0$ & $1$\\
 \hline
 \parbox[c][1.3cm]{1.3cm}{$f_{DE}$} &  $\left\{0,1,\frac{\xi }{3},0,\xi\right\}$ & $-1$ &  $-1$ & $0$ & $0$ & $1$\\
 \hline
 \parbox[c][1.3cm]{1.3cm}{$g_{DE}$} &  $\left\{ 0,1,0,0,0\right\}$ & $-1$ &  $-1$ & $0$ & $0$ & $1$\\
 \hline
 \parbox[c][1.3cm]{1.3cm}{$h_{S}$} &  $\left\{1,0,0,0,\xi\right\}$ & $2$ &  $1$ & $0$ & $0$ & $1$\\
 \hline
    \end{tabular}}
    \caption{Critical points analysis: values of $q$,\, $\omega_{tot}$,\, $\Omega_{r}$,\, $\Omega_{m}$ 
 and $\Omega_{DE}$ for potential \ref{P_2} $(P_2)$.}
    % is used to refer to this table in the text
  \label{P2criticalpoints}
\end{table}
 In this potential function, one thing can be noticed in most of the critical points, the dynamical variable $\lambda$ equals the potential parameter $\xi$. The detailed analysis of the critical point is presented as follows,
\begin{itemize}
\item{\textbf{Radiation-dominated critical points} $a_R, b_R$}: \\
The critical point $a_{R}$ is a standard radiation-dominated critical point with $\Omega_{r}=1$. This critical point describes the early phase of the evolution of the Universe with $\omega_{tot}=\frac{1}{3}$. The standard density parameter for the radiation $\Omega_{r}$ takes the value 1 at this critical point. The critical point $b_{R}$ describes the non-standard radiation-dominated era with $\Omega_{r}=1-\frac{4}{\xi^2}$. This critical point is a scaling solution, and the physical condition on the density parameter $0<\Omega_{DE}<1$ applies the condition on parameter $\xi$, such that $|\xi>2|$. These critical points are considered in \cite{Gonzalez-Espinoza:2020jss}, but are not discussed in \cite{copelandLiddle, Xu_2012}.

\item{\textbf{Matter-dominated critical points} $c_M, d_M$}: \\
The critical point $c_{M}$ is the standard matter dominated critical point with $\Omega_{M}=1$ at which the value of $q=\frac{1}{2}$ and $\omega_{tot}=0$. Another critical point representing the cold DM-dominated era in this case is $d_M$. This critical point is the scaling matter-dominated solution with $\Omega_{DE}=\frac{3}{\xi^2}$. The condition on $0<\Omega_{DE}<1$, applies the condition on the parameter $\xi$ as $|\xi|>\sqrt{3}.$ The critical point $d_{M}$ appeared in the study made in the standard quintessence model and the teleparallel DE model \cite{Xu_2012,copelandLiddle}.

\item{\textbf{DE-dominated critical points} $e_{DE}, f_{DE}, g_{DE}$}: \\
The critical point $e_{DE}$ is the critical point representing the DE-dominated era of the evolution of the Universe. The value of $\omega_{tot}=-1+\frac{\xi^2}{3}$. This critical point describes the late time cosmic acceleration at $-\sqrt{2}<\xi <0\lor 0<\xi <\sqrt{2}$ and a late-time attractor solution. The points $f_{DE}, g_{DE}$ are the cosmological constant solutions with the value of $\omega_{tot}=-1$. So, $e_{DE}$, $f_{DE}$ and $g_{DE}$ represent the standard DE era of the Universe evolution with $\Omega_{DE}=1$.

 \item{\textbf{Critical point representing the stiff DE $h_{S}$:}} 
 This critical point describes the stiff matter-dominated era. During the stiff matter-dominated era, the Universe would expand and cool rapidly. The energy density of stiff matter decreases rapidly with the expansion compared to other forms of matter or radiation.  
\end{itemize}

\begin{figure}[H]
 \centering
  \includegraphics[width=60mm]{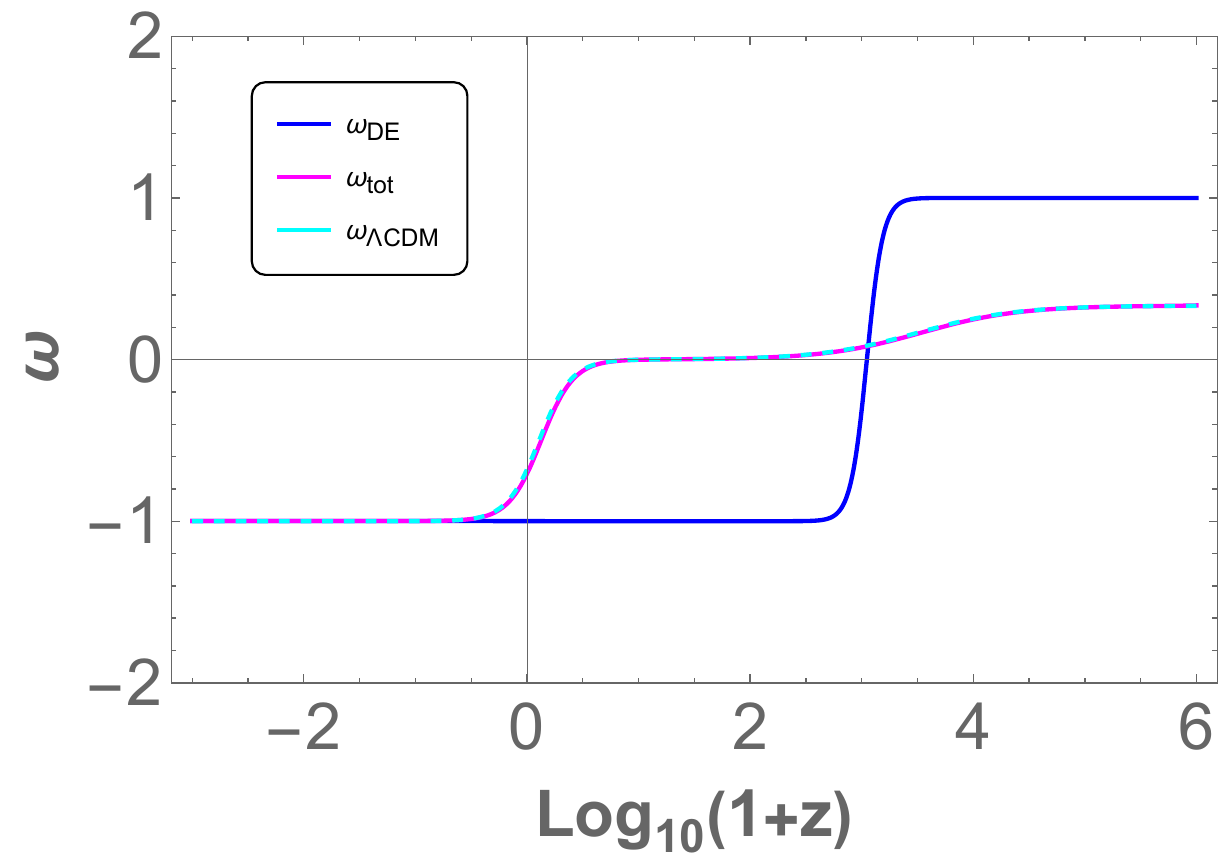}
    \includegraphics[width=60mm]{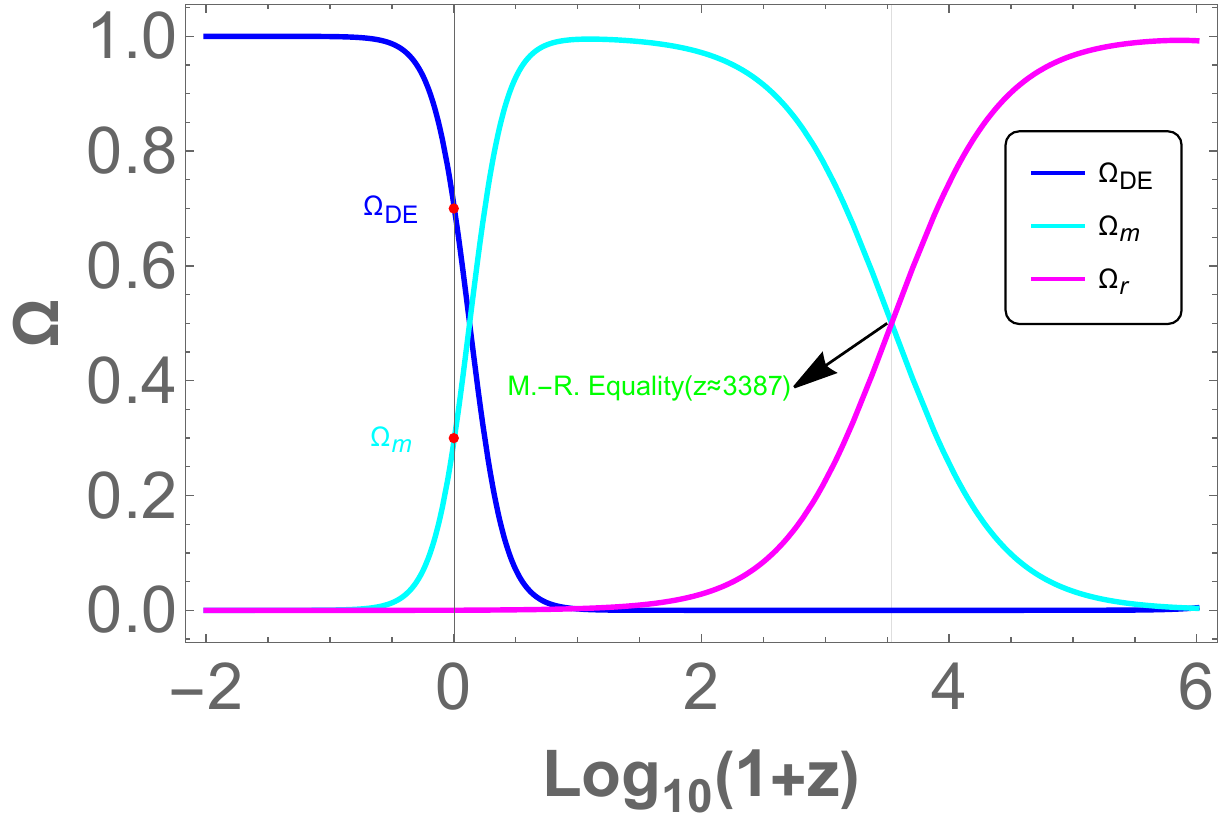}
    \caption{Evolution of EoS and standard density parameters for model \ref{P_2}.} \label{Eosdensitym2}
\end{figure}

\begin{figure}[H]
 \centering
  \includegraphics[width=60mm]{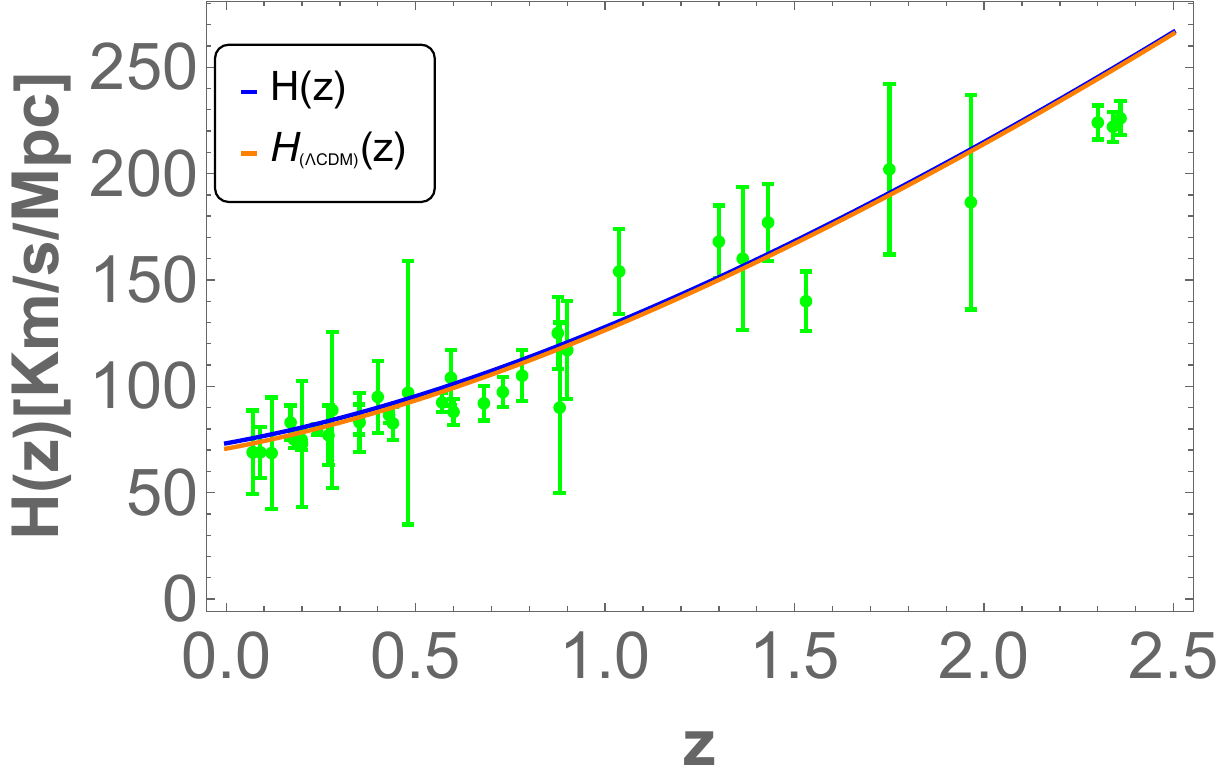}
    \includegraphics[width=60mm]{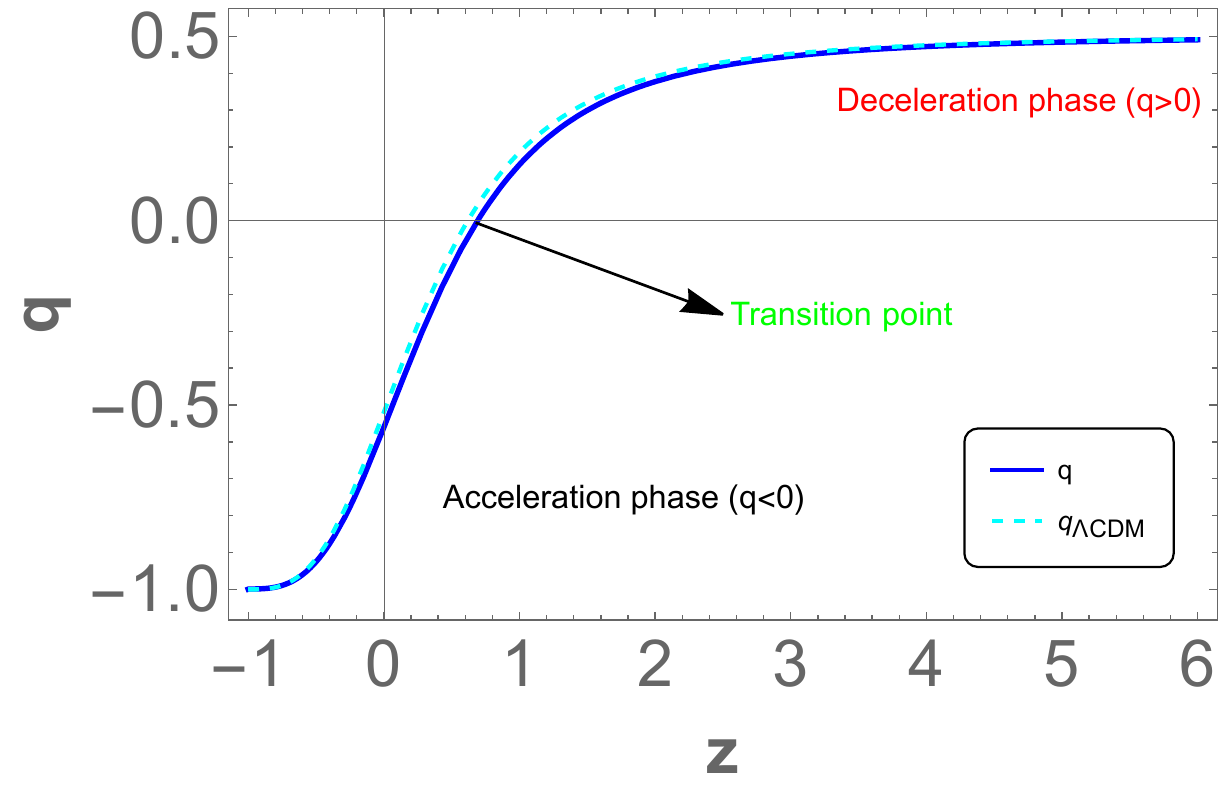}
    \caption{Evolution of Hubble and deceleration parameters for model \ref{P_2}.} \label{h(z)q(z)m2}
\end{figure}

\begin{figure}[H]
    \centering
\includegraphics[width=60mm]{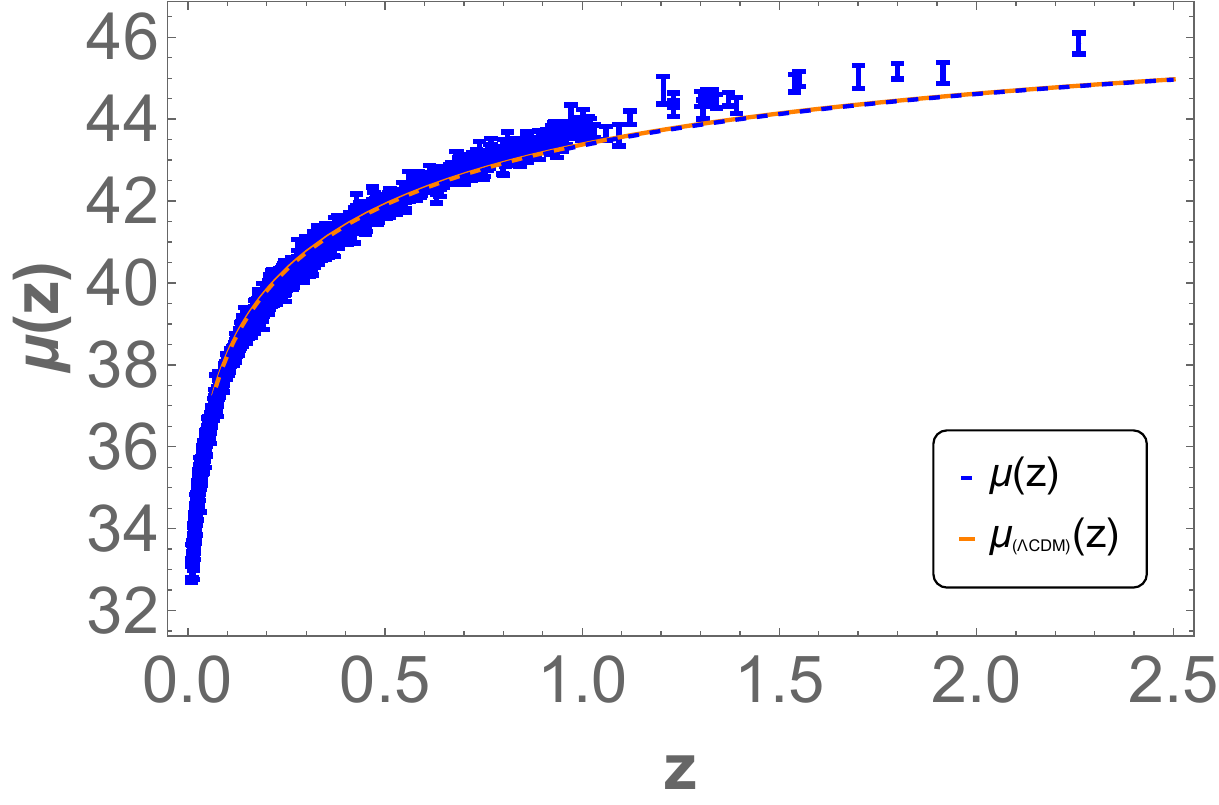}\caption{Plot of the observed distance modulus function $\mu(z)$ and the predicted $\Lambda$CDM model distance modulus function $\mu_{\Lambda CDM}(z)$ for Model \ref{P_2}.} \label{mu(z)m2}
\end{figure}

\textbf{The eigenvalues and the stability conditions:}
\begin{itemize}
\item \textbf{Stability of critical points} $a_{R}, b_{R}:$ \\
The eigenvalues at this critical point are
$\Bigl\{\nu_{1}=2,\, \nu_{2}=-1,\, \nu_{3}=1,\, \nu_{4}=0,\,\nu_{5}=0 \Bigl\}$. The presence of both positive and negative eigenvalues implies the saddle behaviour. The eigenvalues at $b_{R}$ are $\Bigl\{\nu_{1}=-\frac{4 \alpha }{\xi },\, \nu_{2}=1, \, \nu_{3}=4, \, \nu_{4}=-\frac{\sqrt{64 \xi ^4-15 \xi ^6}}{2 \xi ^3}-\frac{1}{2}, \, \nu_{5}=\frac{\sqrt{64 \xi ^4-15 \xi ^6}}{2 \xi ^3}-\frac{1}{2}\Bigl\}$. All the eigenvalues will not take the positive value; this critical point is not an unstable critical point and always shows the saddle point behaviour at $\Bigl\{\alpha <0\land -2<\xi <0\Bigl\}$. 

\item \textbf{Stability of critical points} $c_{M}, d_{M}:$ \\
The eigenvalues at the point $c_{M}$ are $\Bigl\{\nu_{1}=-\frac{3}{2},\,\nu_{2}=\frac{3}{2},\,\nu_{3}=-\frac{1}{2},\,\nu_{4}=0,\,\nu_{5}=0,\,\Bigl\}$, shows the saddle characteristics. The corresponding eigenvalues can be written as, $\Bigl\{\nu_{1}=-\frac{3 \alpha }{\xi },\,\nu_{2}=-\frac{1}{2},\,\nu_{3}=3,\,\nu_{4}=-\frac{3 \left(\xi ^3+\sqrt{24 \xi ^4-7 \xi ^6}\right)}{4 \xi ^3},\,\nu_{5}=\frac{3 \sqrt{24 \xi ^4-7 \xi ^6}}{4 \xi ^3}-\frac{3}{4}\Bigl\}$. This critical point is saddle at $\Bigl\{\alpha <0\land -\sqrt{3}<\xi <0\Bigl\}$.

\item \textbf{Stability of critical points} $e_{DE}, f_{DE}, g_{DE}:$ \\
The eigenvalues of critical point $e_{DE}$ are $\Bigl\{\nu_{1}=-\alpha  \xi,\, \nu_{2}=\xi ^2,\, \nu_{3}=\frac{1}{2} \left(\xi ^2-6\right),\, \nu_{4}=\frac{1}{2} \left(\xi ^2-4\right),\, \nu_{5}= ,\xi ^2-3\Bigl\}$ and it shows stability at $\left(\alpha <0\land -\sqrt{3}<\xi <0\right)\lor \left(\alpha >0\land 0<\xi <\sqrt{3}\right)$ and is saddle at$\left(\alpha <0\land 0<\xi <\sqrt{3}\right)\lor \left(\alpha >0\land -\sqrt{3}<\xi <0\right)$ and is unstable at $\Big(\alpha <0\land \xi >\sqrt{6}\Big)\lor \left(\alpha >0\land \xi <-\sqrt{6}\right)$.\\ The eigenvalues at the critical point $f_{DE}$ are, $\Bigl\{\nu_{1}=0,\, \nu_{2}=-3,\, \nu_{3}=-2,\,$ \\ $ \nu_{4}=-\frac{3 \left(\sqrt{\left(\xi ^2+6\right) \left(8 \alpha  \xi +\xi ^2+6\right)}+\xi ^2+6\right)}{2 \left(\xi ^2+6\right)},\,\nu_{5}= \frac{3}{2} \left(\frac{\sqrt{\left(\xi ^2+6\right) \left(8 \alpha  \xi +\xi ^2+6\right)}}{\xi ^2+6}-1\right)\Bigl\}$. We have obtained the stability at this critical point using the CMT, and this critical point is stable at $(\lambda<0\land \xi >0)\lor (\lambda>0\land \xi >0)$ appendix.

The eigenvalues at critical point $g_{DE}$ are $\Bigl\{\nu_{1}=-3,\, \nu_{2}=-2,\, \nu_{3}=0,\, \nu_{4}=\frac{1}{2} \left(-\sqrt{9-6 \xi ^2}-3\right),\, $ \\ $\nu_{5}=\frac{1}{2} \left(\sqrt{9-6 \xi ^2}-3\right) \Bigl\}$. According to the CMT, this critical point shows stable behaviour for the $\alpha>0$ appendix. \\
\item \textbf{Stability of critical point} $h_{S}:$ \\
The eigenvalues at the critical point $h_{S}$ are $\Bigl\{\nu_{1}=3,\, \nu_{2}=1,\, \nu_{3}=-\sqrt{6} \alpha,\, \nu_{4}=\sqrt{6} \xi,\, \nu_{5}= 3-\sqrt{\frac{3}{2}} \xi\Bigl\}$. This critical point is saddle at $\Bigl\{\alpha >0\land \xi >\sqrt{6}\Bigl\}$ and is unstable at $\Bigl\{\alpha <0\land \xi <\sqrt{6}\Bigl\}$. 
\end{itemize}
\textbf{Numerical results:}
In Fig.  \ref{Eosdensitym2}, we can observe the behavior of the DE and total EoS parameter and the $\Lambda$CDM model. This figure demonstrates the dominance of radiation in the early epoch, followed by a gradual decrease in this dominance and the rise of the DM-dominated era. At the tail end of this sequence, we can observe the present accelerating expansion of the Universe era, with the EoS parameter approaching a value of -1 at late times. Fig.  \ref{Eosdensitym2} also illustrates the EoS parameter $\omega_{tot}$, which starts at $\frac{1}{3}$ for radiation, decreases to 0 during the matter-dominated period and ultimately reaches -1. Both $\omega_{\Lambda CDM}$ and $\omega_{DE}$ (blue) approach -1 at late times, with the value of $\omega_{DE}$ being -1 at present. Additionally, at present, the density parameters are approximately $\Omega_{m}\approx 0.3$ for DM and $\Omega_{DE}\approx 0.7$ for DE. One thing can be noted the matter-radiation equality observed at $z\approx 3387$ in Fig.  \ref{Eosdensitym2}. Fig.  \ref{h(z)q(z)m2} displays the Hubble rate behaviour as a function of redshift $z$, showing that the model closely resembles the standard $\Lambda$CDM model. In Fig.  \ref{h(z)q(z)m2}, the transition from deceleration to acceleration occurs at $z\approx 0.65$, and the present value of the deceleration parameter is noted as approximately -0.56. Furthermore, Fig.  \ref{mu(z)m2} presents the modulus function $\mu_{\Lambda CDM}$ for the $\Lambda$CDM model, along with 1048 pantheon data points and the modulus function $\mu(z)$.\\
In this case, the initial conditions are: $x_C=10^{-8.89} ,\,y_C=10^{-2.89} ,\,u_C=10^{-5.96} ,\,\rho_C=10^{-0.75}, \lambda_{c}=10^{-1.3}, \, \alpha=-5.2, \, \xi= -0.02$.

\begin{figure}[H]
    \centering
\includegraphics[width=60mm]{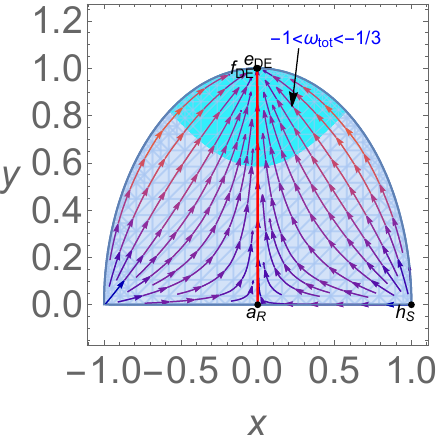}\caption{ 2D phase space for model \ref{P_2}.} \label{2dm2}
\end{figure}

In the 2D phase space portrait represented in Fig.  \ref{2dm2}, the dynamical variables $x$ and $y$ are plotted. The phase space trajectories demonstrate the behavior of critical points $h_{S}$, $a_{R}$, $d_{M}$, and $b_{R}$ as saddle critical points, while the critical points $e_{DE}$ and $f_{DE}$ exhibit stable behavior. These stable critical points are situated within the region of the accelerating expansion phase of the Universe (quintessence) ($-1<\omega_{tot}<-\frac{1}{3}$), which is highlighted by the cyan color in the upper part of the phase-portrait region.

\subsection{Potential \texorpdfstring{$P_3: V_{0} Sinh^{-\eta}(\beta \phi)$}{}}\label{P_3}
This form of potential plays an important role in analysing the generalise form teleparallel tachyonic model \cite{Bahamonde_2019} and to analyse the value of cosmological constant $\Lambda$ using type Ia supernovae \cite{SAHNI2000}. Here we consider this potential form to check its physical viability to discuss the evolutionary epochs of the Universe in the generalized teleparallel scalar tensor formalism. The critical points corresponding to this potential has been presented in the following Table \ref{P3criticalpoints}. In this case, the value of the $\alpha_{1}=\frac{1}{\eta}$ and $\alpha_{2}=-\eta \beta^2$, as presented in Table \ref{Potentialfunctions}, hence the autonomous dynamical system will be of five-dimensions. The detailed analysis of the critical points is presented as follows,
\begin{itemize}
 \item{\textbf{Radiation-dominated critical points} $\mathcal{A}_R, \mathcal{B}_R$}: \\
The critical point $\mathcal{A}_{R}$ is describing the standard radiation-dominated era with $\Omega_{r}=1$, At this critical point, the value of $q=1, \, \omega_{tot}=\frac{1}{3}$. Point $\mathcal{B}_{R}$ describes the non-standard radiation-dominated era with the small contribution of the DE. It is a scaling solution and the physical viability condition on the parameter $0<\Omega_{DE}<1$ applies the condition on the parameters $\beta, \eta$ as $\Bigl\{\beta <0\land \left(\eta <-2 \sqrt{\frac{1}{\beta ^2}}\lor \eta >2 \sqrt{\frac{1}{\beta ^2}}\right)\Big)$ or $\Big(\beta >0\land \left(\eta <-2 \sqrt{\frac{1}{\beta ^2}}\lor \eta >2 \sqrt{\frac{1}{\beta ^2}}\right)\Bigl\}$. This critical point is appeared points are analysed in \cite{Gonzalez-Espinoza:2020jss}.

\item{\textbf{Matter-dominated critical points} $\mathcal{C}_M, \mathcal{D}_M$}: \\
Both of these critical points are in the dark-matter dominated era with $q=\frac{1}{2}, \omega_{tot}=0$. The point $\mathcal{C}_M$ describes the standard matter-dominated era with $\Omega_{m}=1$. The critical point $D_{M}$ represents the scaling matter-dominated solution with $\Omega_{DE}=\frac{3}{\beta^2 \eta^2}$. The physically viable condition applies the condition on model parameters, $\beta, \, \eta$ as $\Bigl\{\beta <0\land \left(\eta <-\sqrt{3} \sqrt{\frac{1}{\beta ^2}}\lor \eta >\sqrt{3} \sqrt{\frac{1}{\beta ^2}}\right)\Bigl\} $ or $ \Bigl\{\beta >0\land \left(\eta <-\sqrt{3} \sqrt{\frac{1}{\beta ^2}}\lor \eta >\sqrt{3} \sqrt{\frac{1}{\beta ^2}}\right)\Bigl\}$.\\ This matter-dominated critical point $\mathcal{D}_{M}$ appeared in the study made in the standard quintessence model and the teleparallel DE model \cite{Xu_2012,copelandLiddle}.

\item{\textbf{DE-dominated critical points} $\mathcal{E}_{DE}, \mathcal{F}_{DE}, \mathcal{G}_{DE}$}: \\
The critical point $\mathcal{E}_{DE}$ is a DE-dominated critical point with $\Omega_{DE}=1$ and is a late-time scaling solution at which $\omega_{tot}=-1+\frac{\beta^2 \eta^2}{3}$. At this point, $\omega_{tot}$ lies in the quintessence region for\\
$\eta \in \mathbb{R}\land \Bigl\{\left(\beta <0\land -\sqrt{2} \sqrt{\frac{1}{\beta ^2}}<\eta <\sqrt{2} \sqrt{\frac{1}{\beta ^2}}\right)\lor \beta =0\lor \left(\beta >0\land -\sqrt{2} \sqrt{\frac{1}{\beta ^2}}<\eta <\sqrt{2} \sqrt{\frac{1}{\beta ^2}}\right)\Bigl\}$.\\
Points $\mathcal{F}_{DE}$ and $G_{DE}$ behave like cosmological constant. These are the de Sitter solutions and represent the standard DE-dominated era of the evolution of the Universe with $\Omega_{DE}=1$.  
 \item{\textbf{Critical point representing the stiff matter $\mathcal{H}_S$:}} \\
 The critical point $\mathcal{H}_{S}$ with $\omega_{tot}=1$ describes the stiff matter-dominated era. Although the value of $\Omega_{DE}=1$, this critical point can not describe the DE-dominated era of the evolution of the Universe. This critical point is also analysed in the \cite{Gonzalez_Espinoza_2020}. 
\end{itemize}
\begin{table}[H]
     % title of Table
    \centering % used for centering table
    \scalebox{0.9}{
    \begin{tabular}{|c |c |c |c| c| c| c|} % centered columns (5 columns)
    \hline\hline %inserts double horizontal lines
    \parbox[c][0.9cm]{2.4cm}{Critical points
    }& $ \{ x, \, y, \, u, \, \rho, \, \lambda \} $ & $q$ &  \textbf{$\omega_{tot}$}& $\Omega_{r}$& $\Omega_{m}$& $\Omega_{DE}$\\ [0.5ex] % inserts table %headin$g$
    \hline\hline % inserts single horizontal line
    \parbox[c][1.3cm]{1.3cm}{$\mathcal{A}_{R}$ } &$\{ 0, 0, 0,  1,  \lambda \}$ & $1$ &  $\frac{1}{3}$& $1$& $0$ & $0$ \\
    \hline
    \parbox[c][1.3cm]{1.3cm}{$\mathcal{B}_{R}$ } & $\left\{\frac{2 \sqrt{\frac{2}{3}}}{\beta  \eta },\frac{2}{\sqrt{3} \beta  \eta },0,\frac{\sqrt{\beta ^2 \eta ^2-4}}{\beta  \eta },\beta  \eta\right\}$ & $1$&  $\frac{1}{3}$& $1-\frac{4}{\beta ^2 \eta ^2}$ & $0$ & $\frac{4}{\beta ^2 \eta ^2}$ \\
    \hline
   \parbox[c][1.3cm]{1.3cm}{$\mathcal{C}_{M}$ } &  $\{0,\, \, 0,\, \, 0, \, \, 0, \, \, \lambda\}$ & $\frac{1}{2}$ &  $0$& $0$ & $1$ & $0$\\
   \hline
   \parbox[c][1.3cm]{1.3cm}{$\mathcal{D}_{M}$} &   $\left\{\frac{\sqrt{\frac{3}{2}}}{\beta  \eta },\frac{\sqrt{\frac{3}{2}}}{\beta  \eta },0,0,\beta  \eta\right\}$ & $\frac{1}{2}$ &  $0$& $0$ & $1-\frac{3}{\beta ^2 \eta ^2}$ & $\frac{3}{\beta ^2 \eta ^2}$\\
   \hline
 \parbox[c][1.3cm]{1.3cm}{$\mathcal{E}_{DE}$} &  $\left\{\frac{\beta  \eta }{\sqrt{6}},\sqrt{1-\frac{\beta ^2 \eta ^2}{6}},0,0,\beta  \eta\right\}$ & $-1+\frac{\beta ^2 \eta ^2}{2}$ & $-1+\frac{\beta ^2 \eta ^2}{3}$ & $0$ & $0$ & $1$\\
 \hline
 \parbox[c][1.3cm]{1.3cm}{$\mathcal{F}_{DE}$} &  $\left\{0,1,\frac{\beta\eta }{3},0,\beta\eta\right\}$ & $-1$ &  $-1$ & $0$ & $0$ & $1$\\
 \hline
 \parbox[c][1.3cm]{1.3cm}{$\mathcal{G}_{DE}$} &  $\left\{ 0,1,0,0,0\right\}$ & $-1$ &  $-1$ & $0$ & $0$ & $1$\\
 \hline
 \parbox[c][1.3cm]{1.3cm}{$\mathcal{H}_{S}$} &  $\left\{1,0,0,0,\beta  \eta\right\}$ & $2$ &  $1$ & $0$ & $0$ & $1$\\
 \hline
    \end{tabular}}
    \caption{Critical points analysis: values of $q$,\, $\omega_{tot}$,\, $\Omega_{r}$,\, $\Omega_{m}$ 
 and $\Omega_{DE}$ for potential \ref{P_3} $(P_3)$.}
    % is used to refer to this table in the text
    \label{P3criticalpoints}
\end{table}

\textbf{The eigenvalues and the stability conditions:}
\begin{itemize}
\item \textbf{Stability of critical points} $\mathcal{A}_{R}, \mathcal{B}_{R}:$ \\
The eigenvalues at $\mathcal{A}_{R}$ are
$\Bigl\{\nu_{1}=2,\, \nu_{2}=-1,\, \nu_{3}=1,\, \nu_{4}=0,\,\nu_{5}=0 \Bigl\}$. The presence of the plus and the minus signatures indicates that this critical is a saddle point. The eigenvalues at critical point $\mathcal{B}_{R}$ are
$\Bigl\{\nu_{1}=-\frac{4 \alpha }{\beta  \eta },\, \nu_{2}=-\frac{8}{\eta }, \nu_{3}=1,\, \nu_{4}=-\frac{\sqrt{64 \beta ^4 \eta ^4-15 \beta ^6 \eta ^6}}{2 \beta ^3 \eta ^3}-\frac{1}{2},\,\nu_{5}=\frac{\sqrt{64 \beta ^4 \eta ^4-15 \beta ^6 \eta ^6}}{2 \beta ^3 \eta ^3}-\frac{1}{2} \Bigl\}$ and is saddle at $\Bigl\{\alpha <0\land \beta >0\land -\frac{2}{\beta }<\eta <0\Bigl\}$.

\item \textbf{Stability of critical points} $\mathcal{C}_{M}, \mathcal{D}_{M}:$\\
The eigenvalues at the critical point $\mathcal{C}_{M}$ are $\Bigl\{\nu_{1}=-\frac{3}{2},\,\nu_{2}=\frac{3}{2},\,\nu_{3}=-\frac{1}{2},\,\nu_{4}=0,\,\nu_{5}=0\Bigl\}$, as there is a presence of positive and the negative eigenvalues, this is a saddle critical point. The eigenvalues at the critical point $\mathcal{D}_{M}$ are $\Bigl\{\nu_{1}=-\frac{3 \alpha }{\beta  \eta },\,\nu_{2}=-\frac{6}{\eta },\,\nu_{3}=-\frac{1}{2},\,\nu_{4}=-\frac{3 \sqrt{24 \beta ^4 \eta ^4-7 \beta ^6 \eta ^6}}{4 \beta ^3 \eta ^3}-\frac{3}{4},\,\nu_{5}=\frac{3 \sqrt{24 \beta ^4 \eta ^4-7 \beta ^6 \eta ^6}}{4 \beta ^3 \eta ^3}-\frac{3}{4}\Bigl\}$. This critical point shows stability within the range of the model parameters as $\alpha \in \mathbb{R}\land \alpha \neq 0$ and is showing saddle behaviour for $\alpha <0$.

\item \textbf{Stability of critical points} $\mathcal{E}_{DE}, \mathcal{F}_{DE}, \mathcal{G}_{DE}:$\\
The eigenvalues at the critical point $\mathcal{E}_{DE}$ are $\Bigl\{\nu_{1}=-\alpha  \beta  \eta,\,\nu_{2}=-2 \beta ^2 \eta ,\,\nu_{3}=\frac{\beta ^2 \eta ^2}{2}-3,\,\nu_{4}=\frac{\beta ^2 \eta ^2}{2}-2,\,\nu_{5}=\beta ^2 \eta ^2-3\Bigl\}$ 
and it shows stability at $\Bigl\{\alpha <0\land \beta <0\land 0<\eta <\sqrt{3} \sqrt{\frac{1}{\beta ^2}}\Bigl\}\lor \Bigl\{\alpha >0\land \beta >0\land 0<\eta <\sqrt{3} \sqrt{\frac{1}{\beta ^2}}\Bigl\}$,
saddle at $\Bigl\{\alpha <0\land \beta >0\land 0<\eta <\sqrt{3} \sqrt{\frac{1}{\beta ^2}}\Bigl\}\lor \Bigl\{\alpha >0\land \beta <0\land 0<\eta <\sqrt{3} \sqrt{\frac{1}{\beta ^2}}\Bigl\}$ and unstable at $\Bigl\{\alpha <0\land \beta <0\land \eta <-\sqrt{6} \sqrt{\frac{1}{\beta ^2}}\Bigl\}\lor \Bigl\{\alpha >0\land \beta >0\land \eta <-\sqrt{6} \sqrt{\frac{1}{\beta ^2}}\Bigl\}$.\\  The eigenvalues at $\mathcal{F}_{DE}$ are $\Bigl\{\nu_{1}=0, \nu_{2}=-3, \nu_{3}=-2, \nu_{4}=-\frac{3 \sqrt{\left(\beta ^2 \eta ^2+6\right) (\beta  \eta  (8 \alpha +\beta  \eta )+6)}}{2 \left(\beta ^2 \eta ^2+6\right)}-\frac{3}{2},\\ \nu_{5}=\frac{3}{2} \Bigl\{\frac{\sqrt{\left(\beta ^2 \eta ^2+6\right) (\beta  \eta  (8 \alpha +\beta  \eta )+6)}}{\beta ^2 \eta ^2+6}-1\Bigl\}\Bigl\}$. The stability at this critical point is analysed using CMT. This critical point shows stable behaviour within the range $\eta \in \mathbb{R}\land ((\lambda<0\land \beta <0)\lor (\lambda>0\land \beta <0))$, and the detailed formalism is presented in the appendix. At the critical point $\mathcal{G}_{DE}$ have the eigenvalues $\Bigl\{\nu_{1}=-3,\,\nu_{2}=-2,\,\nu_{3}=0,\,\nu_{4}=\frac{1}{2} \left(-\sqrt{12 \beta ^2 \eta +9}-3\right),\,\nu_{5}=\frac{1}{2} \left(\sqrt{12 \beta ^2 \eta +9}-3\right)\Bigl\}$. Based upon the CMT, this critical point shows stable behaviour for  $\alpha>0$, where $\dot{u}$ is negative., and the detailed formalism is presented in appendix.

 \item \textbf{Stability of critical point} $\mathcal{H}_{S}$:\\
 The eigenvalues at critical point $\mathcal{H}_{S}$ are $\Bigl\{\nu_{1}=3,\, \nu_{2}=1,\,\nu_{3}=\sqrt{6} \alpha ,\,\nu_{4}=-2 \sqrt{6} \beta ,\,\nu_{5}=3-\sqrt{\frac{3}{2}} \beta  \eta \Bigl\}$
       This critical point is saddle at $\Bigl\{\alpha >0\land \beta >0\land \eta >\frac{\sqrt{6}}{\beta }\Bigl\}$ and is unstable at $\Bigl\{\alpha <0\land \beta <0\land \eta >\frac{\sqrt{6}}{\beta }\Bigl\}$.   
\end{itemize}

\begin{figure}[H]
 \centering
  \includegraphics[width=60mm]{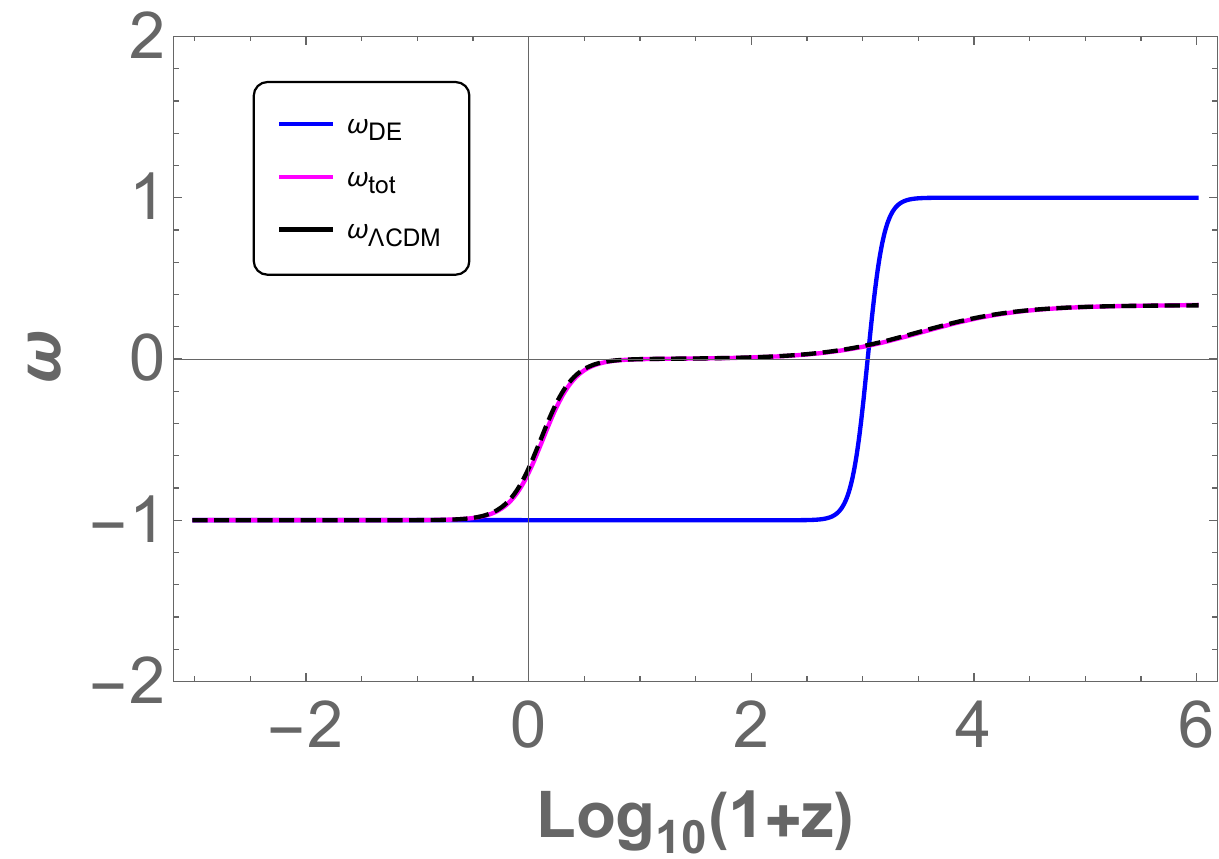}
    \includegraphics[width=60mm]{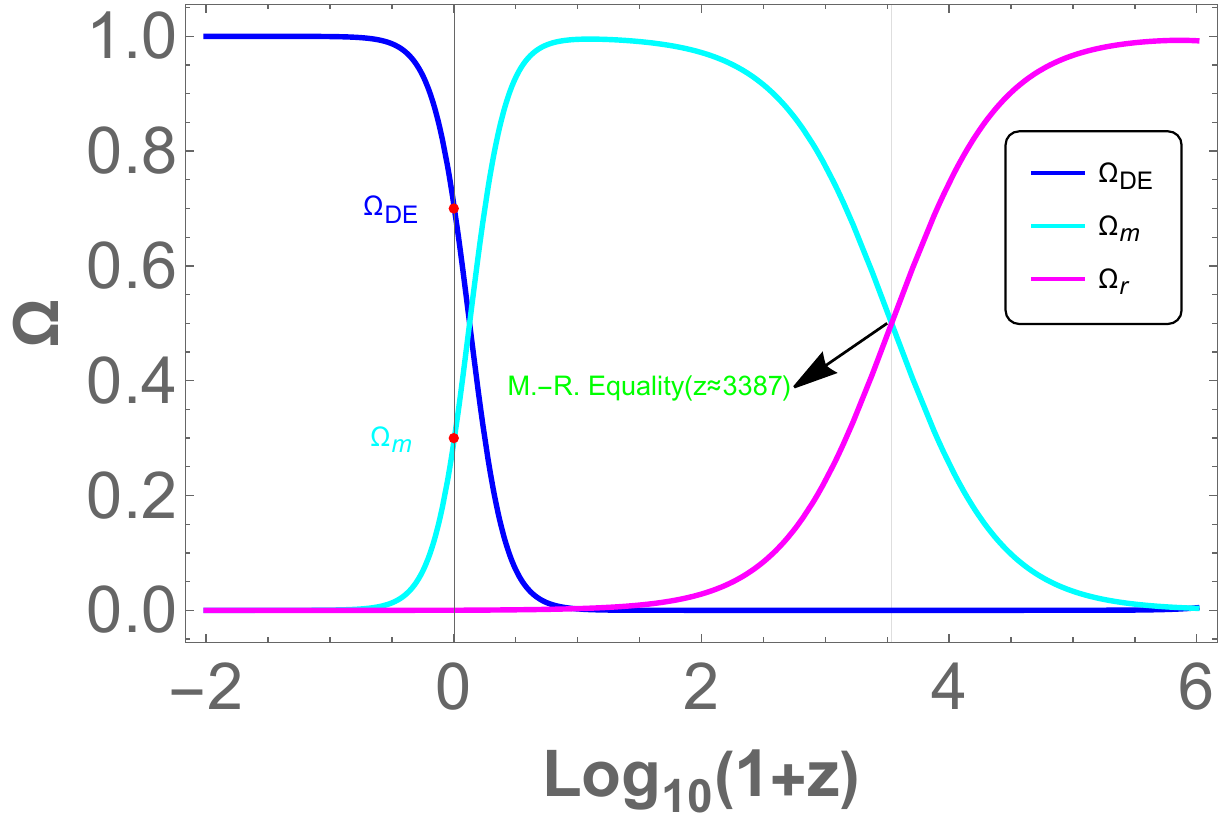}
    \caption{Evolution of EoS and standard density parameters for model \ref{P_3}.} \label{Eosdensitym3}
\end{figure}

\begin{figure}[H]
 \centering
  \includegraphics[width=60mm]{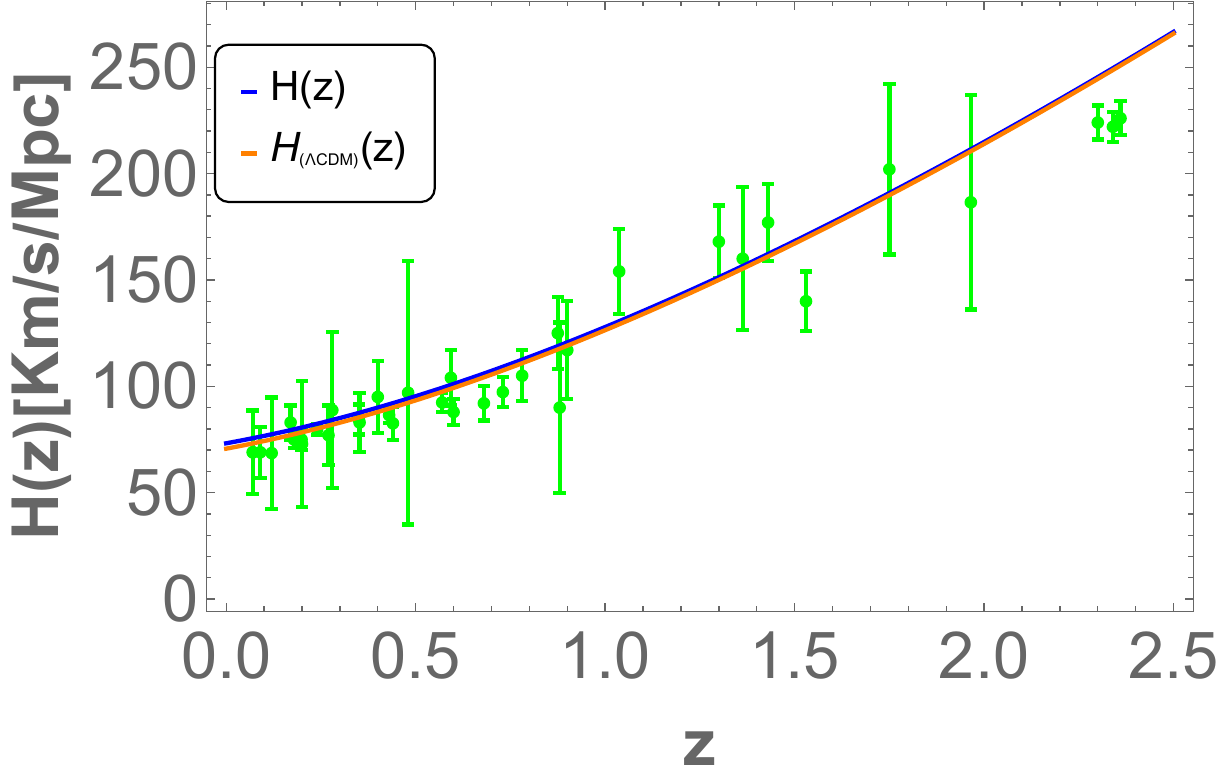}
    \includegraphics[width=60mm]{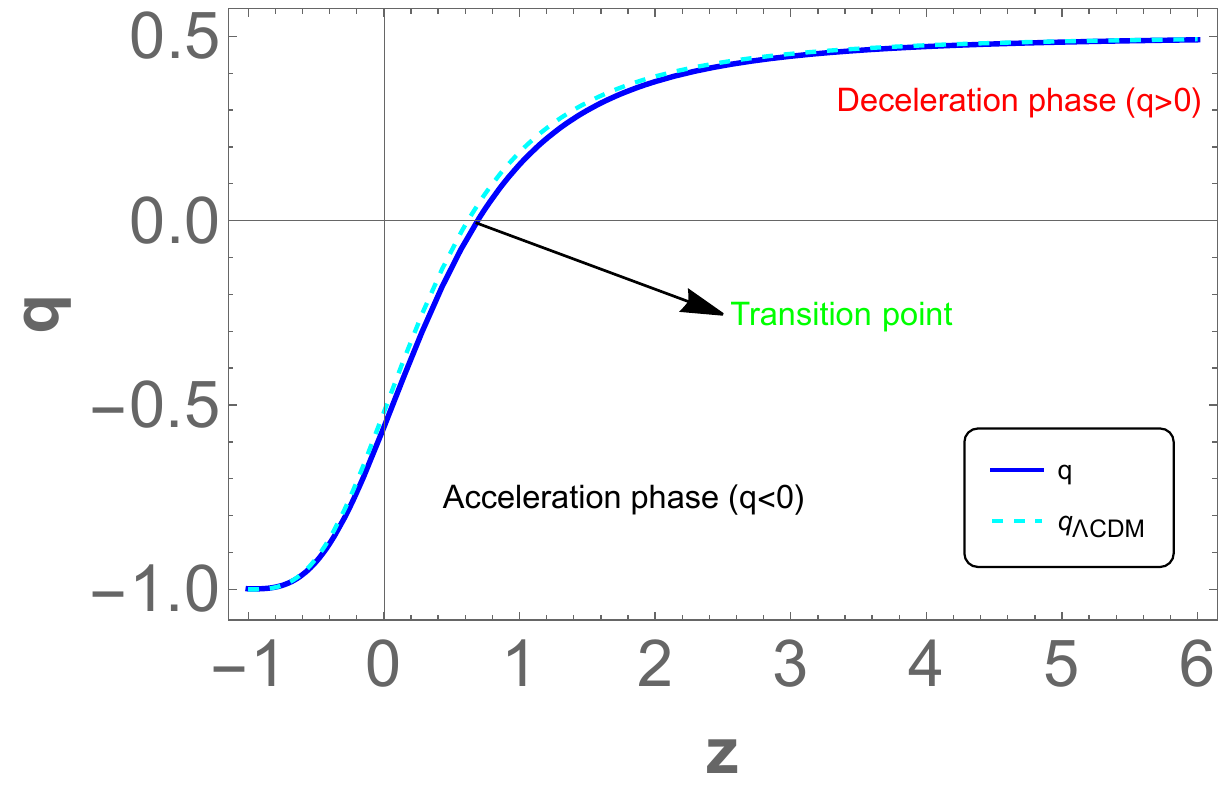}
    \caption{Evolution of Hubble and deceleration parameters for model \ref{P_3}.} \label{h(z)q(z)m3}
\end{figure}

\begin{figure}[H]
    \centering
\includegraphics[width=60mm]{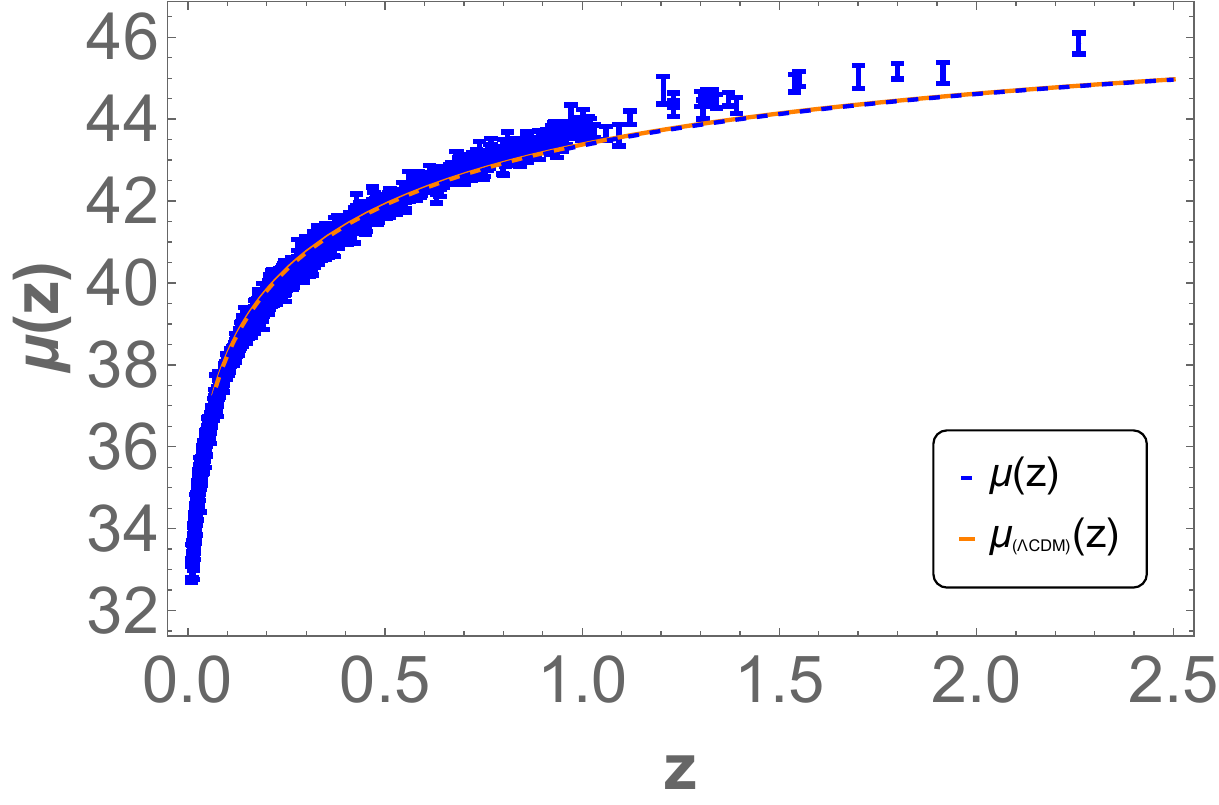}\caption{Plot of the observed distance modulus function $\mu(z)$ and the predicted $\Lambda$CDM model distance modulus function $\mu_{\Lambda CDM}(z)$ for model \ref{P_3}. } \label{mu(z)m3}
\end{figure}

\textbf{Numerical results:}\\
The EoS parameter and standard density parameters are illustrated in Fig.  \ref{Eosdensitym3}. These plots reveal  the sequence of the dominance of radiation in the early epoch, followed by the dominance of DM, and ultimately the dominance of the DE era at present and in the late stage of the evolution. Currently, the value of the standard DM density parameter $\Omega_{m}\approx 0.3$ and the standard DE density parameter $\Omega_{DE}\approx 0.7$. Their equality is observed at $z\approx 3387$ in Fig.  \ref{Eosdensitym3}. The Fig.  \ref{Eosdensitym3} depicts the behavior of the EoS parameter $\omega_{tot}$, which starts from $\frac{1}{3}$ for radiation, approaches 0 during the DM-dominated period, and ultimately tends to $-1$. Both $\omega_{\Lambda CDM}$ and $\omega_{DE}$ approach $-1$ at late times, with the current value of $\omega_{DE}$ being equal to $-1$ at $z=0$. The deceleration parameter $(q)$ and $q_{\Lambda CDM}$ depicted in Fig.  \ref{h(z)q(z)m3}, which describes the transition point for deceleration to acceleration, occurring at $z\approx 0.65$. The present value of the deceleration parameter is observed to be $\approx -0.57$. Additionally, Fig.  \ref{h(z)q(z)m3} displays the Hubble rate evolution as a function of redshift $z$, indicating close agreement between the proposed model and the standard $\Lambda$CDM model. Finally, Fig.  \ref{mu(z)m3} presents the $\Lambda$CDM model modulus function $\mu_{\Lambda CDM}$, 1048 pantheon data points, and the modulus function $\mu(z)$.

In this case the plots are plotted for the initial conditions are: $x_C=10^{-8.89} ,\,y_C=10^{-2.89} ,\,u_C=10^{-5.96} ,\,\rho_C=10^{-0.75}, \lambda_{c}=10^{-1.3}, \, \alpha=-5.2, \, \eta= -0.2, \, \beta=-0.21$.

\begin{figure}[H]
    \centering
\includegraphics[width=60mm]{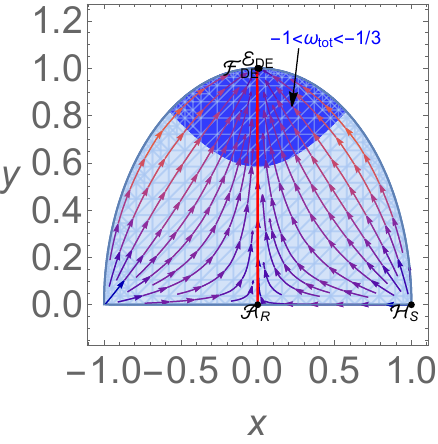}\caption{2D phase space for model \ref{P_3}.} \label{2dm3}
\end{figure}

In Fig.  \ref{2dm3}, we have depicted a 2D phase space diagram illustrating the phase space portrait of the dynamical variables $x$ and $y$. Analysis of trajectories in this plot reveals that the critical points $\mathcal{E}_{DE}$ and $\mathcal{F}_{DE}$ exhibit stable behavior, whereas the critical points $\mathcal{H}_{S}$, $\mathcal{A}_{R}$, $\mathcal{B}_{R}$, and $\mathcal{D}_{M}$ demonstrate saddle behavior. It is noteworthy that the critical points $\mathcal{E}_{DE}$ and $\mathcal{F}_{DE}$ are situated within the region corresponding to the accelerating expansion (quintessence) phase of the Universe, shaded by the blue region in the figure, where ($-1<\omega_{tot}<-\frac{1}{3}$).
\section{Conclusion}\label{Conclusion}
A key approach to understanding the background cosmology is to analyze the dynamical systems. Using this technique, we can identify the critical points of a cosmological viable model and their properties. In addition to being validated by observable physics and cosmology, these predictions can also be hinted at through the expanding Universe. Such tests could be performed on a modified gravity model. By linking critical point analysis with stability and phase space images, specific models or parameter ranges within the selected models can be validated or invalidated. In the present work, we have taken into consideration coupling function in the exponential form $g(\phi)=g_{0}e^{-\alpha \phi \kappa}$ to the teleparallel boundary term $B$ and three different forms of the potential function, which is presented in Table \ref{Potentialfunctions}. 
It is also required for the matter and radiation-dominated eras to occur before the DE period for any DE scenario to succeed. The matter and radiation solutions are explained within the context of dynamical systems as critical points of the autonomous system that are
 unstable for radiation,  or saddle for matter. We have discovered novel scaling solutions for the scaling radiation and matter epochs and the crucial points characterizing the standard radiation and matter eras for the DE model under study in this investigation. 

All three potential functions have stable critical points that explore the accelerated expansion phase of the Universe at a late-time. Scaling solutions have also been obtained for critical points. According to the critical points, non-standard matter and radiation-dominated phases in the Universe have been observed. From our results, we can see that the results match the quintessence model presented in \cite{Xu_2012,copelandLiddle}. We have obtained the matter and DE density parameters at present $z=0$ are found to be $\Omega_{m}^{0}\approx 0.3$ and $\Omega_{de}^{0}\approx 0.7$. Additionally, the matter-radiation equality value is found at $z\approx 3387$. For all three scenarios, the EoS parameter at present and late time have an approach to $-1$ with the present time values aligned with the recent observation \cite{Planck:2018vyg}.
Based on the behavior of Fig.  \ref{h(z)q(z)m1}, we can claim that for the exponential form of $V(\phi)$, the deceleration parameter displays the transition from deceleration to acceleration phase at $z=0.66$. Its current value is $q(z=0)=-0.53$. We obtained the transition point at $z=0.65$ for the $V(\phi)=Cosh(\xi \phi)-1$, and the current value of the deceleration parameter is $q(z=0)=-0.56$ [Fig.  \ref{h(z)q(z)m2}]. We find the transition point at $z=0.61$ for the $V(\phi)=V_{0}Sinh^{-\eta}(\beta \phi)$, and the current value of the deceleration parameter is $q(z=0)=-0.57$ can be observed from Fig.  \ref{h(z)q(z)m3}. In each of the three scenarios, the deceleration parameter value and transition point matched cosmological findings \cite{PhysRevD.90.044016a, PhysRevResearch.2.013028}. We compared our results in all three potential functions with the Hubble 31 data points \cite{Moresco_2022_25} and the Supernovae Ia data 1048 data points \cite{Scolnic_2018}. We may conclude that the outcomes of our model closely resemble those of the conventional $\Lambda$CDM model based on the behaviour of the Figs.  [\ref{h(z)q(z)m1}, \ref{h(z)q(z)m2}, \ref{h(z)q(z)m3}]. The modulus function of our models was shown alongside the 1048 Supernovae Ia data points in Fig.  [\ref{mu(z)m1}, \ref{mu(z)m2}, \ref{mu(z)m3}], using the conventional $\Lambda$CDM model modulus function. The outcomes closely align with the $\Lambda$CDM model. The quintessence region is shaded and is pointed using an arrow in the 2D phase space diagrams presented in Figs.  \ref{2dm1}, \ref{2dm2}, and 
\ref{2dm3}. The phase space trajectories move from the early-time decelerating phase to the stable late-time DE solutions. The present value of different standard density parameters, EoS parameters, and $q$ for all three cases are presented in Table \ref{Potentialfunctionsch6}.

%%%%%%%%%%%%
\begin{table}[H]
     % title of Table
    \centering % used for centering table
    \scalebox{0.8}{
    \begin{tabular}{|c |c |c |c| c| c| c| c|} % centered columns (5 columns)
    \hline 
 \multicolumn{8}{|c|}{\textbf{Results of the potential functions}} \\
    \hline %inserts double horizontal lines
    \parbox[c][0.9cm]{0.9cm}  &\textbf{Potential function} $V(\phi)$ & $\Omega_{DE}^0 \approx$ & $\Omega_{m}^0 \approx$&  \begin{tabular}{@{}c@{}}matter-radiation \\equality $(z_{eq}) \approx$   \end{tabular}&$\omega_{DE}^0\approx$ & $q^{0}$& \begin{tabular}{@{}c@{}}transition point \\$(z_{transi.}) \approx$ \end{tabular} \\ [0.5ex] % inserts table %headin$g$
    \hline % inserts single horizontal line
    \parbox[c][0.9cm]{0.9cm} {$P_1$} &$V_{0}e^{-\kappa \phi}$ & $0.7$&  $0.3$& $3387$ & $-1$ & $-0.53$&0.66 \\
    \hline
   \parbox[c][0.9cm]{0.9cm} {$P_2$} & $Cosh(\xi \phi)-1$ & $0.7$ &  $0.3$& $3387$ & $-1$&$-0.56$&0.65 \\
   \hline
   \parbox[c][0.9cm]{0.9cm }{$P_3$} & $V_{0} Sinh^{-\eta}(\beta \phi)$ & $0.7$ &  $0.3$& $3387$ & $-1$&$-0.57$&0.61 \\
   \hline
    \end{tabular}}
    \caption{Result summary, in the table upper indices $0$ define present time at $z=0$.}
    % is used to refer to this table in the text
    \label{Potentialfunctionsch6}
\end{table}
%%%%%%%%%

 This work demonstrates that the class of potentials may describe the accelerating expansion of the Universe. Thus, the selection of potentials is still arbitrary. Even though three distinct potentials were chosen for this research, one may think of more potentials to perform the same. To keep the analysis from getting too lengthy, we have limited it to just three cases. Despite this, we have determined the stability regions of the parameters of each potential function form. All three potential functions show early-time deceleration and late-time cosmic acceleration through the behavior of the critical points, which indicates that the late-time cosmic acceleration-based critical points are stable. Moreover, we have observed that the exponential form of potential form does not have roots of $f=\lambda (\Gamma-1)$, which means the autonomous system is reduced to four dimensions. However, two other potential function roots are not zero, so in these cases, $\frac{d\lambda}{dN} \neq 0$ indicates a five-dimensional autonomous system.

% Chapter 7
\chapter{Concluding remarks and future
perspectives} % Main chapter title

\label{Chapter7} % For referencing the chapter elsewhere, use \ref{Chapter1} 

\lhead{Chapter 7. \emph{Concluding Remarks and Future
Perspectives}} % This is for the header on each page - perhaps a shortened title

% \chapter{Concluding Remarks and Future
% Perspectives} % Main chapter title

% \label{Chapter7}

\newpage
This thesis highlights the construction and analysis of the autonomous dynamical system in modified teleparallel gravity frameworks. The primary details regarding the construction of the dynamical system approach are presented in chapter \ref{Chapter1}.  Also, different theories that are applicable to derive the stability of the eigenvalues at critical points, like linearisation, Lypnov, and the CMT, are described. The general framework of different teleparallel gravity models, which are considered for detailed analysis, is presented.  Some of these formalisms are related to the scalar field like $f(T, \phi )$ where $\phi$ is the canonical scalar field and the teleparallel analog of the Horndeski theory. Moreover, the general field equations, along with their action formula, are presented for $f(T, B)$, $f(T, T_G)$, and $f(T, B, T_G, B_G)$ gravity. 

Chapter \ref{Chapter2} describes the dynamical system analysis for teleparallel analogous of the Horndeski theory \cite{Bahamonde:2019shr,Bahamonde:2019ipm,Bahamonde:2020cfv,Dialektopoulos:2021ryi,Bahamonde:2021dqn,Bernardo:2021izq,Bernardo:2021bsg}. This formalism includes the non-minimally coupled non-canonical scalar field models.  Two well-motivated models $X^{\alpha} T$ and $X^{\alpha} I_2$ of the teleparallel Horndeski theory have been discussed, including the two particular ($\alpha=1, 2$) cases for these models. The 2D phase space has been demonstrated for the stability ranges of the model parameters. The values of $\omega_{tot}, q$ and the standard density parameters for radiation, matter, and DE are presented at each critical point. The radiation-dominated, matter-dominated, and late-time attractors are demonstrated, which shows the viability of the model. In chapter \ref{Chapter3}, the modification to the teleparallel quintessence model, which has recently come into the study, is $f(T,\phi)$ gravity models \cite{Gonzalez-Espinoza:2021mwr,Gonzalezreconstruction2021,Gonzalez-Espinoza:2020jss}, and the role of power law and the exponential coupling is investigated.  Indeed, when $F(\phi)=0$,
the models reduce to the standard TEGR with a scalar field. The reduction to TEGR with a scalar field does not introduce any new complications; our results and equations derived are consistent in this limit \cite{copelandLiddle}. However, we focused on how the presence of $F(\phi)\ne0$ enriches the dynamics, which is the core novelty of the presented thesis. The two cases, exponential and the power law couplings, are capable of describing critical points that represent radiation, matter, and DE-dominated eras of the evolution of the Universe. These critical points are explained in detail, and the stability conditions are examined. 

To demonstrate the inclusion of the teleparallel boundary terms, chapter \ref{Chapter4} the $f(T, B)$ \cite{Bahamonde:2016grb,Bahamonde:2015zma,bahamonde:2021teleparallel} and $f(T, T_G)$ gravity \cite{Kofinas:2014owa,delaCruz-Dombriz:2017lvj,Kofinas:2014owa} formalism is considered. In both of these teleparallel gravity models, the cosmologically viable models were selected. The critical points have been obtained, and the model parameters are constrained according to the stability conditions of the stable critical points, which can be seen in detail in chapter \ref{Chapter4}. These models are observed to be capable of describing the radiation, matter, and the DE-dominated epoch of the evolution of the Universe. We have also plotted the region plots for the stability ranges of the model parameters. In chapter \ref{Chapter5}, the more generalise form of these two models in teleparallel gravity $f(T, B, T_G, B_G)$ gravity is presented \cite{Bahamonde:2016kba,bahamonde:2021teleparallel}. Here, we have considered the mixed power law and the particular case of the sum of separated power law for the analysis. These models are well-motivated and carry forward from the Noether symmetry approach studied in Ref. \cite{bahamonde2019noether}. We have demonstrated that this model also shows its viability in describing different important phases of the Universe's evolution. The ranges of the model parameters at which these models are cosmologically viable are described in each of these particular studies.

In chapter \ref{Chapter6}, we have performed the dynamical system analysis considering the non-canonical scalar field $\phi$ is coupled to the teleparallel boundary term $B$, the action formula presented in this thesis is novel and is motivated from \cite{Zubair_2017,Bahamonde_2016,Gecim_2018}. In this work, we have presented a detailed analysis of three well-motivated potential functions. This work highlighted that choosing different forms of scalar field potential is supported by comparing the results and testing the effects of the different potential functions. However, after studying qualitative behavior through the dynamical system approach, we conclude that our analysis does not show any favor to a particular form of the potential. It shows that the class of potentials is allowed to describe the accelerated expansion of the Universe. So, the arbitrariness of the choice of potentials remains the same. Though our analysis is done by choosing three different potentials, one can consider more potentials to do the same. The Supernovae Ia and the Hubble data sets are used to compare the obtained results for $H(z), \mu(z)$. The models are showing compatibility with the observation studies, such as contributions from $\Omega_{m}\approx 0.3, \Omega_{DE} \approx 0.7$. It has been observed that the matter radiation equality exists at $z\approx 3387$. These numerical compatible values can be studied in detail in chapter \ref{Chapter6}.

The most crucial future paths of the present work involve using dynamical system analysis, which can be used to set priors for the model parameters for teleparallel gravity models. Precise and reliable data sets such as Hubble, Pantheon Plus, and BAO data can be used to study the cosmological observations. Similar to chapter \ref{Chapter6},  observations of supernovae distance moduli and Hubble parameters using Hubble data sets can be employed to narrow down the parameter space of teleparallel and extended teleparallel gravity models. Viable models can be identified or ruled out by comparing the theoretical predictions from dynamical systems with observational constraints. Some of the key cosmological parameters include the Hubble constant, the DE EoS, and matter density can also be constrained. Modified teleparallel models, particularly those that contain boundary terms like $B$ and $T_G$, offer alternative explanations for the late-time acceleration of the Universe. Future research can concentrate on using modified teleparallel gravity models to explore whether these models can resolve current cosmological tensions, such as the discrepancy in the measured value of the Hubble constant. It may be feasible to identify critical points that correspond to observational signatures by modeling the dynamics of cosmic expansion within these frameworks, offering alternative solutions to standard DE models. Teleparallel gravity models, particularly those that interact with boundary terms, can anticipate distinct gravitational wave signatures that could be investigated by future detectors such as LIGO India, LISA, or the Einstein Telescope.

In the future, this analysis can extend to strong gravity regimes and examine how these models alter the propagation of gravitational waves. By connecting theoretical predictions with observational data, researchers could detect potential deviations from GR and seek new physics in the upcoming era of gravitational wave astronomy.

%-------------------------------------------------------------------------------
%	THESIS CONTENT - APPENDICES
%-------------------------------------------------------------------------------

%\addtocontents{toc}{\vspace{2em}} % Add a gap in the Contents, for aesthetics

%\appendix % Cue to tell LaTeX that the following 'chapters' are Appendices

% Include the appendices of the thesis as separate files from the Appendices
% folder
% Uncomment the lines as you write the Appendices
% \appendix
% \input{Appendices/Appendix}
% %\input{Appendices/AppendixB}
% %\input{Appendices/AppendixC}
% %\appendix
\addtocontents{toc}{\vspace{1em}} % Add a gap in the Contents, for aesthetics

\backmatter

%-------------------------------------------------------------------------------
%	BIBLIOGRAPHY
%-------------------------------------------------------------------------------

%%%\label{Bibliography}
\label{References}
%%%\lhead{\emph{Bibliography}} % Change the page header to say "Bibliography"
\lhead{\emph{References}}
%\input{Chapters/Bibliography}
%%%\input{Bibliography.bib}
%\input{references.bib}
%\input{references1.bib}
%\bibliographystyle{amsplain}
%\bibliographystyle{utphys}
%\printbibliography
%%%\bibliography{Bibliography.bib}
% \bibliographystyle{ieeetr}
% \bibliography{referencesb}
%\bibliographystyle{JHEP}
%\bibliography{referencesb}

% Generated by IEEEtranN.bst, version: 1.14 (2015/08/26)

\cleardoublepage
%%%%%%%%%%%%%%%%%
%%%%%%%%%%%%%%%%%
\pagestyle{fancy}
\lhead{\emph{Appendices}}
\chapter{Appendices}
\section*{Datasets}\label{Datasets}
\subsection*{Hubble data \texorpdfstring{$H(z)$}{}}
In this work, we have taken 31 data points \cite{Moresco_2022_25} to describe the behavior of the Hubble rate in our model. The standard $\Lambda$CDM model will also be compared to our model. We know,
\begin{equation}\label{hubble_LCDM}
H_{\Lambda CDM}= H_{0}\sqrt{(1+z)^3 \Omega_{m}+(1+z)^4 \Omega_{r}+\Omega_{de}} \,,   
\end{equation}

\subsection*{Supernovae Ia}
Another component of our baseline data set is the Pantheon compilation of 1048 SNIa distance measurements spanning $0.01<z<2.3$ redshifts \cite{Scolnic_2018}. This dataset incorporates observations from prominent programs such as PanSTARRS1, Hubble Space Telescope (HST) survey, SNLS and SDSS. By amalgamating data from diverse sources, the Pantheon collection offers valuable insights into the properties and behaviors of Type Ia supernovae and their cosmic implications. Furthermore, it demonstrates the use of stellar luminosity as a means of determining distances in an expanding Universe, with the distance moduli function being a key component of this analysis and is represented as,
\begin{equation}\label{panmoduli}
\mu(z_{i}, \Theta)=5 \log_{10}[D_{L}(z_i, \Theta)]+M    \,.
\end{equation}
Here, $M$ is the nuisance parameter, while $D_{L}$ denotes the luminosity distance. Luminosity distance can be calculated as follows: 
\begin{equation}\label{luminositydistance}
D(z_{i}, \Theta)=c (1+z_{i}) \int^{z_i}_{0} \frac{dz}{H(z, \Theta)} \,,   
\end{equation}

\section*{CMT for critical points in Chapter \ref{Chapter6}}\label{CMT}
The basics for the CMT are well described and presented in Sec \ref{centralmanifoldtheory}.\\
\textbf{CMT for critical point $G_{DE}$:} We have obtained the Jacobian matrix for the critical point \( G_{DE} \) for the autonomous system presented in Eq. (\ref{dynamicalsystem}) as follows:
\[
J(G_{DE}) = 
\begin{bmatrix}
 -3 & 0  &-3\sqrt{\frac{3}{2}}  & 0  \\
 0 & -3 & 0 & 0  \\
 0 & 0 & 0 & 0  \\
 0 & 0 & 0 & -2 
\end{bmatrix}
\]
The eigenvalues of the above Jacobian matrix \( G_{DE} \) are \( \nu_{1} = -2 \), \( \nu_{2} = 0 \), \( \nu_{3} = -3 \) and \( \nu_{4} = -3 \) as obtained in Sec. \ref{P_1}. The eigenvectors corresponding to these eigenvalues are
\begin{math}
\left[0,1,0,0\right]^T
\end{math} \begin{math}
\left[1,0,0,0\right]^T
\end{math} \begin{math}
,\left[0,0,0,1\right]^T
\end{math} \begin{math}
,\left[-\sqrt{\frac{3}{2}},0,1,0\right]^T
\end{math}
Now, applying the CMT, we analyze the stability of this critical point \( G_{DE} \). We use the transformation \( X = x \), \( Y = y-1 \), \( Z = u \) and \( R = \rho \) to shift this critical point to the origin. The resultant equations obtained in the new coordinate system can then be written as,
\begin{align}
\begin{pmatrix}
\dot{X}\\ 
\dot{Y} \\ 
 \dot{R}\\ 
 \dot{Z} 
\end{pmatrix}= 
\begin{pmatrix}
-3 & 0 & 0 & 0 \\
0 & -3 & 0 & 0 \\
0 & 0 & -2 & 0  \\
0 & 0 & 0 & 0 \\
\end{pmatrix} 
\begin{pmatrix}
X\\ 
 Y\\ 
 R\\ 
 Z   
\end{pmatrix}+\begin{pmatrix}
 non\\ linear\\ term   
\end{pmatrix} 
\end{align}
Upon comparing the diagonal matrix with the standard form Eqs. \eqref{dsa1} and \eqref{dsa2}, it is apparent that the variables \( X \), \( Y \) and \( R \) demonstrate stability, while \( Z \) serves as the central variable. At this critical point, the matrices \( \mathcal{A} \) and \( \mathcal{B} \) adopt the following form,
\[
\mathcal{A} =
\begin{bmatrix}
 -3 & 0  &0    \\
  0 & -3  &0   \\
 0 & 0 & -2  \\
 
\end{bmatrix}
\hspace{0.5cm}
\mathcal{B} = 
\begin{bmatrix}
0    
\end{bmatrix}
\]
As per the CMT, the manifold can be expressed using a function that is continuously differentiable. We make the assumption that the stable variables can be represented by the following functions: \( X = g_{1}(Z) \), \( Y = g_{2}(Z) \) and \( R = g_{3}(Z) \). By using Eqs \eqref{dsa1}, \eqref{dsa2} we can derive the zeroth-order approximation of the manifold functions in the following manner:
\begin{eqnarray}
\mathcal{N}(g_1(Z)) = \frac{3 \sqrt{6} Z}{3 Z^2+2}\,, \hspace{0.5cm} \mathcal{N}(g_2(Z)) =-\frac{9 Z^2}{3 Z^2+2}\,, \hspace{0.5cm} \mathcal{N}(g_3(Z)) =0 \,.
\end{eqnarray}
In this case, the center manifold is given by the following expression:
\begin{equation}\label{CMTmanifold1}
\dot{Z}= -\frac{18 \alpha  Z^2}{3 Z^2+2}+ higher \hspace{0.15cm} order \hspace{0.15cm} term \,.  
\end{equation}
By CMT, the critical point $G_{DE}$ exhibits stable behavior for \( Z \neq 0 \) and \( \alpha > 0 \), where \( \dot{Z} \) is negative.

\textbf{CMT for critical point $f_{DE}$:} The Jacobian matrix at the critical point \( f_{DE} \) for the autonomous system (\ref{dynamicalsystem}) is given as,\\
$J(f_{DE}) = 
\left(
\begin{array}{ccccc}
 -\frac{12}{\frac{2 \xi ^2}{3}+4} & \frac{2 \sqrt{6} \xi }{\frac{2 \xi ^2}{3}+4} & -\frac{6 \sqrt{6}}{\frac{2 \xi ^2}{3}+4} & 0 & \frac{2 \sqrt{6}}{\frac{2 \xi ^2}{3}+4} \\
 \frac{\sqrt{6} \xi }{\frac{\xi ^2}{3}+2}-\sqrt{\frac{3}{2}} \xi  & -\frac{2 \xi ^2}{\frac{\xi ^2}{3}+2}-\frac{6}{\frac{\xi ^2}{3}+2} & \frac{3 \xi }{\frac{\xi ^2}{3}+2} & 0 & -\frac{\xi }{\frac{\xi ^2}{3}+2} \\
 -\sqrt{\frac{2}{3}} \alpha  \xi  & 0 & 0 & 0 & 0 \\
 0 & 0 & 0 & -\frac{2 \xi ^2}{3 \left(\frac{\xi ^2}{3}+2\right)}-\frac{4}{\frac{\xi ^2}{3}+2} & 0 \\
 0 & 0 & 0 & 0 & 0 \\
\end{array}
\right)$.\\
The eigenvalues of Jacobian matrix $f_{DE}$ are $\nu_{1}=0,\, \nu_{2}=-3,\, \nu_{3}=-2,\, $ \\ $\nu_{4}=-\frac{3 \left(\sqrt{\left(\xi ^2+6\right) \left(8 \alpha  \xi +\xi ^2+6\right)}+\xi ^2+6\right)}{2 \left(\xi ^2+6\right)},\, \nu_{5}= \frac{3}{2} \left(\frac{\sqrt{\left(\xi ^2+6\right) \left(8 \alpha  \xi +\xi ^2+6\right)}}{\xi ^2+6}-1\right)$.
The corresponding eigenvectors are 
\begin{math}
\left[0,0,\frac{1}{3},0,1\right]^T
\end{math} \begin{math}
,\left[\frac{3 \sqrt{\frac{3}{2}}}{\alpha  \xi },\frac{3 (2 \alpha -\xi )}{2 \alpha  \xi },1,0,0\right]^T
\end{math} \begin{math}
,\left[0,0,0,1,0\right]^T,
\\
\left[\frac{3 \sqrt{\frac{3}{2}} \left(-\sqrt{\left(\xi ^2+6\right) \left(8 \alpha  \xi +\xi ^2+6\right)}+\xi ^2+6\right)}{2 \alpha  \xi  \left(\xi ^2+6\right)},\frac{3 \xi  \left(4 \alpha  \xi ^2+\xi  \sqrt{\left(\xi ^2+6\right) \left(8 \alpha  \xi +\xi ^2+6\right)}+24 \alpha -\xi ^3-6 \xi \right)}{2 \alpha  \left(\xi ^2+6\right) \left(\sqrt{\left(\xi ^2+6\right) \left(8 \alpha  \xi +\xi ^2+6\right)}+3 \xi ^2+6\right)},1,0,0\right]^T
\end{math}
and, \\
\begin{math}
\left[ \frac{3 \sqrt{\frac{3}{2}} \left(\sqrt{\left(\xi ^2+6\right) \left(8 \alpha  \xi +\xi ^2+6\right)}+\xi ^2+6\right)}{2 \alpha  \xi  \left(\xi ^2+6\right)},-\frac{3 \xi  \left(4 \alpha  \xi ^2-\xi  \sqrt{\left(\xi ^2+6\right) \left(8 \alpha  \xi +\xi ^2+6\right)}+24 \alpha -\xi ^3-6 \xi \right)}{2 \alpha  \left(\xi ^2+6\right) \left(\sqrt{\left(\xi ^2+6\right) \left(8 \alpha  \xi +\xi ^2+6\right)}-3 \xi ^2-6\right)},1,0,0 \right]^T  
\end{math}.
To shift the critical points to the origin, the specific transformations we employed here are: \( X = x \), \( Y = y-1 \), \( Z = u - \frac{\xi}{3} \), \( R = \rho \) and \( L = \lambda - \xi \). In this new coordinate system, we expressed the equations in the following form:
\begin{align}
\begin{pmatrix}
\dot{X}\\ 
\dot{Y} \\ 
 \dot{Z}\\ 
 \dot{R} \\
 \dot{L}
\end{pmatrix}= 
\left(
\begin{array}{ccccc}
\tau_{1}&0& 0& 0 &0 \\
 0  &  \tau_{2}& 0 & 0 &0\\
0 & 0 & -3 & 0 & 0 \\
 0 & 0 & 0 & -2 & 0 \\
 0 & 0 & 0 & 0 & 0 \\
\end{array}
\right)
\begin{pmatrix}
X\\ 
 Y\\ 
 Z\\ 
 R\\ 
 L
\end{pmatrix}+\begin{pmatrix}
 non\\ linear\\ term   
\end{pmatrix} 
\end{align}
Where $\tau_{1}=-\frac{3 \left(\sqrt{\left(\xi ^2+6\right) \left(8 \alpha  \xi +\xi ^2+6\right)}+\xi ^2+6\right)}{2 \left(\xi ^2+6\right)} $ and $\tau_{2}=\frac{3}{2} \left(\frac{\sqrt{\left(\xi ^2+6\right) \left(8 \alpha  \xi +\xi ^2+6\right)}}{\xi ^2+6}-1\right)$ we can determine that \( X \), \( Y \), \( Z \) and \( R \) are the stable variables, while \( L \) is the central variable. At this critical point, the matrices \( \mathcal{A} \) and \( \mathcal{B} \) take the following form:
\[
\mathcal{A} =\left(
\begin{array}{ccccc}
-\frac{3 \left(\sqrt{\left(\xi ^2+6\right) \left(8 \alpha  \xi +\xi ^2+6\right)}+\xi ^2+6\right)}{2 \left(\xi ^2+6\right)} &0& 0& 0  \\
 0  &  \frac{3}{2} \left(\frac{\sqrt{\left(\xi ^2+6\right) \left(8 \alpha  \xi +\xi ^2+6\right)}}{\xi ^2+6}-1\right)& 0 & 0 \\
0 & 0 & -3 & 0  \\
 0 & 0 & 0 & -2  \\
\end{array}
\right)
\hspace{0.5cm}
\mathcal{B} = 
\begin{bmatrix}
0    
\end{bmatrix}
\]
According to CMT, we have made specific assumptions that  $X=g_{1}(L)$, $Y=g_{2}(L)$, $Z=g_{3}(L)$ and $R=g_{4}(L)$ are the stable variables. Now By using Eq. \eqref{dsa1} and Eq. \eqref{dsa2}, we have derived the zeroth order approximation of the manifold functions,
\begin{align}
\mathcal{N}(g_1(L)) = -\frac{3 \sqrt{6} L}{\xi ^2+6}\,, \quad  \mathcal{N}(g_2(L)) =\frac{3 L \xi }{\xi ^2+6}\,,  \nonumber\\ \mathcal{N}(g_3(L)) =0 \,, \quad  \mathcal{N}(g_4(L)) =0 \,.
\end{align}
With these, the central manifold can be obtained as 
\begin{equation}\label{CMTmanifold2}
\dot{L}= -\frac{18 L^2 \xi }{\xi ^2+6}+ higher \hspace{0.2cm} order \hspace{0.2cm} term   \,.
\end{equation}
By applying the CMT, this critical point shows stable behavior for  $ \xi >0$, where $\dot{L}$ is negative.

\textbf{CMT for critical point $g_{DE}$:} The Jacobian matrix at the critical point $g_{DE}$ for the autonomous system (\ref{dynamicalsystem} ) is as follows:\\
$J(g_{DE}) = 
\left(
\begin{array}{ccccc}
 -3 & 0 & -3 \sqrt{\frac{3}{2}} & 0 & \sqrt{\frac{3}{2}} \\
 0 & -3 & 0 & 0 & 0 \\
 0 & 0 & 0 & 0 & 0 \\
 0 & 0 & 0 & -2 & 0 \\
 -\sqrt{\frac{3}{2}} \xi ^2 & 0 & 0 & 0 & 0 \\
\end{array}
\right)$.\\
The eigenvalues at critical point $g_{DE}$ are $\nu_{1}=-3,\, \nu_{2}=-2,\, \nu_{3}=0,\, \nu_{4}=\frac{1}{2} \left(-\sqrt{9-6 \xi ^2}-3\right),\, \nu_{5}=\frac{1}{2} \left(\sqrt{9-6 \xi ^2}-3\right) $. The eigenvectors corresponding to these eigenvalues are 
\begin{math}
   \left[0,1,0,0,0\right]^{T} 
\end{math},\\
\begin{math}
   \left[0,0,0,1,0\right]^{T} 
\end{math}
\begin{math}
   \left[0,0,\frac{1}{3},0,1\right]^{T} 
\end{math}
\begin{math}
   \left[-\frac{-\sqrt{2} \sqrt{3-2 \xi ^2}-\sqrt{6}}{2 \xi ^2},0,0,0,1\right]^{T} 
\end{math}
\begin{math}
\left[ -\frac{\sqrt{2} \sqrt{3-2 \xi ^2}-\sqrt{6}}{2 \xi ^2},0,0,0,1\right]^{T} 
\end{math}
To apply CMT, we have shifted the critical point to the origin using shifting transformations $X=x$, $Y=y-1$, $Z=u$, $R=\rho$ and $L=\lambda$. Then we can write equations in the new coordinate system as, 
\begin{align}
\begin{pmatrix}
\dot{X}\\ 
\dot{Y} \\ 
 \dot{R}\\ 
 \dot{L} \\
 \dot{Z}
\end{pmatrix}= 
\left(
\begin{array}{ccccc}
\frac{1}{2} \left(-\sqrt{9-6 \xi ^2}-3\right) &0& 0& 0 &0 \\
 0  &  \frac{1}{2} \left(\sqrt{9-6 \xi ^2}-3\right)& 0 & 0 &0\\
0 & 0 & -3 & 0 & 0 \\
 0 & 0 & 0 & 0-2 & 0 \\
 0 & 0 & 0 & 0 & 0 \\
\end{array}
\right)
\begin{pmatrix}
X\\ 
 Y\\ 
 R\\ 
 L\\ 
 Z
\end{pmatrix}+\begin{pmatrix}
 non\\ linear\\ term   
\end{pmatrix} \nonumber
\end{align}
The variables $X, Y, R$ and $L$ are stable, whereas $Z$ is the central variable. At this critical point, the matrices $\mathcal{A}$ and $\mathcal{B}$ take on the following form:
\[
\mathcal{A} =\left(
\begin{array}{ccccc}
\frac{1}{2} \left(-\sqrt{9-6 \xi ^2}-3\right) &0& 0& 0  \\
 0  &  \frac{1}{2} \left(\sqrt{9-6 \xi ^2}-3\right)& 0 & 0 \\
0 & 0 & -3 & 0  \\
 0 & 0 & 0 & 0-2  \\
\end{array}
\right)
\hspace{0.5cm}
\mathcal{B} = 
\begin{bmatrix}
0    
\end{bmatrix}
\]\,.
Using CMT, we assume the: \( X = g_{1}(Z) \), \( Y = g_{2}(Z) \), \( R = g_{3}(Z) \) and \( L = g_{4}(Z) \) are the stable variables and the zeroth approximation of the manifold functions are as follows:
\begin{eqnarray}
\mathcal{N}(g_1(Z)) = \frac{3 \sqrt{6} Z}{3 Z^2+2}\,, \quad  \mathcal{N}(g_2(Z)) =-\frac{9 Z^2}{3 Z^2+2}\,, \nonumber\\ \mathcal{N}(g_3(Z)) =0 \,, \quad \mathcal{N}(g_4(Z)) =0 \,.
\end{eqnarray}
With these assumptions, the central manifold can be obtained as follows:
\begin{equation}\label{CMTmanifold3}
\dot{Z}=-\frac{18 \alpha  Z^2}{3 Z^2+2}+ higher \hspace{0.2cm} order \hspace{0.2cm} term   \,.
\end{equation}
According to CMT, this critical point demonstrates stable behavior for \( Z \ne 0 \) and \( \alpha > 0 \), where \( \dot{Z} \) is negative.

\textbf{CMT for critical point $\mathcal{F}_{DE}$:} The Jacobian matrix at the critical point $\mathcal{F}_{DE}$ for the autonomous system (\ref{dynamicalsystem} ) is,\\
$J(\mathcal{F}_{DE}) = 
\left(
\begin{array}{ccccc}
 -\frac{12}{\frac{2 \beta ^2 \eta ^2}{3}+4} & \frac{2 \sqrt{6} \beta  \eta }{\frac{2 \beta ^2 \eta ^2}{3}+4} & -\frac{6 \sqrt{6}}{\frac{2 \beta ^2 \eta ^2}{3}+4} & 0 & \frac{2 \sqrt{6}}{\frac{2 \beta ^2 \eta ^2}{3}+4} \\
 \frac{\sqrt{6} \beta  \eta }{\frac{\beta ^2 \eta ^2}{3}+2}-\sqrt{\frac{3}{2}} \beta  \eta  & -\frac{2 \beta ^2 \eta ^2}{\frac{\beta ^2 \eta ^2}{3}+2}-\frac{6}{\frac{\beta ^2 \eta ^2}{3}+2} & \frac{3 \beta  \eta }{\frac{\beta ^2 \eta ^2}{3}+2} & 0 & -\frac{\beta  \eta }{\frac{\beta ^2 \eta ^2}{3}+2} \\
 -\sqrt{\frac{2}{3}} \alpha  \beta  \eta  & 0 & 0 & 0 & 0 \\
 0 & 0 & 0 & -\frac{2 \beta ^2 \eta ^2}{3 \left(\frac{\beta ^2 \eta ^2}{3}+2\right)}-\frac{4}{\frac{\beta ^2 \eta ^2}{3}+2} & 0 \\
 0 & 0 & 0 & 0 & 0 \\
\end{array}
\right)$
The eigenvalues at $\mathcal{F}_{DE}$ are $\nu_{1}=0, \nu_{2}=-3, \nu_{3}=-2, \nu_{4}=\Big[-\frac{3 \sqrt{\left(\beta ^2 \eta ^2+6\right) (\beta  \eta  (8 \alpha +\beta  \eta )+6)}}{2 \left(\beta ^2 \eta ^2+6\right)}-\frac{3}{2}\Big],\\ \nu_{5}=\frac{3}{2} \Big[\frac{\sqrt{\left(\beta ^2 \eta ^2+6\right) (\beta  \eta  (8 \alpha +\beta  \eta )+6)}}{\beta ^2 \eta ^2+6}-1\Big]$.
The eigenvectors corresponding to these eigenvalues are 
\begin{math}
   \left[0,0,\frac{1}{3},0,1\right]^{T} 
\end{math},
\begin{math}
   \left[\frac{3 \sqrt{\frac{3}{2}}}{\alpha  \beta  \eta },\frac{3 (2 \alpha -\beta  \eta )}{2 \alpha  \beta  \eta },1,0,0\right]^{T} 
\end{math}
\begin{math}
   \left[0,0,0,1,0\right]^{T} 
\end{math}\\
\begin{math}
   \left[\frac{3 \sqrt{\frac{3}{2}} \left(-\sqrt{\left(\beta ^2 \eta ^2+6\right) \left(8 \alpha  \beta  \eta +\beta ^2 \eta ^2+6\right)}+\beta ^2 \eta ^2+6\right)}{2 \alpha  \beta  \eta  \left(\beta ^2 \eta ^2+6\right)},\frac{3 \beta  \eta  \left(4 \alpha  \beta ^2 \eta ^2+\beta  \eta  \sqrt{\left(\beta ^2 \eta ^2+6\right) \left(8 \alpha  \beta  \eta +\beta ^2 \eta ^2+6\right)}+24 \alpha -\beta ^3 \eta ^3-6 \beta  \eta \right)}{2 \alpha  \left(\beta ^2 \eta ^2+6\right) \left(\sqrt{\left(\beta ^2 \eta ^2+6\right) \left(8 \alpha  \beta  \eta +\beta ^2 \eta ^2+6\right)}+3 \beta ^2 \eta ^2+6\right)},1,0,0\right]^{T} 
\end{math}
\begin{math}
\left[\frac{3 \sqrt{\frac{3}{2}} \left(\sqrt{\left(\beta ^2 \eta ^2+6\right) \left(8 \alpha  \beta  \eta +\beta ^2 \eta ^2+6\right)}+\beta ^2 \eta ^2+6\right)}{2 \alpha  \beta  \eta  \left(\beta ^2 \eta ^2+6\right)},-\frac{3 \beta  \eta  \left(4 \alpha  \beta ^2 \eta ^2-\beta  \eta  \sqrt{\left(\beta ^2 \eta ^2+6\right) \left(8 \alpha  \beta  \eta +\beta ^2 \eta ^2+6\right)}+24 \alpha -\beta ^3 \eta ^3-6 \beta  \eta \right)}{2 \alpha  \left(\beta ^2 \eta ^2+6\right) \left(\sqrt{\left(\beta ^2 \eta ^2+6\right) \left(8 \alpha  \beta  \eta +\beta ^2 \eta ^2+6\right)}-3 \beta ^2 \eta ^2-6\right)},1,0,0\right]^{T} 
\end{math}
The transformations used in this case are: \( X = x \), \( Y = y-1 \), \( Z = u - \frac{\beta \eta}{3} \), \( R = \rho \) and \( L = \lambda - \beta \eta \) and in the new coordinate system, the equations can be written as,
\begin{align}
\begin{pmatrix}
\dot{X}\\ 
\dot{Y} \\ 
 \dot{Z}\\ 
 \dot{R} \\
 \dot{L}
\end{pmatrix}= 
\left(
\begin{array}{ccccc}
\tau_{11}&0& 0& 0 &0 \\
 0  & \tau_{22} & 0 & 0 &0\\
0 & 0 & -3 & 0 & 0 \\
 0 & 0 & 0 & 0-2 & 0 \\
 0 & 0 & 0 & 0 & 0 \\
\end{array}
\right)
\begin{pmatrix}
X\\ 
 Y\\ 
 Z\\ 
 R\\ 
 L
\end{pmatrix}+\begin{pmatrix}
 non\\ linear\\ term   
\end{pmatrix} 
\end{align}
Where $\tau_{11}=-\frac{3 \sqrt{\left(\beta ^2 \eta ^2+6\right) (\beta  \eta  (8 \alpha +\beta  \eta )+6)}}{2 \left(\beta ^2 \eta ^2+6\right)}-\frac{3}{2}, \tau_{22}=\frac{3}{2} \Big[\frac{\sqrt{\left(\beta ^2 \eta ^2+6\right) (\beta  \eta  (8 \alpha +\beta  \eta )+6)}}{\beta ^2 \eta ^2+6}-1\Big] $. In this case, using CMT, we can conclude that \( X \), \( Y \), \( Z \) and \( R \) are the stable variables, while \( L \) is the central variable. At this critical point, the matrices \( \mathcal{A} \) and \( \mathcal{B} \) appear as follows:
\[
\mathcal{A} =\left(
\begin{array}{ccccc}
-\frac{3 \sqrt{\left(\beta ^2 \eta ^2+6\right) (\beta  \eta  (8 \alpha +\beta  \eta )+6)}}{2 \left(\beta ^2 \eta ^2+6\right)}-\frac{3}{2} &0& 0& 0  \\
 0  &  \frac{3}{2} \Big[\frac{\sqrt{\left(\beta ^2 \eta ^2+6\right) (\beta  \eta  (8 \alpha +\beta  \eta )+6)}}{\beta ^2 \eta ^2+6}-1\Big]& 0 & 0 \\
0 & 0 & -3 & 0  \\
 0 & 0 & 0 & 0-2  \\
\end{array}
\right)
\hspace{0.5cm}
\mathcal{B} = 
\begin{bmatrix}
0    
\end{bmatrix}
\]
In this case the stable variables are \( X = g_{1}(L) \), \( Y = g_{2}(L) \), \( Z = g_{3}(L) \) and \( R = g_{4}(L) \), the zeroth approximation of the manifold functions as follows:
 \begin{eqnarray}
\mathcal{N}(g_1(L)) = -\frac{3 \sqrt{6} L}{\beta ^2 \eta ^2+6}\,,\quad  \mathcal{N}(g_2(L)) =\frac{3 \beta  \eta  L}{\beta ^2 \eta ^2+6}\,,  \nonumber\\ \mathcal{N}(g_3(L)) =0 \,,\quad \mathcal{N}(g_4(L)) =0 \,.
\end{eqnarray}
With these assumptions, the central manifold can be obtained as follows: 
\begin{equation}\label{CMTmanifold4}
\dot{L}=\frac{36 \beta  L^2}{\beta ^2 \eta ^2+6}+ higher \hspace{0.2cm} order \hspace{0.2cm} term   
\end{equation}
According to CMT, this critical point exhibits stable behavior for \( \eta \in \mathbb{R} \) and \( \beta < 0 \), where \( \dot{L} \) is negative.

\textbf{CMT for critical point $\mathcal{G}_{DE}$:} The Jacobian matrix at the critical point $\mathcal{G}_{DE}$ for the autonomous system (\ref{dynamicalsystem} ) is as follows:\\
$J(\mathcal{G}_{DE}) = 
\left(
\begin{array}{ccccc}
 -3 & 0 & -3 \sqrt{\frac{3}{2}} & 0 & \sqrt{\frac{3}{2}} \\
 0 & -3 & 0 & 0 & 0 \\
 0 & 0 & 0 & 0 & 0 \\
 0 & 0 & 0 & -2 & 0 \\
 \sqrt{6} \beta ^2 \eta  & 0 & 0 & 0 & 0 \\
\end{array}
\right)$\\
At the critical point $\mathcal{G}_{DE}$ we have the eigenvalues $\nu_{1}=-3,\,\nu_{2}=-2,\,\nu_{3}=0,\,$\\$\nu_{4}=\frac{1}{2} \left(-\sqrt{12 \beta ^2 \eta +9}-3\right),\,\nu_{5}=\frac{1}{2} \left(\sqrt{12 \beta ^2 \eta +9}-3\right)$ The eigenvectors corresponding to these eigenvalues are 
\begin{math}
   \left[0,1,0,0,0\right]^{T} 
\end{math},
\begin{math}
   \left[0,0,0,1,0\right]^{T} 
\end{math}
\begin{math}
   \left[0,0,\frac{1}{3},0,1\right]^{T} 
\end{math}
\begin{math}
   \left[-\frac{\sqrt{2} \sqrt{4 \beta ^2 \eta +3}+\sqrt{6}}{4 \beta ^2 \eta },0,0,0,1\right]^{T} 
\end{math}
\begin{math}
\left[-\frac{\sqrt{6}-\sqrt{2} \sqrt{4 \beta ^2 \eta +3}}{4 \beta ^2 \eta },0,0,0,1\right]^{T} 
\end{math}
In this case, to shift the critical point towards the origin, the transformations we used are: \( X = x \), \( Y = y-1 \), \( Z = u \), \( R = \rho \) and \( L = \lambda \). The equations in the new co-ordinate system can be written as:
\begin{align}
\begin{pmatrix}
\dot{X}\\ 
\dot{Y} \\ 
 \dot{R}\\ 
 \dot{L} \\
 \dot{Z}
\end{pmatrix}= 
\left(
\begin{array}{ccccc}
\frac{1}{2} \left(-\sqrt{12 \beta ^2 \eta +9}-3\right)&0& 0& 0 &0 \\
 0  & \frac{1}{2} \left(\sqrt{12 \beta ^2 \eta +9}-3\right)& 0 & 0 &0\\
0 & 0 & -3 & 0 & 0 \\
 0 & 0 & 0 & 0-2 & 0 \\
 0 & 0 & 0 & 0 & 0 \\
\end{array}
\right)
\begin{pmatrix}
X\\ 
 Y\\ 
 R\\ 
 L\\ 
 Z
\end{pmatrix}+\begin{pmatrix}
 non\\ linear\\ term   
\end{pmatrix} \nonumber
\end{align}
In this case the stable variables are \( X \), \( Y \), \( R \) and \( L \) , while \( Z \) is the central variable. At this critical point, the matrices \( \mathcal{A} \) and \( \mathcal{B} \) appear as  \[
\mathcal{A} =\left(
\begin{array}{ccccc}
\frac{1}{2} \left(-\sqrt{12 \beta ^2 \eta +9}-3\right) &0& 0& 0  \\
 0  &  \frac{1}{2} \left(\sqrt{12 \beta ^2 \eta +9}-3\right)& 0 & 0 \\
0 & 0 & -3 & 0  \\
 0 & 0 & 0 & 0-2  \\
\end{array}
\right)
\hspace{0.5cm}
\mathcal{B} = 
\begin{bmatrix}
0    
\end{bmatrix}
\]
    We have assumed the following functions for the stable variables: \( X = g_{1}(Z) \), \( Y = g_{2}(Z) \), \( R = g_{3}(Z) \) and \( L = g_{4}(Z) \). The zeroth approximation of the manifold functions can be obtained as follows:
        \begin{eqnarray}
            \mathcal{N}(g_1(Z)) =\frac{3 \sqrt{6} Z}{3 Z^2+2}\,,\quad \mathcal{N}(g_2(Z)) =-\frac{9 Z^2}{3 Z^2+2}\,,\nonumber\\  \mathcal{N}(g_3(Z)) =0 \,,\quad \mathcal{N}(g_4(Z)) =0 \, ,
        \end{eqnarray}
with these, the central manifold can be obtained as follows:
\begin{equation}\label{CMT-manifold}
\dot{Z}=-\frac{18 \alpha  Z^2}{3 Z^2+2}+ higher \hspace{0.2cm} order \hspace{0.2cm} term \,.  
\end{equation}
According to CMT, this critical point exhibits stable behavior for \( Z \ne 0 \) and \( \alpha > 0 \), where \( \dot{Z} \) is negative.

\cleardoublepage
%%%%%%%%%%%%%%%%%%
%%%%%%%%%%%%%%%%%%
%%\backmatter
\pagestyle{fancy}
\lhead{\emph{List of publications and presentations}}
\chapter{List of publications and presentations}
\justifying
% \pagestyle{fancy}
% \label{Publications}
% \lhead{\emph{List of publications}}
\section*{Publications included in this thesis}
\begin{enumerate}[itemsep=2pt]
 \item \textbf{S. A. Kadam}, B. Mishra, and J. L. Said, ``Teleparallel Scalar-Tensor Gravity Through Cosmological Dynamical Systems", \textit{European Physical Journal C}, \textbf{82}, 680 (2022).
 \item \textbf{S. A. Kadam}, Ananya Sahu, S. K. Tripathy, B. Mishra, ``Dynamical system analysis for scalar field potential in teleparallel gravity", \textit{European Physical Journal C}, \textbf{84}, 1088 (2024).
  \item \textbf{S. A. Kadam}, Ninaad P. Thakkar and B. Mishra, ``Dynamical System Analysis in Teleparallel Gravity with Boundary Term", \textit{European Physical Journal C}, \textbf{83}, 809 (2023)
  \item \textbf{S. A. Kadam}, Santosh V. Lohakare and B. Mishra, ``Dynamical Complexity in Teleparallel Gauss-Bonnet Gravity", \textit{Annals of Physics}, \textbf{460}, 169563 (2024).
   \item \textbf{S. A. Kadam} and B. Mishra, ``Constraining $f(T, B, T_G, B_G)$ gravity by dynamical system analysis", \textit{Physics of the Dark Universe}, \textbf{46}, 101693 (2024). 
  \item \textbf{S. A. Kadam}, L. K. Duchaniya, B. Mishra, ``Teleparallel Gravity and Quintessence: The Role of Nonminimal Boundary Couplings", \textit{Annals of Physics}, \textbf{470}, 169808 (2024).
\end{enumerate}
\section*{Other Publications}
\begin{enumerate}[itemsep=4pt, topsep=5pt]
   \item \textbf{S. A. Kadam}, B. Mishra, and S. K. Tripathy, ``Dynamical Features of $f (T, B)$ Gravity", \textit{Modern Physical Letter A}, \textbf{32}, 2250104 (2022).

    \item Laxmipriya Pati, \textbf{S.A. Kadam}, S. K. Tripathy and B. Mishra, ``Rip Cosmological Models in Extended Symmetric Teleparallel Gravity", \textit{Physics of the Dark Universe}, \textbf{35}, 100925 (2022).

    \item \textbf{S. A. Kadam}, J. L. Said, and B. Mishra, ``Accelerating Cosmological Models in $f (T, B)$ Gravitational Theory", \textit{International Journal of Geometric Methods in Modern Physics}, \textbf{20}, 2350083 (2023).

    \item L. K. Duchaniya, \textbf{S. A. Kadam}, J. L. Said, and B. Mishra, ``Dynamical Systems Analysis in $f (T, \phi)$ Gravity", \textit{European Physical Journal C}, \textbf{83}, 27 (2023). 

    \item \textbf{S. A. Kadam}, B. Mishra and J. L. Said, ``Noether Symmetries in $f (T, T_G)$ Cosmology", \textit{Physica Scripta}, \textbf{98}, 045017 (2023).

    \item B. Mishra, \textbf{S. A. Kadam},  and S. K. Tripathy, ``Scalar Field Induced Evolution in Teleparallel Gravity", \textit{Physics Letter B}, \textbf{857}, 138968 (2024). 
\end{enumerate}
\section*{Conferences and talks}  
\begin{enumerate}[itemsep=4pt, topsep=5pt]
     \item Attend Cosmology from Home 2021, held from July 5 –16, 2021, and present a talk entitled “Late-time Cosmic Acceleration Model In $f(T, B)$ Gravity.”
     
    \item	Attend the conference “Advances in Relativity and Cosmology (PARC-2021)” (October 26-28, 2021) Organized by the Department of Mathematics, BITS-Pilani, Hyderabad Campus, and present a talk entitled “Late-time cosmic acceleration model in extended theory of gravity.”
\item	Participated in Physical Interpretations of Relativity Theory (PIRT) – 2021 conference, organized by Bauman Moscow State Technical University during the period Monday 5 July – Friday 9 July 2021, and presented a talk entitled “Late-time Cosmic Acceleration Model In $f (T, B)$ Gravity”.

\item	Delivered an invited talk in the International Webinar on Recent Advances in Science and Technology (RAST-2021) held at Indira Gandhi Institute of Technology, Sarang, Odisha, INDIA on the topic.

\item	Participated in Recent Developments in Modified Gravity and Cosmology organized by the Department of Mathematics, Birla Institute of Technology and Science – Pilani, Hyderabad Campus, Hyderabad, India, from March 09-11, 2021.

\item	Attended the GRAVITEX international conference on Gravity: Theory and Experiment on 9-12 August 2021 and delivered a talk. Organized by the University of KwaZulu Natal, Durban, Republic of South Africa. 

\item	Attended the ``Workshop on Tensions in Cosmology”, held in Corfu, Greece, from the 7th to the 12th of September 2022.

\item	Attended the Cosmology from Home 2022 conference, held from July 4-15, 2022, and contributed a talk on Dynamical features of $f(T, B)$ Cosmology.

\item	Attended the 23rd International Conference on General Relativity and Gravitation 2022 annual meeting of the Division of Gravitation and Relativistic Astrophysics and gave a talk entitled “Dynamical behavior of $f(T, B)$ Cosmology.”

\item	Participated in the 32nd meeting of the Indian Association for General Relativity and Gravitation (IAGRG32) from 19-21 December 2022 at ISSER Kolkata and presented a poster entitled ``Teleparallel Scalar-tensor gravity through cosmological dynamical systems.”

\item	Participated in ``International Conference on Mathematical Sciences $\&$ its Applications (ICMSA - 2022)" organized by the School of Mathematical Sciences, Swami Ramanand Teerth Marathwada University, Nanded, Maharashtra, India, and has presented a paper entitled `` Energy Conditions and Dynamical Study in Modified Teleparallel Cosmology.”

\item	Participated in Prof. P. C. Vaidya's National Conference on Mathematical Sciences held from 14-15 March 2022 and presented a paper entitled ``Rip Cosmological Models in Extended Symmetric Teleparallel Gravity”. Organized by the Department of Mathematics, Sardar Patel University, Vallabh Vidyanagar, Gujarat, India, and Gujarat Ganit Mandal from 14-15 March 2022.

\item	Participated in ``XXIII International Scientific Conference ``Physical Interpretations of Relativity Theory PIRT – 2023”, organized by Bauman Moscow State Technical University during the period Monday 3 July – Thursday 6 July 2023. Delivered the talk on ``Teleparallel scalar-tensor gravity through cosmological dynamical systems.”

\item	 Participated in Metric Affine Framework for Gravity 2022 and presented a paper entitled ``Dynamical features of $f(T, B)$ Cosmology.” Organized by the Institute of Physics, University of Tartu, Estonia, from 27 June to 1 July 2022.

\item	Attend Cosmology from Home 2023, held from July 3 –14, 2023, and present a talk entitled ``Noether Symmetries in Extended Teleparallel Gauss-Bonnet Cosmology.”

\item	Participated in ``Teacher's Enrichment Workshop" at BITS-Pilani Hyderabad Campus" (From 09.01.2023 to 14.01.2023).

\item	Participated in 89th Annual Conference of Indian Mathematical Society An International Meet BITS Pilani - Hyderabad Campus, Hyderabad, December 22 – 25, 2023, and presented a talk entitled “Dynamical System Analysis in Teleparallel Gravity with Boundary Term.”
\end{enumerate}
\subsection*{Research visit/ International conference (In person) }  
\begin{enumerate}[itemsep=4pt, topsep=5pt]
\item A research visit to the Inter-University Centre for Astronomy and Astrophysics (IUCAA) Pune from June 16 – 23, 2024.
\item Attended the International Conference on Beyond Standard Model: From Theory to Experiment BSM-2023 Conference in Hurghada, Egypt, and presented a talk entitled “Dynamical system analysis in teleparallel scalar-tensor gravity.” This visit is funded by the Science and Engineering Research Board in India.
\end{enumerate}
%\clearpage
\cleardoublepage
%\addtocontents{toc}{\vspace{2em}} % Add a gap in the Contents, for aesthetics

%\backmatter
\pagestyle{fancy}
\lhead{\emph{Biography}}
\chapter{Biography}
\textbf{Brief Biography of Candidate}\\
\textbf{Mr. Siddheshwar Atmaram Kadam} completed his B.Sc. in 2015 and M.Sc. in 2017 from N.E.S. Science College Nanded, Maharashtra. He achieved a UGC NET-JRF all-India rank of 153 in December 2019, cleared the MH-SET exam and has published twelve research papers during his Ph.D., in national and international journals. He presented his work at various renowned national and international conferences. He was selected to present his work at ``BSM-2023" in Egypt, held in November 2023, with financial support from DST-SERB India. He also has an academic visit to IUCAA Pune in June 2024 for his research work.

\textbf{Brief Biography of Supervisor}\\
\textbf{Prof. Bivudutta Mishra} received his Ph.D. degree from Sambalpur University, Odisha, India, in 2003. His main research areas are Geometrically Modified Theories of Gravity, Theoretical Aspects of Dark Energy, and Wormhole Geometry. He has published over 160 research papers in national and international journals, presented papers at conferences in India and abroad, supervised four Ph.D. students, and is currently guiding eight more. He has also organized academic and scientific events in the department. He has become a member of the scientific advisory committee of national and international academic events. He has successfully completed multiple sponsored projects funded by Government Funding agencies and is at present working on three projects funded by CSIR, SERB-DST (MATRICS), and SERB-DST(CRG-ANRF).  He is also an awardee of DAAD-RISE, 2019,2022. He has also reviewed several research papers in highly reputed journals, is a Ph.D. examiner, and is a BoS member of several universities. He has been invited by many foreign universities to share his research in scientific events, some of which are Canada, Germany, the Republic of China, Russia, Australia, Switzerland, Japan, the UK, and Poland. As an academic administrator, he was Head of the Department of Mathematics from September 2012 to October 2016 and was Associate Dean of International Programmes and Collaborations from August 2018 to September 2024. He is also a visiting professor at Bauman Moscow State Technical University, Moscow, a visiting associate at Inter-University Centre for Astronomy and Astrophysics, Pune, a Fellow of the Royal Astronomical Society, UK, and a Fellow of the Institute of Mathematics and Applications, UK.
Foreign member of the Russian Gravitational Society, Moscow. He is among the top $2\%$ scientists in the world.
                         
\end{document}